%% file: main.tex
\documentclass[pdflatex,sn-basic]{sn-jnl}

\usepackage{graphicx}%
\usepackage{multirow}%
\usepackage{amsmath,amssymb,amsfonts}%
\usepackage{amsthm}%
\usepackage{mathrsfs}%
\usepackage[title]{appendix}%
\usepackage{xcolor}%
\usepackage{textcomp}%
\usepackage{manyfoot}%
\usepackage{booktabs}%
\usepackage{algorithm}%
\usepackage{algorithmicx}%
\usepackage{algpseudocode}%
\usepackage{listings}%

\makeatletter
\input{aas_macros.sty}
\def\ref@jnl#1{{\jnl@style#1\ }}
\makeatother

\usepackage{esvect}
\usepackage[T1]{fontenc}
\usepackage{ae, aecompl}
\usepackage[math]{cellspace}
\usepackage{csquotes}
\usepackage{bm}
\usepackage{lscape}

\input{own_latex_commands_i}

\newcommand{\SPHI}{\texttt{SPHINCS\_BSSN }}

\begin{document}

\title[SPH methods in the modelling of compact objects]{SPH methods in the modelling of compact objects}
\footnotetext{
This article is a revised version of \url{ https://doi.org/10.1007/lrca-2015-1}\\
\textbf{Change summary} Major revision, updated and substantially expanded.\\
\textbf{Change details} 
The review has been re-structured and in large parts re-written. In more detail: introduction updated, instead of 
the earlier Sec. 2 \enquote{Kernel Approximation} and Sec. 3 \enquote{SPH formulations of Ideal fluid dynamics}
there are now two  new sections, Sec.2 \enquote{SPH as a numerical method for Newtonian gas dynamics} and Sec. 3 \enquote{General relativistic SPH}
which are  all new text with new subsections. The Sec.3 \enquote{General relativistic SPH} now contains special
relativistic SPH as a limit  of general relativity. Now, there is a section on fully general relativistic SPH in dynamical spacetimes
      which did not exist yet at the time of writing the first version of this review. In Sec. 4.1 \enquote{Encounters of two white dwarfs}
      all sections have been updated. Sec. 4.2 \enquote{Encounters between two neutron stars and neutron stars and black holes}:
       Sec. 4.2.1 \enquote{Relevance} has been rewritten, new Sec. 4.2.2 \enquote{Gravitational-wave detections: the first decade}, 
       Sec.4.2.3 \enquote{The modeling challenge} has been re-written, Sec. 4.2.4  \enquote{Double neutron star mergers} all parts updated
       and streamlined, new paragraph on  \enquote{The first SPH simulations in full General Relativity}. New Appendix A  \enquote{Common SPH kernels},
       New Appendix B \enquote{Coefficients of the linearly reproducing....}. New figures:  1,2, 4, 5, 6, 7, 8, 9, 10, 11, 12, 14, 15, 16, 17, 18,  1
       9, 20, 21, 26, 28, 29, A1, A2, A3, A4, A5.}

\author*[1,2]{\fnm{Stephan} \sur{Rosswog}}\email{stephan.rosswog@uni-hamburg.de}

\affil[1]{\orgdiv{Hamburg Observatory}, \orgname{University of Hamburg}, \orgaddress{\street{Gojenbergsweg 112}, \city{Hamburg}, \postcode{21029}, \country{Germany}}}

\affil[2]{\orgdiv{The Oskar Klein Centre for Cosmoparticle Physics, Department of Astronomy}, \orgname{AlbaNova University Centre}, \orgaddress{\city{Stockholm}, \postcode{106 91}, \country{Sweden}}}

\abstract{
We review the current status of compact object simulations that are based on the 
Smoothed Particle Hydrodynamics (SPH) method.  The first section of this review is dedicated
to SPH as a numerical method for Newtonian, ideal gas dynamics and it should be fairly self-contained. It begins with
the basics of the method, but also describes recent advances including
various meshless derivatives or methods for treating shocks. A separate chapter 
summarizes general relativistic SPH, including its special relativistic limit, and it explains 
in some detail the recent development of full numerical relativity in SPH where  matter is 
evolved together with a dynamical spacetime. The remainder of the 
review has an astrophysical focus, here we discuss the status of the simulations of white dwarf--white dwarf, 
neutron star--neutron star and neutron star--black hole systems. For each type 
of system the emphasis is on gravitational-wave-driven mergers, but we also briefly summarize
dynamical collisions that can occur in locations with large stellar densities. 
}


\maketitle

\newpage

\tableofcontents

\newpage

\section*{List of abbreviations}
\label{sec:abbreviations}

\begin{tabbing}
    Abbreviation \hspace*{3cm} \= meaning\kill
    AIC   \> accretion-induced collapse\\
    ALE  \>  arbitrary Lagrangian Eulerian\\
    AMR \> adaptive mesh refinement\\
    APM  \> Artificial Pressure Method\\
    BH    \> black hole\\
    CF    \> computing frame\\
    CFA  \> conformal flatness approximation\\
    DoF  \> degrees of freedom\\
    EM   \>  electromagnetic \\
    GR  \> general relativity \\
    GRB \> Gamma-ray burst\\
    GWTC-4.0 \> 4th Gravitational Wave Transient Catalogue\\
    sGRB \> short gamma ray burst\\
    HLL  \> Harten--Lax--van Leer\\
    HLLC \> Harten--Lax--van Leer-contact\\
    h.o.t. \> higher order terms\\
    ISCO \> innermost stable circular orbit\\
    KH    \> Kelvin--Helmholtz\\
    LLF   \> local Lax--Friedrich\\
    LRE  \> local regression estimate\\
    MFN \> meshless Finite Mass\\
    MFV \> meshless Finite Volume\\
    MOOD \> multi-dimensional optimal order detection\\
    NLTE \> non-local thermodynamic equilibrium\\
    NS    \>  neutron star\\
    RHS  \> right hand side\\
    SPH  \> Smoothed Particle Hydrodynamics\\
    WD   \>  white dwarf \\
    
  \end{tabbing}

\newpage

\section{Introduction}
\label{sec:intro}

\subsection{Relevance of compact object encounters}
The vast majority of stars in the Universe will finally become  compact stellar objects, either 
white dwarfs (WDs),  neutron stars (NSs) or black holes (BHs). Our Galaxy therefore harbors large numbers of them,
probably $\sim 10^{10}$ white dwarfs, several $10^8$ neutron stars and tens of millions of 
stellar-mass black holes. These objects stretch the physics that is known 
from terrestrial laboratories to extreme limits. The structure of white dwarfs, for example, 
is governed by electron degeneracy pressure, they are therefore Earth-sized manifestations 
of the quantum mechanical Pauli-principle. Neutron stars, on the other hand, reach in their 
cores multiples of the nuclear saturation density ($= 2.6 \times 10^{14}$ \gcc) which makes
them excellent probes for nuclear matter theories. The dimensionless compactness parameter
$\mathcal{C}= (G/c^2) (M/R)= R_\mathrm{S}/(2R)$, where $M$, $R$ and $R_\mathrm{S}$ are mass, radius 
and Schwarzschild radius of the object,
can be considered  as  a measure of the strength of  a gravitational field. It is proportional to
the Newtonian gravitational potential and directly related to the gravitational redshift. For
(non-rotating) black holes, the parameter takes the value of 0.5 at the Schwarzschild radius and for neutron stars
it is only moderately smaller, $\mathcal{C}\approx 0.2$, both are gigantic in comparison
to the solar value of $\sim 10^{-6}$.  Neutron 
stars and black holes, therefore, offer the possibility to test gravity in the 
strong field regime \citep{psaltis08,yunes16,berti18a,berti18b}.

Compact objects that have had time to settle into an equilibrium state possess a high degree of symmetry 
and are essentially perfectly spherically symmetric. Moreover, they are cold enough to be excellently 
described in a $T=0$ approximation (since for all involved particle species $i$ the thermal energies are 
much smaller than the involved Fermi-energies, $k T_i \ll E_{{\rm F},i}$) and they are in chemical 
equilibrium. Such highly symmetric, zero temperature systems have the advantage that results can 
be obtained by (semi-) analytical methods, but due to their very low electromagnetic (or other)
emission such objects are very difficult to detect.

Compact objects still possess -- at least in principle -- very large energy reservoirs and in
cases where these reservoirs can be tapped, they can produce electromagnetic emission that is so
bright that it can serve as a cosmic beacon. For example, each nucleon inside of a carbon-oxygen 
white dwarf  can potentially still release $\approx$~0.9~MeV via nuclear burning to the most stable 
elements, or approximately $1.7 \times 10^{51} $ erg per solar mass. The gravitational binding energy 
of a neutron star or black hole is even  larger, $E_\mathrm{grav} \sim G M^2/R = \mathcal{C} M c^2 = 3.6 \times 10^{53} \; 
{\rm erg} \; (\mathcal{C}/0.2)  (M/$\msun). Tapping these gigantic energy reservoirs, however, usually 
requires special, often catastrophic circumstances, such as the collision or merger with yet another compact 
object. For example, the merger of a neutron star with another neutron star or with a black hole can
power gamma-ray bursts (GRBs) and mergers of two white dwarfs are believed to trigger type Ia supernovae.
Such events are highly dynamical and do not possess enough symmetries to be accurately described by 
(semi-) analytical methods. They require a careful, three-dimensional numerical modeling of gravity,
the hydrodynamics  and the relevant ``microphysical'' ingredients such as neutrino processes, nuclear
burning or a nuclear equation of state. For such simulations it is crucial to choose a numerical method that 
is well suited for the problem at hand.

\subsection{When/why SPH?}
Each numerical method has particular merits and challenges and it is usually pivotal in terms 
of work efficiency to choose the best method for the problem one is interested in. Therefore, we want to 
collect here the major strengths of SPH, but also point out challenges for the method.

\emph{Numerical conservation.} The probably most outstanding feature of SPH is that it allows in a straight forward way
to \emph{exactly conserve} mass, energy, momentum and angular momentum \emph{by construction}, 
i.e., independent of the numerical resolution. This property is ensured by appropriate (anti-)symmetries in the 
SPH evolution equations, see Sect.~\ref{sec:conservation}. Since robustness and numerical conservation were 
key design goals in SPH, particular choices were made: 
\begin{itemize}
\item [i)] densities are calculated via summations over
positive-definite, radial kernel functions, which are guaranteed to yield a strictly positive density estimate
(therefore it is always safe to divide by the density at a particle position), 
\item [ii)]  the momentum equation is
a numerical representation of Newton's 3rd law (\enquote{actio = reactio}) and
\item [iii)] the energy equation is a straight 
forward translation of the first law of thermodynamics. 
\end{itemize}
While the exact conservation is a major asset of SPH,
these hardwired design choices come at a price. For example, radial kernels ensure  straight-forwardly
angular momentum conservation, but they restrict the ability to enforce low-order consistency relations, see 
Sects.~\ref{sec:discrete_approx} and \ref{sec:RPK}. Positive definite kernels, on the other hand, guarantee
a positive density estimate, but they prevent the exact reproduction  of functions beyond linear degree \citep{monaghan92}.
Last but not least, the hydrodynamics equations can be thought of as a conservation law for a state vector
${\bf q}$, $\partial_t \bf{q} + \nabla \cdot \bf{F}(\bf{q})= s,$ where $\bf{F}$ are fluxes and 
$\bf{s}$ source terms, see Sect.~\ref{sec:SPH_landscape}. But if one ensures mass conservation via 
constant particle masses and density kernel summations, and energy conservation via a straight-forward 
translation of the first law of thermodynamics, one treats
different components of the same conservation law in different ways. This implies, for example,
that the density $\rho$ comes from a smoothing process, while the specific internal energy $u$ does not. 
One needs to be aware of this, for example, when setting up contact discontinuities across which the 
product $\rho u$ should be continuous (for a polytropic equation of state). Thus,  there is a tradeoff between 
straight forward conservation and the accuracy of function approximation.

\emph{Adaptivity.} Apart from its superb numerical conservation properties,  another benefit that comes essentially for free 
(i.e. without a need for additional computational infrastructure) is the natural adaptivity of SPH. Since the particles move
with the local fluid velocity, they naturally trace the flow motion. As a corollary, simulations are 
not bound, like usually in Eulerian simulations, to a pre-defined ``simulation volume'', but instead they can 
follow whatever the astrophysical system decides to do, be it to collapse  or to expand rapidly
or both in different parts of the flow.  This automatic ``refinement on density'' is also closely related
to the fact that vacuum does not pose a problem: such regions are simply devoid of SPH particles and no
computational resources are wasted on regions without matter. In Eulerian methods, in contrast, 
vacuum is usually treated as a background fluid (\enquote{atmosphere}) in which the \enquote{fluid of interest} moves
and special attention needs to be payed to avoid artefacts caused by the interaction of these two fluids. 
For example, the neutron star surface of a binary neutron star system close to merger moves
with an orbital velocity of $> 0.1c$ against the ``vacuum'' background medium. This can cause 
strong shocks (e.g. \citealt{gittins25}) and it may become challenging to disentangle, say, physical neutrino emission from 
the one that is entirely due to the artificial interaction with the background medium. There is however
progress been made on the Eulerian treatment of vacuum, see e.g.  \citet{duez02,poudel20}.

SPH's ``natural tendency to ignore vacuum'' may also become a disadvantage in cases 
where the major interest of the investigation is a tenuous medium close to a dense one, say, for gas that 
is accreted onto the surface of a compact star. Such cases are probably more efficiently handled with an adaptive 
Eulerian method that can refine on physical quantities that are different from density. Several examples of 
hybrid approaches between SPH and Eulerian methods are discussed in Sect.~\ref{sec:WDWD_SNIa}.

\emph{Galilean invariance.} SPH is Galilean invariant and thus independent of the chosen computing 
frame. The lack of this property can cause serious artefacts for Eulerian schemes if the simulation is
performed in an inappropriate reference frame, see \citet{springel10b} for a number of examples.
Particular examples related to binary mergers have been discussed in \citet{new97} and \citet{swesty00}:
simulating the orbital motion of a binary system in a space-fixed frame can lead to an entirely 
spurious inspiral and merger, while simulations in a corotating frame may yield accurate results.
For SPH, in contrast, it does not matter in which frame the calculation is performed.

\emph{Advection.} Another strong asset of SPH is its exact advection of fluid properties: an attribute attached to an 
SPH particle is simply carried along as the particle moves. This makes it particularly easy to, say, 
post-process the nucleosynthesis from a particle trajectory, without any need for additional ``tracer particles''. 
In Eulerian methods high velocities with respect to the computational grid can seriously compromise the
accuracy of a simulation. For SPH, this is essentially a ``free lunch'', see for example Figure~\ref{fig:advection},
where a high-density wedge is advected with a velocity of $0.9999~c$ through the computational domain without 
any visible artefact.

\emph{Adaptive gravity and coupling to n-body.} The particle nature of SPH also allows for a natural transition to n-body methods.
For example, if ejected material from a stellar encounter becomes ballistic so that hydrodynamic
forces are negligible, one may decide to simply follow the long-term evolution of  point masses (rather than fluid parcels)
in a gravitational potential. Such a treatment can make time scales accessible 
that cannot be reached within a hydrodynamic approach
\citep{faber06b,rosswog07a,lee07,ramirezruiz09,lee10a}.
SPH can also straight forwardly be combined with highly flexible and accurate
gravity solvers such as tree methods.  Such approaches are extremely powerful for problems in which a 
fragmentation with complicated geometry due to the interplay of gas dynamics and self-gravity occurs, 
say in star or planet formation. 
Many successful examples of couplings of SPH with trees exist in the literature. Maybe the first one was the use the
Barnes-Hut oct-tree algorithm \citep{barnes86} within the TREESPH code \citep{hernquist89}, closely
followed by the implementation of a mutual nearest neighbour tree due to Press for the simulation 
of white dwarf encounters \citep{benz90b}. Today a number of very fast and flexible
tree variants are routinely used in SPH, see e.g., \citet{dave97,carraro98a,springel01a,wadsley04,springel05a,wetzstein09,nelson09},
and for a long time SPH-tree combinations were at the leading edge ahead of Eulerian approaches
that only have caught up later, see for example \citet{kravtsov99,tessier02}. More recently, also ideas 
from the fast multipole method have been borrowed \citep{dehnen00,dehnen02,gafton11,dehnen14,springel21,schaller24}
that allow for a better  scaling than $O(N \log N)$ with the particle number  $N$.\\
The ease of adaptive gravity/coupling to n-body methods has also made  SPH a primary choice for
planetary impact simulations, see for example \cite{benz89,cameron91,benz94,benz95,asphaug96,canup01,schaefer16,reinhardt17,chau18,golabek18,reinhardt20}.
Such simulations need to go beyond ideal hydrodynamics,  since they have to include effects 
such as elasticity or fracture of the planetary material.

\emph{Fluid instabilities.} Traditional SPH formulations have been shown to struggle with weakly
triggered fluid instabilities \citep{agertz07,springel10a,read10,mcnally12}. As we discussed before and will cover
in more detail in Sect.~\ref{sec:volume_elements}, the different treatment of different components of the same 
conservation law can cause surface tension forces near contact discontinuities. This insight triggered
a flurry of suggestions on how to improve this aspect of SPH 
\citep{price08a,cha10,read10,hess10,valcke10,junk10,abel11,gaburov11,murante11,read12,hopkins13,saitoh13}
and modern versions of SPH, e.g. \citet{frontiere17,rosswog20a,sandnes24}, obtain good results at essentially all 
the \enquote{hard problems}.

\emph{Magnetic fields.}  Magnetic fields are ubiquitous in astrophysics and there have been many attempts to implement them into
SPH \citep{boerve01,boerve04,price04a,price04b,price05,price06,boerve06,rosswog07a,dolag09,buerzle11a,buerzle11b,tricco12,tricco13}. 
 This is, however, notoriously difficult with  one of the major challenges being to preserve the $\nabla \cdot \vec{B}=0$ constraint during the MHD 
evolution. To date, this can only approximately be enforced by hyperbolic divergence cleaning methods,  as originally suggested by \cite{dedner02}
and adapted for SPH e.g. in \cite{tricco12} and \cite{tricco16}. There is no equivalent to the Eulerian \enquote{constrained transport} method  \citep{evans88},
which can enforce the $\nabla \cdot \vec{B}=0$ constraint to machine precision. Until recently, the employed SPMHD discretization performed
well on some problems, but seriously struggled, for example, in reproducing grid-based results for the magnetorotational instability (MRI),
see for example \cite{deng19}. There has, however, been very substantial progress recently: \cite{wissing22} managed to successfully simulate
the MRI in accretion disks for over 100 orbits. Major improvements came here (as is often the case in numerics) from seemingly small changes,
including the use of noise-suppressing Wendland kernels, see Sec.~\ref{sec:Wendland_kernel}, and in particular the use of a so-called GDF-gradient formulation that is described
in more detail in Sec.~\ref{sec:Alternative_SPH}.\\
One challenge, however, remains for SPH and what is usually considered
an advantage for SPH can turn into the opposite: since the particles follow the density evolution, they are mostly concentrated in high-density regions.
The magnetic field, however, can also be very strong in regions of low density (e.g. above a neutron star merger remnant) and in such potentially important
regions there would only be a low number density of SPH particles, i.e. very poor resolution. \\
For a more complete overview of this exciting, but technically complicated field of magnetic fields in SPH, we refer to  the 
literature \citep{iwasaki11,price12a,tsukamoto13,price18a} and in particular to the recent review of \cite{tricco23}.

\emph{Shocks.} Artificial dissipation is also often considered as a major drawback. However, if dissipation is steered properly,
see Sect.~\ref{sec:AV}, the performance should be very similar to the one of approximate 
Riemann solver approaches. 
A Riemann solver approach may, from an aesthetical point of view, be more appealing, though, and a number of such
approaches have been successfully implemented and tested  \citep{inutsuka02,cha03,cha10,murante11,puri14,rosswog25a,kitajima26}. In fact,
techniques such as slope-limited reconstruction (applied within the artificial dissipation terms) 
\citep{christensen90,frontiere17,rosswog20a,rosswog21a,rosswog25a,rosswog26a} that have been borrowed from Finite Volume methods
have turned out to be very effective in reducing dissipation where it is not needed, but more improvements should be
achievable \citep{garcia_senz26}.

\emph{Computational efficiency.} Contrary to what was often claimed in the early literature, however, SPH is not necessarily a very efficient 
method. It is true that if only the bulk matter distribution is of interest, one can often obtain robust
results already  with  an astonishingly small number of SPH particles. To obtain accurate results for the 
thermodynamics of the flow, however, still usually requires large particle numbers. In large SPH simulations
it becomes a serious challenge to maintain cache-coherence since particles that were initially close 
in space and computer memory can later on become completely scattered throughout different (mostly slow) 
parts of the memory.
This can be improved by using cache-friendly numerical variables and/or
by various sorting techniques to re-order particles in memory according to their spatial location.
This can be done --in the simplest case-- by using a uniform hash grid, but in many astrophysical
applications hierarchical structures such as trees are highly preferred for sorting the particles, see e.g., 
\citet{warren95,springel05a,nelson09,gafton11}. While such measures can easily improve the performance
by factors of a few, they come with some bookkeeping overhead which, together with, say, individual
time steps and particle sinks or sources, can make codes rather unreadable and error-prone.\\
In recent years much effort has been invested in achieving massive parallelism and in porting codes
to GPUs. Examples of such SPH codes include ChaNGa \citep{menon15}, PKDGRAV3 \citep{potter17}, SWIFT \citep{schaller24} and SPH-EXA  \citep{cabezon25}.

\subsection{Roadmap through this review}
Since this review is rather long, we would like to provide some pointers
to useful starting points:
\begin{itemize}
\item Readers that are only interested in SPH as a numerical method can focus entirely on Sect.~\ref{sec:SPH_numerics}.
   The basics of the method, such as kernel interpolation, basic SPH formulations of the ideal gas dyamics
   equations, numerical conservation etc. are outlined in Sect.~\ref{sec:kernel_interpolation}--\ref{sec:SPH_from_Lagrangian}.
   More advanced topics such as alternative SPH formulations, various meshless derivatives and the treatment of shocks are 
   discussed in Sects.~\ref{sec:Alternative_SPH}--\ref{sec:shocks}.  Sect.~\ref{sec:SPH_numerics} closes by making 
   broad-brush connections to various particle methods that have been developed over, roughly,  the last decade, and it stresses 
   that, in addition to accurate evolution schemes,  the initial conditions of a simulation are a very crucial ingredient for the accuracy of
   a simulation. Finally, we show some examples involving different SPH discretizations.
\item Sect.~\ref{sec:GR_SPH} is dedicated to general relativistic SPH. It explains general relativistic formulations of SPH
   both in  fixed and dynamical spacetimes and provides the special relativistic limit. A fair part of this section is dedicated to the recent
   development of fully dynamical, general relativistic SPH which also makes ample use of the more advanced numerical techniques
   described in Sect.~\ref{sec:SPH_numerics}.
\item The more astrophysics-inclined reader may want to focus instead on the remaining Sect.~\ref{sec:astro_mergers}. Here we summarize
    the current status of SPH simulations in white dwarf--white dwarf, in neutron star--neutron star and in neutron star--black hole
    collisions. Unfortunately, we cannot comprehensively summarize the extremely broad areas themselves, instead we have to focus on those
    parts that involve SPH. In this section, we give brief references to the technicalities discussed in Sect.~\ref{sec:SPH_numerics} and \ref{sec:GR_SPH}.
    The modular structure of these sections should allow for a selective consultation on the more technical issues of SPH.
\end{itemize}

\section{SPH as a numerical method for Newtonian gas dynamics}
\label{sec:SPH_numerics}
\begin{figure}[ht]
     \centerline{\includegraphics[width=0.5\textwidth]{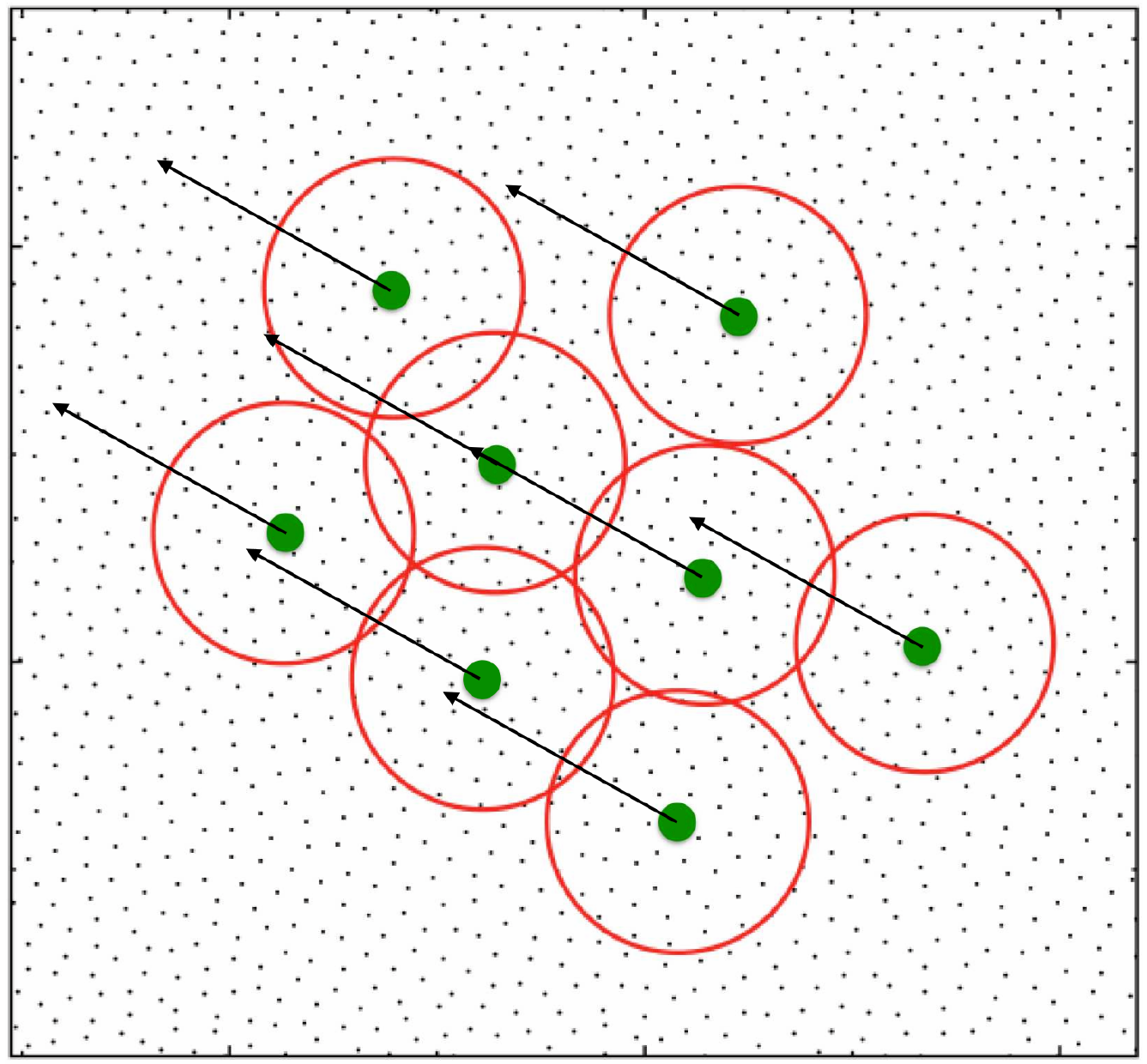}}
    \caption{Sampling of a fluid flow by SPH particles: the continuum is represented by a finite
                  number of sampling points (``particles'') that move with the local fluid velocity.
                  The particles share the total mass and move in a way that they conserve energy, 
                  momentum and angular momentum. For some particles (green, filled circles)
                  also their support (``sphere of influence'') is indicated by red circles.}
    \label{fig:sph_flow}
\end{figure}

The first suggestions of SPH date back to 1977 \citep{gingold77,lucy77}.
The main idea of SPH is to model a fluid via spherical 
particles of finite size and overlapping support, see Fig.~\ref{fig:sph_flow}. These particles
move with the local fluid velocity, i.e. SPH is a fully \emph{Lagrangian} meshless method,
as opposed to Eulerian or Arbitrary Lagrangian Eulerian (\enquote{ALE}) methodologies.
The overlapping nature of the particles distinguishes them, for example,
from (usually quasi-) Lagrangian Moving Mesh methods, see e.g. 
\citet{springel10b} and \citet{duffell11}.  The latter methods instead tessellate space into non-overlapping 
cells which have a sharp interface surface between them. The task is now to 
define the interactions between the numerical particles in a way so that the
underlying continuum equations and their properties are as accurately as possible
reproduced by a finite set of discrete particles. Ideally, this should include the
exact conservation of nature's conservation laws \emph{on the discrete level}.
This is indeed possible in SPH and it is one of its major assets.

\subsection{Basics of kernel interpolation}
\label{sec:kernel_interpolation}
At the heart of SPH is a kernel interpolation. To obtain SPH evolution
equations for  Lagrangian fluid dynamics involves two steps: 
a smooth kernel-interpolation in the continuum limit and a subsequent
discretization of the related integrals. This latter step  represents the
transition to a discrete system of \enquote{particles}.

\subsubsection{Continuum limit}
\begin{figure}[ht]
   \centering
   \includegraphics[width=8cm]{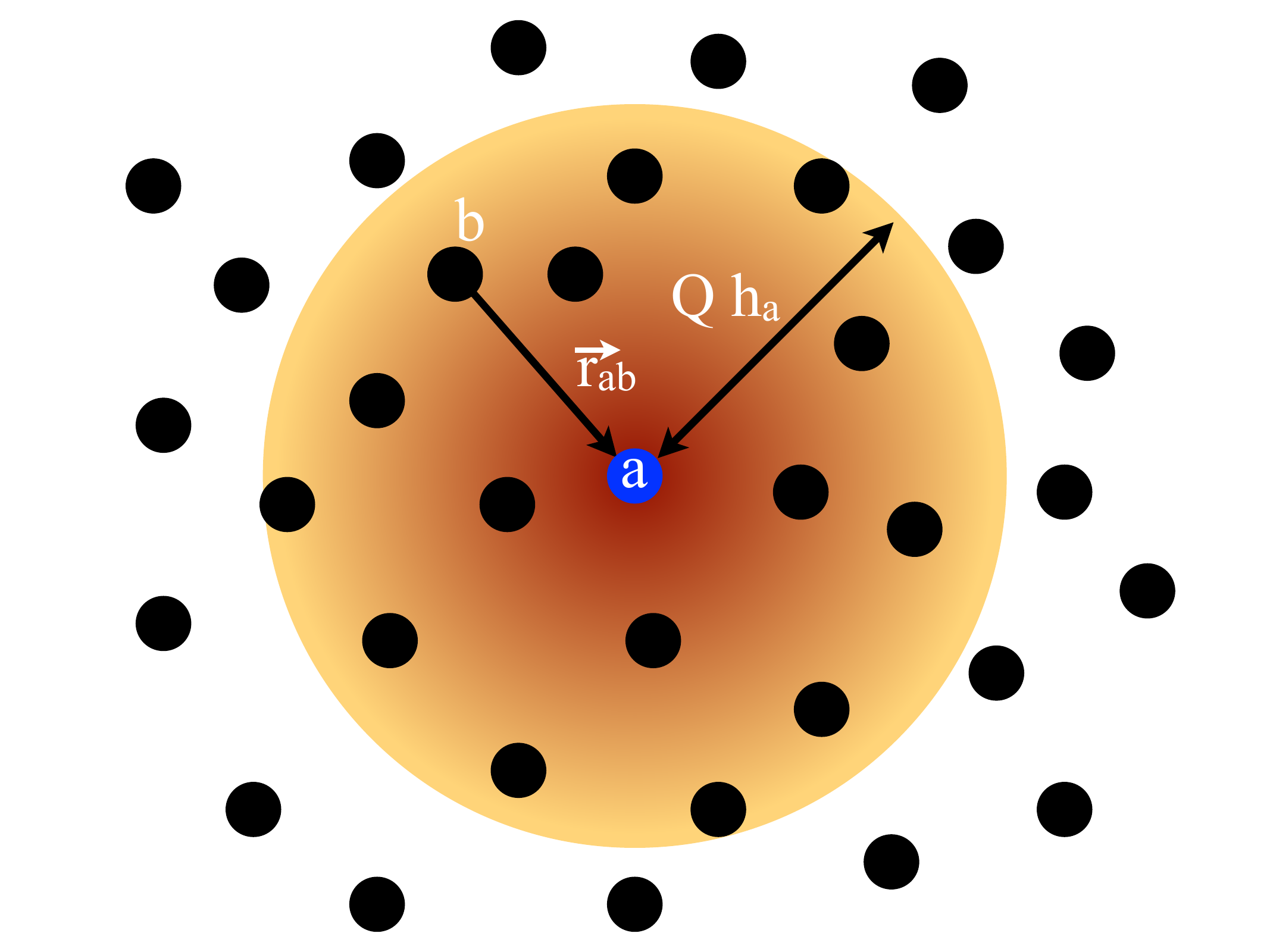} 
   \caption{Sketch of the interaction of a particle $a$ with its
     neighboring particles. To avoid a computationally expensive
     interaction of each particle with all other particles, kernels
     with a finite support (indicated by the shaded region) are
     usually used. The support size of a particle $a$ is set as a
     multiple, $Q h_a$, of its smoothing length, $h_a$. Many
     frequently used kernels have $Q=2$.}
   \label{fig:sketch_support}
 \end{figure}

The integral interpolant of a function $f$ reads
\be
\tilde{f}(\vr) \equiv \int f(\vr') W_h(\vr - \vr') dV', \label{eq:integ_interpol}
\ee
where the quantity $h$ determines the width of the kernel function
$W_h$. In an SPH context, $h$ is called the \enquote{smoothing length}.
Obviously, the kernel function should be normalized
\be
\int W_h(\vr - \vr') dV' = 1 \label{eq:kernel_normalization}
\ee
and, for $\tilde{f}$ to reproduce the original function $f$ in the
limit of vanishing smoothing length, the
kernel must have the \enquote{delta-property'}
\be
\lim_{h \rightarrow 0} W_h(\vr - \vr') \rightarrow \delta(\vr - \vr'),
\label{eq:delta_prop}
\ee
where $\delta(\vr)$ is the Dirac delta distribution.
To keep the computational effort under control, one is usually
interested in a \emph{local} smoothing and therefore uses kernels
with a compact support, i.e. kernels that vanish outside of a cut-off
radius, which is usually a multiple of the smoothing length $h$, see
Fig.~\ref{fig:sketch_support}. For reasons 
related to numerical conservation, in particular of angular momentum,
see Sect.~\ref{sec:conservation} below, \emph{radial kernels} with the property
\be
W_h(\vr - \vr') = W_h(|\vr - \vr'|) 
\ee
are usually preferred. This choice, however, imposes a substantial
restriction on the kernel function and, depending on the intended
context, one may want to relax this condition (but then should monitor that
the degree of angular momentum conservation is acceptable in practice). Another obvious
implication of Eq.~(\ref{eq:integ_interpol}) is that the kernel must
have the dimension \enquote{1/volume}. A number of SPH kernels that are
frequently used are
summarized in Appendix \ref{sec:kernels}.

For a clearer understanding of the interpolation accuracy, one may now
expand the function $f(\vr')$ in Eq.~(\ref{eq:integ_interpol}) around the
position $\vr$
\bea
\tilde{f}(\vr) &=&  \hspace*{-0.3cm} \int\left\{ f(\vr) + (\p_i f)_{\vr} \; (\vr' - \vr
)^i + \frac{1}{2!} (\p_j \p_k f)_{\vr} \; (\vr' - \vr
)^j  \; (\vr' - \vr
)^k + \mathrm{h.o.t.} \right\}  W_h(\vr - \vr') dV' \nonumber \\
&=& f(\vr) \int  W_h(\vr - \vr') dV' +  (\p_i f)_{\vr} \; \int (\vr' -
\vr)^i W_h(\vr - \vr') dV' + \nonumber  \\
&& \frac{1}{2!} (\p_j \p_k f)_{\vr}   \int  (\vr' - \vr
)^j  \; (\vr' - \vr
)^k  W_h(\vr - \vr') dV' + \mathrm{h.o.t},
\eea
where higher order terms in the Taylor expansion are abbreviated as
``h.o.t''.
At this point one can introduce the continuous moments of the kernel
at position $\vr$
\bea
(\mathcal{M}_0)(\vr) &=& \int W_h(\vr -\vr') dV'\\
(\mathcal{M}_1^i)(\vr) &=& \int  (\vr' - \vr)^i \; W_h(\vr -\vr') dV'\\
(\mathcal{M}_2^{ij})(\vr) &=& \int  (\vr' - \vr)^i   (\vr' - \vr)^j
\; W_h(\vr -\vr') dV' \\
...\nonumber
\eea
So apart from fulfilling the normalization condition
Eq.~(\ref{eq:kernel_normalization}), or, the zeroth moment being equal to
unity, the higher order moments of the kernel
function should vanish for a good interpolation accuracy in the
continuum limit, so that all error terms vanish individually. If the kernel is
symmetric, $W_h(\vr' - \vr)= W_h(\vr - \vr')$, the first moment vanishes and
therefore, for such kernels, the integral approximation is second
order accurate. One can, of course, enforce the vanishing of higher
order moments, but this has implications on the conservation
properties and usually destroys the exact conservation that otherwise
can be enforced in SPH in a very simple way. So there is a tradeoff 
between exact conservation and the accuracy with which functions can 
be reproduced.

%
 %
 
 \subsubsection{Discrete approximation}
 \label{sec:discrete_approx}
To introduce the concept of ``particles'', one discretizes the above
integrals by introducing the volume $V_b$ of a particle labelled $b$.
The most straight forward and  most commonly used way is to use the mass
of a particle $b$, $m_b$ and its mass density, $\rho_b$, to construct
a volume element
\be
V_b= \frac{m_b}{\rho_b}.\label{eq:std_vol_element'}
\ee
This is a \emph{choice} and
--unless noted otherwise-- we use this expression, but other
choices are possible, see e.g. \citet{saitoh13,hopkins13,rosswog15b,cabezon17}.
With a one-point quadrature to discretize the integral approximation,
Eq.~(\ref{eq:integ_interpol}), evaluated at the position
of a particle $a$, one finds 
\be
\tilde{f}_a= \tilde{f}(\ra)= \sum_b f_b \; V_b W_h(\rab) \equiv \sum_b
f_b \; \Phi_b^h(\ra),
\label{eq:disc_func_approx}
\ee
where we have abbreviated the weight, sometimes referred to as ``shape
functions'', as
\be
\Phi_b^h(\ra)\equiv V_b W_h(\rab)
\ee
and $\vec{r}_{ab}\equiv \ra - \rb$.
Note, that so far
we have not yet specified which smoothing length $h$ to use, one could
use, for example, the smoothing length of particle $a$, $h_a$ or
particle $b$, $h_b$, or an
average of both such as $(h_a+h_b)/2$.

\begin{figure}[ht]
\centering
 \includegraphics[width=11cm]{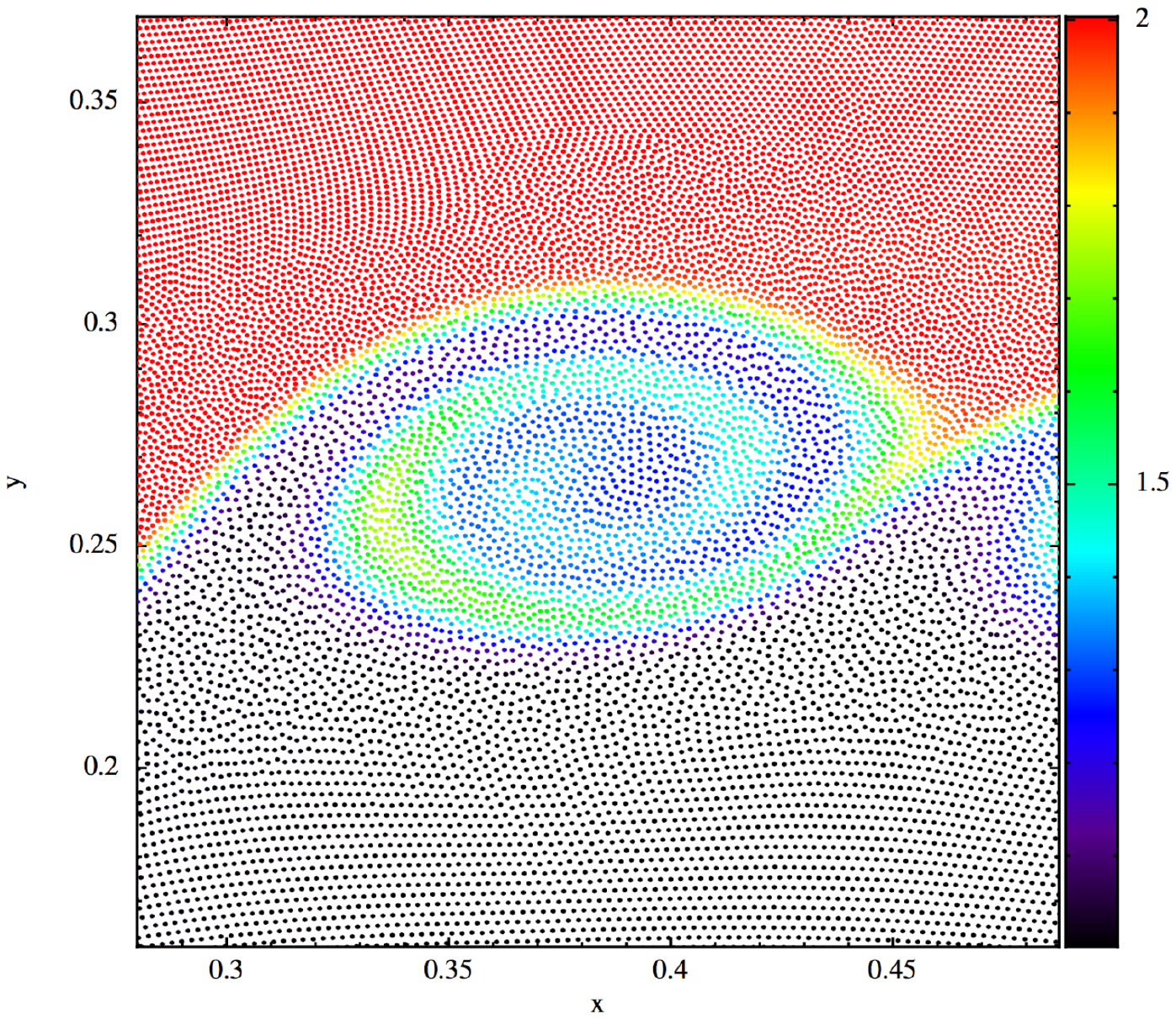}
 \caption{SPH particle distribution in a Kelvin--Helmholtz instability
   test, colour-coded is the density.
               Note that the particles are
               \emph{not} distributed randomly, but instead show a high degree of regularity. This
               is crucial for the accuracy of the SPH kernel interpolation, see the quality
               indicators $\mathcal{Q}_1 ... \mathcal{Q}_4$ in Eqs.~(\ref{eq:quality_int}) and
               (\ref{eq:Q3_Q4}). This example uses  a high-order Wendland kernel,
               see Appendix~\ref{sec:kernels},  which enforces  a
               particularly regular particle distribution. }
   \label{fig:particle_dist}
\end{figure}

For a more quantitative understanding of the quality of the above discrete function
interpolation, one can, similar to the previous section, Taylor-expand the
function $f$ around $\vec{r}_a$, so that its value at $\rb$ becomes
\be
f_b= f_a +  (\p_i f)_a  \rba^i + \frac{1}{2!}
(\p_{jk}  f)_a \rba^j \rba^k  + O(r_{ba}^3),
\label{eq:Taylor_fb}
\ee
where the Einstein summation convention has been
used. Here --and in the rest of the review-- we will label the
particle of interest with $a$, neighbour particles that are summed
over are usually labelled $b$.
Inserting the expansion into the discrete function approximation
Eq.~(\ref{eq:disc_func_approx}) one finds
\be
\tilde{f}_a= f_a \; \sum_b \Phi_b^h(\ra) +  (\p_i f)_a \; \sum_b
\rba^i \Phi_b^h(\ra)  + \frac{1}{2!}  (\p_{jk} f)_a \; \sum_b
\rba^j \rba^k \Phi_b^h(\ra) + ...,
\ee
so similar to the continuous case, the zeroth discrete moment should be equal
to unity and the discrete higher order moments at position $a$ should vanish:
\bea
(M_0)_a       &=& \sum_b \Phi_b^h(\ra) \stackrel{!}{=} 1 \label{eq:zeroth_moment}\\
(M_1^i)_a    &=& \sum_b \rba^i \Phi_b^h(\ra) \stackrel{!}{=} 0 \label{eq:first_moment}\\
(M_2^{jk})_a &=& \sum_b \rba^j \rba^k  \Phi_b^h(\ra) \stackrel{!}{=} 0 \label{eq:second_moment}\\
...\nonumber
\eea
 Thus, the two lowest order  \enquote{interpolation quality
indicators} read
\be
\mathcal{Q}_1: \;  (M_0)_a = 1 \quad \text{and} \quad
\mathcal{Q}_2: \; (M_1^i)_a = 0.
\label{eq:quality_int}
\ee
The first relation simply states that the particles should provide a good approximation to a partition of
unity, i.e. the applied weights should add up to unity. The second relation indicates that 
there is no preferred direction in the particle distribution.
If fixed, prescribed kernels are used --as is the common
approach in SPH-- these relations are only approximately fulfilled and
it depends on the chosen kernel function, the number of
contributing particles in the kernel support (\enquote{neighbour particles}
or simply \enquote{neighbours}) and the actual particle
distribution  how well these relations
are fulfilled. A particle distribution from a Kelvin--Helmholtz test is
shown in Fig.~\ref{fig:particle_dist}, which shows that the particles
are far from randomly distributed, their distribution is 
``disordered, but in an \emph{orderly} way'' \citep{monaghan05}.
One can, however, also enforce consistency
relations by construction, this is, for example,  the basic idea behind the
``reproducing kernel method'' \citep{liu95} which we summarize in Sect.~\ref{sec:RPK}. 
This usually comes at the price of a non-negligible computational effort to calculate discrete 
moment satisfying kernels and of sacrificing exact numerical angular momentum conservation.
The latter, however, seems to be often acceptably small.

%
%
To estimate the gradient of a function one can straight
forwardly apply the nabla operator to Eq.~(\ref{eq:disc_func_approx})\footnote{From now
on we drop the distinction between a function and its numerical approximation.}
\be
\nabla f(\vec{r}) = \sum_b V_b \; f_b \; \nabla_r
W_h(\vec{r_b} - \vec{r}),
\label{eq:nabla_f}
\ee
where the index $r$ at the nabla operator indicates that the
derivative is taken with respect to the variable $\vec{r}$ (rather
than $\vec{r}_b$). In SPH, the derivatives are usually needed at the position
of a particle (here labelled $a$) and abbreviated as
\be
(\nabla f)^{(0)}_a= \sum_b V_b  f_b \nabla_a W_h(\vec{r}_{ab}),
\label{eq:std_grad}
\ee
where  the label at the nabla operator is again a reminder that the derivative
is taken with respect to $\vec{r}_a$ (and not $\vec{r}_b$). This is the simplest possible choice
and since we will discuss more sophisticated choices below, especially in Sec.~\ref{sec:accurate_gradients}, 
we label this gradient prescription with superscript "(0)".
To get a quantitative condition for the quality of the gradient
estimate, one can insert Eq.~(\ref{eq:Taylor_fb}) into
(\ref{eq:std_grad}) and finds
\be
\p_i f(\ra)= f_a \left[ \sum_b V_b \p_i W_h\rba \right] + (\p_j f)_a
\left[ \sum_b V_b \rba^j \p_i W_h \rba \right] + ... ,
\ee
so for good gradient accuracy the first bracket should vanish, and the
second one should give  a Kronecker delta, so we have two more quality
indicators
\be
\mathcal{Q}_3: \sum_b V_b \p_i W_h\rba = 0 \quad \text{and} \quad
\mathcal{Q}_4: \sum_b V_b \rba^j \p_i W_h
\rba = \delta^j_i.
\label{eq:Q3_Q4}
\ee
$\mathcal{Q}_3$ is, of course, just the gradient of the quality
indicator $\mathcal{Q}_1$ which expresses the partition of unity. It
is therefore sometimes called ``partition of nullity''. 
As mentioned above, in standard SPH one only has an approximate
partition of unity / partition of nullity
\bea
\sum_b \Phi^h_b(\vr) = \sum_b \frac{m_b}{\rho_b} \;
W_h(\vec{r_b} - \vec{r}) &\approx& 1\\
\sum_b \frac{m_b}{\rho_b} \; \nabla_r W_h(\vec{r_b} -
\vec{r})  &\approx& 0,
\eea
where the second equation is simply the result of applying the nabla
operator to the first one. This has the effect, that when applying Eq.~(\ref{eq:nabla_f})
to  a constant function $f=f_0$, one only approximately recovers the
theoretical value of zero
\be
\nabla_r \left(\sum_b \frac {m_b}{\rho_b} f_b W_h(\vec{r_b} -
  \vec{r})\right)= f_0 \sum_b \frac {m_b}{\rho_b} \nabla_r W_h(\vec{r_b} -
  \vec{r}) \approx 0.
\ee
This, however, can be easily corrected by simply ``subtracting a numerical
zero'', so that a better gradient estimate reads
\begin{align}
\left( \nabla f\right)^{(1)}_a  &= \sum_b \frac{m_b}{\rho_b} f_b \nabla_a W_{h_a}(\vec{r}_{ab})
- \left(f_a \sum_b \frac{m_b}{\rho_b} \nabla_a W_{h_a}(\vec{r}_{ab})\right) \nonumber \\
 & = - \sum_b \frac{m_b}{\rho_b} f_{ab} \nabla_a
   W_{h_a}(\vec{r}_{ab}),
\label{eq:better_grad}
\end{align}
where $f_{ab}= f_a - f_b$ and this expression vanishes by construction
when all $f_k$ are the same. Further meshless derivatives are
discussed in Sect.~\ref{sec:accurate_gradients}.

\subsection{\enquote{Vanilla Ice} Newtonian SPH}
\label{sec:vanilla_ice}
In the simplest case of ideal, non-relativistic fluids the equations read
\bea
\frac{d \rho}{dt}   &=& - \rho \nabla \cdot \vec{v}\label{eq:drhodt}\\
\frac{d \vec{v}}{dt} &=& - \frac{\nabla P}{\rho} + \vec{f}\label{eq:dvdt}\\
\frac{d u}{dt}&=& \frac{dq}{dt} + \frac{P}{\rho^2} \frac{d \rho}{dt}
\label{eq:dudt}
\eea
and they express the conservation of mass, momentum and energy.
Here,
$\rho$ is the mass density, $\vec{v}$ the fluid velocity, $P$ the pressure,
$\vec{f}$ denotes other ``body forces'' e.g. from gravity, $u$ is the specific 
internal energy (i.e. energy per mass) and $dq/dt$ denotes
potential sources of heating and/or cooling. The Lagrangian time derivative, $d/dt$,
is related to the Eulerian time derivative $\p/\p t$ by
\be
\frac{d}{dt} (.)= \left(\frac{\p}{\p t} + \vec{v} \cdot\nabla\right) (.).
\label{eq:Lag_dt}
\ee
Note that Eq.~(\ref{eq:dudt}) is simply the first law of
thermodynamics written ``on a per mass basis", i.e. the involved
quantities are e.g. energy per mass or volume per mass (=1/density).\\
%
%
One has to decide now whether one wants to integrate 
the continuity equation (\ref{eq:drhodt}) or not. While
some recent studies \citep{meng21,sandnes24} find such an approach
beneficial, the most common approach in SPH
is a summation over nearby particles. The corresponding
``density-by-summation'' can be found by simply applying
Eq.~(\ref{eq:disc_func_approx})
to the density
\be
\rho_a= \sum_b m_b W_h(\rab).\label{eq:rho_sum}
\ee
In other words: the density can be calculated by summing over the
local neighborhood of a particle  $a$ and thereby weighting each particle’s
mass according to how far away it is from particle $a$. If one keeps
each particle’s mass fixed in time, one has enforced \emph{exact mass
conservation} and one does not need to solve the mass conservation
equation  explicitly. 

Keep in mind, however, that the density summation
introduces some smoothing: even when the masses are distributed with a
sharp jump from one particle to the next, Eq.~(\ref{eq:rho_sum}) will
return a smooth approximation 
of the jump in the density. While this may seem trivial, it can have
non-trivial consequences. For example, it can introduce a surface
tension effects which can suppress 
weakly triggered Kelvin--Helmholtz instabilities \citep{agertz07}.
This will be discussed in more detail in
Sect.~\ref{sec:volume_elements}.

If one decides to explicitly solve the evolution equation
for the density, Eq.~(\ref{eq:drhodt}), one can apply the gradient
prescription Eq.~(\ref{eq:better_grad}) to the velocity
\be
\left(\nabla \cdot \vec{v}\right)_a = - \sum_b \frac{m_b}{\rho_b} \vec{v}_{ab} \cdot
\nabla_a W_h(\rab)\label{eq:divv_v1},
\ee
where
\be
\vec{v}_{ab}= \vec{v}_a - \vec{v}_b.
\ee
An alternative discretization of the velocity divergence can be found by taking Eq.~(\ref{eq:drhodt})
and by calculating the Lagrangian time derivative of
Eq.~(\ref{eq:rho_sum}) directly via
\be
(\nabla \cdot \vec{v})_a= -\frac{1}{\rho_a} \frac{d\rho_a}{dt}= -\frac{1}{\rho_a} \sum_b m_b \frac{dW_h(\rab)}{dt}
= \frac{1}{\rho_a} \sum_b m_b \; \vec{v}_{ba} \cdot \nabla_a W_h(\rab),
\label{eq:divv_v2}
\ee
where we have taken the straight-forward Lagrangian time derivative of
the kernel function, see Eq.~(\ref{eq:k5}) in Appendix \ref{sec:kernels}. Thus, we have 
derived an alternative numerical discretization for
$\nabla \cdot \vec{v}$, where the difference between the two estimates is whether
the local density ("$\rho_a$") or the weighted nearby densities ("$\rho_b$") are used. 
Both are equally valid and both vanish 
for the uniform velocity case.

Both of these $\nabla \cdot \vec{v}$ prescriptions can used to evolve
the density. With Eq.~(\ref{eq:divv_v2}), the continuity equation becomes
\be
\frac{d\rho_a}{dt}= \sum_b m_b \; \vec{v}_{ab} \cdot \nabla_a
W_h(\rab). \label{eq:drho_dt}
\ee
Note that if the density is integrated, mass conservation needs to be ensured
by the time stepping scheme rather than being explicitly guaranteed as
in the summation approach, Eq.~(\ref{eq:rho_sum}).

%
Eq.~(\ref{eq:dvdt}) expresses the conservation of momentum in the
continuum limit. For reasons of numerical conservation, see Sect.~\ref{sec:conservation} below,
one does not start by straight forwardly discretizing $\nabla P/\rho$,
but instead one uses a mathematically 
equivalent expression, that leads to a different discrete
approximation. If one applies
\be
\nabla \left( \frac{P}{\rho}\right)= \frac{\nabla P}{\rho} - P
\frac{\nabla \rho} {\rho^2},
\ee
to Eq.~(\ref{eq:dvdt}), evaluated at particle $a$ and applies the
standard gradient Eq.~(\ref{eq:std_grad}) one finds (ignoring other
body forces)
\be
\frac{d\vec{v}_a}{dt}= - \sum_b m_b \left\{ \frac{P_a}{\rho_a^2} + \frac{P_b}{\rho_b^2} \right\} \nabla_a W_{h}(r_{ab}).\label{eq:dvdt_van}
\ee
A discrete energy equation can be found by using
Eq.~(\ref{eq:drho_dt}) in the first law of thermodynamics,
Eq.~(\ref{eq:dudt}),
\be
\frac{du_a}{dt}= \frac{P_a}{\rho_a^2} \sum_b m_b \vec{v}_{ab} \cdot
\nabla_a W_h(\vec{r}_{ab}). \label{eq:dudt_van}
\ee
Eqs.~(\ref{eq:rho_sum}), (\ref{eq:dvdt_van}) and (\ref{eq:dudt_van})
form a valid set of conservative SPH equations.

As in other numerical methods, one has a fair amount of freedom which variables
one would like to evolve. Instead of evolving $u$, one could also
evolve the ``thermokinetic energy'' $e \equiv  u + \frac{1}{2}
\vec{v}^2$, which is connected to the internal energy and momentum
equation by $de/dt=du/dt + \vec{v} \cdot d\vec{v}/dt$. Its continuous
evolution equation reads 
\be
\frac{d e}{dt}= - \frac{1}{\rho} \nabla \cdot \left( P
\vec{v}\right)= - \frac{P}{\rho^2} \nabla \cdot\left( \rho \vec{v}
\right) - \vec{v} \cdot \nabla \left( \frac{P}{\rho} \right).
\ee
If one straight forwardly applies Eq.~(\ref{eq:std_grad}) to both
terms on the right hand side (RHS), one finds
\be
\frac{d e_a}{dt}= - \sum_b m_b \left( \frac{P_a
\vec{v}_b}{\rho_a^2} +  \frac{P_b
\vec{v}_a}{\rho_b^2} \right) \nabla_a W_h (r_{ab}).
\label{eq:dedt}
\ee
This form of the equation is very similar to the special- and
general-relativistic energy equations based on the canonical energy
from a Lagrangian, see Eqs.~(\ref{eq:GR_energy_evolution}) and (\ref{eq:SR_energy_evolution}).

Yet another alternative is to integrate an evolution equation for the
specific entropy, see e.g. \citet{springel02} and \citet{liptai19}. This has the
advantage that one has full control over the entropy evolution which,
in the absence of shocks, is strictly conserved in ideal fluid
dynamics.

\subsection{Conservation in standard SPH}
\label{sec:conservation}
Not evolving the masses of SPH particles and using
Eq.~(\ref{eq:rho_sum}) is a straight forward way to
ensure \emph{exact mass conservation}. The  conservation of momentum and
energy results in standard SPH from the combination of terms that are
symmetric and anti-symmetric with respect to exchanging the particle
indices $a \leftrightarrow b$. 
For example, according to Eq.~(\ref{eq:dvdt_van}) the force from particle $b$ on $a$ is
\be
\vec{F}_{b\rightarrow a}= -m_a m_b \left\{ \frac{P_a}{\rho_a^2} +
  \frac{P_b}{\rho_b^2}\right\} \nabla_a W_{h}(r_{ab})= f_{(ab)}
  g_{[ab]}= - \vec{F}_{a\rightarrow b} \label{eq:F_anti_sym}
\ee
where $f_{(ab)}/g_{[ab]}$ denote  functions that are
symmetric/antisymmetric under $a \leftrightarrow b$. The
anti-symmetry of the term $\nabla_a W_{h}(r_{ab})= - \nabla_b W_{h}(r_{ab})$, see
Eq.~(\ref{eq:k4}), stems from assuming radial kernels which we always do
unless explicitly noted otherwise.
Therefore, the total change of momentum is
\be
\sum_a m_a \frac{d\vec{v}_a}{dt}= \sum_{ab} f_{(ab)} g_{[ab]}= 0,
\ee
which cancels since each term has a matching term of equal absolute
value and opposite sign, so that we have \emph{exact momentum conservation}.
The \emph{exact conservation of energy} follows in a similar way by
using Eq.(\ref{eq:dedt})
\be
\frac{dE}{dt}= \sum_a m_a \frac{d \tilde{e}_a}{dt} = - \sum_{a,b} m_a
m_b \left( \frac{P_a
\vec{v}_b}{\rho_a^2} +  \frac{P_b
\vec{v}_a}{\rho_b^2} \right) \nabla_a W_h (r_{ab})= \sum_{ab} f_{(ab)} g_{[ab]}= 0,
\ee
where the cancellation occurs for the same reasons as before.
For the angular momentum change we sum up all torques on the
particles,
\begin{align}
\frac{d\vec{L}}{dt}= \sum_{a,b} \vec{r}_a \times
\vec{F}_{b \rightarrow a} = & \frac{1}{2} \left( \sum_{a,b} \vec{r}_a \times
\vec{F}_{b \rightarrow a} + \sum_{b,a} \vec{r}_b \times
\vec{F}_{a \rightarrow b} \right)= \nonumber\\
& \frac{1}{2} \left( \sum_{a,b}
\left( \vec{r}_a - \vec{r}_b \right) \times \vec{F}_{b \rightarrow a}
\right)= 0,
\end{align}
where we have split the first double-sum into two parts, echanged the
summation order ($a \leftrightarrow b$) and used the anti-symmetry of
$\vec{F}_{b \rightarrow a}$, see Eq.~(\ref{eq:F_anti_sym}). The total
angular momentum change finally vanishes, because $\vec{F}_{b
\rightarrow a} \propto \nabla_a W_h(r_{ab}) \propto \vec{r}_a -
\vec{r}_b$, see Eq.~(\ref{eq:k4}) in Appendix~\ref{sec:kernels}. This clearly illustrates the
advantages of using \emph{radial} kernels, $W_h(\vec{r}_{ab})= W_h(|\vec{r}_{ab}|)$.

 \subsection{Adaptive resolution and \enquote{grad-h} terms}
 \label{sec:adapt_res}
Up to here we have not discussed how to (dynamically) choose the size
of the smoothing lengths. In many engineering contexts, where 
essentially incompressible flows are simulated (with $\nabla \cdot \vec{v} = 0$ or
$\rho=$ const, see Eq.~(\ref{eq:drhodt})), one may get
away with keeping the smoothing lengths at constant values, but in most
astrophysical contexts the densities can change by orders of magnitude,
and therefore one needs to adapt the smoothing lengths dynamically. While
this may seem like a small technical detail, these choices have a fair
impact on the kernel approximation accuracy and on the particle
distribution/the noise in a simulation.\\ 
Usually the density is used to steer the smoothing length evolution,
e.g. by starting from
\be
\frac{h(t)}{h_0}= \left( \frac{\rho_0}{\rho(t)}\right)^{1/3},
\ee
taking the Lagrangian time derivative and using the continuity
equation (\ref{eq:drhodt}), to obtain an evolution equation
\citep{benz90a},  
\be
\frac{dh_a}{dt}= \frac{1}{3}h_a \left(\nabla \cdot \vec{v} \right)_a
\label{eq:h_update1}
\ee
that can be integrated forward in time together with  the hydrodynamic equations.

More frequently, the smoothing length is updated according to \citet{gingold82}
\be
h_a= \eta \left( \frac{m_a}{\rho_a} \right)^{1/3}, \label{eq:h_update2}
\ee
where the parameter $\eta$ is typically chosen in the range between
1.2 and 1.5.\footnote{The best $\eta$-value depends on the exact kernel that is
  used, see Appendix~\ref{sec:kernels}.}
Note, however, that for this prescription the density
\be
\rho_a= \sum_b m_b W_{h_a}(\rab) \label{eq:rho_ha}
\ee
and smoothing lengths, Eq.~(\ref{eq:h_update2})
depend on each other, so that an iteration is necessary to find consistent
$\rho$/$h$-values.

If non-constant smoothing lengths are used one should (at least 
in principle)  account for additional terms in the SPH equations
\citep{nelson94,serna96,springel02,monaghan02}. 
If we take the changes in the smoothing length into account, we
find by straight forwardly taking the Lagrangian time derivative of
Eq.~(\ref{eq:rho_ha}) and by applying the chain rule
\be
\frac{d \rho_a}{dt}= \sum_b m_b \left\{ \frac{\p W_{h_a}(\vec{r}_{ab})}{\p
    \vec{r}_{ab}} \frac{d \vec{r}_{ab}}{dt} +
  \frac{\p W_{h_a}(\vec{r}_{ab})}{\p h_a} \frac{\p h_a}{\p\rho_a} \frac{d \rho_a}{dt}\right\},
\ee
which, after collecting the $d\rho_a/dt$-terms and applying Eq.~(\ref{eq:k5}), yields
\be
\frac{d \rho_a}{dt}= \frac{1}{\Omega_a} \sum_b m_b \vec{v}_{ab} \cdot \nabla_a W_{h_a}(\vec{r}_{ab})
\label{eq:drhodt_gradh}
\ee
with the ``grad-h'' correction
\be
\Omega_a=  1 -  \frac{\p h_a}{\p\rho_a} \sum_b m_b \frac{\p W_{h_a}(\vec{r}_{ab})}{\p h_a}.
\label{eq:omega_grad_h}
\ee
In a similar way, the spatial derivatives can be calculated
\be
\frac{\p\rho_b}{\p\vec{r}_a}
=  \sum_k m_k \left\{ \nabla_a W_{h_b}(\vec{r}_{bk})
   + \frac{\partial W_{h_b}(\vec{r}_{bk})}{\partial h_b} \frac{\p h_b}{\p \rho_b} 
      \frac{\p \rho_b}{\p\vec{r}_a} \right\}
    = \frac{1}{\Omega_b} \sum_k m_k \nabla_a W_{h_b}(\vec{r}_{bk}).
\label{eq:nabla_rho_grad_h}
\ee
To summarize: if derivatives of the smoothing length $h$ are accounted for,
the ``standard'' SPH expressions for the spatial and temporal density
derivatives have to be corrected by factors $1/\Omega$.
In \citet{rosswog07c} the ``grad-h'' terms were found to further improve the
(already very good) numerical conservation, but overall these terms
did not appear to be crucial and were therefore disregarded.

Another way to update the smoothing lengths, similar in spirit to
Eq.~(\ref{eq:h_update2}), but without involving masses, is to use
\be
h_a= \eta \; n_a^{-1/3}, 
\ee
where
\be
n_a= \sum_b W_h(\vec{r}_{ab})
\ee
is the number density of the SPH particles at position $a$. Also here
one needs an iteration for consistent values.

%
\begin{figure}[ht]
   \centering
   \includegraphics[width=8cm]{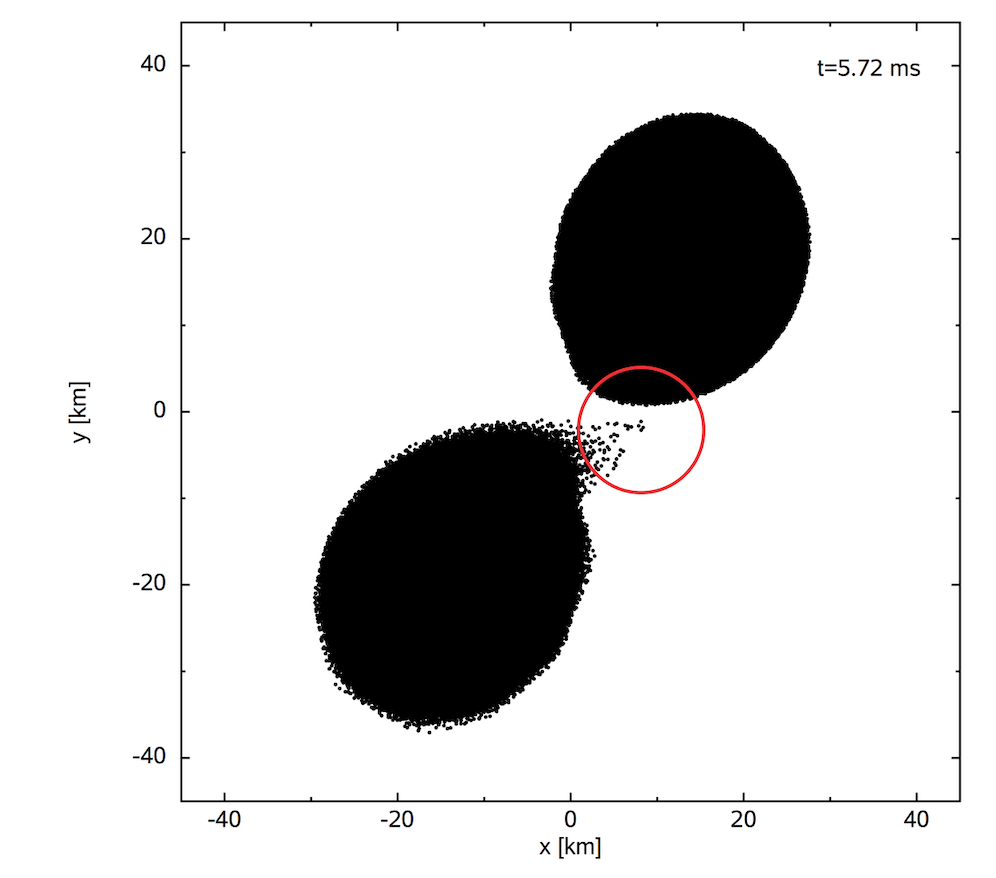} 
   \caption{Merger between a 1.3 and a 1.4 \msun merger of two neutron stars. 
   Image reproduced with permission from \citet{rosswog20a}, copyright by the author.
}
   \label{fig:h_adaption_NSM}
 \end{figure}
 %

Yet another way to update the smoothing lengths is motivated by
more practical considerations. If high-order kernels are used, the
SPH-summations should run over a large (typically $>100$) number of
neighbors, see Appendix~\ref{sec:kernels}. If now an expanding gas
stream encounters a sharp surface, like in the situation sketched in
Fig.~\ref{fig:h_adaption_NSM} from \citet{rosswog20a}, a $\rho_a-h_a$
iteration may temporarily require prohibitively long neighbor lists.
For that reason one may want to adjust the smoothing lengths
so that each particle has \emph{exactly} the desired number of
neighbors \citep{read10,rosswog20a}. While this may appear to be
prohibitively computationally expensive, this approach works at an 
acceptable speed in the \Ma code \citep{rosswog20a}: first a tree walk
is performed with smoothing lengths increased by 10\% to find a
generous list of potential neighbor particles. From this candidate list, the distance to
the $N_\mathrm{nei}+1$th closest particle is determined via a fast
partitioning algorithm \citep{press92}, and this distance is chosen as
the support size of the kernel. In this way, each particle has
exactly $N_\mathrm{nei}$ neighbors with non-zero contribution and throughout the
simulation smoothing lengths change very gently and produces very
little noise compared to smoothing length updates where the number of
contributing neighbors is kept only approximately constant.

\subsection{SPH derived from a Lagrangian}
\label{sec:SPH_from_Lagrangian}
As we have seen in Sect.~\ref{sec:vanilla_ice}, one can enforce exact
conservation in SPH by insightful choices in the discretization
process. There is, however, a more elegant road to fully conservative
equations by starting from a (discretized) Lagrangian and by applying
a variational principle
\citep{gingold82,speith98,monaghan01,springel02,rosswog09b,price12a}. Apart
from avoiding ``by hand choices'',  in this approach the particle
system also inherits much of the geometric structure of the continuum
system, e.g. Liouville's theorem and Poincare invariants, and
additional physics can be consistently incorporated in the evolution
equations by starting from a modified Lagrangian, see, for example,
\citet{price07a} for a consistent inclusion of self-gravity.

If we just consider ideal fluid dynamics without additional physics, a
Lagrangian is given by \citep{eckart60,salmon88,morrison98} 
\be
L= \int \rho \left(\frac{v^2}{2} - u(\rho,s) \right) dV,
\ee
where $s$ is the specific entropy and the discrete SPH-form reads
\be
L= \sum_b m_b \left( \frac{v_b^2}{2} - u_b \right)
\label{eq:Newtonian_Lagrangian}.
\ee
By applying the Euler-Lagrange equations
\be
\frac{d}{dt} \frac{\p L}{\p \vec{v}_a} - \frac{\p L}{\p \vec{r}_a}= 0
\ee
and keeping the particle masses fixed, one finds the momentum equation
of a particle $a$. Here $\p L/\p \vec{v}_a$ is  
the canonical momentum, which for Eq.~(\ref{eq:Newtonian_Lagrangian})
simply becomes $m_a \vec{v}_a$. By applying the adiabatic first law of thermodynamics,
$du= -P d(1/\rho)$, one finds
\be
\left(\frac{\p u}{\p \rho}\right)_s= \frac{P}{\rho^2}
\ee 
and 
\bea
\hspace*{-0.5cm}\frac{d \vec{v}_a}{dt}&=& \frac{1}{m_a}  \frac{\p
L}{\p \vec{r}_a}=  - \frac{1}{m_a} \sum_b m_b \frac{\p u_b}{\p \rho_b}
\frac{\p \rho_b}{\p \vec{r}_a} = - \frac{1}{m_a}\sum_b m_b
\frac{P_b}{\rho_b^2} \left[\frac{1}{\Omega_b} \sum_k m_k \nabla_a W_{h_b}(r_{bk})\right]
\nonumber  \\
&=& - \sum_b m_b \left( \frac{P_a}{\Omega_a \rho_a^2} \nabla_a
W_{h_a}(r_{ab}) + \frac{P_b}{\Omega_b \rho_b^2} \nabla_a W_{h_b}(r_{ab}) \right),
\label{eq:dvdt_NSPH}
\eea
where we have used $\nabla_a W_{bk}= \nabla_b W_{kb} (\delta_{ba}
-\delta_{ka})$, see Eq.~(\ref{eq:k3}) in Appendix \ref{sec:kernels}. 
For the energy equation, a straight forward insertion of
Eq.~(\ref{eq:drhodt_gradh})
into Eq.~(\ref{eq:dudt}) (without the heating/cooling term) yields
\be
\frac{du_a}{dt}= \frac{P_a}{\Omega_a \rho_a^2} \sum_b m_b \vec{v}_{ab}\cdot \nabla_a W_{h_a}(r_{ab}), 
\label{eq:dudt_NSPH}
\ee
and with Eqs.~(\ref{eq:rho_sum}), (\ref{eq:dvdt_NSPH}) and (\ref{eq:dudt_NSPH}) we now
have a complete set of discrete hydrodynamics equations.

\subsection{Alternative SPH formulations}
\label{sec:Alternative_SPH}
\subsubsection{Insightful, conservative ``ad-hoc'' choices}
A large family of alternative formulations of SPH may be found by
realizing  that the continuity equation can be written as
\citep{price04b}
\be
\frac{d\rho}{dt}= \psi \left[ \vec{v}\cdot \nabla \left(
\frac{\rho}{\psi}\right) - \nabla \cdot \left( \frac{\rho \vec{v}}{\psi}\right)\right],
\ee
where $\psi$ is an arbitrary scalar function defined on the particle field,
which can, by straight forwardly applying Eq.~(\ref{eq:std_grad}), be
translated into
\be
\frac{d\rho_a}{dt}=  \sum_b m_b \frac{\psi_a}{\psi_b} \vec{v}_{ab} \cdot
\nabla_a W_h(r_{ab}).
\ee
A consistent momentum equation is
\be
\frac{d\vec{v}_a}{dt}= - \sum_b m_b \left( \frac{P_a}{\rho_a^2} \frac{\psi_a}{\psi_b} +
\frac{P_b}{\rho_b^2} \frac{\psi_b}{\psi_a}  \right) \nabla_a W_h(r_{ab})
\ee
and the corresponding equations for the specific internal energy and
the thermokinetic energy read
\be
\frac{du_a}{dt}= \frac{P_a}{\rho_a^2} \sum_b m_b \frac{\psi_a}{\psi_b}
\vec{v}_{ab} \cdot \nabla_a W_h(r_{ab})
\ee
and
\be
\frac{d \tilde{e}_a}{dt}= - \sum_b \left( \frac{\psi_a}{\psi_b}\frac{P_a
\vec{v}_b}{\rho_a^2} +  \frac{\psi_b}{\psi_a} \frac{P_b
\vec{v}_a}{\rho_b^2} \right) \nabla_a W_h (r_{ab}),
\ee
respectively.

For the specific choice $\psi= \rho^{2 - \sigma}$ one recovers a generalization
earlier suggested by \citet{monaghan92}
\bea
\frac{d\rho_a}{dt}&=& \rho_a^{2-\sigma} \sum_b
\frac{\vec{v}_{ab}}{\rho_b^{2-\sigma}}  \cdot \nabla_a
                                                                  W_{h}(r_{ab})\\
\frac{d\vec{v}_a}{dt}&=& - \sum_b m_b \left[ \frac{P_a}{\rho_a^\sigma \rho_b^{2-\sigma}} + 
                                                                  \frac{P_b}{\rho_a^{2-\sigma}
                                                                  \rho_b^{\sigma}}
                                                                  \right]
                                                                  \nabla_a
                                                                  W_{h}(r_{ab})\\
\frac{du_a}{dt}&=& \frac{P_a}{\rho_a^\sigma} \sum_b m_b
\frac{\vec{v}_{ab}}{\rho_b^{2-\sigma}} \nabla_a W_{h}(r_{ab}),                                                                
\eea
where $\sigma$ is a free parameter.

For the choice $\psi=1$ we recover the vanilla ice SPH formulation of
Eqs.~(\ref{eq:drho_dt}), (\ref{eq:dvdt_van}) and
(\ref{eq:dudt_van}). If one instead chooses $\psi= \rho$, one obtains
the sometimes called ``Geometric Density Average Force (GDF)''
formulation of SPH
\bea
\frac{d\rho_a}{dt}    &=& \sum_b m_b \frac{\rho_a}{\rho_b}
\vec{v}_{ab} \cdot \nabla_a W_{h}(r_{ab})\label{eq:dens_GDF}\\
\frac{d\vec{v}_a}{dt}&=& - \sum_b m_b  \left( \frac{P_a + P_b}{\rho_a
    \rho_b} \right) \nabla_a W_{h}(r_{ab}) \label{eq:mom_GDF}\\
\frac{d u_a}{dt}        &=& \sum_b m_b \left(\frac{P_a}{\rho_a \rho_b}\right)
\vec{v}_{ab} \cdot \nabla_a W_{h}(r_{ab})\label{eq:en_GDF}.
\eea
Several studies found such GDF formulations advantageous
compared to the ``vanilla ice'' SPH formulation, both for hydrodynamics
\citep{oger07,read10,wadsley17,rosswog20a,rosswog26a} and 
magnetohydrodynamics \citep{wissing20,wissing22}.
For the choice $\psi= \rho/\sqrt{P}$ one finds
\bea
\frac{d\rho_a}{dt}    &=& \sum_b m_b \left(\frac{\rho_a}{\rho_b} \sqrt{\frac{P_b}{P_a}}
\right) \vec{v}_{ab} \cdot \nabla_a W_{h}(r_{ab})\\
\frac{d\vec{v}_a}{dt}&=& - \sum_b m_b  \left( 2\frac{\sqrt{P_a  P_b}}{\rho_a
    \rho_b} \right) \nabla_a W_{h}(r_{ab}) \\
\frac{d u_a}{dt}        &=& \sum_b m_b \left(\frac{\sqrt{P_a P_b}}{\rho_a \rho_b}\right)
\vec{v}_{ab} \cdot \nabla_a W_{h}(r_{ab}),
\eea
which is the SPH formulation used in \citet{hernquist89}.

\subsubsection{Formulations based on generalized volume elements}
\label{sec:volume_elements}
\begin{figure}[ht]
   \centering
   \includegraphics[width=10cm]{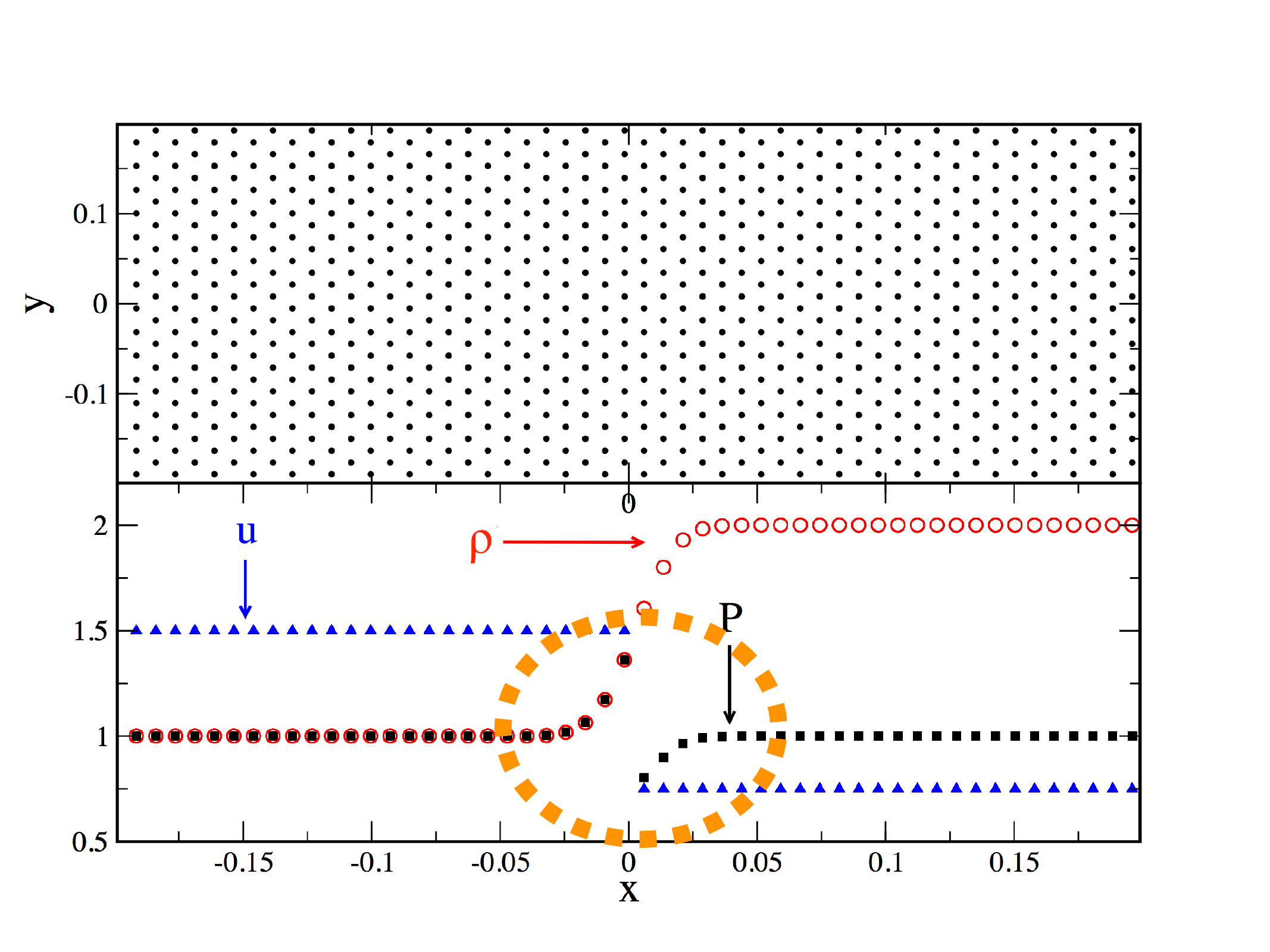} 
   \caption{The density summation introduces some smoothing. If
     the internal energy is set up as a sharp transition, this leads
     to
     ``pressure blips'' that can introduce surface tension effects.
     Figure adapted from \citet{rosswog15b}.
}
   \label{fig:contact_discontinuity}
 \end{figure}

The \enquote{standard} choice of volume element,  $V_b= m_b/\rho_b$,  can
introduce spurious surface tension forces 
near contact discontinuities \citep{saitoh13}. 
For a polytropic equation of state, $P= (\Gamma-1) u \rho$, the
product of density and  internal energy must be the same on both sides
to ensure a single value of  $P$ at the discontinuity, i.e. $\rho_1
u_1= \rho_2 u_2$, where the subscripts label the two sides of the
discontinuity.  So the jumps in $u$ and $\rho$ need to be consistent
with each other,  otherwise the mismatch results in a ``pressure
blip'', see Fig.~\ref{fig:contact_discontinuity}, which can cause
spurious forces that have an effect like a surface tension and can
suppress weakly triggered fluid instabilities, see for example,
\citet{agertz07,springel10a,read10}. In the ``standard'' SPH formulation
such a mismatch does occur because the density estimate is smooth, but
the internal energy enters the SPH equations as an un-smoothed quantity.

%
 %
 
 The following measures have been suggested to alleviate the problem
 \begin{itemize}
   \item apply thermal conductivity, see Sec.~\ref{sec:AV}, to smooth the internal energy
            transition \citep{price08a}  which has been shown to improve Kelvin-Helmholtz instabilities and also helps in some shock problems
            \citep{noh87,rosswog07c}. 
    \item use good kernels with appropriately large neighbour numbers, see Appendix \ref{sec:kernels}, to have good interpolation quality, see Sec.~\ref{sec:discrete_approx}
    \item use the GDF-symmetry rather than the "vanilla ice SPH" form, see Sec.~\ref{sec:Alternative_SPH}
    \item use reproducing kernels, see Sec.~\ref{sec:RPK}, which produce by construction a consistent interpolation across contact discontinuities and
    \item apply different volume elements, as we describe in the following.
  \end{itemize}
            
In \citet{saitoh13} it was pointed out
that SPH formulations that do not  include density explicitly in the
equations of motion do avoid the pressure becoming multi-valued  
at contact discontinuities. Since the density usually enters the
equation of motion via the  choice of the volume element, a different
choice can possibly avoid the problem  altogether. This observation is
consistent with the findings of \citet{hess10} who used a  
particle hydrodynamics method, but calculated the volumes via a
Voronoi tessellation rather than via smooth density sum. Closer to the
original SPH spirit is the class of  kernel-based particle volume
estimates that have been suggested \citep{hopkins13}  
as a generalization of the approach from \citet{saitoh13} .

A straight forward estimate of the volume element at particle
position $b$ is to use the inverse of the local SPH particle number
density
\be
V_b= \frac{1}{\sum_k W_{h_b}(\vr_{kb})}.
\ee
While this approach is straight forward, it can be generalized
by weighting the kernel with a scalar property $X$  \citep{hopkins13},
so that the volume becomes 
\be
V_b^{(X)}= \frac{X_b}{\sum_k X_k W_{h_b}(\vr_{kb})} \equiv \frac{X_b}{\kappa_{X, b}} 
\label{eq:gen_vol_element}
\ee
and one can calculate the density via
\be
\rho_b^{(X)}= \frac{m_b}{V_b^{(X)}}= \frac{m_b}{X_b} \kappa_{X, b}.
\label{eq:density}
\ee
Then, the smoothing length update reads
\be
h_b= \eta \left(V_b^{(X)}\right)^{1/d},
\ee
where  $d$ is the number of spatial dimensions. The derivatives of the  
quantity $\kappa_{X,b}$ become\footnote{The derivation follows closely
  the steps explained in detail  in Sect.~3 of \citet{rosswog09b}.}
\be
\nabla_a \kappa_{X,b}= \frac{1}{\Omega_b} \sum_k X_k \nabla_a W_{h_b}(\vr_{bk})
\quad \text{and} \quad 
\frac{d \kappa_{X,b}}{dt}= \frac{1}{\Omega_b} \sum_k X_k \vec{v}_{bk} \cdot \nabla_b W_{h_b}(\vr_{bk}), 
\label{eq:kappa_derivs}
\ee
with the generalized ``grad-h terms'' being
\be
\Omega_b= 1 - \frac{m_b}{X_b} \frac{\p h_b}{\p \rho_b} \sum_k X_k \frac{\p W_{h_b}(\vr_{kb})}{\p h_b}.
\ee
The derivatives of the volume elements then become
\be
\nabla_a V_b
= -\frac{V_b^2}{X_b \Omega_b} 
\sum_k X_k \nabla_a W_{h_b}(\vr_{bk})
\label{eq:nabla_V}
\ee
and 
\be
\frac{d V_b}{dt} 
= - \frac{V_b^2}{X_b \Omega_b} \sum_k X_k \vec{v}_{bk} \cdot \nabla_b W_{h_b}(\vr_{bk})
\label{eq:ddt_V}
\ee
and the SPH equations for the general volume element of Eq.~(\ref{eq:gen_vol_element})
read
\bea
V_b                   &=& \frac{X_b}{\kappa_{X,b}}\label{eq:gen_vol_N}\\
m_a \frac{d\vec{v}_a}{dt} &=& - \sum_{b} X_a X_b \left\{  \frac{P_a}{\kappa_a^2 \Omega_a} \nabla_a W_{h_a}(\vr_{ab}) 
+ \frac{P_b}{\kappa_b^2 \Omega_b} \nabla_a W_{h_b}(\vr_{ab})  \right\}\label{eq:gen_mom_N} \\
m_a \frac{du_a}{dt}       &=& \frac{P_a X_a}{\kappa_a^2\Omega_a} \sum_b X_b \vec{v}_{ab} \cdot \nabla_a W_{h_a}(\vr_{ab}).\label{eq:gen_en_N}
\eea
For the choice $X=m$ one recovers the commonly used equation set
\citep{monaghan01,price04c,rosswog07c}. Explicit forms of the equations, also for other choices of $X$, 
are provided  in Table~\ref{tab:explicit_forms_Newt_SPH}.

\begin{table}[htp]
\caption{Explicit forms of the equations for special choices of the
  weight $X$ for Newtonian and special-relativistic SPH. In the
  relativistic case we have to specify whether a density is measured
in the fluid rest frame or a chosen \enquote{computing frame},
abbreviated as CF.}
\label{tab:explicit_forms_Newt_SPH}
\centering
{\small


\begin{tabular}{ l l}
\multicolumn{2}{c}{\textbf{\underline{Newtonian SPH}}}\\
\addlinespace[2ex]
\multicolumn{2}{c}{\textbf{Weight $X= m$}}\\
\midrule
\\
 density      & $\rho_a \; \,= \sum_b m_b W_{ab}(h_a)$ \\
momentum      & $\frac{d\vec{v}_a}{dt}= - \sum_b m_b \left\{ \frac{P_a}{\Omega_a \rho_a^2} \nabla_a W_{ab}(h_a) 
                + \frac{P_b}{\Omega_b \rho_b^2} \nabla_a W_{ab}(h_b) \right\}$   \\
energy        & $\frac{du_a}{dt}=  \frac{P_a}{\Omega_a \rho_a^2} \sum_b m_b \vec{v}_{ab} \cdot \nabla_a W_{ab}(h_a)$
\end{tabular}

\vspace*{0.5cm}

\begin{tabular}{ l l}

\multicolumn{2}{c}{\textbf{Weight $X=1$}}\\
\midrule
\\
 density      & $\rho_a \; \,= m_a \sum_b W_{ab}(h_a)$ \\
momentum      & $\frac{d\vec{v}_a}{dt} = - 
                                   \sum_b m_b \left\{ \frac{m_a}{m_b} \frac{P_a}{\Omega_a \rho_a^2} \nabla_a W_{ab}(h_a) 
                                   + \frac{m_b}{m_a} \frac{P_b}{\Omega_b \rho_b^2}   \nabla_a W_{ab}(h_b) \right\}$   \\
energy        & $\frac{du_a}{dt} = \frac{P_a m_a}{\Omega_a \rho_a^2} \sum_b \vec{v}_{ab} 
                \cdot \nabla_a W_{ab}(h_a)$
\end{tabular}

\vspace*{0.5cm}

\begin{tabular}{ l l}
\multicolumn{2}{c}{\textbf{Weight $X= P^k$}}\\
\midrule
\\
 density      & $\rho_a \; \,= m_a \sum_b \left( \frac{P_b}{P_a} \right)^k W_{ab}(h_a)$ \\
momentum      & $\frac{d\vec{v}_a}{dt}= - \sum_b m_b \left\{ \frac{m_a}{m_b}\frac{P_a^{1-k} P_b^k }{\Omega_a^2 \rho_a^2} \nabla_a W_{ab}(h_a) 
                + \frac{m_b}{m_a} \frac{P_a^k P_b^{1-k}}{\Omega_b \rho_b^2} \nabla_a W_{ab}(h_b) \right\}$   \\
energy        & $\frac{du_a}{dt}=  \frac{P_a m_a}{\Omega_a \rho_a^2} \sum_b \left( \frac{P_b}{P_a} \right)^k \vec{v}_{ab} \cdot \nabla_a W_{ab}(h_a)$
\end{tabular}




\begin{tabular}{ l l}
\addlinespace[6ex]
\multicolumn{2}{c}{\textbf{\underline{Special-relativistic SPH}}}\\
\addlinespace[2ex]
\multicolumn{2}{c}{\textbf{Weight $X= \nu$}}\\
\midrule
\\
CF bar. num. density       & $N_a \; \,= \sum_b \nu_b W_{ab}(h_a)$ \\
spec. canon. momentum & $\frac{d\vec{S}_a}{dt}= - \sum_b \nu_b 
                                          \left\{ \frac{P_a}{\Omega_a N_a^2} \nabla_a W_{ab}(h_a) 
                                        + \frac{P_b}{\Omega_b N_b^2} \nabla_a W_{ab}(h_b) \right\}$   \\
spec. canon. energy        &  $\frac{d \epsilon_a}{dt}= - \sum_b \nu_b
                                           \left\{ \frac{P_a }{\tilde{\Omega}_a N_a^2}  \vec{v}_b \cdot 
                                           \nabla_a W_{ab}(h_a)
                                        + \frac{P_b }{\tilde{\Omega}_b}   \vec{v}_a \cdot 
                                            \nabla_a W_{ab}(h_b)\right\}$ 
\end{tabular}

\vspace*{0.5cm}

\begin{tabular}{ l l}
\multicolumn{2}{c}{\textbf{Weight $X=1$}}\\
\midrule
\\
CF bar. num. density            & $N_a \; \,= \nu_a \sum_b W_{ab}(h_a)$ \\
spec. canon. momentum      & $\frac{d\vec{S}_a}{dt} = - 
                                   \sum_b \nu_b \left\{ \frac{\nu_a}{\nu_b} \frac{P_a}{\Omega_a N_a^2} \nabla_a W_{ab}(h_a) 
                                   + \frac{\nu_b}{\nu_a} \frac{P_b}{\Omega_b N_b^2}   \nabla_a W_{ab}(h_b) \right\}$   \\
spec. canon. energy             & $\frac{d \epsilon_a}{dt}= - \sum_b \nu_b
                                           \left\{ \frac{\nu_a}{\nu_b} \frac{P_a}{\tilde{\Omega}_a N_a^2}  \vec{v}_b \cdot 
                                           \nabla_a W_{ab}(h_a)
                                        +\frac{\nu_b}{\nu_a}  \frac{P_b}{\tilde{\Omega}_b}   \vec{v}_a \cdot 
                                            \nabla_a W_{ab}(h_b)\right\}$ 
\end{tabular}

\vspace*{0.5cm}

\begin{tabular}{ l l}
\multicolumn{2}{c}{\textbf{Weight $X= P^k$}}\\
\midrule
\\
CF bar. num. density       & $N_a \; \,= \nu_a \sum_b \left( \frac{P_b}{P_a} \right)^k W_{ab}(h_a)$ \\
spec. canon. momentum      & $\frac{d\vec{S}_a}{dt}= - \sum_b \nu_b \left\{ \frac{\nu_a}{\nu_b}\frac{P_a^{1-k}P_b^k}{\Omega_a^2 N_a^2} \nabla_a W_{ab}(h_a) 
                + \frac{\nu_b}{\nu_a} \frac{P_a^k P_b^{1-k}}{\Omega_b N_b^2} \nabla_a W_{ab}(h_b) \right\}$   \\
spec. canon.  energy        &  $\frac{d \epsilon_a}{dt}= - \sum_b \nu_b
                                           \left\{ \frac{\nu_a}{\nu_b} \frac{P_a^{1-k} P_b^{k}}{\tilde{\Omega}_a N_a^2}  \vec{v}_b \cdot 
                                           \nabla_a W_{ab}(h_a)
                                        +\frac{\nu_b}{\nu_a}  \frac{P_a^k P_b^{1-k}}{\tilde{\Omega}_b}  \vec{v}_a \cdot 
                                            \nabla_a W_{ab}(h_b)\right\}$ 
\end{tabular}}
\end{table}

\subsection{More meshless derivatives}
\label{sec:accurate_gradients}
We focus here mostly on first-order derivatives as they are needed for
basic SPH formulations. Higher order derivatives (e.g. for modelling
physical viscosity) can be straight forwardly be calculated with the
LRE approach described in Sect.~\ref{sec:LRE}. For more on 
higher-order derivatives we refer the interested reader to the
literature \citep{espagnol03,monaghan05,price12a}.  There are  
various ways to calculate gradients at particle positions beyond the
most straight-forward Eqs.~(\ref{eq:std_grad}) and
(\ref{eq:better_grad}). One concern is the gradient accuracy, but one
of SPH's most salient features is its excellent numerical
conservation, which is, in the standard formulation, due to the
anti-symmetry of kernel gradients with respect to the exchange 
of particle indices, see Sect.~\ref{sec:conservation} and
Eq.~(\ref{eq:k4}), and due to the SPH kernel gradient being radial,
see Eq.~(\ref{eq:k4}). An accurate gradient estimate 
without built-in conservation can be less useful in practice than a
less accurate estimate that ensures Nature's conservation laws, see
Sect.~5 in \citet{price12a} for a striking example of how a seemingly
good  gradient without built-in conservation can lead to pathological
particle distributions. The challenge is to  combine exact (or at
least very good) conservation with an accurate gradient estimate.

As discussed in Sect.~\ref{sec:conservation}, the anti-symmetry with
respect to exchanging particle labels $a \leftrightarrow b$ is
crucial for conservation, but even without this anti-symmetry accurate
gradient estimates can be very useful, e.g. in the reconstruction
of a Riemann problem or for limiters to suppress unnecessary
dissipation, see Sect.~\ref{sec:shocks}.

\subsubsection{Linear exact gradients}
Exact gradients of linear functions can be constructed in the
following way \citep{price04c}. Start with the RHS of
Eq.~(\ref{eq:std_grad})  at $\vec{r}_a$ and insert the Taylor
expansion of $f_b$ around $\vec{r}_a$ like in Eq.~(\ref{eq:Taylor_fb})
\be
\sum_b V_b \;  f_b \nabla_a W_{h_a}(\vec{r}_{ab})= \sum_b V_b \;  \left\{ f_a +
  (\nabla f)_a \cdot (\vec{r}_{b}-\vec{r}_{a}) + ... \right\} \nabla_a W_{h_a}(\vec{r}_{ab})
    \nabla_a, 
\ee
re-arrange into (summation over $k$)
\be
\sum_b V_b \;  (f_b - f_a) \nabla^i_a W_{h_a}(\vec{r}_{ab})= \nabla^k f_a \; M_a^{ki}
\quad {\rm with} \quad M_a^{ki}= \sum_b V_b (\vec{r}_b - \vec{r}_a)^k \nabla^i_a W_{h_a}(\vec{r}_{ab}).
\ee
Solving this equation for the gradient yields
\be
\nabla^i f_a^{(2)}= C_{(2),a}^{ik} \sum_b V_b (f_b-f_a)
\nabla_a^k W_{h_a}(\vec{r}_{ab})= - \sum_b V_b f_{ab} \widetilde{\nabla^i_a W_{ab}},
\label{eq:lin_exact_gradient}
\ee
where the ``corrected kernel gradient'' reads
\be
\widetilde{\nabla^i_a W_{ab}}=  C_{(2),a}^{ik}  \; \nabla_a^k W_{h_a}(\vec{r}_{ab}).
\ee
Here we have introduced the ``correction matrix'' $C_{(2)}^{ik}$ which
is the inverse of the matrix $M^{ik}$. Several of the subsequent
meshless gradient prescriptions involve such correction matrices,
therefore we add a label $(n)$ indicating that the correction matrix
belongs to gradient prescription $\nabla f^{(n)}$.
Note that the sum is the same as in the gradient
estimate Eq.~(\ref{eq:better_grad}), but corrected by the matrix
${\bf C}$ which accounts for the local particle distribution. Obviously, this
comes at the price of inverting a $d \times d$ matrix in $d$ dimensions,
but since the inversion can be calculated analytically, this is not
a major computational burden.

Comparing $\nabla f^{(2)}$, Eq.~(\ref{eq:lin_exact_gradient}), with
$\nabla f^{(1)}$, Eq.~(\ref{eq:better_grad}) suggests the
correspondence
\be
\nabla^i W_h(\vr -\vr_b) \leftrightarrow  C^{ik}_{(2)}(\vr) \; \nabla^k
W_h(\vr-\vr_b)\equiv G_{(2),b}^i(\vr,h(\vr)).
\ee

An SPH scheme that is based on these \enquote{corrected gradients} can be constructed, see
\citet{rosswog26a} for more details, by using
\be
\nabla^i_a \bar{W}_{ab} \rightarrow \left(\widetilde{\nabla W}\right)^i_{ab} \equiv 
\frac{C_{(2),a}^{ik} \nabla_a^k W_{h_a}(r_{ab}) + C_{(2),b}^{ik} \nabla_a^k W_{h_b}(r_{ab})}{2},
\ee
so that energy and momentum equation read
\bea
\left(\frac{d\vec{v}}{dt}\right)_a &=&  - \sum_b m_b \left(\frac{P_a + P_b}{\rho_a \rho_b} \right) \left(\widetilde{\nabla W}\right)_{ab}\\
\left(\frac{du}{dt}\right)_a &=&\sum_b m_b \left(\frac{P_a}{\rho_a \rho_b} \right) \vec{v}_{ab} \cdot  \left(\widetilde{\nabla W}\right)_{ab} .
\label{eq:V2_set}
\eea
This corrected-gradient scheme achieves very good accuracy in benchmark tests \citep{rosswog26a}.

Since the above gradient estimate requires the knowledge of the
volume/density, one needs to have performed at least one previous loop
over the particle distribution, if the density prescription
Eq.~(\ref{eq:rho_sum}) is used.  If this is a concern, one may apply a
slight variant \citep{price04c,rosswog07c}. Start from a linear
Taylor-expansion of $f_b$  like Eq.~(\ref{eq:Taylor_fb}) and insert it in
\be
\sum_b m_b \left\{f_b - f_a\right\} \nabla_a^k W_{h_a}(r_{ab}) \approx \sum_b m_b
\left\{ \nabla_a f \cdot (\vr_b - \vr_a)\right\}  \nabla_a^k W_{h_a}(r_{ab}) 
\ee
to find
\begin{align}
\nabla^i f_a^{(3)}=&  C_{(3)}^{ik}\sum_b m_b \; (f_b - f_a) \; 
\nabla_a^k W_{h_a}(r_{ab})  \nonumber\\ \text{with} \quad
C_{(3)}^{ik}=& \left(\sum_b m_b
  (\vec{r}_{ba})^i \; \nabla_a^k W_{h_a}(r_{ab}) \right)^{-1}.
\label{eq:mod_lin_exact_gradient}
\end{align}
While this is very similar to Eq.~(\ref{eq:lin_exact_gradient}), this
gradient estimate only involves masses in the sums (instead of
densities/volumes) and can therefore be conveniently calculated
alongside the density loop.

\subsubsection{Integral-based approximation}
Equation~(\ref{eq:lin_exact_gradient}) is equivalent to an approach
that was independently derived by discretizing an integral in
\citet{garcia_senz12,cabezon12a}. To see this, start from
\be
\nabla_a W_{h_a}(\vec{r}_{ab})= \frac{\p}{\p \vr_a}W\left(\frac{|\vr_a
    - \vr_b|}{h_a}\right)= \frac{\p W}{\p u} \frac{\p u}{\p \vr_a}=
\frac{\p W}{\p u}  \frac{\vr_a -\vr_b}{h_a|\vr_a - \vr_b|},
\ee
where $u\equiv |\vr_a - \vr_b|/h_a$. Since $\p W/\p u <0$, we can
write this as
\be
\nabla_a W_{h_a}(\vec{r}_{ab})= - \frac{\p W}{\p u} \frac{\vr_b
  -\vr_a}{h_a|\vr_a - \vr_b|} \equiv (\vr_b -\vr_a)
\tilde{W}_{h_a}(\vr_{ab}),
\label{eq:gradW_tildeW}
\ee
where $\tilde{W}$ is another valid, positive definitive kernel with finite
support. We can now re-write Eq.~(\ref{eq:lin_exact_gradient})
\be
\nabla_a^i f^{(2)}= \left( \sum_l V_l (\vr_b
  -\vr_a)^i \nabla_a^k W_{h_a}(
\vr_{ab})\right)^{-1} \left\{ \sum_b V_b (f_b - f_a) \nabla_a^k W_{h_a}(\vr_{ab})\right\}
\ee
and by inserting Eq.~(\ref{eq:gradW_tildeW}) we find
\begin{align}
\nabla_a^i f^{(4)}= & \left( \sum_l V_l (\vr_b
  -\vr_a)^i (\vr_b - \vr_a)^k \tilde{W}_{h_a}(
  \vr_{ab})\right)^{-1} \nonumber\\
  & \left\{ \sum_b V_b (f_b - f_a) (\vr_b -
  \vr_a)^k \tilde{W}_{h_a}(\vr_{ab})\right\},
\label{eq:der_fIA}
\end{align}
which is (apart from $\tilde{W}$ instead of $W$) the gradient expression found
in \citet{garcia_senz12,cabezon12a} {bf in another way}.
The ``correction matrix'' 
\be
C_{(4)}^{ik}(\vr)= \left(\sum_b   (\vr - \vr_b)^i (\vr - \vr_b)^k V_b W_h(\vr-\vr_b)\right)^{-1}
\ee
is, of course, the discrete moment matrix from Eq.~(\ref{eq:first_moment})
  in Sect.~\ref{sec:kernel_interpolation}.
The comparison between Eqs.~(\ref{eq:der_fIA}) and
(\ref{eq:better_grad}) suggests the correspondence
\be
\nabla^i W_h(\vr -\vr_b) \leftrightarrow  C^{ik}_{(4)}(\vr) \; (\vr_b - \vr)^k
\; W_h(\vr-\vr_b)\equiv G_{(3),b}^i(\vr,h(\vr))
\ee
If we write the gradient at $\vr=\vr_a$
\begin{align}
\nabla^i f_a  =& - \sum_b V_b \;
f_{ab} \;  G^i_{(3),b}(\vr_a,h_a) =
- f_a C_{(4),a}^{ik}\sum_b  \; (\vr_{ab})^k V_b W(\vr_{ab}) \nonumber\\
&+ \sum_b V_b \;
f_b \;  G^i_{(3),b}(\vr_a,h_a)\\
\nabla^i f_a ^{(5)} \approx& \sum_b V_b \; f_b \;  G^i_{(3),b}(\vr_a,h_a), \label{eq:der_IA}
\end{align}
where we have assumed that the first moment, 
Eq.~(\ref{eq:first_moment}), vanishes to a good approximation and we have therefore
neglected the first term. Whether this is a good approximation or not, is
difficult to know a priori, but very extensive tests with the
gradient from Eq.~(\ref{eq:der_IA}) have shown substantially improved results
compared to the standard SPH gradient while achieving the same level
of numerical conservation \citep{rosswog15b,rosswog20a}.

\subsubsection{Error-minimizing gradients}
Yet another meshless gradient estimate can be obtained by starting
from a linear approximation to a function $f$ attached at particle $a$ 
\be
\tilde{f}_a(\vr)= f_a + (\nabla f)_a \cdot (\vr - \vr_a)
\ee
and determining the gradient $(\nabla f)_a$ so that $\tilde{f}_a$ is
the best representation of the surrounding particles. In other words,
one searches the gradient that minimizes, at the position of particle
$a$, the functional  
\be
E_a \equiv \sum_b \Psi_b(\vr_a) \left\{ \tilde{f}_a(\vr_b)  - f_b
\right\}^2= \sum_b \Psi_b(\vr_a) \left\{ f_{ab} + \nabla^k f_a
  (\vr_a - \vr_b)^k \right\}^2,
\ee
where $\Psi_b$ is a suitable weight function that gives more weight to
nearby particles. Minimizing the error functional via 
\be
\frac{\partial E_a}{\partial (\nabla f)_a^j} \stackrel{!}{=} 0
\ee
results in
\be
\nabla^j f_a^{(6)} = C^{jk}_{(6),a} \sum_b f_{ab} \Psi_b(\vr_a) (\vr_a - \vr_b)^k 
\ee
with
\be
C^{jk}_{(6),a}= \sum_b \Psi_b(\vr_a) (\vr_a -\vr_b)^j  (\vr_a -\vr_b)^k. 
\ee
If we choose for the weight function
\be
\Psi_b(\vr_a)= V_b W_{h_a}(\vr_{ab}),
\ee
we recover exactly the linear-exact gradient approximation,
see Eqs.~(\ref{eq:lin_exact_gradient}) or (\ref{eq:der_fIA}). If one
uses a weight function $\Psi_b(\vr)$ that is an exact partition of
unity, one reproduces the meshless derivative of 
\citet{lanson08a} which is also frequently used in meshless finite
volume methods, see e.g. \citet{gaburov11,hopkins15a}.

If instead one chooses
\be
\Psi_b(\vr_a)= W_{h_a}(\vr_{ab}),
\ee
one obtains
\be
\nabla_a^i f^{(7)}= \left( \sum_l  (\vr_{ab})^i (\vr_{ab})^k W_{h_a}(
  \vr_{ab})\right)^{-1} \left\{ \sum_b f_{ab} (\vr_{ab})^k W_{h_a}(\vr_{ab})\right\}.
\label{eq:something}
\ee

\subsubsection{Linearly reproducing kernel gradients}
\label{sec:RPK}
One of the main criticisms of SPH is that the standard
SPH-approximation  neither exactly reproduces 
constant nor linear functions. This, however, can be enforced, by enhancing the kernel functions with
additional parameters $A$ and $B^i$,
\be
\mathcal{W}_{ab}(\vec{r}_{ab})= A_a \left[1 + B_a^i \; (\vec{r}_{ab})^i\right] \bar{W}_{ab},
\ee 
where $\bar{W}_{ab}$ is a symmetrized kernel function, e.g. 
$\bar{W}_{ab}= 0.5 \left[W(r_{ab},h_a )+ W(r_{ab},h_b)\right]$.
The parameters $A$ and $B^i$ are determined at every point (here labelled
$a$) so that $\mathcal{W}$ \emph{exactly reproduces the discrete first-order consistency relations}
\bea
\sum_b V_b \; \mathcal{W}_{ab} &=& 1 \nonumber\\
\sum_b (\vec{r}_{ab})^i \;  V_b \; \mathcal{W}_{ab} &=& 0,
\label{eq:consistency_relations}
\eea
see Eqs.~(\ref{eq:zeroth_moment}) and (\ref{eq:first_moment}).  The
price for this exact reproduction is that the kernel is no longer
guaranteed to be radial due to the $B^i$-term. This property is in
standard SPH responsible for exact angular momentum conservation.  It
is, however, possible to relatively straight-forwardly write a set of
equations that conserves energy, momentum and mass, if density
summation is used, but not necessarily angular momentum, since
the mutual accelerations  can no longer be guaranteed to be along
the line connecting two particles, see
Sect.~\ref{sec:conservation}. In practice, 
however, this does not have to be a major concern, \citet{frontiere17} find, in their SPH formulation based on
reproducing kernels, in typical tests violations of exact angular momentum
conservation on the sub-percent level.

The gradient of the kernel $\mathcal{W}$ with respect to position $\ra$ reads
\begin{align}
\p_k \mathcal{W}_{ab}\equiv (\nabla_a)^k  \mathcal{W}_{ab} =& A_a \;
B_a^k \; \bar{W}_{ab} + A_a \left(1+ B_a^i (\vec{r}_{ab})^i\right)
\nabla_a^k \bar{W}_{ab} + \nonumber \\ 
&
\left(  1+  B_a^i  (\vec{r}_{ab})^i \right)
\bar{W}_{ab} (\nabla_a^k A)_a + A_a \;
(\vec{r}_{ab})^i \; (\nabla_a^k B)_a^i \;
\bar{W}_{ab}.
\label{eq:RPK_kernel_gradient1}
\end{align}
Taking the gradient of $\mathcal{W}_{ba}$ with respect to $\vr_b$
results in
\begin{align}
\p_k \mathcal{W}_{ba}\equiv (\nabla_b)^k  \mathcal{W}_{ba} =& A_b
\; B_b^k \; \bar{W}_{ab} - A_b \left(1 - B_b^i (\vec{r}_{ab})^i\right) \nabla_a^k \bar{W}_{ab} + \nonumber \\
& \left( 1- B_b^i (\vec{r}_{ab})^i \right) \bar{W}_{ab} (\nabla_b^k A)_b - A_b \; (\vec{r}_{ab})^i \; (\nabla_b^k B)_b^i  \; \bar{W}_{ab} ,                                    \end{align}
where we have used $\bar{W}_{ab}= \bar{W}_{ba}$, $\vec{r}_{ba}=
-\vec{r}_{ab}$ and $\nabla_b^k \bar{W}_{ba}= - \nabla_a^k
\bar{W}_{ab}$.
The derivations that lead to the explicit expressions for
$A_a$, $B_a$ and their derivatives are straight-forward, but lengthy.
We provide the resulting expressions in Appendix \ref{sec:RPK_expressions}. 

With the linearly reproducing kernels at hand, we can now approximate a function $f$ via
\be
\tilde{f}(\ra)= \sum_b V_b \; f_b \; \mathcal{W}_{ab}
\ee
and its derivative via (dropping the tilde for ease of notation)
\be
\nabla^k f_a^{(8)}= \sum_b V_b \; F_b \; \p_k \mathcal{W}_{ab}.
\ee
While these expressions look very similar to standard SPH approximations, they 
exactly reproduce linear functions on a discrete level, which the standard SPH equations do not.

The gradient of $\bar{W}_{ab}$ can be written as
\be
\nabla_a \bar{W}_{ab}= \frac{1}{2} \left[ \nabla_a W_{h_a}(\vr_{ab}) + \nabla_a
  W_{h_b}(\vr_{ab}) \right] = \frac{1}{2} \left[ \nabla_a W_{h_a}(\vr_{ab}) - \nabla_b
  W_{h_b}(\vr_{ab}) \right], 
\ee
which suggests the replacement
\be
\nabla_a \bar{W}_{ab} \rightarrow (\nabla \mathcal{W})_{ab} \equiv
\frac{1}{2} \left[ \nabla_a \mathcal{W}_{ab} - \nabla_b
  \mathcal{W}_{ba}\right].
\label{eq:RPK_kernel_gradient}
\ee
Various particle hydrodynamics schemes based on reproducing
kernels have recently been developed, for example the artificial
viscosity-based  codes \texttt{CRKSPH} 
\citep{frontiere17} and \texttt{REMIX} \citep{sandnes24} or the
Riemann solver-based approach with reproducing kernels developed in \citet{rosswog25a}. 
All of them have shown excellent results. 

\subsubsection{Local regression estimate (LRE) derivatives}
\label{sec:LRE}
We want to summarize here a general method that we have used to
map particle properties to a mesh in the context of fully dynamical
general-relativistic SPH \citep{rosswog23a}, see
Sect.~\ref{sec:fullGR_SPH}. This method is somewhat different from
the previous ones, since it is not exclusively designed to obtain accurate
gradient approximations, but it also delivers the function value
that is consistent with the gradients, all of this for some specifiable point 
$\vec{R}$ that does not need to be a particle position. For higher polynomial 
orders it also delivers higher order derivatives.

\begin{figure}[ht]
\centering
 \includegraphics[width=13cm]{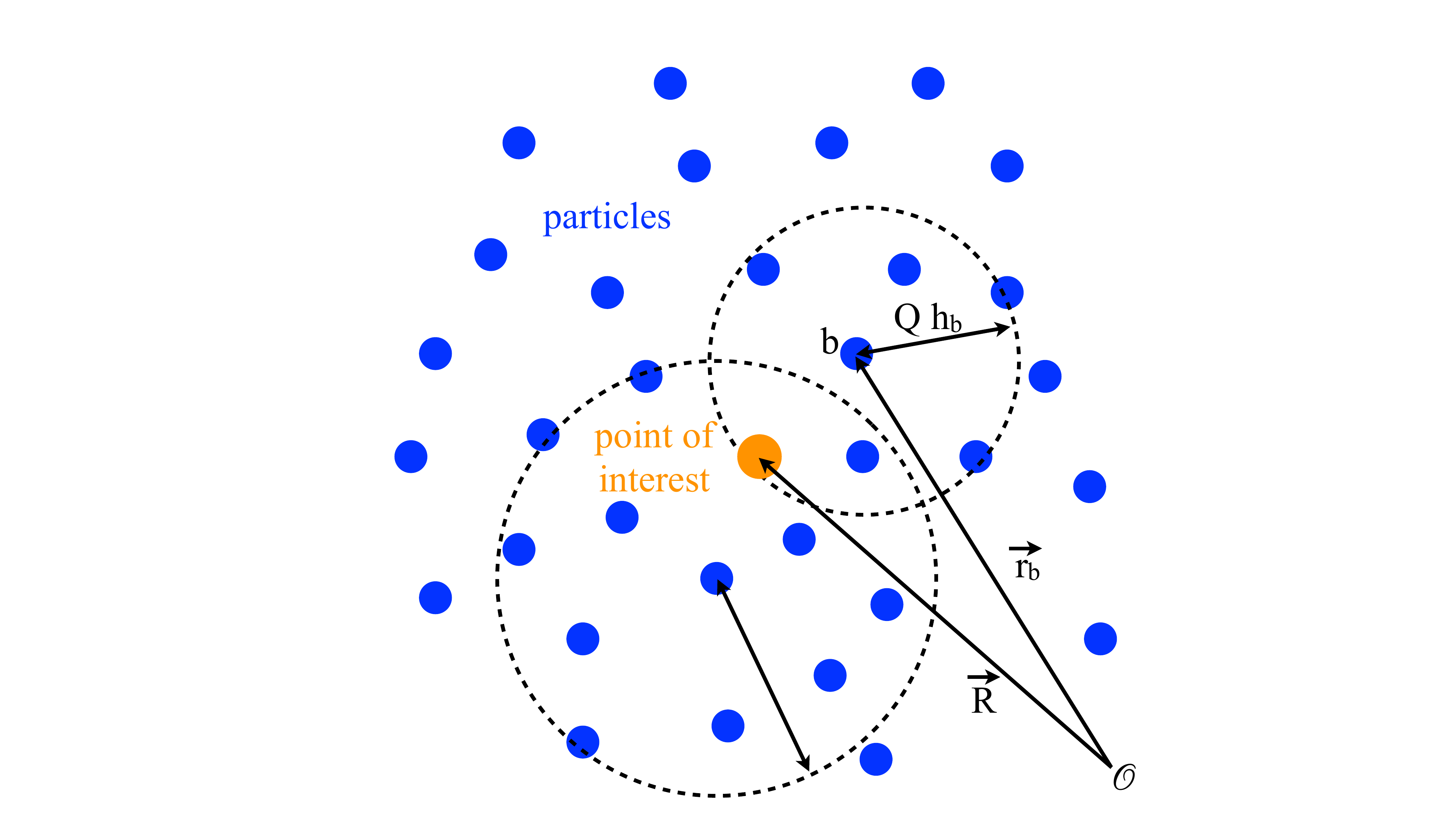}
 \caption{Estimating function properties at a point of interest
   (in orange; either particle or grid point) based on the surrounding particles.}
   \label{fig:LRE_config}
\end{figure}

Assume that we have some specifiable ``point of interest''
$\vec{R}=(X,Y,Z)^T$, it could be 
either a particle position, or, say, the position of a grid point,
surrounded by other particles labelled $b$, see
Fig.~\ref{fig:LRE_config}.

If we assume that the function values at the particle positions $\{ f_b\}$ are described by a function $f(\vec{r})$,
we can Taylor-expand this function around the point of interest $\vec{R}$:
\bea
f(\vec{r})=&& f(\vec{R}) + (\p_i f)_{\vec{R}} (\vec{r} - \vec{R})^i + \frac{1}{2!} (\p_{ij} f)_{\vec{R}} \; (\vec{r} - \vec{R})^i  (\vec{r} - \vec{R})^j \nonumber \\
                &+& \frac{1}{3!} (\p_{ijk} f)_{\vec{R}} \; (\vec{r} - \vec{R})^i  (\vec{r} - \vec{R})^j  (\vec{r} - \vec{R})^k \nonumber \\
                &+& \frac{1}{4!} (\p_{ijkl}  f)_{\vec{R}} \; (\vec{r} - \vec{R})^i  (\vec{r} - \vec{R})^j  (\vec{r} - \vec{R})^k  (\vec{r} - \vec{R})^l  \nonumber\\
                &+& \mbox{higher order terms.}
\eea
This Taylor expansion can be interpreted as a polynomial approximation of a given order, where the basis functions have been
shifted to the point of interest $\vec{R}$. The local approximation $\tilde{f}(\vec{r})$ of $f(\vec{r})$ around the point $\vec{R}$ then reads
\be
\tilde{f}^R(\vr)= \vec{\beta}^R \cdot \vec{P}^R(\vr),
\label{eq:func_approx}
\ee 
where the coefficient vector (containing the function and its derivatives) reads
\begin{align}
   \vec{\beta}^R = \Big[ f_{\vec{R}},
           \p_x f_{\vec{R}},
           \p_y f_{\vec{R}},
           \p_z f_{\vec{R}},
           \frac{1}{2}\p_{xx} f_{\vec{R}},
           \p_{xy} f_{\vec{R}},
           \p_{xz} f_{\vec{R}},
           \frac{1}{2}\p_{yy} f_{\vec{R}},
           \p_{yz} f_{\vec{R}},
           \frac{1}{2}\p_{zz} f_{\vec{R}},
           \cdots\Big],\label{eq:coeff}
  \end{align}
and the ``shifted basis functions'' read
 \begin{align}
   \vec{P}^R(\vec{r}) = \Big [ 
           1, \;\;
           \Delta x,\,
           \Delta y,\,
           \Delta z, \,
            \Delta x \Delta x,\,
           \Delta x \Delta y,\,
           \Delta x \Delta z,\,
           \Delta y \Delta y,\,
           \Delta y \Delta z, \,
           \Delta z \Delta z,\;\;
        \cdots\Big]^T,
   \end{align}
where $\Delta \vec{r}=\vec{r}-\vec{R}= [\Delta x, \Delta y,\Delta z]^T$,
and $\Delta x= x - X$ (similarly for the other components).
The degrees of freedom ($\mathrm{DoF}$) for a given number of dimensions $d$ and a maximum polynomial order of the basis $m$ are given by
\be
\mathrm{DoF}= \frac{(d+m)!}{d!\,m!}.
\ee
So in 3D, we have 1, 4, 10, 20 and 35 DoF for constant, linear,
quadratic, cubic and quartic polynomials.

To find the coefficients, i.e. function estimates
and derivatives, that are optimal at $\vec{R}$, one has to minimize an error functional
\be
\epsilon^R \equiv  \sum_b [f_b - \tilde{f}^R(\vec{r}_b)]^2 \; \Psi_{bR} = \sum_b \left[f_b - { \sum_{i=1}^{\mathrm{DoF}} \beta^R_i P^R_i(\vec{r}_b)} \right]^2 \; \Psi_{bR},
\ee
where $\Psi_{bR}$ is a weighting function that depends on the distance
between the particle $b$ and the point of interest $\vec{R}$ and that
gives more weight to nearby particles. Obvious choices for $\Psi$ would
be an SPH-kernel $W$ or the above used shape functions $\Phi$ that form
a partition of unity. The optimal coefficients that minimize
the error at $\vec{R}$ are determined via
\be
\left(\frac{\p \epsilon^R}{\p \beta_i^R }\right)_{\vec{R}} \stackrel{!}{=} 0
\ee
and one finds
\begin{align}
\beta^R_i = \left(M_{ik}\right)^{-1}  B_k \quad \text{with} \quad
M_{ik} =&  \sum_b P^R_i(\vec{r}_b) \; P^R_k(\vec{r}_b) \, \Psi_{bR} \nonumber\\
\quad \text{and} \quad B_k =& \sum_b f_b \; P^R_k(\vec{r}_b) \, \Psi_{bR}
\label{eq:coeff2},
\end{align}
where the ``moment matrix''  $M_{ik}$, very similar to
Sect.~\ref{sec:kernel_interpolation}, is independent of the function
$f$. The function $f$ only enters in the vector $B_k$, therefore, $M_{ik}$ only
depends on the particle distribution and, once determined, it can be
used to approximate several functions.

\subsubsection{Accuracy of different gradient prescriptions}
\begin{figure}[ht]
  \centerline{
    \includegraphics[width=0.49\textwidth]{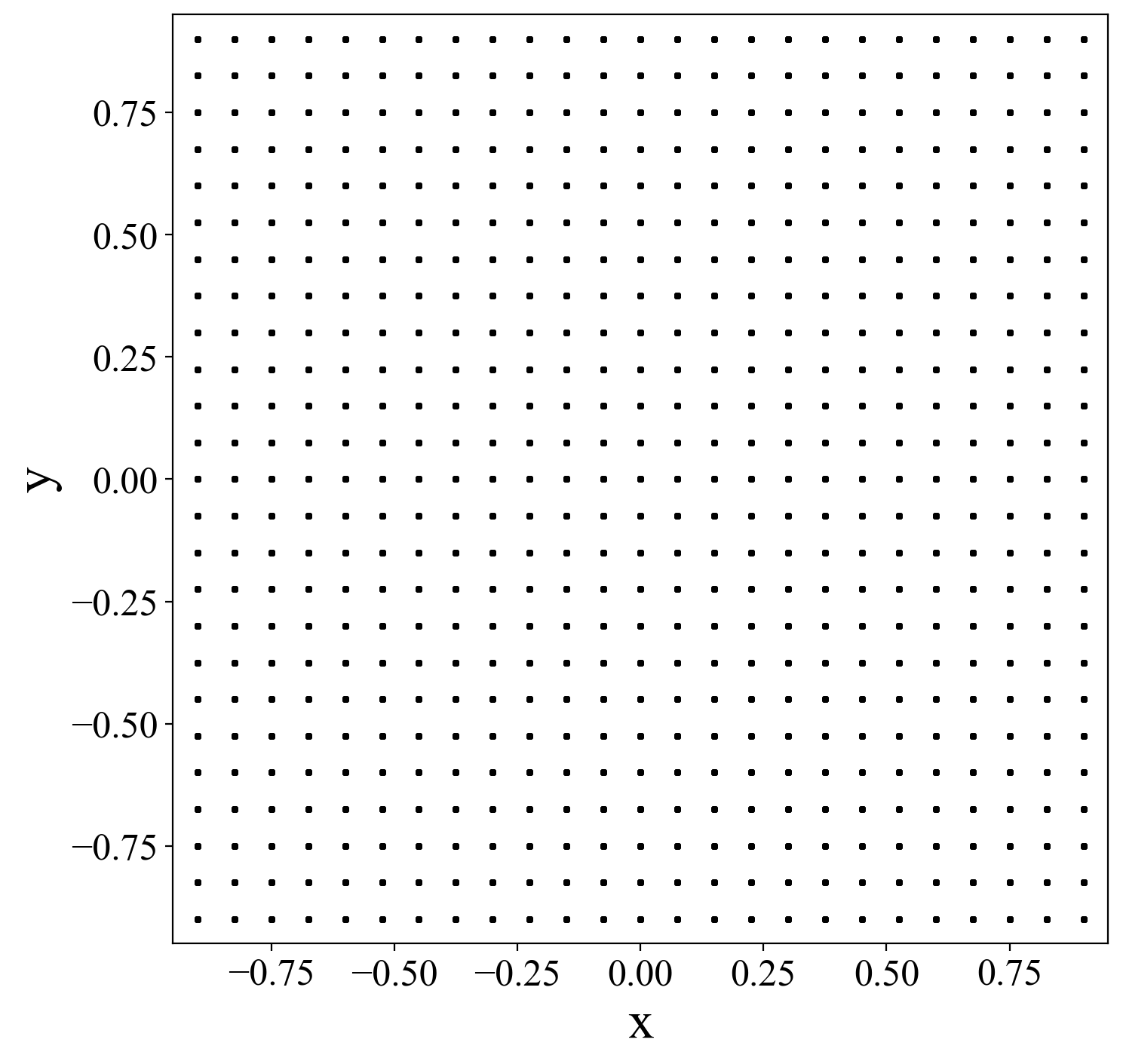}
    \includegraphics[width=0.49\textwidth]{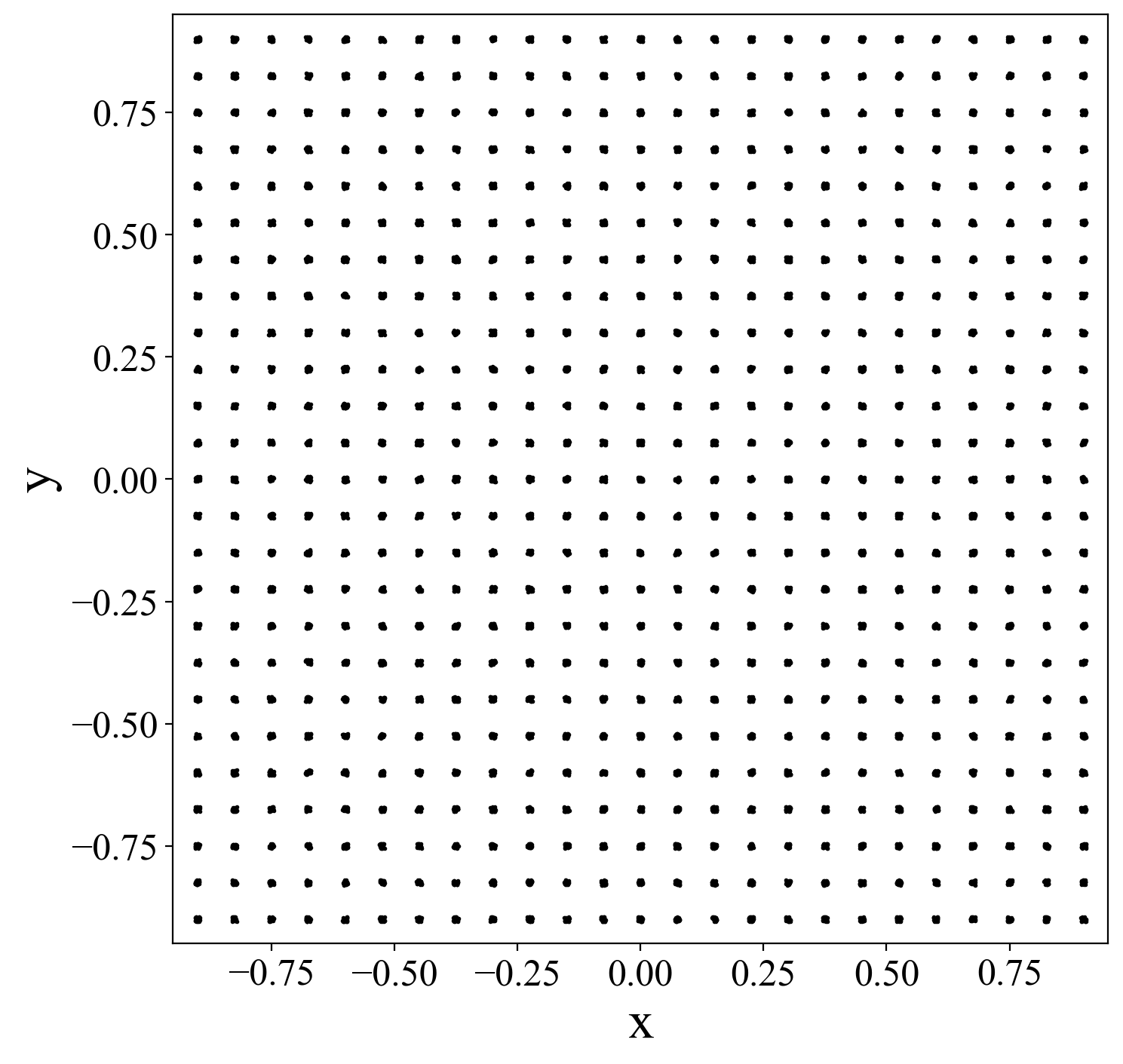}
  }
  \centerline{
    \includegraphics[width=0.49\textwidth]{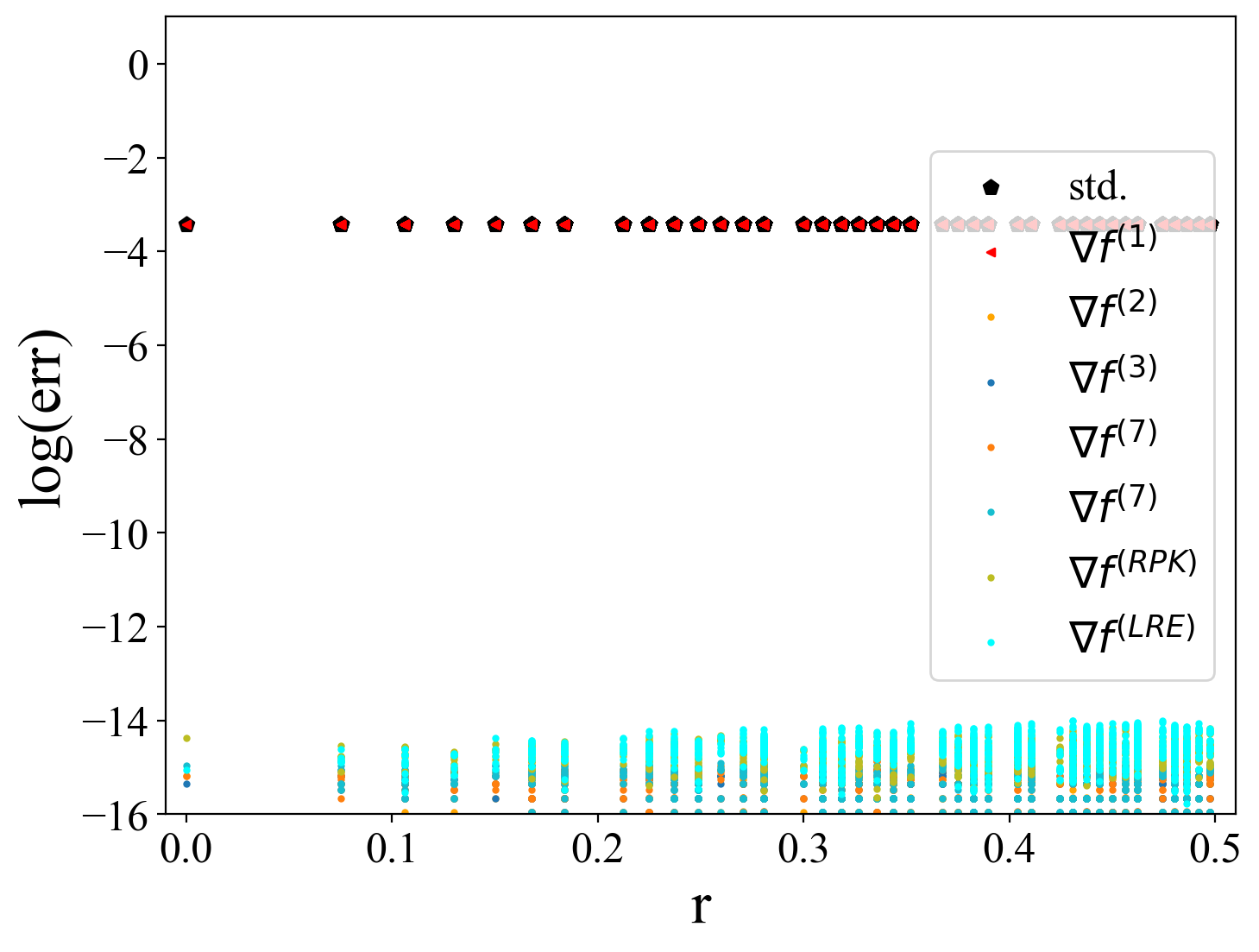}
    \includegraphics[width=0.49\textwidth]{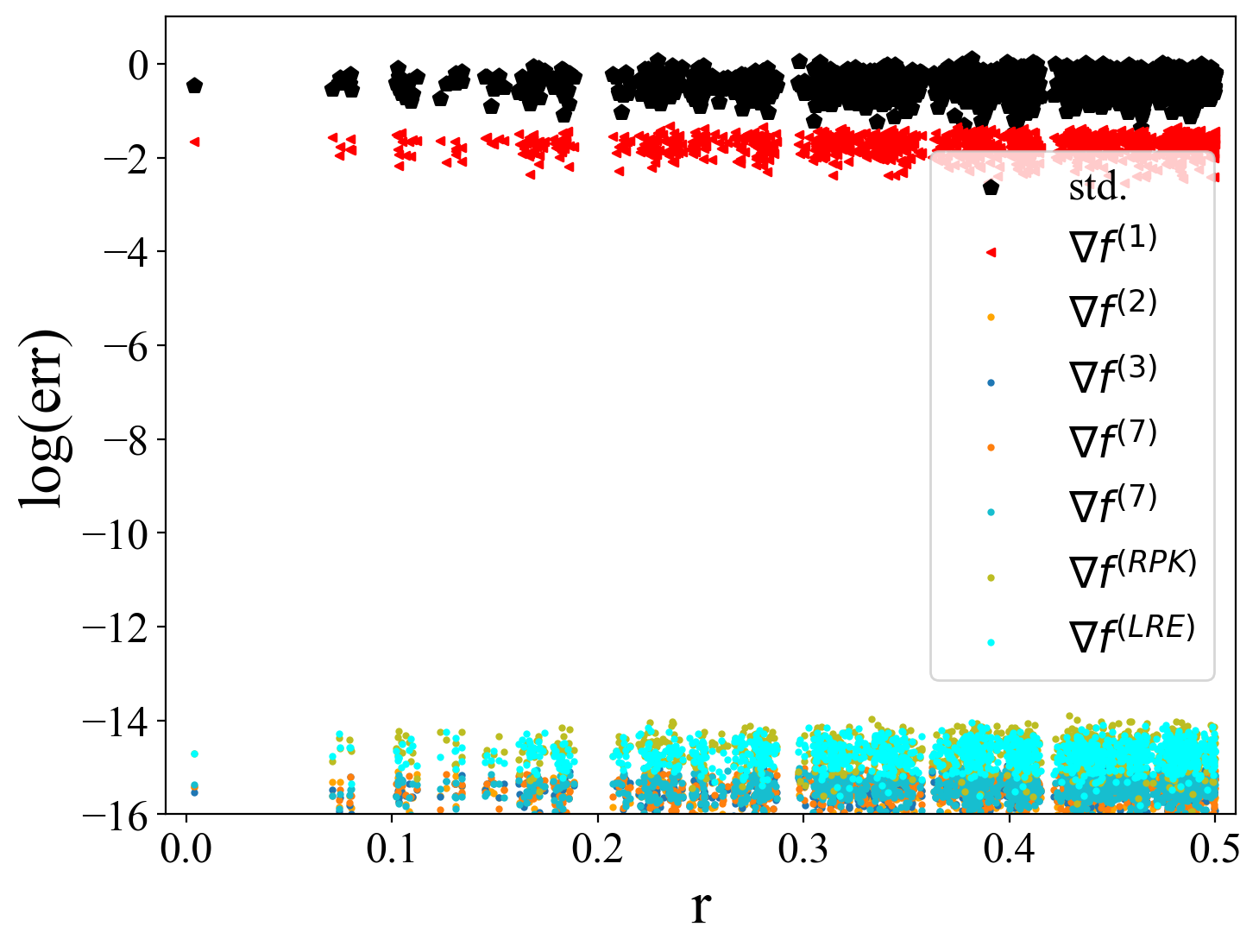}
  }
 \caption{Numerical experiment to measure gradient accuracy. The upper
 row shows the particle positions, exactly arranged on a cubic
 lattice (left), and the same lattice, but with each particle coordinate
shifted by a small random number chosen uniformly from the interval
[-0.005,0.005].
The second row shows the gradient errors for both particle
configurations for the different gradient gradient prescription. Note
that the hardly visible randomization of the positions leads to a serious
gradient deterioration for the standard gradient $\nabla f^{(0)}$ and
for $\nabla f^{(1)}$. Please see the main text for the exact
definitions of the gradient estimates.}
   \label{fig:gradient_experiment}
\end{figure}

We perform here a small experiment to scrutinize the gradient
approximations. We use first particles that are exactly arranged in
a cubic lattice and then compare against a case where such this lattice has
been randomly disturbed by a small amplitude, see the upper two panels of
Fig.~\ref{fig:gradient_experiment}.

In the randomized case, right column Fig.~\ref{fig:gradient_experiment},
each particle coordinate was displaced by a random number chosen 
uniformly from the interval [-0.005,0.005]. Each particle is assigned
a function $\f(\vr)= (-x,0,0)$ and we measure the error between the
numerically determined gradient and the exact result, $\epsilon\equiv
|\nabla f^\mathrm{num} - \nabla f^\mathrm{ex}|/|\nabla f^\mathrm{ex}|$.
For the LRE-approximation, the ``point of interest'' is each time a
particle position, and for the shown plots we use polynomial order
$m=1$. Higher polynomial orders show a slightly different level of
accuracy (due to different floating point operations), but all have
average errors $<10^{-14}$. Note that the hardly visible perturbation
(right panel in Fig.~\ref{fig:gradient_experiment}) has a serious impact
on the gradient estimates $\nabla f^{(0)}$ and $\nabla f^{(1)}$, while
the other gradient estimates reproduce the theoretical result to
within machine precision as expected from their construction.

\subsection{Treatment of shocks}
\label{sec:shocks}
In gas dynamics, 
even smooth initial conditions can steepen  into shocks 
\citep{landau59,whitham74,shu92}. At discontinuities
the differential form of the fluid equations is no longer valid, instead
their integral form needs to be used, which, at shocks, translates into
the {\em Rankine--Hugoniot conditions} \citep{landau59}, which relate the upstream 
and downstream properties of the flow. They show in particular, that the entropy 
increases in shocks, i.e. that dissipation occurs inside the shock front. Numerical 
methods should  mimick this behaviour, i.e. in smooth parts of the flow they should 
(ideally) be dissipationless, but they need to have a built-in mechanism which 
produces entropy in a shock front. For this reason the inviscid SPH equations 
need to be augmented by further measures that become active near shocks.

The most common approach in SPH is to add artificial dissipation terms, in the 
simplest case only involving \enquote{artificial viscosity} (acting on  velocity jumps), but 
sometimes also \enquote{artificial conductivity} (acting on jumps in the internal energy). 
Note that the comparison with Riemann solver approaches \citep{monaghan97} 
suggests to  apply dissipative terms also to the continuity equation, but to date 
there are only few such implementations, see \citet{read12} and \citet{sandnes24}. While 
artificial viscosity is one of the oldest techniques in computational  fluid dynamics, 
it is still being developed further and modern artificial dissipation schemes perform 
--and actually are-- very similar to approximate Riemann solver approaches. We 
will discuss artificial dissipation  in Sect.~\ref{sec:AV} and Riemann solver 
approaches in Sect.~\ref{sec:Riemann_SPH}.

\subsubsection{Artificial dissipation}
\label{sec:AV}
\begin{figure}[t]
\centerline{
\includegraphics[width=0.5\textwidth]{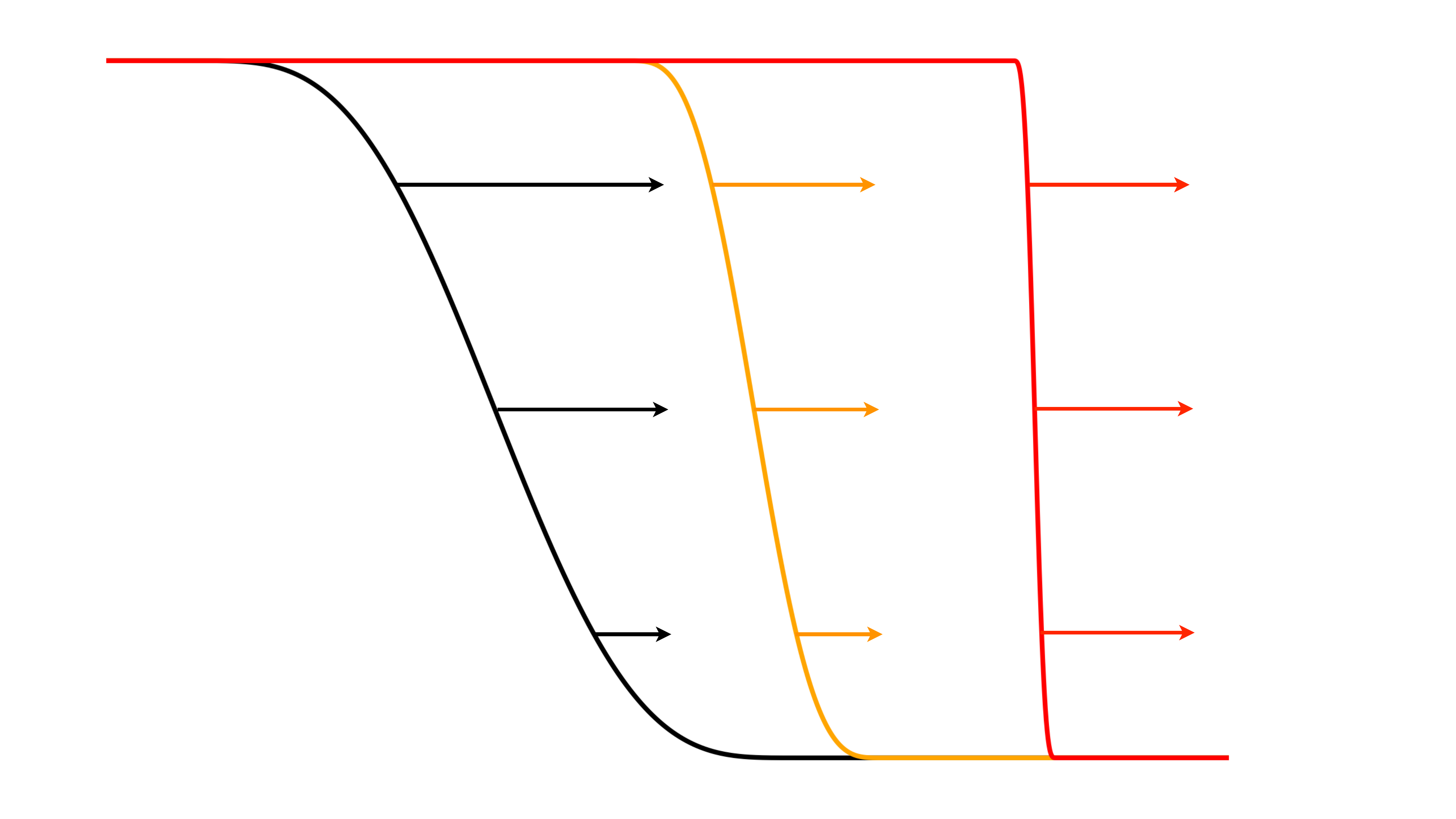}}
\caption{Sketch of a sound wave steepening into a shock.}
\label{fig:AV}   
\end{figure}

Seemingly harmless sound waves can steepen into shocks during the
hydrodynamic evolution, when high-density parts with larger sound
speeds  catch up with the slower low-density parts, see the sketch in the left panel of
Fig.~\ref{fig:AV}. Without any dissipation this steepening will continue
until the wave becomes mathematically discontinuous. 
The main idea of artificial viscosity is to add an ``artificial
pressure'' $Q$ to the physical pressure $P$ (wherever it occurs)
\be
P \rightarrow P + Q
\ee
in order to prevent the wave from becoming infinitely steep and instead 
keep it  at a numerically treatable level.
The suggestion of \citet{vonneumann50} was
to compose the artificial pressure $Q$, for converging flows, out of
two terms $P_\mathrm{B} \propto - \rho l \divv$, and a second term,
$P_\mathrm{NR} \propto \rho l^2 (\nabla \cdot \vec{v})^2$, where $l$ is a
local resolution length. The first term mimics  physical bulk
viscosity while the second term is designed to suppress spurious
oscillations in the post-shock region. 
Keep in mind that $\divv= -(d \rho/dt) / \rho$, see
Eq.~(\ref{eq:drhodt}), is a measure of the local relative volume 
change of the fluid, with negative sign for compression and positive
sign for expansion. The term $l (\divv)$ is thereby a measure of the velocity
jump between two adjacent entities (either cells or particles).

In SPH, the expression for the viscous pressure at particle position $a$
is often written as \citep{monaghan83}
\be
Q_a= \rho_a \left(- \alpha c_{{\rm s},a} \mu_a + \beta \mu_a^2\right),
\label{eq:Qvis}
\ee
where the velocity jump estimate is
\be
\mu_a= {\rm min} \left(0,\frac{(\vec{v}_{ab} \cdot \vec{r}_{ab})h_a}{r_{ab}^2 +
\epsilon^2 h_a^2} \right).
\label{eq:mu_vis}
\ee
Here, $\vec{v}_{ab}= \vec{v}_a - \vec{v}_b$, $\vec{r}_{ab}= \vec{r}_a
- \vec{r}_b$, $r_{ab}^2= \vec{r}_{ab} \cdot \vec{r}_{ab}$ and $\alpha,
\beta$ and $\epsilon$ are numerical parameters with typical
values  1, 2 and $0.1$ and $c_{{\rm s}}$ is the sound speed\footnote{In our own experiments we have not found the $\epsilon$-term to be necessary as long as there is no pairing instability, e.g. because Wendland kernels are used.}. The
min-function switches off the artificial pressure for receding
particles ($\vec{v}_{ab} \cdot \vec{r}_{ab}>0$) and the
$\epsilon$-term in the denominator avoids divergence. Enhancing
physical pressures at the particle positions with such artificial
pressures keeps the equations Galilean invariant, they are still
conservative since none of  the symmetries in the particle indices, see
Sect.~\ref{sec:conservation}, was changed, and the
dissipation vanishes for rigid rotation. Variants of these expressions
exist, but the differences between them in hydrodynamic tests are usually
small. More important is that dissipation is avoided where it is not
needed, see below.

%
%
Comparison with Riemann solver approaches \citep{monaghan97} suggests an alternative form 
that involves signal velocities and jumps in variables across 
characteristics. The main idea of these ``discontinuity capturing terms'' is that 
for any conserved scalar variable $A$ with $\sum_a m_a dA_a/dt=0$ a dissipative term 
of the form
\be
\left( \frac{dA_a}{dt}\right)_\mathrm{diss}= \sum_b m_b \frac{\alpha_{A,b} v_\mathrm{sig}}{\bar{\rho}_{ab}} 
(A_a-A_b)\hat{e}_{ab} \cdot \nabla_a W_{ab}
\ee
should be added, where the parameter $\alpha_{A,b}$ determines the exact amount of dissipation and $v_\mathrm{sig}$ 
is a signal velocity between particle $a$ and $b$. Applied to the velocity and the thermokinetic energy
$e= u + v^2/2$ this yields
\bea
\left( \frac{d \vec{v}_a}{dt} \right)_\mathrm{diss}&=& \sum_b m_b \frac{\alpha v_\mathrm{sig}(\vec{v}_a-\vec{v}_b)\cdot 
\hat{e}_{ab}}{\bar{\rho}_{ab}} \nabla_a W_{ab}\label{basic:eq:v_diss}\\
\left( \frac{d e_a}{dt} \right)_\mathrm{diss}&=& \sum_b m_b \frac{e^\ast_a - e^\ast_b}{\bar{\rho}_{ab}} 
\hat{e}_{ab}\cdot \nabla_a W_{ab},\label{basic:eq:e_diss}
\eea
where, following \citet{price08a}, the energy $e^\ast$ includes velocity components along the line 
joining particles $a$ and $b$, $e^\ast_a= \frac{1}{2} \alpha v_\mathrm{sig} (\vec{v}_a\cdot\hat{e}_{ab})^2 
+ \alpha_u v_\mathrm{sig}^u u_a$. Note that in this equation different signal velocities and dissipation
parameters can be used for the velocities and the thermal energy terms. Using
$du_a/dt= de_a/dt -\vec{v}_a \cdot d\vec{v}_a/dt$, this translates into 
\be
\left( \frac{du_a}{dt} \right)_\mathrm{diss}= - \sum_b \frac{m_b}{\bar{\rho}_{ab}} \left[ \alpha v_\mathrm{sig} \;  
\frac{1}{2} (\vec{v}_{ab}\cdot\hat{e}_{ab})^2 + \alpha_u v^u_\mathrm{sig}
(u_a-u_b) \right] \hat{e}_{ab} \cdot \nabla_a W_{ab}
\label{eq:du_diss}
\ee
for the thermal energy equation.
The first term in this equation bears similarities with the
``standard'' \emph{artificial viscosity} prescription, see
Eq.~(\ref{eq:mu_vis}).  The second one expresses the exchange of thermal energy between particles
and therefore represents an artificial thermal conductivity which
smoothes discontinuities in the specific energy. Artificial
conductivity had been suggested earlier \citep{noh87}  to cure the so-called ``wall heating problem''. 
Tests have shown that \emph{artificial conductivity} substantially improves SPH's performance
in simulating Sedov blast waves \citep{rosswog07c} and in the treatment of Kelvin--Helmholtz instabilities \citep{price08a}.
Note that the general strategy suggests to use dissipative terms also for the continuity equation so that particles
can exchange mass in a conservative way. This has so far been rarely applied, but \citet{read12,sandnes24} find good results
with this strategy in  multi-particle-mass simulations.

\paragraph{How to avoid dissipation where it is not needed?}

A major concern is to not have dissipation where it is not
needed. There are three strategies to avoid unwanted dissipation:
1.) apply ``limiters'' that identify where dissipation should be suppressed,
2.) use time-dependent dissipation  parameters so that dissipation
decreases if not needed and
--maybe most effectively-- 3.) perform a slope-limited
reconstruction in the dissipative terms, very similar to Finite Volume
methods. Combinations of these three approaches are also possible.\\

\begin{enumerate}
\item {\bf ``Limiters''}

Here the idea is to multiply the dissipation parameter $\alpha$ with a limiter that is smart
enough to decide, based on the local fluid properties, whether a particle is
near a shock or not and, in the latter case, the limiter should have a
very small or zero value to avoid dissipation. Various versions of
such limiters have been suggested, see
e.g. \citet{balsara95,cullen10,rosswog15c}.

\citet{balsara95} suggested to 
distinguish between shock and shear motion based on the ratio
\be
\xi_a^{\rm B}= \frac{|\nabla \cdot \vec{v}_a|}{|\nabla \cdot \vec{v}_a| 
+ |\nabla \times \vec{v}_a| + 0.0001 c_{s,a}/h_a}.
\label{eq:Balsara}
\ee
Dissipation is suppressed, $\xi_{a}^{\rm B} \rightarrow 0$, where $|\nabla \times \vec{v}_a| \gg |\nabla \cdot \vec{v}_a|$, 
whereas $\xi_{a}^{\rm B}$ tends to unity in the opposite limit. 
If symmetrized appropriately, e.g. by multiplying the $\xi_a^{\rm B}$
with the artificial pressure $Q_a$, exact conservation is guaranteed.
This limiter has been found  useful in many applications \citep{steinmetz96,navarro97,rosswog00}, but 
it can be challenged if shocks occur in a shearing environment like an accretion disk \citep{owen04}. Part of this
challenge comes from the finite accuracy of the standard SPH-derivatives. It is, however,
straight forward \citep{cullen10,read12,rosswog14a} to use more accurate derivatives,
see Sect.~\ref{sec:accurate_gradients}, in the limiters. Suppression of dissipation in shear flows can
then be obtained by simply replacing the SPH-gradient operators in Eq.~(\ref{eq:Balsara}) by more accurate
expressions, or by simply adding terms proportional to $|\nabla \times \vec{v}|$ 
in the denominators of Eqs.~(\ref{eq:trigger_CD}) and (\ref{eq:trigger_RH}) (instead of multiplying $\alpha_\mathrm{des}$
with a limiter). 

Another limiter was proposed by \citet{cullen10}:
\be
\xi_{a}^{\rm C}= \frac{|2(1 - R_a)^4 (\nabla \cdot \vec{v})_a|^2}{|2(1 - R_a)^4 
           (\nabla \cdot \vec{v})_a|^2 + tr({\bf S}_a \cdot {\bf
             S}_a^t)},
         \label{eq:CD_shock_trigger}
\ee
where
\be
R_a= \frac{1}{\rho_a} \sum_b {\rm sign}(\divv)_b \; m_b \; W_{h_a}(r_{ab}).
\ee
Note that $R_a$ is simply the ratio of a density summation where each term is weighted by the sign of $\divv$
and the normal density, Eq.~(\ref{eq:density}). Therefore, near a shock $R_a \rightarrow -1$. The matrix
${\bf S}$ is the traceless symmetric  part of the velocity gradient matrix ${\mathcal V}= (\p_i v_j)$ and a measure of
the local shear. Similar to the Balsara factor, $\xi^{\rm C}$ approaches unity if compression clearly dominates over
shear and it vanishes in the opposite limit.\\

\item {\bf Time-dependent dissipation  parameters}

\citet{morris97} suggested an approach with time-dependent parameters 
for each particle and with triggers that indicate where the
dissipation needs to be increased. Using  $\beta_a= 2\alpha_a$ in
Eq.~(\ref{eq:Qvis}),  they evolved $\alpha_a$ according  
to 
\be
\frac{d\alpha_a}{dt}= \mathcal{A}_a^+ - \mathcal{A}_a^- \quad {\rm with} \quad 
\mathcal{A}_a^+= \max\left(-(\nabla\cdot\vec{v})_a,0\right) \quad \text{and} \quad
\mathcal{A}_a^- = \frac{\alpha_a(t) - \alpha_{\min}}{\tau_a},
\label{eq:alpha_steering_MM}
\ee
where $\alpha_{\min}$ represents a minimum, ``floor'' value 
for the viscosity parameter and $\tau_a\sim h_a/c_{{\rm s},a}$ is the
individual decay time scale at particle $a$. 
This approach (or slight modifications of it) has been shown to substantially 
reduce unwanted effects in practical simulations \citep{rosswog00,dolag05,wetzstein09}
but, as already realized in the original publication, triggering on the velocity
divergence, see $\mathcal{A}_a^+$ in Eq.~(\ref{eq:alpha_steering_MM}), also raises the viscosity in adiabatic compression with
$(\nabla\cdot\vec{v})=$ const, where it is actually not needed. 

\paragraph{Dissipation triggers}

Several improvements of the basic Morris and Monaghan idea were suggested by \citet{cullen10}.
First, the authors argued that the floor value can be safely set to zero, provided 
that $\alpha$ can grow fast enough. To ensure the latter, the current value of $\alpha_a$ 
is compared to values indicated by triggers and, if necessary, $\alpha_a$ is increased 
{\em instantaneously} rather than by solving Eq.~(\ref{eq:alpha_steering_MM}). Second, 
instead of using $(\nabla\cdot\vec{v})_a$ one triggers on its time derivative 
to find the locally desired dissipation parameter:
\be
A_a= \xi_a \; \max\left[-\frac{d (\nabla\cdot\vec{v})_a}{dt},0\right] \quad \text{and} \quad 
\alpha_{a, \rm des}= \alpha_{\max} \frac{A_a}{A_a + c^2_a/h_a^2},
\label{eq:trigger_CD}
\ee
where $c_a$ is a signal velocity, $\xi_a$ the limiter of Eq.~(\ref{eq:CD_shock_trigger}) and $\alpha_{\max} $ a maximally
admissible dissipation value.
If $\alpha_{a, \rm des} > \alpha_\mathrm{a}$ then $\alpha_\mathrm{a}$ is raised immediately to 
$\alpha_{a, \rm des}$, otherwise it decays according to the $\mathcal{A}_a^-$-term 
in Eq.~(\ref{eq:alpha_steering_MM}). Apart from overall substantially reducing dissipation, 
this approach possesses the additional virtue that $\alpha$ peaks $\sim 2$ smoothing lengths ahead 
of a shock front and decays immediately thereafter. 

\citet{read12} suggest a similar strategy, but trigger on the \emph{spatial }
change of the compression
\be
A_a= \xi_a \; h_a^2 |\nabla (\nabla \cdot \vec{v})| \quad \text{and} \quad 
\alpha_{a,\rm des}= \alpha_{\max} \frac{A_a}{A_a + h_a |\nabla \cdot \vec{v}|_a + 0.05 \; c_a},
\label{eq:trigger_RH}
\ee 
where $\nabla \cdot \vec{v} < 0$ and $\alpha_\mathrm{des}= 0$ otherwise. 
The major idea is to detect convergence \emph{before} it actually occurs. Particular care is taken
to ensure that all fluid properties remain single valued as particles approach each other and
higher-order gradient estimators are used. With this approach they 
find good results with only very little numerical noise.

In \citet{rosswog15b} a trigger that combines both spatial and temporal changes was explored:
\be
A_a=  \max \left[ h_a \; D_{v} \left( \nabla \cdot \vec{v} \right)_a \;   
\frac{d (\nabla\cdot\vec{v})_a}{dt},0 \right],
\quad 
{\rm where} \quad  D_{v} \left( \nabla \cdot \vec{v} \right)_a=  \hat{e}_{v_a} \cdot 
\nabla \left( \nabla \cdot \vec{v} \right)_a
\label{eq:SR_shock_trigger}
\ee
is the directional derivative, $\hat{e}_{v_a}$ is the unit vector in direction $\vec{v}_a$ and  all 
involved velocity derivatives are evaluated with the accurate gradient estimates of Eq.~(\ref{eq:der_fIA}).
The desired shock parameter is
\be
\alpha_\mathrm{a, shock}= \alpha_{\max} \frac{A_a}{A_a  + 0.1 \; c_\mathrm{s,a}^3/h_a^3}.
\ee
An additional trigger is suggested to measure local sign fluctuations in $\nabla \cdot \vec{v}$ via
\be
A_{a, \rm noise}= \left| \frac{\tilde{S}_{1,a}}{S_{2,a}} - 1\right|,
\ee
as an indication for noise. Here
\be
S_{1,a}= \sum_b (\divv)_{b} \quad \text{and} \quad S_{2,a}= \sum_b |\divv|_{b}
\ee
and $\tilde{S}_{1,a}$ is equal to $-S_{1,a}$ if $(\nabla \cdot \vec{v})_{a} < 0$ and equal
to $S_{1,a}$ otherwise\footnote{Recent work by \citet{rosswog26a} find that weighting the contribution 
by the kernel function, i.e. using $S_{1,a}= \sum_b W_{ab} (\divv)_{b}/\sum_b W_{ab}$  and 
$S_{2,a}= \sum_b  W_{ab}|\divv|_{b}/\sum_b W_{ab}$, further improves the noise identification 
and avoids false triggers.}.
 So if all neighbouring particles have the same sign in $\divv$, i.e.~all are
either expanding or all are being compressed, $A_{a, \rm noise}$ vanishes, while finite values indicate
sign fluctuations, in other words noise. The noise dissipation parameter then follows from  
\be
\alpha_\mathrm{a, noise}=  \alpha_{\max} \frac{A_{a,\rm noise}}{A_\mathrm{ref, noise} + A_{a,\rm noise}},
\label{eq:SR_noise_trigger}
\ee
where $A_\mathrm{ref,  noise}$ determines the level of noise that is
considered tolerable. The desired dissipation parameter is
chosen as $\alpha_{a, \rm  des}= \max(\alpha_\mathrm{a, shock},\alpha_\mathrm{a, noise})$ and, again, 
$\alpha_\mathrm{a}$ is raised immediately if $\alpha_{a, \rm  des} > \alpha_\mathrm{a}$. The addition of this noise
trigger has been shown \citep{rosswog15b} to significantly increase the accuracy of SPH in the Gresho--Chan 
vortex test and in Schulz-Rinne tests \citep{rosswog26a}.

\citet{rosswog20b} suggested to trigger on the local
non-conservation of entropy to identify ``troubled particles''. In an
ideal fluid, the entropy should be strictly conserved, but if the
internal energy is evolved,  this conservation is not enforced on a
numerical level. Therefore one can monitor the entropy conservation
as a measure for the numerical quality of the flow and the entropy
non-conservation (either due to a shock or numerical noise)
indicates how much dissipation should be applied. 

\begin{figure}[t]
\centerline{
\includegraphics[width=0.7\textwidth]{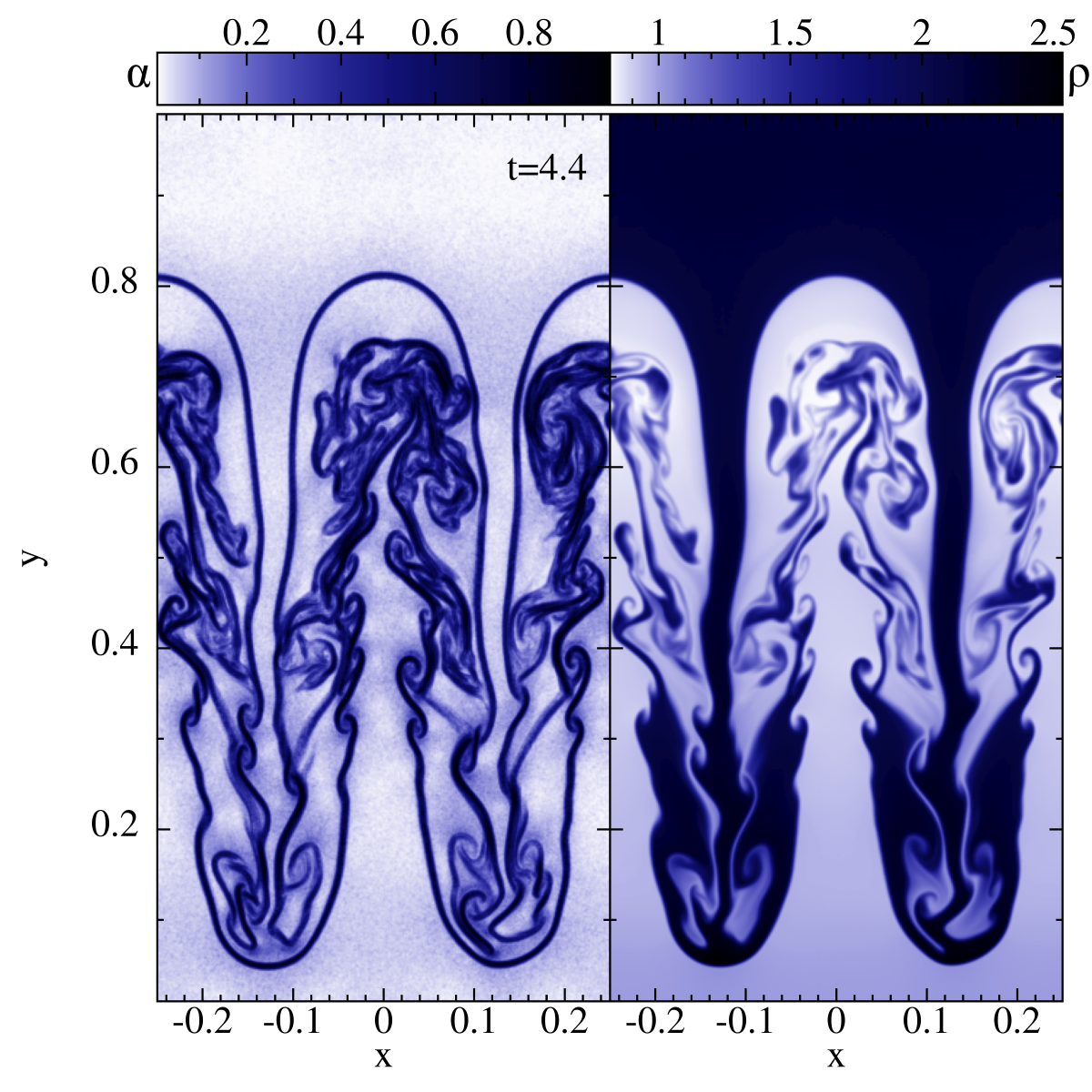}}
\caption{Rayleigh-Taylor instability where the dissipation is steered
by entropy monitoring. The right panel shows the density evolution,
while the left panel shows the value of the dissipation parameter
$\alpha$. Dissipation is only triggered at the interface between high
and low densities, otherwise it is essentially absent. For more
details see \citet{rosswog20b}.}
\label{fig:entropy_trigger}   
\end{figure}

One can use the non-dimensionalized relative
entropy change rate of a particle $a$ between time step $t^{n-1}$ and
time step $t^{n}$ ($\Delta t= t^{n-1} - t^n$) 
\be
\dot{\epsilon}_a^n \equiv \frac{ |s_a^{n} - s_a^{n-1}|}{s_a^{n-1}} \frac{\tau_a}{\Delta t},
\ee
as a measure of how much dissipation is needed.  Here $s$ is the
specific entropy of a particle, $\tau_a= h_a/c_{s,a}$ is the particle's
dynamical time scale and $c_{s,a}$ its sound speed.
Based on the logarithm of the entropy change rate
$l_a^n=\log(\dot{\epsilon}_a^n)$ as a trigger, the desired value  of
the dissipation parameter is chosen as
\be
\alpha_{a,\rm des}^n= \alpha_{\max} \; \mathcal{S}(l_a^n),
\ee
where the smooth ``switch-on'' function 
\be
\mathcal{S}(x)= 6x^5 - 15x^4 + 10x^3,
\ee
with
\be
x= \rm min\left[max\left(\frac{l_a^n - l_0}{l_1-l_0},0\right),1\right]
\ee
was used.
Here, $l_0$ is a numerically acceptable level of the entropy change
rate below which no dissipation is applied and $l_1$ is the rate where
the maximum dissipation is needed. Rates in between are steered by the
smooth switch-on function. \citet{rosswog20b} finds good
results for $l_0= \log_{10}(10^{-4})$ and $l_1= \log_{10}(5 \times
10^{-2})$. With this prescription dissipation is 
rapidly and robustly switched on in strong shocks, but in gentle flows
dissipation is applied only very locally. For example,
Fig.~\ref{fig:entropy_trigger} shows a snapshot of a 
Rayleigh--Taylor instability simulation. It demonstrates that
dissipation is essentially absent apart from the interface between 
the high- and the low-density fluid, see the left panel.\\

\item {\bf Slope-limited reconstruction in the dissipative terms}

\begin{figure}[t]
\centerline{
\includegraphics[width=0.9\textwidth]{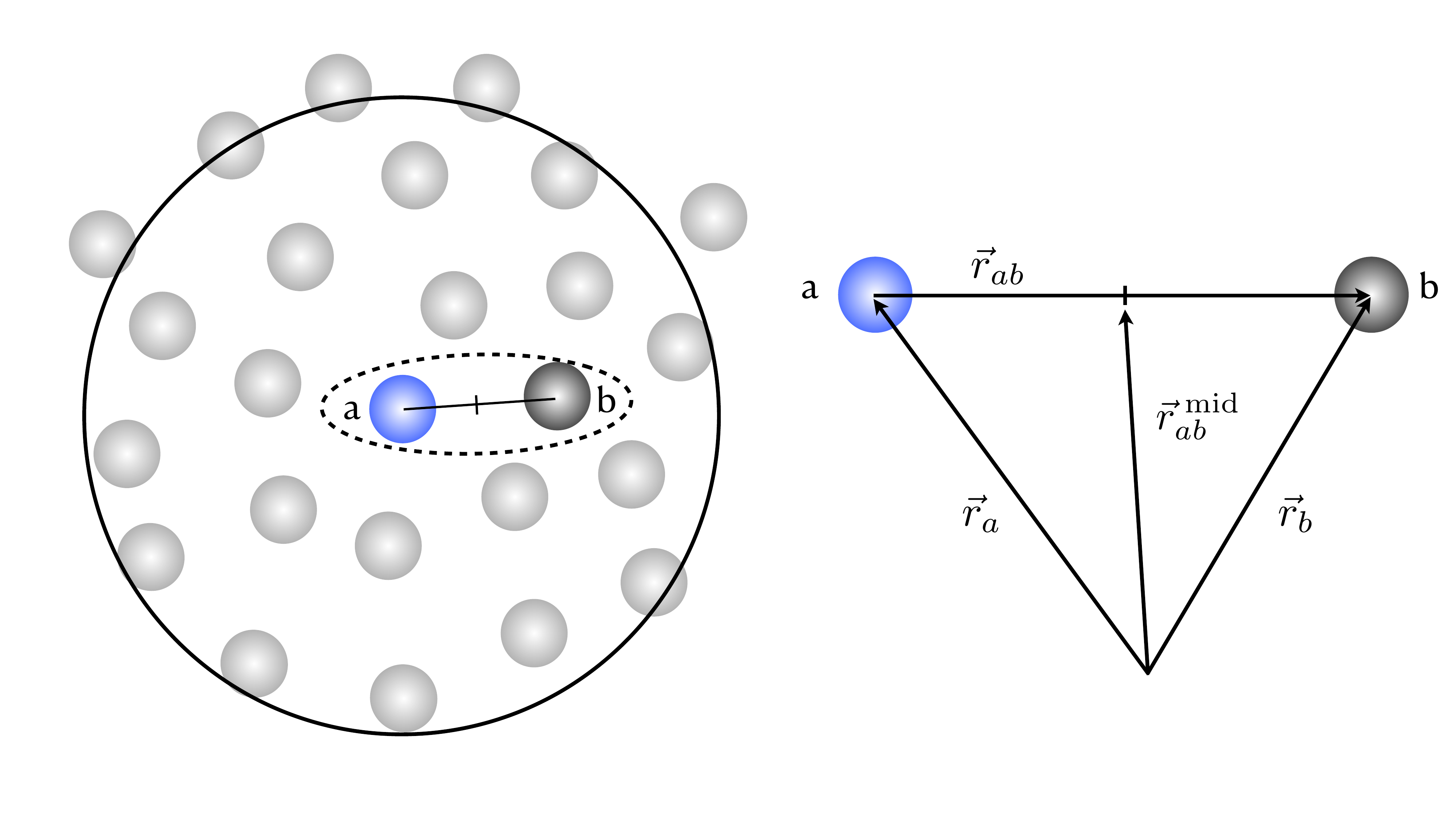}}
\caption{Sketch of the inter-particle Riemann problem between
particles $a$ and $b$. The jumping quantities, in either artificial
dissipation or Riemann solver approaches, can be constructed to the
mid-point of each particle pair and one then applies either artificial
viscosity or Riemann solver to the jump in the \emph{reconstructed} values.
Image reproduced with permission from \citet{rosswog25a}, copyright by the author.}
\label{fig:interparticle_riemann}   
\end{figure}

An, in retrospect maybe obvious, approach is to use slope-limited
reconstruction \emph{in the artificial dissipation}
\citep{christensen90}, very similar to what is 
done in Riemann problems in  Finite Volume methods
\citep{toro99}. This approach has found its way into astrophysics only
recently \citep{frontiere17,rosswog20a,price20,sandnes24}, but it has turned out to be
very powerful in getting rid off excess dissipation, actually
even when large, constant dissipation parameters (e.g. $\alpha=1$ and $\beta=2\alpha$) 
are used \citep{rosswog20a}.

All of the dissipative terms discussed above,
Eqs.~(\ref{eq:mu_vis})-(\ref{eq:du_diss}), act on jumps between
particle properties, e.g. $\vec{v}_a - \vec{v}_b$ or $u_a - u_b$,
i.e. on differences \emph{at the particle positions}. This approach
corresponds to a ``zeroth order reconstruction'' in a Finite Volume
scheme. Such approaches are known to be very dissipative. The solution in Finite
Volume schemes is to solve a Riemann problem at the interfaces between
computational cells, but not with the differences of the average cell values, but instead
with values that have been reconstructed to the interface. To avoid
potential overshoots/oscillations, one uses typically slope limiters
\citep{toro99}. 

Exactly the same can be done in SPH: one can reconstruct
quantities (e.g. velocities or specific internal energies) to the
midpoint between each particle pair,  $\vec{r}_{ab}^{\rm mid}=
0.5(\vec{r}_a + \vec{r}_b)$, see Fig.~\ref{fig:interparticle_riemann},
and then use these reconstructed values in the jumps of
Eqs.~(\ref{eq:mu_vis})--(\ref{eq:du_diss}). Explicitely, if one wants
\emph{linearly reconstructed} values of velocities and internal
energies, both from $a$- and the $b$-side, one can use
\be
v_a^{i, \rm rec} = v^i_a - \frac{1}{2} \Psi(\nabla v^i_a,\nabla v^i_b)
\cdot \vec{r}_{ab} \quad \text{and} \quad
v_b^{i, \rm rec} = v^i_b + \frac{1}{2} \Psi(\nabla v^i_a,\nabla v^i_b)
\cdot \vec{r}_{ab}, \label{eq:rec1}
\ee
and
\be
u_a^{\rm rec} = u_a - \frac{1}{2} \Psi(\nabla u_a,\nabla u_b) \cdot
\vec{r}_{ab} \quad \text{and} \quad
u_b^{\rm rec} = u_b + \frac{1}{2} \Psi(\nabla u_a,\nabla u_b) \cdot
\vec{r}_{ab} \label{eq:rec3}.
\ee
Here $\psi$ is a suitable slope limiter function such as 
 \texttt{minmod} 
\be
\Psi_\mathrm{mm}(x,y)= \frac{1}{2}\left[\mathrm{SGN}(x) + \mathrm{SGN}(y)\right] \; \mathrm{MIN}(|x|,|y|),
\ee
the \texttt{vanLeer limiter} \citep{vanLeer77}
\be
\Psi_\mathrm{vL}(x,y)= \begin{cases}\frac{2 x y}{x+ y} \quad & \text{if} \; x y > 0\\
                                               0 & \text{otherwise},
                         \end{cases}                      
\ee
the vanLeer monotonized Central (\texttt{vanLeerMC}) \citep{vanLeer77}
\be
\Psi_\mathrm{vLMC}(x,y)= \begin{cases} \mathrm{SGN}(x) \; \mathrm{MIN}\left[\frac{|x+y|}{2}, 2 |x|, 2 |y|\right] \quad & \text{if} \; x y > 0 \\
                                                      0 & \text{otherwise},
                              \end{cases}                        
\ee
or the \texttt{vanAlbada limiter} \citep{vanAlbada82}
\be
\Psi_\mathrm{vA}(x,y)= \begin{cases} \frac{(x^2 + \epsilon^2) y + (y^2 + \epsilon^2) x}{x^2 + y^2 + 2 \epsilon^2} \quad & {\rm if } \; x y > 0,\\
                                                 0 & \text{otherwise}.
                                               \end{cases}
                                               \label{eq:vanAlbada}
\ee
The latter is insensitive to the exact value of $\epsilon$, good results are found for
$\epsilon^2= 10^{-6}$.  In Eq.~(\ref{eq:rec1}) the slope limiter $\Psi$ is to be applied to
each component of the gradients.

The strategy of using reconstructed values has turned out to be very
successful \citep{frontiere17,rosswog20a,price20,sandnes24}. For example, weakly
triggered Kelvin--Helmholtz instabilities that do not grow with
standard large dissipation parameters ($\alpha=1$ and $\beta=2\alpha$)
show a healthy growth if the only change is to switch to reconstructed
values in the dissipative terms, see Fig.~20  in
\citet{rosswog20a}. The test bench in \citet{rosswog20a} shows excellent
results even for large, constant dissipation parameters, which may
suggest that reconstruction alone is good enough to get rid of most
unwanted dissipation.
\end{enumerate}

We conclude this section on artificial dissipation (viscosity and conductivity) with a short summary.
Artificial dissipation bears many similarities with approximate Riemann solvers, and, not too surprisingly,
many of the results are similar. If just naively applied everywhere, artificial dissipation can lead to flows that are way more dissipative
than what is desired. Such unwanted dissipation can also artificially transport angular momentum, so that the total
angular momentum is still conserved, but transported like in a dissipative fluid rather than an ideal one 
and this can substantially impact simulation results, e.g. in accretion disks \citep{deng17}.
Of course, these artificial dissipation effects die out with resolution, see Eq.~(\ref{eq:mu_vis}), and
for some (but not all) problems just increasing the resolution may cure all worries \citep{meier26}. 
In practice, however, one usually has to deal with finite resolution and therefore much effort
in the last two decades has been invested in reducing unwanted dissipation by the means discussed 
above. These approaches,  however, usually contain free parameters that need to be gauged in large test batteries, but
still, while this leads in general to great improvements, it is not guaranteed to work in an optimal way every time 
and everywhere \citep{garcia_senz26}. The most robust improvement, however, is the reconstruction in the dissipative terms and we are 
not aware of any reasonable argument against it. \\
Closely related to potential overdamping is the behaviour of traditional SPH in simulations of subsonic turbulence, which was
found to be problematic in \cite{bauer12}. It seems, however, that the problematic behaviour  was 
only partially due too excessive dissipation, the other part resulting from the finite gradient accuracy 
of the standard SPH gradient, Eq.~(\ref{eq:std_grad}). 
A recent study \citep{cabezon25} which includes better gradient estimates, time dependent artificial
viscosity with reconstruction and improved volume elements finds much improved results that
are very similar to those found with AREPO \citep{springel10b} and GIZMO \citep{hopkins15a}.
\\
What we have described above is the standard approach with a linear and quadratic contribution to
the viscous pressure, see Eq.~(\ref{eq:Qvis}). There are, however,  indications that the quadratic term alone, though with a
larger prefactor, may be enough \citep{reisner13,price24b}. Such possibilities should be explored systematically
in the future.

\subsubsection{Riemann solver approaches}
\label{sec:Riemann_SPH}
Another way to deal with shocks is --similar to Eulerian methods--  the use of Riemann solvers \citep{toro99}.
The numerical treatment of shocks is a difficult numerical problem and also Riemann solver approaches are not free
of drawbacks. The use of Riemann solvers in an SPH context is sometimes advertised to be beneficial for SPH,
since it gets rid of free artificial dissipation parameters that need to be gauged in large sets of benchmark tests.
It needs to be stressed, however,  that also Riemann solvers can introduce a fair amount of dissipation, similar
to non-steered artificial dissipation. For such cases limiters have been suggested, e.g. \citet{meng21,fang25},
to detect and suppress unwanted dissipation from the Riemann solver, actually very similar to the dissipation
limiters for artificial viscosity discussed in Sect.~\ref{sec:AV}. Moreover, Riemann solver approaches (e.g. the
Roe and the HLLC solver) can suffer from instabilities, e.g. the ``carbuncle phenomenon'' \citep{peery88}. While
a Riemann solver is an elegant and mathematical sound approach, it is not guaranteed to always yield better
results that a well-designed artificial viscosity scheme. In fact, given that many modern artificial viscosity
schemes are guided by the use of Riemann solvers in Eulerian hydrodynamics one should expect them to
perform similarly.

\citet{vila99} has pioneered the use of Riemann solvers in an SPH-ALE context,
where ALE stands for \enquote{Arbitrary Lagrangian Eulerian} which is discussed in more detail in the next
section. \citet{parshikov02}, in contrast, restricted themselves in their
\enquote{interparticle contact algorithm} to a purely Lagrangian SPH approach. In astrophysics, the use of
Riemann solvers in SPH has been pioneered by \citet{inutsuka02,iwasaki11,kitajima25}.
Contrary to \citet{parshikov02} who designed a dissipative first-order method, Inutsuka applied, already in the
first paper, a reconstruction approach to a particle interface. They started from a proper partition of unity/nullity, 
see Sect.~\ref{sec:kernel_interpolation}, of the form
\bea
1&=& \sum_b \frac{m_b}{\rho(\vr)} W_{h}(\vr-\vr_b)\\
0&=& \sum_b m_b \nabla \left[ \frac{W_{h}(\vr-\vr_b)}{\rho(\vr)} \right].
\eea
Their SPH approximation involves a convolution integral
\be
\tilde{f}(\vr)= \int \sum_b \frac{f(\vr')}{\rho(\vr')} W_h(\vr' - \vr) W_h(\vr' - \vr_b) dV'
\ee
that can only be solved approximately. \citet{cha03,cha10} derived and explored several Riemann-SPH versions.
\citet{sirotkin13} derived an alternative, first-order accurate
Riemann-SPH approach which was (exclusively) tested in shocks for
a Local-Lax--Friedrich (LLF) \citep{vanLeer77,davis88}, a Harten--Lax--van Leer (HLL) \citep{rider94} and an exact Riemann solver. They found no obvious advantage
of using an exact Riemann solver, in fact, the exact solver in their scheme always resulted in spikes
at the contact discontinuity.  More recent work, has focused on more accurate reconstruction schemes, see for example \citet{avesani14,avesani21}.

While various Riemann-solver SPH versions exist, we want to
summarize here a particularly simple formulation that is based on the original
Inutsuka approach \citep{inutsuka02}, but with several approximations to get rid
of the convolution integrals \citep{cha03,iwasaki11,puri14}. In Puri and Ramachandran's \citeyearpar{puri14} final equation set the density is calculated via summation and the  momentum and energy are evolved according to
\bea
\frac{d\vec{v}_a}{dt}&=& - \sum_b m_b P^\ast_{ab} \left( \frac{\nabla_a W_{h_a}(r_{ab})}  {\rho_a^2} + \frac{\nabla_a W_{h_b}(r_{ab})}  {\rho_b^2} \right)\\
\frac{du_a}{dt}&=& - \sum_b  m_b P^\ast_{ab} \left[ \vec{v}_{ab}^\ast - \dot{\vec{r}}_a^\ast\right] \cdot \left( \frac{\nabla_a W_{h_a}(r_{ab})}  {\rho_a^2} + \frac{\nabla_a W_{h_b}(r_{ab})}  {\rho_b^2} \right),
\eea
where $P^\ast_{ab}$ is the Riemann-solver solution for the pressure at the contact discontinuity,
\be
\vec{v}_{ab}^\ast= v_{ab}^\ast \hat{e}_{ab} + \frac{1}{2} 
\left[ \vec{v}_a + \vec{v}_b - (\tilde{v}_a + \tilde{v}_b) \hat{e}_{ab} \right],
\ee
$v^\ast_{ab}$ is the Riemann-solver solution for the velocity at the contact discontinuity and the tilde indicates a projection on the direction connecting the two particles
\be
\tilde{v}_k = \vec{v}_k \cdot \hat{e}_{ab},
\ee
where $k$ stands for either $a$ or $b$. The intermediate velocity $\dot{\vec{r}}_a^\ast$
is given by
\be
\dot{\vec{r}}_a^\ast= \dot{\vr}_a + \frac{\Delta t}{2} \ddot{\vr}_a.
\ee
For the Riemann solver either exact or approximate Riemann solvers, e.g. as the Local-Lax--Friedrich \citep{vanLeer77,davis88}, Roe's solver \citep{roe81}  or the HLLC\footnote{The HLLC Riemann solver is a variant of the HLL solver, but restores the contact discontinuity. It stands for Harten-Lax-van Leer-Contact.} solver \citep{toro94,toro99} have been explored in \citet{puri14}, where the authors find no obvious advantage of using the exact Riemann solver.

More recently, \citet{rosswog25a} has started from the ``interparticle contact algorithm''  \citep{parshikov02} and extended the approach by i) reproducing kernels $\mathcal{W}$, see Sect.~\ref{sec:RPK}, ii) linear reconstruction with various slope limiters and iii) used the intermediate state values (labelled by a $^\ast$) of Roe's approximate Riemann solver \citep{roe81}. Their final equation set reads
\bea
\rho_a&=&   \sum_b m_b \bar{W}_{ab}\label{eq:dens_sum3}\\
\frac{d \vec{v}_a}{dt} &=& - \frac{2}{\rho_a} \sum_b V_b P_{ab}^\ast  (\nabla \mathcal{W})_{ab} \label{eq:dvdt3}\\
\frac{d u_a}{dt}&=&  \frac{2}{\rho_a}  \sum_b  V_b  P_{ab}^\ast  (\vec{v}_a - \vec{v}^\ast_{ab}) \cdot (\nabla \mathcal{W})_{ab} \label{eq:dudt3}.
\eea
In their tests, where they used a high order harmonic-like kernel, see Appendix \ref{sec:kernels} and found very good results in particular for the \texttt{vanAlbada limiter}, see Eq.~(\ref{eq:vanAlbada}). A follow-up study \citep{rosswog26a} compared this approach with modern artificial viscosity approach, also with reproducing kernels.
This study found only small advantages for the Riemann solver approach for the shock tests, but clearly slower growth of instabilities,
thus giving overall preference to the artificial viscosity approaches. This result, however, may be specific to the tested formulations and there
is likely room for improvements on the Riemann-solver + SPH side.

\subsection{SPH in the landscape of particle hydrodynamics methods}
\label{sec:SPH_landscape}
SPH is the archetype of a mesh-free particle method, but many other
particle methods have emerged over time for solving fluid equations.
The last two decades have seen ample cross-fertilization between
different methods, the use of slope-limited reconstruction in
artificial dissipation terms, see Sect.~\ref{sec:AV},
is just one of many examples. It has, for example, been realized that
Finite Volume methods are not tied to meshes, but they can
also be formulated on a particle basis
\citep{vila99,benmoussa00,junk03,gaburov11,hopkins15a}.
The corresponding codes have a ``look-and-feel'' similar to SPH, one
deals with particles, kernels, smoothing lengths etc., but care is
taken to systematically average the evolution equations over particle
volumes. In SPH, in contrast, very much focus is on exact conservation
and in a standard approach one uses a density summation, the momentum
equation from a Lagrangian and the internal energy evolves according
to a straight forward translation of the first law of
thermodynamics. While such an approach enforces exact conservation it
comes with its own subtleties, for example a potentially different
smoothness of density and internal energy which can lead, for example,
to surface tension effects \citep{agertz07}, as explained in
Sect.~\ref{sec:volume_elements}.\\ 
Once formulated in a ``Finite volume language'', many techniques that have
been developed over decades in a mesh-based finite Volume context can
be translated to the particle-based context with comparatively little
effort. As of today, there is a plethora of ``hybrid methods'' that
combine various elements in different ways and that come under many
different names, which we will not try to summarize here. 

\begin{figure}[ht]
\centering
 \includegraphics[width=7cm]{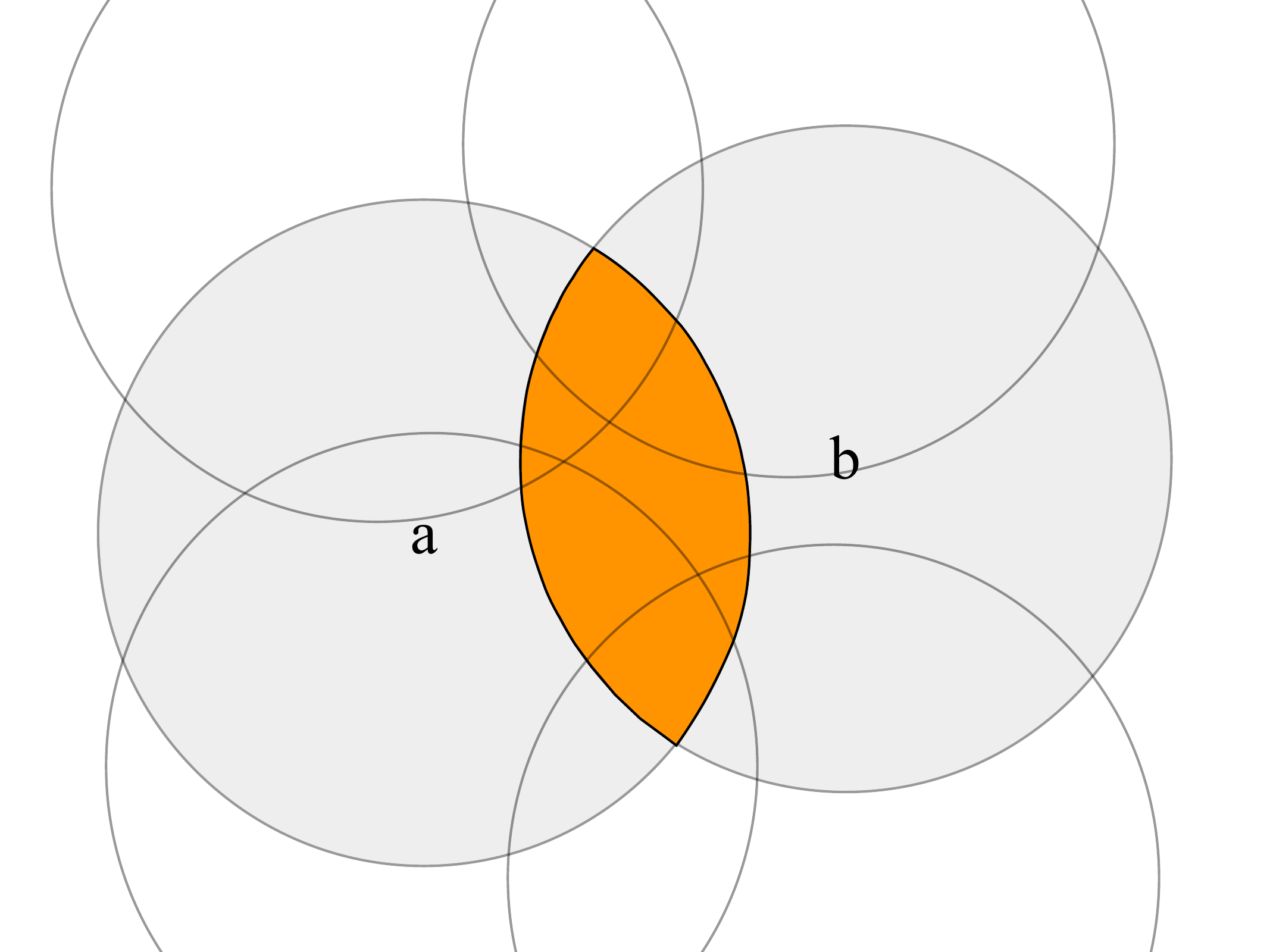}
 \caption{In a particle-based Finite Volume method the ``effective
   area vectors'' $\vec{A}_{ab}$ are calculated over the overlapping
   regions of the shape functions.
 }
   \label{fig:overlapping_area}
\end{figure}

We will  start here with the ideal hydrodynamics equations in conservative
form, we will then perform a general Finite Volume discretization
within an ALE (``Arbitrary Lagrangian Eulerian'') framework, derive a
general formulation and then explore how specific approximations lead
to different particle discretizations. Clearly, there are similarities with 
moving mesh methods, as implemented for example in the AREPO code 
\citep{springel10b} which also use Finite Volume discretizations in an ALE
framework. The major difference is that moving mesh codes tesselate space
into disjunct cells with sharp interfaces, while Finite Volume particle methods
work, similar to SPH, with overlapping particle volumes, see 
Fig.~\ref{fig:overlapping_area}.

We start from a general conservation law
\be
\partial_t \bf{q} + \nabla \cdot \bf{F}_E(\bf{q})= s,
\label{eq:cons_law}
\ee
where $\bf{q}$ is the state vector and $\bf{F}_E$ and $\bf s$ are the
corresponding fluxes and source terms. While the procedure is
generally valid for different conservation laws (e.g. also including
viscous fluxes) we are here mostly interested in ideal fluids for
which the state vector reads
\be
\bf{q}= \left(
\begin{array}{c}
\rho\\
\rho \vec{v}\\
e
\end{array}
\right)
\ee
with $e= \rho(u + v^2/2)$ being the total energy \emph{per volume} and $u$  the specific
internal energy. The corresponding flux tensor reads
\bea
{\bf{F}_E}
&=& \left(
\begin{array}{c}
\rho \vec{v} \\
\rho \vec{v} \; \otimes \vec{v}  + P \bf{I}\\
(e + P) \vec{v}
\end{array}
\right),
\eea
where $\vec{v}$ is the fluid velocity.

To be more general, one can write the above conservation law in a
frame moving with velocity $\vec{w}$ so that it reads
\be
\p_t {\bf q} + \nabla \cdot  {\bf F_w} = {\bf s}
\label{eq:ALE_cons_law}
\ee
where the ALE flux is given by
\be
{\bf F_w} = {\bf F}_E - {\bf w} \otimes {\bf q}.
\label{eq:ALE_flux}
\ee
\noindent To find a particle-based Finite Volume discretization, 
we now assume that we have shape functions $\Phi(\vr,h(\vr))$
whose support is determined by $h$ and that, attached to a particle $b$,
read $\Phi_b(\vr)= \Phi(\vr -\vr_b,h_b)$. These shape functions should
form a partition of unity and their gradients a partition of nullity
(since $\nabla 1= 0$), see Sect.~\ref{sec:discrete_approx},
\be
\sum_b \Phi_b(\vr)= 1 \quad \text{and} \quad \sum_b \nabla \Phi_b(\vr)= 0,
\label{eq:PU}
\ee
so that approximations to functions $f$ and their gradients can be represented as
\be
\tilde{f}(\vr)=\sum_b f_b \Phi_b(\vec{x}) \quad \text{and} \quad
\tilde{(\nabla f)}(\vr)= \sum_b f_b \nabla \Phi_b(\vec{x}).
\label{eq:PU_function_approx}
\ee
By ``inserting a 1'' according to Eq.~(\ref{eq:PU}) one finds for the
total volume
\be
V= \int 1 dV= \int \sum_b \Phi_b(\vr) dV= \sum_b \int \Phi_b(\vr) dV=
\sum_b V_b,
\ee
so that we have
\be
V_b= \int \Phi_b(\vr) dV.
\ee
If we define the average of a function $f$ over the volume of a
particle $a$
\be
\bar{f}_a \equiv \frac{\int f(\vr) \Phi_a(\vr) dV}{\int \Phi_a(\vr) dV},
\ee
we have
\be
\int f(\vr) \Phi_a(\vr) dV= V_a \bar{f}_a.
\label{eq:one_point_quad}
\ee
Multiplying the conservation law Eq.~(\ref{eq:ALE_cons_law}) with
shape function $\Phi_a(\vr)$ and integrating over the volume one finds
\be
\frac{dQ_a}{dt} + \int \left( \nabla \cdot {\bf F_w}\right)
\Phi_a(\vr) dV = \int {\bf s}(\vr) \Phi_a(\vr) dV, 
\ee
where we have abbreviated $\int {\bf q}(\vr) \Phi_a(\vr) dV$ as $Q_a=
V_a \bar{q}_a$, which are the particle-volume integrated conserved
quantities
\be
\bf{Q}_a= \left(
\begin{array}{c}
m_a\\
\vec{P}_a\\
E_a
\end{array}
\right),
\ee
corresponding to mass, momentum and total energy of particle $a$.
If we now insert the discrete approximation Eq.
(\ref{eq:PU_function_approx}) for flux divergence $\nabla \cdot {\bf
  F_w}= \sum_b {\bf F}_b \cdot \nabla \Phi_b(\vr)$, one finds
\be
\frac{dQ_a}{dt} + \sum_b {\bf F}_b \cdot \int \Phi_a(\vr) \; \nabla
\Phi_b(\vr) dV= \bar{\bf s}_a V_a. \label{eq:n1}
\ee
The integral in this equation can be integrated by parts, so  that it
reads
\be
\oint \Phi_a(\vr)\Phi_b(\vr) dA - \int \Phi_b(\vr) \nabla \Phi_a(\vr) dV.
\ee
The surface integral would be the natural place to incorporate
boundary conditions, but since such surface terms are unimportant in
most astrophysical applications and are also ignored in standard SPH, we also
ignore them here. If we insert this approximation in
Eq.~(\ref{eq:n1}), we find
\be
\frac{dQ_a}{dt} - \sum_b {\bf F}_b \cdot \int \Phi_b(\vr) \; \nabla
\Phi_a(\vr) dV= \bar{\bf s}_a V_a. \label{eq:n2}
\ee
Taking the average of Eqs.~(\ref{eq:n1}) and (\ref{eq:n2})
yields
\be
\frac{dQ_a}{dt} + \frac{1}{2} \sum_b {\bf F}_b \int \left[
  \Phi_a(\vr)  \nabla \Phi_b(\vr) -  \Phi_b(\vr)  \nabla \Phi_a(\vr)
\right] dV = \bar{\bf s}_a V_a.
\label{eq:n3}
\ee
The integral
\be
\vec{A}_{ab} \equiv \int \left[  \Phi_a(\vr)  \nabla \Phi_b(\vr) -  \Phi_b(\vr)  \nabla \Phi_a(\vr)
\right] dV
\label{eq:area_vector}
\ee
is calculated over the overlapping shape functions
$\Phi_k$, see Fig.~\ref{fig:overlapping_area}, and has the dimension of a surface.
It is therefore often called ``effective area vector'' and it is the
overlapping-particle equivalent to the area-vector in a mesh-based
Finite Volume scheme. The area vector
obviously is anti-symmetric,
\be
\vec{A}_{ab}= - \vec{A}_{ba}
\ee
and also fulfills the ``closure condition''
\be
\sum_b \vec{A}_{ab} =0. \label{eq:A_ab_closure}
\ee
This can be seen by using the partition of nullity, Eq.~(\ref{eq:PU}),
\be
\sum_b \int \Phi_a(\vr)  \nabla \Phi_b(\vr) dV = 0,
\label{eq:n4}
\ee
integrating it by parts  and dropping again the surface term. This yields
\be
\sum_b \left\{ \oint \Phi_a(\vr)  \Phi_b(\vr)
  d\vec{A} - \int \Phi_b(\vr)  \nabla \Phi_a(\vr) dV   \right\}
\simeq - \sum_b \int \Phi_b(\vr)  \nabla \Phi_a(\vr) dV   = 0
\label{eq:n5}
\ee
and by adding  Eqs.~(\ref{eq:n4}) and (\ref{eq:n5}) we find
\be
0= \sum_b \int \left\{ \Phi_a(\vr)  \nabla \Phi_b(\vr) - \Phi_b(\vr)
  \nabla \Phi_a(\vr) \right\} dV= \sum_b \vec{A}_{ab}.
\ee
Now back to the conservation law Eq.~(\ref{eq:n3}), we can subtract
$\frac{1}{2} {\bf F}_a \cdot \sum_b   \vec{A}_{ab}= 0$ from it to find
\be
\frac{dQ_a}{dt} + \sum_b {\bf \bar{F}}_{ab} \cdot \vec{A}_{ab}=
\bar{\bf s}_a V_a,
\label{eq:FVP_cons_law}
\ee
where we have abbreviated the average flux as
\be
{\bf \bar{F}}_{ab} = \frac{1}{2} ({\bf F}_{a} + {\bf F}_{b}).
\ee
Equation (\ref{eq:FVP_cons_law}) looks exactly like a mesh-based
Finite Volume evolution equation with the difference, that the area
vector $\vec{A}_{ab}$ does not refer to sharp cell
interfaces, but instead it is calculated from the integral over
overlapping shape functions according to Eq.~(\ref{eq:area_vector}).
Due to the anti-symmetry, the fluxes from particle $a$ to $b$ are the
negative of the fluxes from particle $b$ to $a$. Therefore the
discrete equations are manifestly conservative.
To proceed to a concrete particle-based finite volume scheme, one needs
to find suitable ways to calculate the area vector $\vec{A}_{ab}$ and
to evaluate the interparticle fluxes ${\bf \bar{F}}_{ab}$, and here
obviously different choices and approximations are possible.

Since such Finite Volume evolution equations are well studied in a
mesh-based context, one can rather straight-forwardly apply the
corresponding techniques to a particle context, for example,
apply slope-limited reconstructions to interfaces between particles,
see also Sect.~\ref{sec:AV},
and use the corresponding values as inputs to  solve inter-particle
Riemann problems, e.g., via approximate Riemann solvers
\citep{toro99}. Since only the projection of the flux $\bar{\bf
  F}_{ab}$ to the effective area
vector $\vec{A}_{ab}$ is needed, one only needs to solve a
one-dimensional, unsplit Riemann problem.

As derived above, the method is ``Arbitrary Lagrangian Eulerian'',
i.e. by specifying the velocity $\vec{w}$, one can set the mode in
which the code is run. The most obvious choice is to use the
``Lagrangian mode'', i.e. $\vec{w}$ is set equal to the local fluid
velocity, but one could as well set $\vec{w}=0$ to run in a ``Eulerian
mode'' where the particles do not move (although this would not have
any obvious advantage). At least in principle (but to our knowledge
not much used), one could actually use a velocity different from zero
and from the local fluid velocity, for example, to refine on some
specific region, but one could also allow the particles to move with a
small non-Lagrangian component, e.g. to allow them to locally optimize
their position.

The first Finite Volume approach with particles in astrophysics is due
to Gaburov and Nitadori who formulated an
``astrophysical weighted particle magnetohydrodynamics''
method \citep{gaburov11}. Somewhat later, Hopkins presented GIZMO \citep{hopkins15a}, a
Finite Volume particle method implemented in the infrastructure of the
GADGET-3 code \citep{springel01a,springel05a}. GIZMO distinguishes between
two sub-variants, the ``Meshless Finite Volume (MFV)'' method and the
``Meshless Finite Mass (MFM)'' method which differ only at second
order. In the MFV method the Riemann problem is solved in a stationary
frame with effective frame velocity $\vec{v}_\mathrm{eff}^{\rm \; frame}=
0$, while in the MFM method the frame is assumed to move with the
velocity of the contact discontinuity (``the star state''). Since
there is no mass flux through contact discontinuities, this means
that the particles do not exchange mass, i.e. they keep their original
mass as in standard SPH. By now, there are further implementations of
Finite Volume methods in astrophysical fluid dynamics, see for example
\citet{hubber18}  or \citet{groth23}.

\subsection{Importance of initial conditions}
\label{sec:IC}
It is not always sufficiently appreciated  how important the initial conditions are for the accuracy of SPH simulations.
Unfortunately, it can become rather non-trivial to construct them. One obvious requirement is the regularity 
of  a particle distribution so that the quality
indicators $\mathcal{Q}_1 - \mathcal{Q}_4$, Eqs.~(\ref{eq:quality_int}) and (\ref{eq:Q3_Q4}), 
are fulfilled with high accuracy. This suggests placing the particles initially on some type of lattice.
However, these lattices should ideally possess at least two more properties: a) they should not contain preferred directions
b) they should represent a \emph{stable} minimum energy configuration for the applied SPH formulation.
Condition a) is desirable since otherwise shocks propagating along a symmetry axis of a lattice will 
continuously ``collect'' particles in this direction and this can lead
to ``ringing effects'' \citep{morris96b,lombardi99}. Condition b)
is important since a regular lattice does by no means guarantee that the configuration is actually an
equilibrium of the SPH particles. If it is not, particles will start to move ``off the lattice'' and this
will introduce noise.

The above discussion suggests the use of some ``glass-like'' particle distribution: regular, but without
preferred directions. A heuristic approach to construct good, low-noise initial conditions, say for a star
in hydrostatic equilibrium, is to apply a velocity-dependent damping term,
\be
\vec{f}_\mathrm{damp} \propto - \frac{\vec{v}_a}{\tau_a},
\ee
with a suitably chosen damping time scale $\tau_a$
to the momentum equation. This ``relaxation process'' can be applied
until some convergence criterion is met (say, the kinetic 
energy has dropped below some threshold). 
This procedure, however, becomes difficult to apply for more
complicated initial conditions.

\citet{diehl12} suggested a variant of a ``weighted
Voronoi tessellation'' to construct more general initial
conditions. Assuming a desired average distance to neighbor particles, $\delta(\vr)$,
is known at each position $\vr$, they start from a guessed particle
distribution and iteratively improve the positions. The motion
amplitude is reduced with each iteration so that the
process slowly ``freezes'' to the final particle distribution. The authors
find very smooth particle distributions that perform very well,
e.g. in a Sedov explosion test, compared to various prescribed lattice
setups. We refer to the original paper for more details on the methods
and results.

\citet{kaltenborn23} suggest what they call a ``halted
pendulum relaxation'' to construct, e.g. single and binary stars. The
main idea is to find the minimum of a potential energy (here of the
particle configuration) and to monitor the global kinetic
energy. Motivated by a swinging pendulum where the maximum kinetic
energy is reached when the potential energy is minimal, they monitor
the kinetic energy in the to-be-relaxed system and once it reaches a
maximum, they set all velocities to zero and then continue the
evolution. In this simple way, they find good initial conditions for
astrophysically relevant configurations such as mass transferring
binary systems.

Another approach to generate arbitrary particle distributions of good interpolation quality is
the so-called ``Artificial Pressure Method'' (APM) \citep{rosswog20a}.
The aim is to reproduce, as accurately as possible, a given density profile
with a particle distribution where each particle has a given, prescribed mass.
One may wonder ``why not use a uniform particle distribution and enforce the
desired density distribution by choosing the particle masses?''. This may actually
be an option for simple cases with small density contrasts, so that the resulting
differences in the SPH particle masses are small. But if the density
contrasts are large, such an approach may produce a lot of noise when
particles of very low mass interact with high-mass particles (``ping
pong ball vs cannon ball effect'') and this should be absolutely avoided.

\begin{figure}[ht]
   \centering
   \includegraphics[width=10cm]{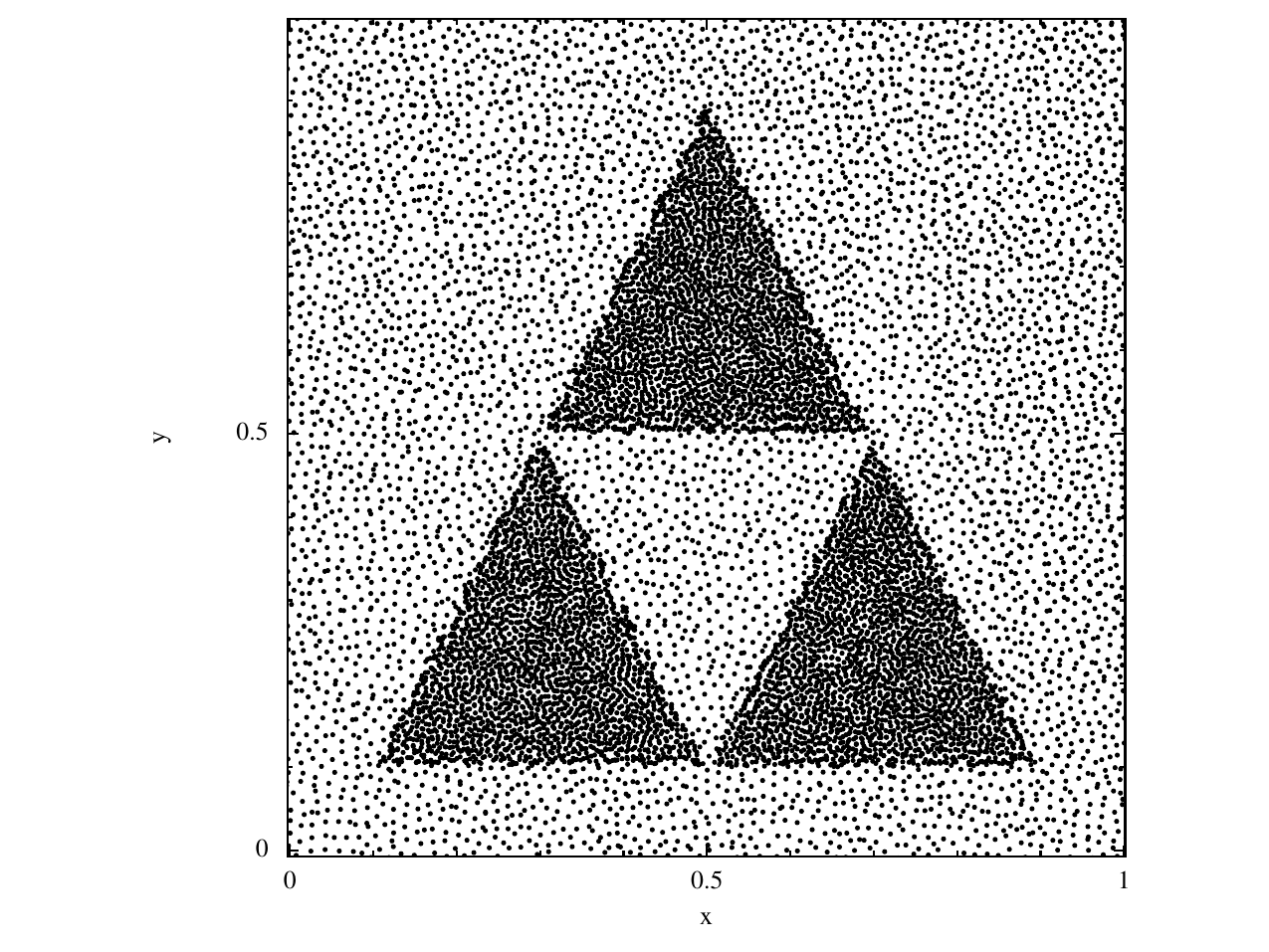} 
   \caption{Illustration of setting up particles according to the ``Artificial Pressure Method'' \citep{rosswog20a}.}
   \label{fig:APM_triangles}
\end{figure}

The main idea of the APM is to start from a guess distribution of
particles, measure their local error compared to the desired result
and translate this error into an “artificial pressure”. This  pressure
is then used in an equation very similar to a hydrodynamic momentum
equation to push the particle into a position where it reduces its error.
Instead of integrating this ``pseudo-momentum equation'' one performs
iterations, each time driving a particle into a better location. As an example,
we show a configuration with a rather complicated
density profile with three high density triangles that are embedded in a uniform background,
that we try to reproduce with equal mass particles, see
Fig.~\ref{fig:APM_triangles} \citep{rosswog20a}.
The method has also been translated to the general relativistic case where it was used
to construct both single and binary neutron stars \citep{rosswog21a,diener22a,rosswog22b,
rosswog23a}.
An iterative method in a similar spirit of the APM was recently devised by \cite{wissing26a}.

\subsection{Some numerical examples}
\label{sec:numerical_examples}
We want to show here some numerical examples to illustrate the impact of different methodological 
choices. The first test (``Advection'') is insensitive to specific methodological choices, here we pick a special-relativistic SPH
formulation from \citet{rosswog15b} since we want to demonstrate the
excellent advection properties even at very large velocities.
For the subsequent tests we show the results for  five
different SPH versions. In all of them we use reproducing kernel gradients
wherever reconstructions are needed and we always use the van Albada
slope limiter, see Eq.~(\ref{eq:vanAlbada}) in Sect.~\ref{sec:AV}. The
density is calculated in all cases via summation.
\begin{itemize}
\item{\bf Version $\mathcal{V}_0$}: 
This is an ``old school'' SPH formulation where we use the ``vanilla ice'' SPH version, see
  Sect.~\ref{sec:vanilla_ice}, together with constant artificial
  dissipation, see Eq.~(\ref{eq:Qvis}), with $\beta=2 \alpha= 2=$
  const. without reconstruction and a cubic spline kernel with $50$ neighbours.
  This version is hardly used today, here it serves the pedagogical purpose
  to illustrate which effects of a combination of poor SPH choices can have.
\item {\bf Version $\mathcal{V}_1$}: This is the same as $\mathcal{V}_0$, the only change is
that we use the Wendland C4 kernel with 250 neighbours. So comparing $\mathcal{V}_1$ with 
$\mathcal{V}_0$ illustrates the {\em impact of the kernel choice}.
 \item {\bf Version $\mathcal{V}_2$}:  We apply here
 slope-limited reconstruction and we steer the parameters
 via a Cullen--Dehnen approach \citep{cullen10} with one modification:
 we use
 \be
\alpha_{a, \rm des}= \alpha_{\max} \frac{A_a}{A_a + 0.1 (c_a/h_a)^2}
\ee
instead of Eq.~(\ref{eq:trigger_CD}). The additional factor 0.1 leads
to an earlier switch-on of the dissipation which we find useful to damp
post-shock oscillations in Schulz-Rinne tests \citep{schulzrinne93a}. For the
decay time scale in Eq.~(\ref{eq:alpha_steering_MM}) we use $\tau_a=30$.
Steering dissipation via entropy conservation monitoring
\citep{rosswog20b}, see Sect.~\ref{sec:AV}, yields similar results. 
We further apply a small amount of conductivity using Eq.~(\ref{eq:du_diss})
with $v_{\rm sig}= \sqrt{(|P_a - P_b|)/\bar{\rho}_{ab}}$ and $\alpha_u= 0.05$.
We use this dissipation prescription in all shown experiments.
So comparing $\mathcal{V}_2$ with $\mathcal{V}_1$
 illustrates the {\em effect of reduced dissipation}.
 \item  {\bf Version $\mathcal{V}_3$}: Similar to $\mathcal{V}_2$, but
 here we change the symmetries of the SPH equations  to the ``GDF'' form, see Eqs.~(\ref{eq:mom_GDF}) and
 (\ref{eq:en_GDF}), which has been found advantageous in several
 studies  \citep{oger07,read10,wadsley17,rosswog20a,wissing20,wissing22} 
 without any known drawbacks. Comparing version $\mathcal{V}_3$ with version $\mathcal{V}_2$
 illustrates the {\em effect of different SPH symmetrizations}.
\item  {\bf Version $\mathcal{V}_4$}: This version is the same as version
$\mathcal{V}_3$ apart from using the reproducing kernel gradient
$\nabla \mathcal{W}_{ab}$, see
Eq.~(\ref{eq:RPK_kernel_gradient}). Comparing  version $\mathcal{V}_4$  
with  version $\mathcal{V}_3$ therefore illustrates the {\em effect of improved
gradient accuracy and the exact partition of unity, see Eq.~(\ref{eq:consistency_relations}).}
\end{itemize}

\subsubsection{Advection}
\label{sec:advection_test}

\begin{figure}[ht]
   \centerline{\includegraphics[width=0.7\textwidth]{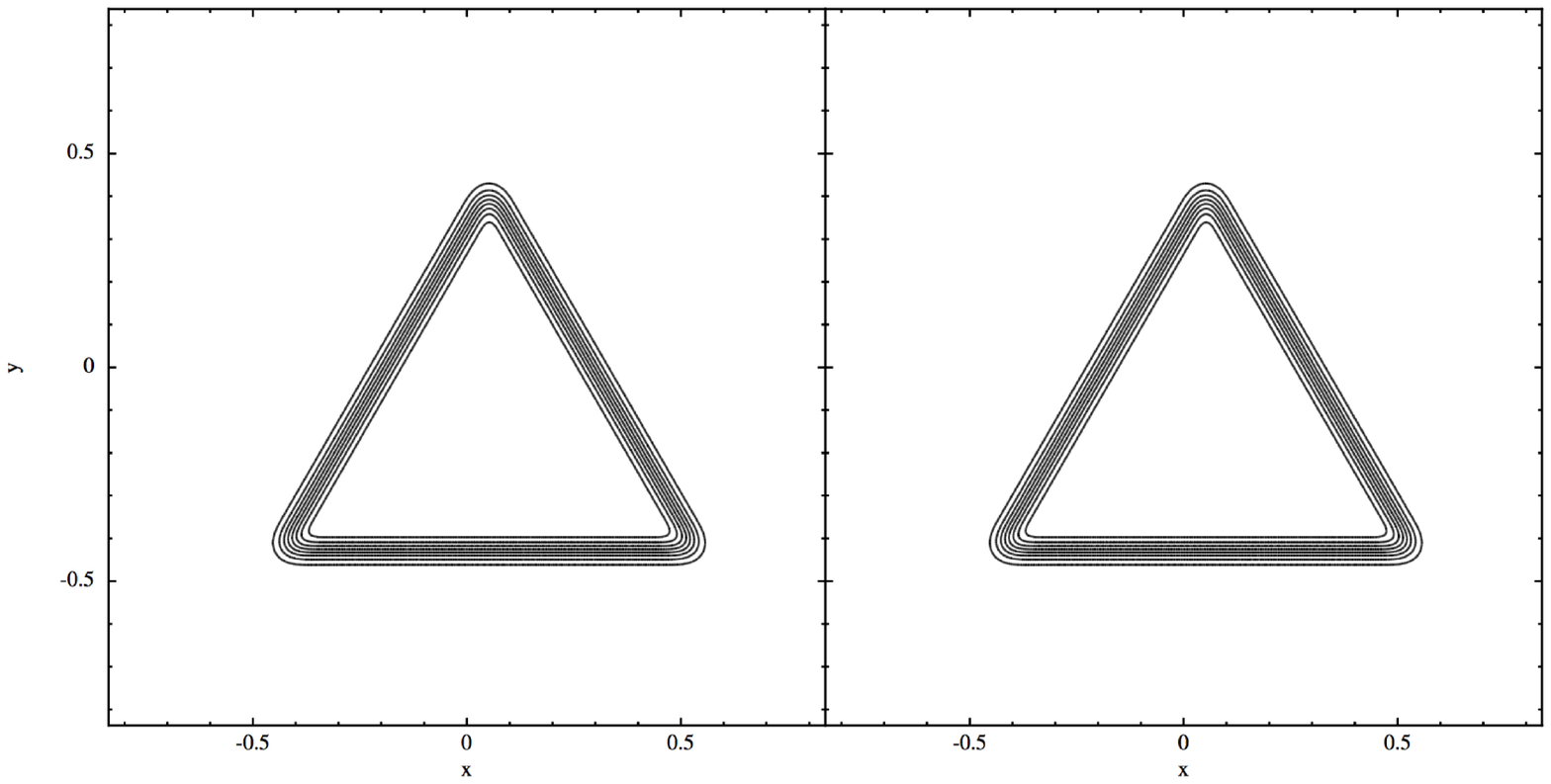}}
    \caption{Demonstration of SPH's superb advection properties. A high-density wedge ($N=2$) 
in pressure equilibrium with the background ($N=1$) is advected with velocity $v_x= 0.9999$, 
corresponding to a Lorentz factor of $\Gamma= 70.7$, through a box with periodic boundaries. 
For this test only 7 000 particles are used. After crossing the box 10 times (or more than 23\,000 
time steps; right panel) no deterioration of the shape with respect to the initial condition (left panel) 
is noticeable.}
    \label{fig:advection}
\end{figure}

SPH is essentially perfect in terms of advection: if a particle carries a certain property,
say some electron fraction, it simply carries this property along as
it moves (unless the property is changed by physical processes, of
course). The advection accuracy is, contrary to Eulerian schemes, 
independent of the numerical resolution and just governed by the time
integration accuracy of the involved ordinary differential equations.
We briefly want to illustrate the advection accuracy  in an example
that is a very serious challenge for Eulerian schemes, but essentially
a ``free lunch'' for SPH: the highly supersonic advection of a density 
pattern through the computational domain with periodic boundary
conditions.  To this end, we set up  (only!) 7 000 SPH particles inside the domain
$[-1,1] \times [-1,1]$. Each fluid element is assigned a velocity in the x-direction of 
$v_x= 0.9999$ corresponding to a Lorentz factor of $\gamma\approx 70.7$ and 
periodic boundary conditions are applied. The simulation is performed
with the special relativistic \texttt{SPHINCS\_SR} code
\citep{rosswog15b}.\footnote{More specifically, the SPH formulation
  $\mathcal{F}_1$ is used, see Sect.~7.5 in \citet{rosswog15b} for the specifics.}
The result of this experiment is shown in 
Fig.~\ref{fig:advection}: after crossing the computational domain ten times (more than 
23 000 time steps; right panel) the shape of the triangle has not changed in any 
noticeable way with respect to the initial condition (left panel). 

\subsubsection{Sedov explosion}
\label{sec:Sedov}
The Sedov--Taylor test, a strong initial point-like explosion
expanding into a low-density environment, is one of the standard
benchmark tests for hydrodynamics codes \citep{sedov59,taylor50}.
For an explosion energy $E$ and a density of the ambient medium
$\rho$, the blast wave propagates to the radius
$r(t)= \beta (E t^2/\rho)^{1/5}$ at  time $t$, where $\beta$ depends on the
adiabatic exponent of the gas ($\approx 1.15$ in 3D  for the
$\Gamma=5/3$ we use).  In the strong explosion limit, the density
jumps by a factor of  $\rho_2/\rho_1= (\Gamma + 1)/(\Gamma-1)= 4$ at
the shock front, where the numerical value refers to our chosen 
value of $\Gamma$. We place $128^3$ particles according to a
Centroidal Voronoi Tessellation \citep{du99}  in the volume
[-0.5,0.5]$^3$. For an even better particle distribution we perform
additional sweeps according to our ``Artificial Pressure Method'' (APM),
see Sect.~\ref{sec:IC} and \citet{rosswog20a}.

\begin{figure}[ht]
   \centering
   \includegraphics[width=\textwidth]{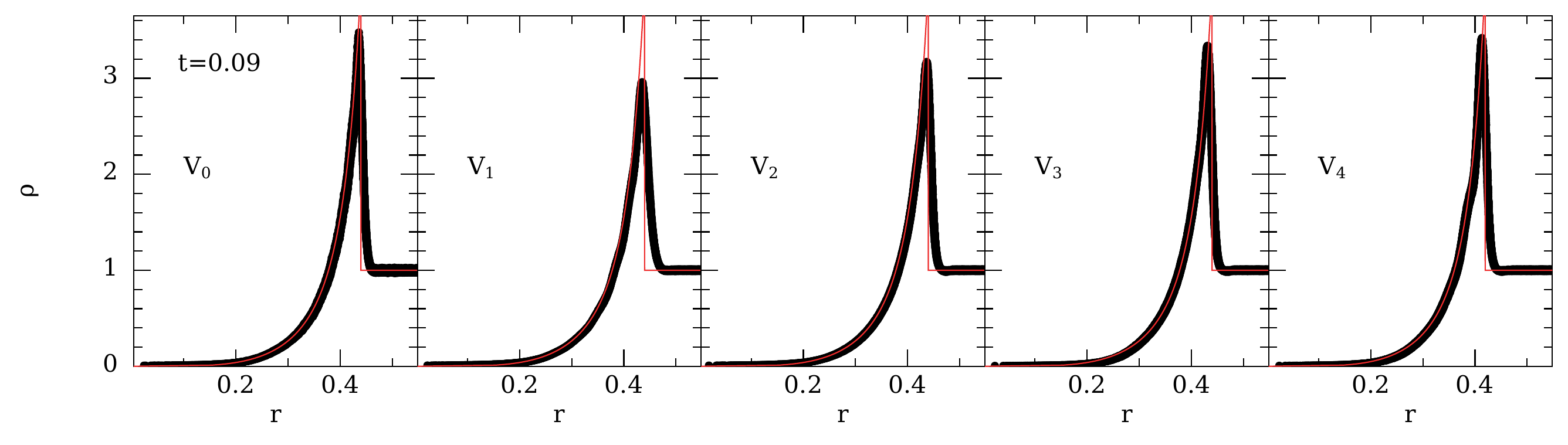}
   \caption{Comparison between different methodological choices in
      the Sedov explosion test. Left panel: ``old school'' choices ($\mathcal{V}_0$), 
      in particular using the cubic spline kernel with 50 neighbours and full dissipation everywhere; second
    panel: like $\mathcal{V}_0$, but using the Wendland C4 kernel with 250 neighbours instead; 
    third panel: like $\mathcal{V}_1$ but with reduced dissipation
    ($\mathcal{V}_2$); fourth panel: like $\mathcal{V}_2$, but
    different SPH discretization ($\mathcal{V}_3$); fifth panel: like
    $\mathcal{V}_3$, but with linearly reproducing kernels
    ($\mathcal{V}_4$). See main text for full details.}
    \label{fig:Sedov}
\end{figure}
We show in  Fig.~\ref{fig:Sedov} the results for $\mathcal{V}_0$ to
$\mathcal{V}_4$. Overall, all results agree very well with the analytic
solution (red), the major difference is in the  peak heights and in some 
cases minor post-shock oscillations are visible. The "old school" version $\mathcal{V}_0$
performs here pretty well: since it only samples over 50 neighbours, it reaches the 
largest peak density (= 3.44) and the large, constant dissipation keeps the post-shock region
well-behaved. Part of the good behaviour in the post-shock region is also be due to the carefully prepared initial conditions. 
Changing to $\mathcal{V}_1$, where now the Wendland C4 kernel with
250 neighbours is used, reduces the peak height  (2.96). Reducing the dissipation
($\mathcal{V}_2$) increases the peak height again (to 3.17) without any visible artifacts.
The only change in going from $\mathcal{V}_2$ to $\mathcal{V}_3$ is the 
use of the GDF-form of the SPH equations. This leads to a better resolution of sharp gradients,
 which is consistent with observations in the literature 
\citep{oger07,read10,wadsley17,rosswog20a,wissing20,wissing22}, here the peak height is 3.34.
Yet another improvement comes from using linearly reproducing kernels in $\mathcal{V}_4$
(peak at 3.42) though at the price of very small post-shock oscillations. So with
the combination of all improvements ($\mathcal{V}_4$) one recovers with 250 neighbours
the peak height of the 50 neighbour result ($\mathcal{V}_0$).

\subsubsection{Kelvin--Helmholtz instability}
\label{sec:KeHe_test}
\begin{figure}[ht]
   \centering
   \hspace*{-1cm}\includegraphics[width=1.1\textwidth]{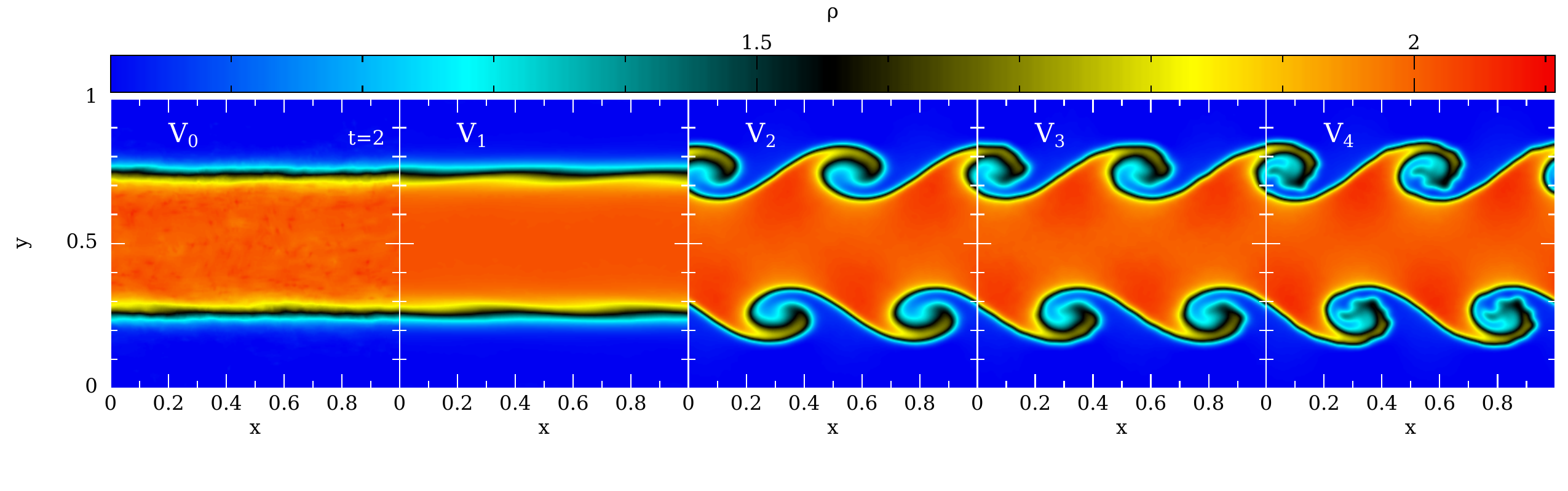}
   \vspace*{-0.4cm}
   \caption{\bf Kelvin-Helmholtz instability test ($128^2$, $t=2.0$) for $\mathcal{V}_0$ to $\mathcal{V}_4$. Here, the full-dissipation
   versions ($\mathcal{V}_0, \mathcal{V}_1$) do not grow, reducing the dissipation make the decisive difference.
   The GDF-version ($\mathcal{V}_3$) grows slightly faster than the vanilla ice version and improving gradient
   accuracy ($\mathcal{V}_4$)  helps further.}
    \label{fig:KeHe}
\end{figure}

Traditional versions of SPH have been shown to struggle with weakly triggered Kelvin-Helmholtz  instabilities \citep{agertz07,mcnally12}.
We want to show here the impact of different methodological choices on this test.
We use a test setup in which traditional SPH has been shown to fail, even at a rather high resolution  in 2D, see \cite{mcnally12}.
We follow the latter paper, but 
we use the full 3D code and set up the "2D" test as a thin 3D slice with $N \times N \times 20$ particles (referred to as $"N^2"$).
For simplicity, the particles are initially placed on a cubic lattice. Periodic boundary conditions are enforced by placing 
appropriate particle copies outside of the "core" volume. The test is initialized as:
\be
\rho(y)=
 \left\{
\begin{array}{l}
\rho_1 - \rho_m e^{(y - 0.25)/\Delta} \quad {\rm for \; \; 0.00 \le y < 0.25}\\
\rho_2 + \rho_m e^{(0.25 - y)/\Delta} \quad {\rm for \; \; 0.25 \le y < 0.50}\\
\rho_2 + \rho_m e^{(y - 0.75)/\Delta} \quad {\rm for \; \; 0.50 \le y < 0.75}\\
\rho_1 - \rho_m e^{(0.75 - y)/\Delta} \quad {\rm for \; \; 0.75 \le y < 1.00},\\
\end{array}
\right.
\ee
where $\rho_1= 1$, $\rho_2= 2$, $\rho_m= (\rho_1 - \rho_2)/2$ and $\Delta= 0.025$.
The velocity is set up as
\be
v_x(y)=
 \left\{
\begin{array}{l}
v_1 - v_m e^{(y - 0.25)/\Delta} \quad {\rm for \; \; 0.00 \le y < 0.25}\\
v_2 + v_m e^{(0.25 - y)/\Delta} \quad {\rm for \; \; 0.25 \le y < 0.50}\\
v_2 + v_m e^{(y - 0.75)/\Delta} \quad {\rm for \; \; 0.50 \le y < 0.75}\\
v_1 - v_m e^{(0.75 - y)/\Delta} \quad {\rm for \; \; 0.75 \le y < 1.00}\\
\end{array}
\right.
\ee
with $v_1$= 0.5, $v_2= -0.5$, $v_m= (v_1-v_2)/2$ and a small velocity perturbation in 
$y$-direction is introduced as $v_y=  0.01 \sin(2\pi x/\lambda)$ with the perturbation 
wave length $\lambda= 0.5$. In the linear regime, a Kelvin-Helmholtz instability grows 
in the incompressible limit on a characteristic time scale of
\be
\tau_{\rm KH}= \frac{(\rho_1+\rho_2) \lambda}{\sqrt{\rho_1 \rho_2} |v_1-v_2|},
\ee
with $\tau_{\rm KH}\approx 1.06$ for the chosen parameters.
The tests use a polytropic equation of state with
exponent $\Gamma=5/3$ and a deliberately low resolution ($128^2$) 
to see  clear distinctions between the different versions, see Fig.~\ref{fig:KeHe}. 

\begin{figure}[ht]
   \centering
   \includegraphics[width=0.6\textwidth]{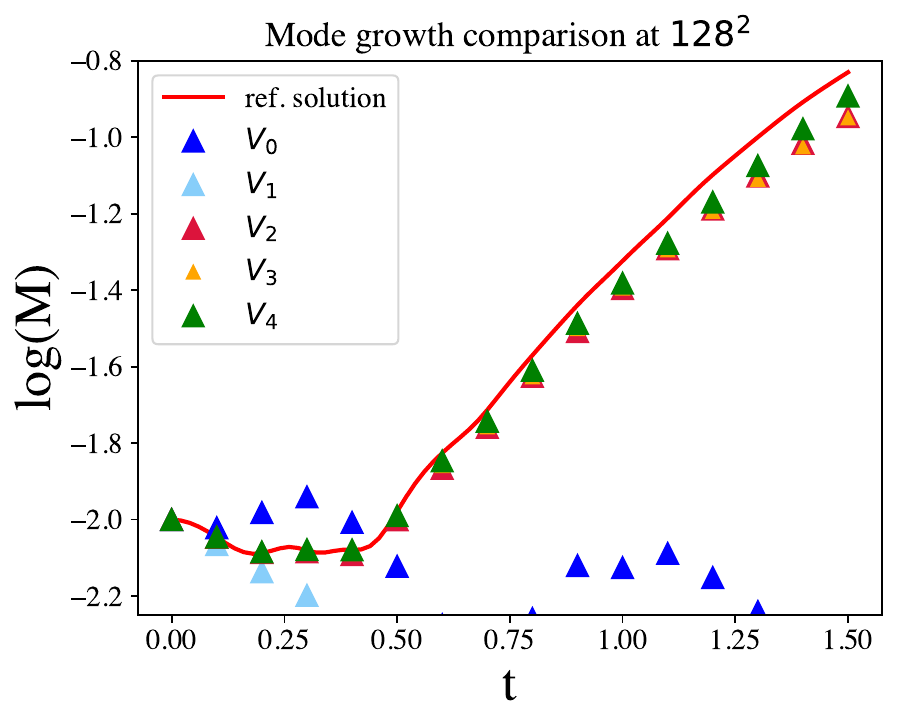}
   \caption{Growth rates of Kelvin-Helmholtz instability tests ($128^2$) for versions $\mathcal{V}_0$ to $\mathcal{V}_4$. 
   Note that the marker size of $\mathcal{V}_3$ has been kept intentionally smaller to keep $\mathcal{V}_2$ visible.
   Several points of $\mathcal{V}_0$ and $\mathcal{V}_1$ are below the shown ordinate range.}
    \label{fig:KH_growth_rate}
\end{figure}

The two full-dissipation versions  ($\mathcal{V}_0, \mathcal{V}_1$)
do not show noticeable growth, not for the cubic spline, but neither for the Wendland kernel. The major improvement
comes from reducing the (un-needed) dissipation ($\mathcal{V}_2$). The GDF-  ($\mathcal{V}_3$) and 
the "vanilla ice" ($\mathcal{V}_2$) versions grow at similar speeds, but more accurate gradients ($\mathcal{V}_4$) 
further accelerate the growth of the instability.\\
For a more quantitative assessment, we compare the growth rates of the instability (calculated exactly as in \cite{mcnally12})
 to a high-resolution ($4096^2$) simulation obtained with the PENCIL code \citep{brandenburg02}, 
see Fig.~\ref{fig:KH_growth_rate}. These growth rates confirm the visual impression with $\mathcal{V}_2$ and $\mathcal{V}_3$
growing very similarly, with $\mathcal{V}_4$ having the closest agreement with the high-resolution solution. 

\subsubsection{Rayleigh--Taylor instability}
In a Rayleigh--Taylor instability test,  a density layer $\rho_t$ rests on top of a layer with
density $\rho_b < \rho_t$ in a constant acceleration field, e.g. due
to gravity.   While the denser fluid sinks down, it develops a
characteristic ``mushroom-like'' pattern.  Simulations with traditional
SPH implementations have  shown only retarded growth or even complete
suppression of the instability \citep{abel11,saitoh13}. 
We adopt a quasi-2D setup, but evolve the fluid with a full 3D
code. For simplicity, we place particles on a cubic lattice in the
$xy$-domain $[-0.25,0.25] \times [0,1]$ with a spacing $\Delta x= 0.5/400= 0.00125$
and use 20 layers of particles
in the $z$-direction, and also place 20 layers of particles as
boundaries around this core region. The properties of the particles below
$y=0$ are ``frozen'' at the values of the initial conditions, and we
multiply the temporal derivatives of particles with $y_a > 1$
with a damping factor 
\be
\xi= \exp\left\{-\left(\frac{y_a - 1}{0.05}\right)^2\right\},
\ee 
so that any evolution in this upper region is strongly suppressed. We
apply periodic boundaries in the $x-$direction 
at $x= \pm 0.25$. Similar to \citet{frontiere17} we use  $\rho_t=2$,
$\rho_b=1$, a constant acceleration $\vec{g}= -0.5 \hat{e}_y$ and 
\be
\rho(y)= \rho_b + \frac{\rho_t-\rho_b}{1+\exp[-(y-y_t)/\Delta]}
\ee
with transition width $\Delta=0.025$ and transition coordinate
$y_t=0.5$. We apply a small velocity perturbation to the interface 
\be
v_y(x,y)= \delta v_{y,0} [1 + \cos(8\pi x)][1 + \cos(5\pi(y-y_t))]
\ee
for $y$ in $[0.3,0.7]$ with an initial amplitude $\delta
v_{y,0}=0.025$, and we use a polytropic equation of state with
exponent $\Gamma=1.4$. The equilibrium pressure profile is given by
\be
P (y) = P_0 -  g \rho(y) [y - y_t]
\ee
with $P_0= \rho_t/\Gamma$, so that the sound speed is near unity in
the transition region. 

\begin{figure}[ht]
    \centerline{\includegraphics[width=\textwidth]{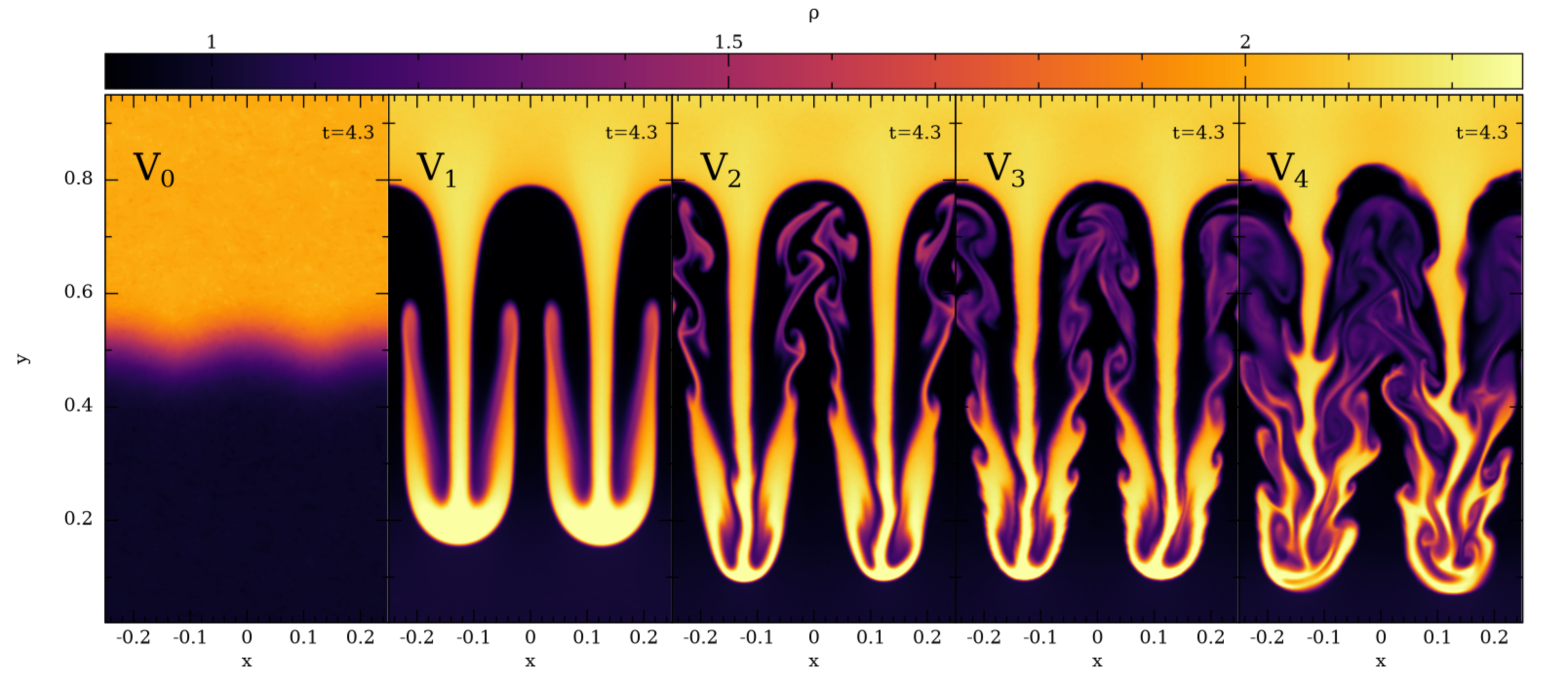} }  
        \caption{Comparison between different methodological choices in
      the Rayleigh--Taylor test. Left panel: ``old school'' choice ($\mathcal{V}_0$; full dissipation, cubic spline kernel with 50 neighbours); second
    panel: like $\mathcal{V}_0$, but with a better kernel (Wendland $C4$, 250 neighbours;
    $\mathcal{V}_1$); third panel: like $\mathcal{V}_1$, but with reduced dissipation
     ($\mathcal{V}_2$); fourth panel: like
    $\mathcal{V}_2$, but with GDF symmetry of the SPH equations ($\mathcal{V}_3$), and $\mathcal{V}_4$ which is like 
    $\mathcal{V}_3$, but with linearly reproducing kernels. See main text for full details.}
    \label{fig:RaTa}
\end{figure}

A comparison for the different SPH versions
$\mathcal{V}_0..\mathcal{V}_4$ is shown in Fig.~\ref{fig:RaTa}. In the
 ``old school'' version $\mathcal{V}_0$  the instability does not
develop at all and the gas is at $t= 4.3$ still very close to its initial
condition. Contrary to the previous Kelvin-Helmholtz test, just 
using a better kernel with more neighbours
($\mathcal{V}_1$) leads to the growth of the instability, though at a
reduced pace and without smaller scale structures developing. Reducing
the dissipation ($\mathcal{V}_2$) leads to much improved results,
though with little substructure developing. Using the same
dissipation, but changing the form of the equations from ``vanilla
ice'' to GDF ($\mathcal{V}_3$) makes the flow more susceptible to 
the instability, and last but not least, using accurate reproducing
kernel gradients ($\mathcal{V}_4$) improve the results even further.

\subsubsection{Schulz-Rinne tests}

Schulz-Rinne tests \citep{schulzrinne93a} are very challenging 2D
benchmarks \citep{schulzrinne93a,lax98,kurganov02,liska03}.
They are constructed so that four constant states meet at one corner,
and the initial values are chosen so that an elementary wave, either a
shock, a rarefaction, or a contact discontinuity, appears at each
interface.  During subsequent evolution, complex wave patterns
emerge for which exact solutions are not known. \\
These tests are considered as very difficult and although no analytical solutions exist,
the comparison with other state-of-art-methods can be very insightful and reveal weaknesses.
Schulz-Rinne tests have not often been shown for SPH, we are only aware 
of the work of \citet{puri14}, applying Godunov SPH with approximate Riemann solvers
with some success,
the tests in the \Ma code \citep{rosswog20a} and in a Riemann solver
approach with reproducing kernels \citep{rosswog25a}. As we illustrate below,
"old school"-type SPH ($\mathcal{V}_0$ and $\mathcal{V}_1$) performs not 
particularly well in these tests, but the other, more modern SPH versions
obtain results that are very similar to good mesh-based methods.

Here, we show three such tests. We place particles on a cubic
lattice in a 3D slice thick enough so that the midplane is unaffected
by edge effects (we use 20 particle layers in $z$-direction),   
so that 200 x 200 particles are within  $[x_c-0.3,x_c+0.3] \times [y_c-0.3,y_c+0.3]$, 
where $(x_c,y_c)$ is the contact point of the quadrants, and we use
a polytropic exponent $\Gamma=1.4$ in all tests. We refer to these
Schulz-Rinne type problems as SR1-SR3 \footnote{In 
  \citet{kurganov02} these are tests 3, 11, and 12.} and give their initial
parameters for each quadrant in Table~\ref{tab:SchulzRinne}. 

\begin{table}[ht]
\caption{Initial data for the Schulz-Rinne-type 2D Riemann problems with vorticity creation.}
\begin{tabular}{ l c | c | c | c | c | c | c |}
\hline
 & & SR1; contact point: $(0.3,0.3)$ & &\\
  \hline	
  \hline		
  variable & NW & NE & SW & SE \\ \hline
  $\rho$ &  0.5323 &  1.5000  & 0.1380 &  0.5323 \\
  $v_x$ &  1.2060 &  0.0000   & 1.2060 & 0.0000  \\
  $v_y$ &   0.0000 & 0.0000   & 1.2060 & 1.2060 \\
  $P$    &   0.3000 & 1.5000   & 0.0290 & 0.3000    \\
  \hline  
   & & SR2; contact point: $(0.0,0.0)$ & &\\
  \hline	
  \hline		
  variable & NW & NE & SW & SE \\ \hline
  $\rho$ &  0.5313 &  1.0000 & 0.8000  &  0.5313 \\
  $v_x$ &   0.8276 &  0.1000 & 0.1000 & 0.1000  \\
  $v_y$ &   0.0000 & 0.0000  & 0.0000  &  0.7276\\
  $P$    &   0.4000  & 1.0000  & 0.4000 & 0.4000    \\
  \hline  
   & & SR3; contact point: $(0.0,0.0)$ & &\\
  \hline	
  variable & NW & NE & SW & SE \\ \hline
  $\rho$ &  1.0000 &  0.5313 & 0.8000  &  1.000 \\
  $v_x$ &   0.7276 &  0.0000 & 0.0000 & 0.0000  \\
  $v_y$ &   0.0000 & 0.0000  & 0.0000  &  0.7262\\
  $P$    &   1.0000  & 0.4000  & 1.0000 & 1.0000    \\
  \hline  
\end{tabular}
\label{tab:SchulzRinne}
\end{table}

\begin{figure}[ht]
    \centerline{\includegraphics[width=1.1\textwidth]{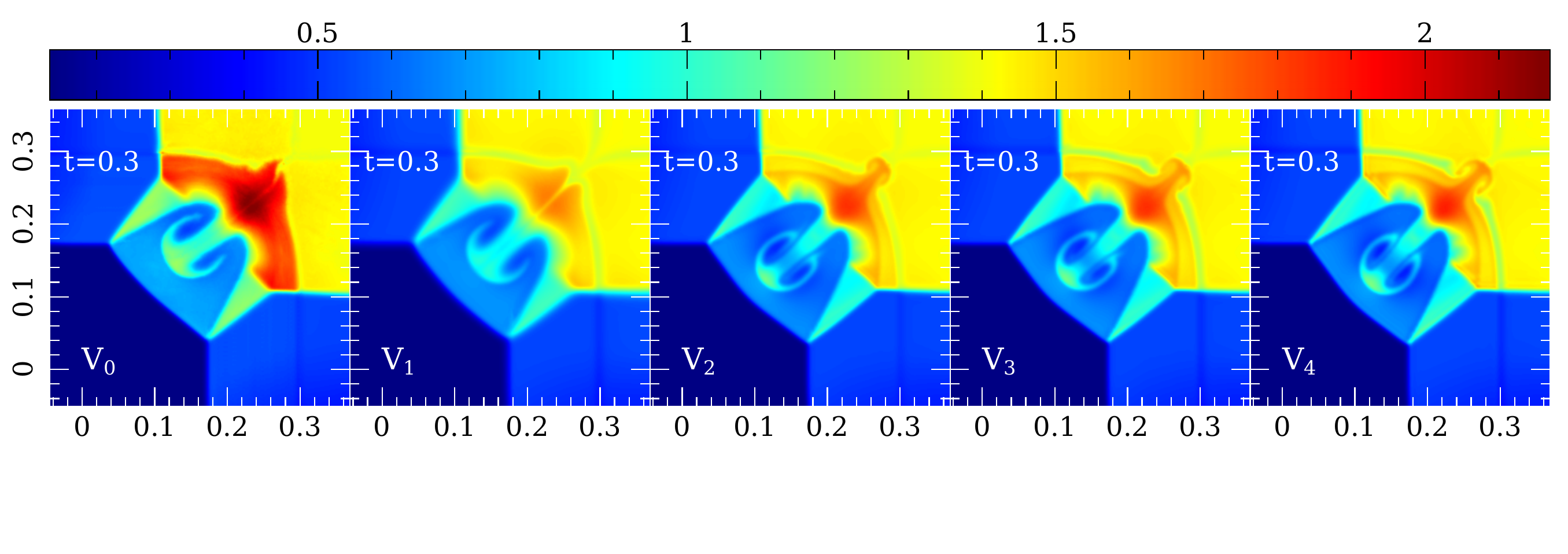}} \vspace*{-0.7cm}
   \centerline{\includegraphics[width=1.1\textwidth]{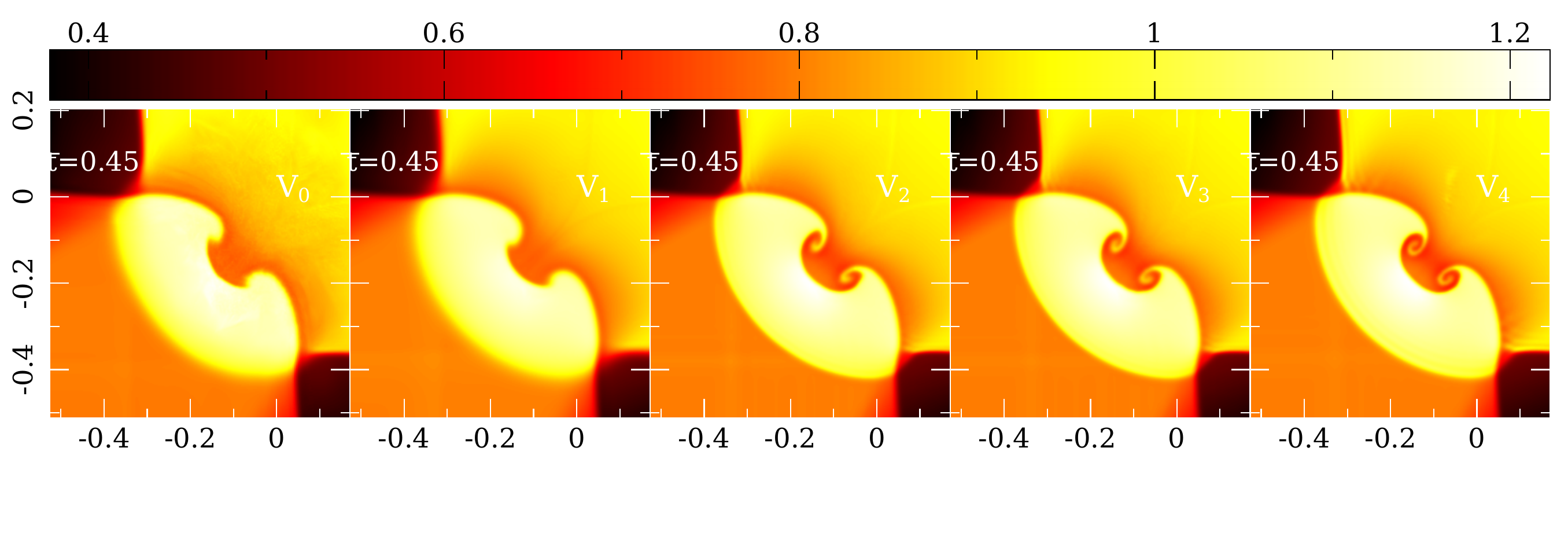}}        \vspace*{-0.7cm}   
   \centerline{\includegraphics[width=1.1\textwidth]{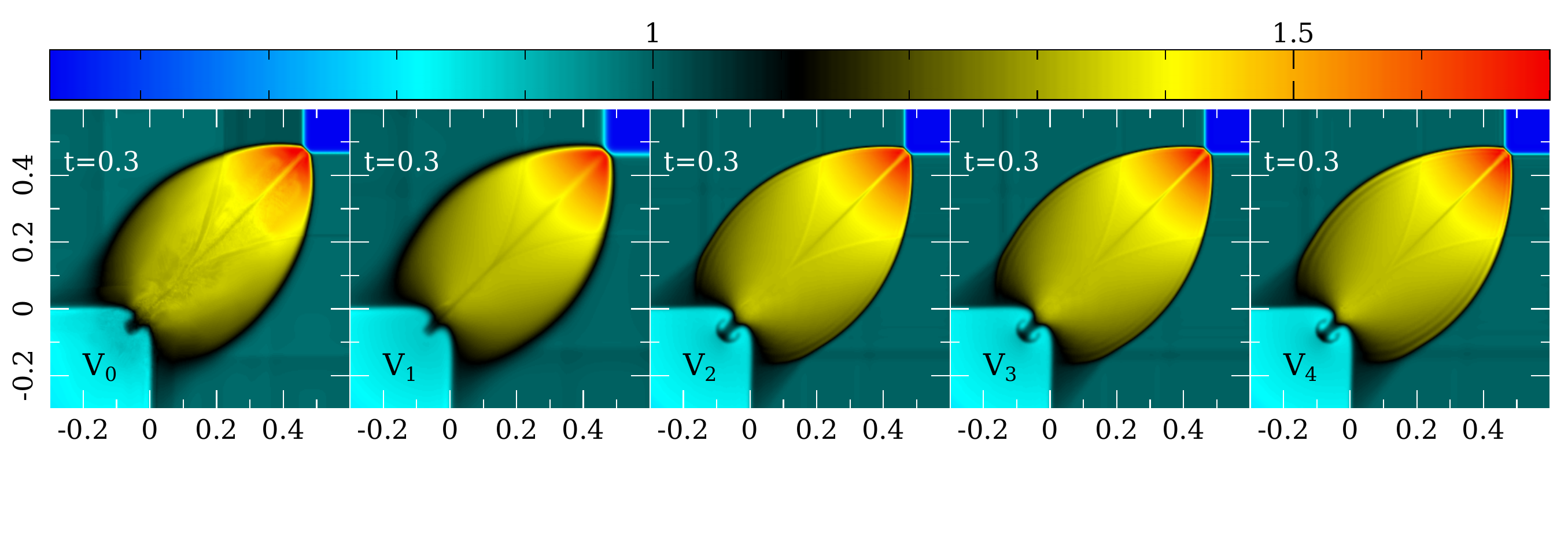}}. \vspace*{-0.7cm}       
   \caption{Comparison between different methodological choices in
      the chosen set of Schulz-Rinne tests. First row: SR1, second
      row: SR2, third row: SR3. Left column: ``old school'' choice ($\mathcal{V}_0$) with cubic spline kernel and 50 neighbours; second column:
     like $\mathcal{V}_0$, but with Wendland C6 kernel and 250 neighbours
    ($\mathcal{V}_1$); third column: like $\mathcal{V}_1$, but with reduced dissipation; fourth column: like
    $\mathcal{V}_2$, but with GDF symmetrization; fifth column: like $\mathcal{V}_3$, but with linearly reproducing kernels
    ($\mathcal{V}_4$). See text for full details.}
    \label{fig:SchulzRinne}
  \end{figure}

For all three  tests, the most advanced SPH version ($\mathcal{V}_4$; last panel in each row) yields 
good agreement with the results published in the literature
\citep{schulzrinne93a,lax98,kurganov02,liska03,rosswog20a}. We see
similar tendencies as in the previous tests. For example, excessive,
constant dissipation ($\mathcal{V}_0$ and $\mathcal{V}_1$) suppresses the formation of
``mushroom-like'' structures in all three tests, see the first two panels in each row of
Fig.~\ref{fig:SchulzRinne}. The cases with cubic spline kernels and only 50 neighbours,
in addition, struggle in maintaining the diagonal symmetry.
The reduction of dissipation (third
panels) already massively improves the test results. The change of the
SPH equations to a GDF-symmetrization has in these tests only a minor impact, 
it enhances the \enquote{mushroom structure} in SR1 (first column) near [0.25,0.25] and also the mushroom is more
curled in comparison to $\mathcal{V}_1$ in SR2. Finally, the enhanced
gradient accuracy from the reproducing kernels helps with curling in of the
mushroom structure, but it also enhances the symmetry compared to the
standard SPH kernel gradients, compare  for example the last panel of SR1 with the
previous ones.

To summarize these tests: the bulk of improvements is achieved by switching to a better kernel
and by reducing dissipation ($\mathcal{V}_2$), but additionally changing to the GDF symmetry and
applying reproducing kernels further improves the results.

\section{General relativistic SPH}
\label{sec:GR_SPH}
For binaries that contain a neutron or a black hole general-relativistic effects are  important. 
The first relativistic SPH formulations were developed by \citet{kheyfets90}
and \citet{mann91,mann93}.
Shortly after, \citet{laguna93a} developed a 3D, general-relativistic SPH code in a fixed background metric 
that was subsequently applied to the tidal disruption of stars by massive black holes \citep{laguna93b}.
Their SPH formulation is complicated by several issues: the continuity 
equation contains a gravitational source term that requires SPH kernels
for curved space-times. Moreover, due to their choice of variables, the equations contain 
time derivatives of Lorentz factors that are treated by finite difference approximations and restrict
the ability to handle shocks to only moderate Lorentz factors. The Laguna et al. formulation has been 
extended by \citet{rantsiou08} and applied to neutron star black hole binaries. \citet{rosswog10a} derived
the general relativistic SPH equations from a Lagrangian of an ideal fluid, thereby accounting 
for \enquote{grad-h}-terms, see Sect.~\ref{sec:adapt_res}.
More recently, a modern SPH version in a fixed metric has been 
developed by \citet{liptai19} within the framework of the \texttt{PHANTOM} code \citep{price18a}, 
which has  been applied to study tidal disruption events \citep{liptai19b,toscani22,price24}.

We focus in Sect.~\ref{sec:fixed_metric_GR} on SPH formulations in a known background metric
and take in Sect.~\ref{sec:SR_SPH} the special-relativistic limit of this formulation. Major additional technical
challenges need to be overcome for full numerical relativity, where also the spacetime is
dynamically evolved, in tight coupling with the matter evolution. This challenging topic is addressed
in Sect.~\ref{sec:fullGR_SPH}.

\subsection{General relativistic SPH on a fixed background metric}
\label{sec:fixed_metric_GR}
Special- and general-relativistic versions of SPH can be derived  similarly to
the Newtonian case
\citep{monaghan01,rosswog09b,rosswog10b,rosswog10a}. We use 
geometric units with $G=c=1$ and adopt (-,+,+,+) for the metric
signature. Greek letters are reserved for space-time indices from 0...3 
with 0 being the temporal component and we use $i$ and $j$ for spatial
components. SPH particles are labeled by $a,b$ and
$k$. Contravariant spatial indices of a vector quantity $\vec{V}$ at
particle $a$ are denoted as $V^i_a$, while covariant ones will be
written as $(V_i)_a$. The line element and proper time are given by
$ds^2= g_{\mu \nu} \; dx^\mu \; dx^\nu$ and $d\tau^2= - ds^2$. The
proper time is related to the coordinate time $t$ by 
\be
\Theta d\tau = dt,
\label{eq:theta_t}
\ee
where we have introduced a generalization of the Lorentz-factor 
\be
\Theta\equiv \frac{1}{\sqrt{-g_{\mu\nu} v^\mu v^\nu}}.
\label{eq:theta}
\ee
The coordinate velocity components are given by
\be
v^\mu= \frac{dx^\mu}{dt}= \frac{dx^\mu}{d\tau} \frac{d\tau}{dt}= \frac{U^\mu}{\Theta}= \frac{U^\mu}{U^0},
\label{eq:v_mu}
\ee
where we have used Eq.~(\ref{eq:theta_t}) and $U^\mu$ is the
four-velocity which is normalized as
\be
U^\mu U_\mu= -1. \label{eq:U_norm}
\ee
The Lagrangian of a relativistic fluid can be written as \citep{fock64}
\be
L= - \int T^{\mu \nu} U_\mu U_\nu \sqrt{-g} dV,
\ee
where $g= {\rm det}(g_{\mu \nu})$ and $T^{\mu\nu}$ denotes the energy-momentum tensor which, for an ideal fluid, reads
\be
T^{\mu \nu}= (\tilde{\rho}+P)U^\mu U^\nu + P g^{\mu \nu}.
\ee
Here $\tilde{\rho}$ is the energy density measured in the local rest frame of the fluid and 
$P$ is its pressure. For clarity, we will in the following lines write
out factors of $c$ explicitly.  The energy density possesses a contribution
from  the rest mass and one from the thermal energy:
\be
\tilde{\rho}= \rho + u \rho/c^2= n m_0 c^2 (1+u/c^2).
\ee
Here $n$ is the baryon number density measured in the local fluid rest frame, $m_0$ is the baryon mass and $u= u(n,s)$ the specific energy, with $s$ being the specific entropy.\footnote{Here we simply use $m_0= m_u$ for the average baryon mass, where $m_u$ is the atomic mass unit. In reality, $m_0$ depends on the exact composition, but
even in extreme cases the deviations from $m_u$ are only a small fraction of a percent. See Sect.~2.1 in \citet{diener22a} for a more detailed discussion.}
Here, we 
measure all energies in units of $m_0 c^2$.  With this convention (and now using again $c=1$) the energy 
momentum tensor reads
\be
T^{\mu \nu}= \left\{ n (1+u) + P \right\} U^\mu U^\nu + P g^{\mu \nu}.\label{eq:EM_tensor}
\ee
To perform practical simulations, we give up general covariance and choose
a particular frame (``computing frame'') in which the simulations are performed. Like in special-relativistic hydrodynamics,
one must clearly distinguish between quantities that are measured in the computing frame
and those measured in the local rest frame of the fluid.
With the normalization of the four-velocity, Eq.~(\ref{eq:U_norm}), the Lagrangian 
can be written as
\be
L= - \int n(1+u)\sqrt{-g} dV\label{eq:Lagrangian_cont}.
\ee
Local baryon number conservation, $(U^\mu n);_\mu= 0$,
can be expressed as 
\be 
\frac{1}{\sqrt{-g}}\p_\mu (\sqrt{-g} U^\mu n)= 0,
\ee
where we have used an identity for the covariant derivative, see e.g. \cite{schutz89}.
More explicitly, we can write 
\be 
\p_t (N) + \p_i(N v^i)= 0, \label{eq:continuity_N}
\ee
where we have made use of Eq.~(\ref{eq:v_mu}) and have introduced the computing frame 
baryon number density
\be
N= \sqrt{-g} \Theta n.\label{eq:N_n}
\ee

The total conserved baryon number, $\mathcal{N}$, can then be expressed as a sum over fluid parcels
with volume $\Delta V_b$ located at $\vec{r}_b$, where each parcel carries a baryon 
number\footnote{Be careful not to confuse the baryon number $\nu_b$ with the velocity component of a particle $v^i_b$.} $\nu_b$ 
\be 
\mathcal{N}= \int N dV \simeq \sum_b N_b \Delta V_b = \sum_b \nu_b,\label{eq:parcel_volumes}
\ee
therefore the particle volume in the computing frame reads $\Delta V_b= \nu_b/N_b$.
Eq.~(\ref{eq:continuity_N}) looks like the Newtonian continuity equation
and we will use it for the SPH discretization process. Similar to
Eq.~(\ref{eq:disc_func_approx}), a 
quantity $f$ can now be approximated by 
\be
\tilde{f} (\vec{r}) \simeq \sum_b \frac{\nu_b}{N_b} \; f_b \; W(\vec{r}-\vec{r}_b,h)\label{eq:SPH_discretization},
\ee
where  the particle's baryon number $\nu_b$ replaces the mass and the Newtonian mass density $\rho$ is 
replaced by the computing frame baryon number density $N$. The latter can be calculated,
very similarly to the Newtonian case, as
\be
N_a= \sum_b \nu_b W_{ab}(h_a).
\ee
If each particle keeps its baryon number constant, we have
exact baryon number conservation. The fluid Lagrangian, Eq.~(\ref{eq:Lagrangian_cont}), then becomes
\be
L= - \int \frac{1+u}{\Theta} \; N \; dV \approx  - \sum_b  \left( \frac{1+u}{\Theta} \right)_b N_b \Delta V_b = - \sum_b \nu_b \left( \frac{1+u}{\Theta} \right)_b,
\ee
where in the last step we have made use of the volume element $\Delta V_b= \nu_b/N_b$, as suggested
by Eq.~(\ref{eq:parcel_volumes}).

Again, as in the Newtonian case, one can use the Euler-Lagrange equations
\be
\frac{d}{dt} \frac{\p L}{\p v^i_a} - \frac{\p L}{\p x^i_a}= 0, \label{eq:EL}
\ee
to derive an evolution equation for the canonical momentum. From
the Lagrangian one can define a canonical momentum 
\be
(p_i)_a\equiv \frac{\p L}{\p v^i_a}= - \frac{\p}{\p v^i_a}\sum_b \nu_b \left( \frac{1+u}{\Theta} \right)_b.
\ee
Note that here $\p u_b/\p v^i_a$ has a non-zero value, since
\be
\frac{\p u_b}{\p v^i_a}= \frac{\p u_b}{\p n_b} \frac{\p n_b}{\p v^i_a} = 
\frac{P_b}{n_b^2} \frac{\p}{\p v^i_a} \left( \frac{N_b}{\sqrt{-g} \; \Theta_b} \right),
\ee
where we have used the first law of thermodynamics and the
relation Eq.~(\ref{eq:N_n}) and the velocity dependence
comes in via the generalized Lorentz factor $\Theta$, see
Eq.~(\ref{eq:theta}). This yields \citep{rosswog10a} 
\be
\frac{\p}{\p v^i_a} \left( \frac{1}{\Theta_b}\right)= - \Theta_b (g_{i\mu} v^\mu)_a \delta_{ab}.
\ee 
The \emph{canonical momentum per baryon} reads
\be
(S_i)_a\equiv \frac{1}{\nu_a} \frac{\p L}{\p v^{i}_{a}}= \Theta_a \left(1+u_a+\frac{P_a}{n_a} \right) (g_{i\mu} v^\mu)_a= \mathcal{E}_a (U_i)_a,
\ee
where
\be
\mathcal{E}\equiv  1 + u + \frac{P}{n}
\ee
is the specific enthalpy. Similarly, one can use the canonical energy 
\be
E\equiv \sum_a \frac{\p L}{\p v^i_a} v^i_a - L= \sum_a \nu_a \left( v^i_a (S_i)_a + \frac{1+u_a}{\Theta_a}\right)
\ee
to identify the \emph{canonical energy per baryon}
\be
e_a \equiv v^i_a (S_i)_a + \frac{1+u_a}{\Theta_a}.
\label{eq:can_energy}
\ee
The evolution equation for the canonical momentum per baryon follows, according to the 
Euler-Lagrange equations (\ref{eq:EL}), from $\frac{d (S_i)_a}{dt}= \frac{1}{\nu_a} \frac{\p L}{\p x^i_a}$, 
and after some algebra \citep{rosswog10a} one  finds
\bea
\frac{d (S_i)_a}{dt} = - \sum_b  \nu_b 
\left\{
\frac{P_a}{\Omega_a N_a^2} D_i^a  + \frac{P_b}{\Omega_b N_b^2} D_i^b
\right\}
 + \left(\frac{\sqrt{-g}}{2 N} T^{\mu\nu} 
\frac{\p g_{\mu\nu}}{\p x^i}\right)_a,\label{eq:GR_momentum_evolution}
\eea
where the general relativistic ``grad-h terms'' read \citep{rosswog10a}
\be
\Omega_k= 1 - \frac{\p h_k}{\p N_k} \sum_l \nu_l  \frac{\p
  W_{kl}(h_k)}{\p h_k}
\ee
and we have introduced the abbreviations
\be
D^a_i \equiv   \sqrt{-g_a} \;  \frac{\p W_{ab}(h_a)}{\p x_a^i} \quad \text{and} \quad 
D^b_i \equiv    \sqrt{-g_b} \; \frac{\p W_{ab}(h_b)}{\p x_a^i}
\label{eq:kernel_grad}.
\ee
The first term in Eq.~(\ref{eq:GR_momentum_evolution}) (including the
summation) is due to hydrodynamic accelerations and it is formally
very similar to the Newtonian momentum equation (\ref{eq:dvdt_NSPH}),
although the involved quantities have a different meaning. The second
term represents the accelerations due to spacetime curvature and it depends
on the spatial derivatives and the determinant of the metric.

The evolution equation of the canonical energy per baryon
follows from straight forwardly taking the Lagrangian time derivative
of Eq.~(\ref{eq:can_energy}), applying the first law of thermodynamics
and after some algebra, see \citet{rosswog10a}, one obtains
\bea
\frac{de_a}{dt}&=& - \sum_b \nu_b \left\{
\frac{P_a v^i_b}{\Omega_a N_a^2} \; D_i^a  + 
\frac{P_b v^i_a}{\Omega_b N_b^2} \; D_i^b   
\right\}
 - \left( \frac{\sqrt{-g}}{2 N} T^{\mu\nu} \p_t g_{\mu\nu}\right)_a.\label{eq:GR_energy_evolution}
\eea
Similar to the momentum equation, there is a hydrodynamic contribution
(involving the summation over neighbour particles) and a gravitational
contribution that is proportional to the time derivative of the
metric tensor.

In order to calculate the RHSs of  Eqs.~(\ref{eq:GR_momentum_evolution}) and
(\ref{eq:GR_energy_evolution})  one needs to convert the numerical (“conservative”)
variables ($N, S^i ,e$), that are evolved forward in  time, back to
the corresponding physical (“primitive”) variables ($v^i , n, u, P$).
This is actually not an entirely trivial problem
(sometimes called ``recovery problem'', ``recovery'' or
``conservative-to-primitive transformation'') that requires 
numerical root finding. The exact algorithm depends on which
equations are evolved and on the equation of state that is used. The
strategy in the \SPHI code is to express $n$ and $u$ in terms of the evolved
variables $N,S_i$ and $e$ and the pressure $P$, substitute them into
the (polytropic or piecewise polytropic) equation of state and use
numerical rootfinding (e.g. Ridders' method, \citealt{press92}) to find
the pressure value that solves this equation. Once this pressure value
is found, all physical quantities can be found by a simple
backsubstitution. See Sect.~2.2.4 in \citet{rosswog21a} and Appendix A
in \citet{rosswog22b} for a detailed description of the polytropic and
the piecewise polytropic case, respectively. For a generalization of
the recovery procedure for Fermi liquid theory thermal contributions
see \cite{biswas26a} and for recovery algorithms for tabulated nuclear equations 
of state see \cite{shankar26a}.\\
The GR-SPH equations (\ref{eq:GR_momentum_evolution}) and 
(\ref{eq:GR_energy_evolution}) still need to be augmented by a shock 
capturing mechanism. For explicit dissipation terms, please  see \cite{liptai19} and
\cite{rosswog22b}.

\subsection{The special relativity limit}
\label{sec:SR_SPH}
In the special-relativistic limit we can neglect the gravitational
terms in  Eqs.~(\ref{eq:GR_momentum_evolution}) and
(\ref{eq:GR_energy_evolution}). In  flat space-time with Cartesian
coordinates one has $\sqrt{-g} \rightarrow 1$ and $\Theta \rightarrow
\gamma$, see Eq.~(\ref{eq:theta}), and Eq.~(\ref{eq:N_n}) becomes  
$N= \gamma n$, which simply expresses the increase in the computing
frame number density  $N$ with respect to the local fluid rest frame
density $n$ due the Lorentz contraction, because the volume appears to
be Lorentz-contracted by a factor $\gamma$ when seen from the
computing frame. The momentum and energy equations reduce in this
limit to 
\bea
\left(\frac{d \vec{S}_a}{dt}\right)_\mathrm{SR}&=& - \sum_b \nu_b 
\left\{ 
\frac{P_a}{\Omega_a N_a^2} \frac{\p W_{ab}(h_a)}{\p \vec{r}_a} +
\frac{P_b}{\Omega_b N_b^2} \frac{\p W_{ab}(h_b)}{\p \vec{r}_a}
\right\}\label{eq:SR_momentum_evolution}
\eea
and
\bea
\left(\frac{de_a}{dt}\right)_\mathrm{SR}&=& - \sum_b \nu_b \left\{ 
\frac{P_a \vec{v}_b}{\Omega_a N_a^2 } \;  \cdot \frac{\p W_{ab}(h_a)}{\p \vec{r}_a} + 
\frac{P_b \vec{v}_a}{\Omega_b N_b^2 } \; \cdot \frac{\p W_{ab}(h_b)}{\p \vec{r}_a}  
\right\}\label{eq:SR_energy_evolution}
\eea
which are the equations derived and successfully tested in
\citet{rosswog10b,rosswog11a}.

Note that by choosing the canonical energy and momentum as numerical
variables, one avoids complications such as time derivatives of
Lorentz factors, that have plagued earlier SPH formulations
\citep{laguna93a}. The price one has to pay is that the physical
variables (such as $u$ and $\vec{v}$) need to be recovered at every
time step from $N$, $\epsilon$ and $\vec{S}$ by solving a non-linear
equation, see \citet{chow97,rosswog10b}.

As in the non-relativistic case, these equations need to be augmented
by extra measures to handle shocks. The dissipative terms  \citep{chow97}
\be
\left(\frac{d\vec{S}_a}{dt}\right)_\mathrm{diss}= - \sum_b \nu_b \Pi_{ab} 
\overline{\nabla_a W_{ab}} \;\; {\rm with} \;\; \Pi_{ab}= - 
\frac{K v_\mathrm{sig}}{\bar{N}_{ab}} (\vec{S}_a^\ast-\vec{S}_b^\ast) \cdot\hat{e}_{ab}
\label{eq:diss_mom}
\ee
\be
\left(\frac{d\epsilon_a}{dt}\right)_\mathrm{diss}=  - \sum_b \nu_b \vec{\Omega}_{ab} \cdot
\overline{\nabla_a W_{ab}} \;\; {\rm with} \;\; \vec{\Omega}_{ab} = -
\frac{K v_\mathrm{sig}}{\bar{N}_{ab}} (\epsilon_a^\ast-\epsilon_b^\ast)\hat{e}_{ab}
\label{eq:diss_en}
\ee
with symmetrized kernel gradients
\be
\overline{\nabla_a W_{ab}} = \frac{1}{2}\left[\nabla_a W_{ab}(h_a) +  \nabla_a W_{ab}(h_b) \right],
\ee
\be
\gamma_k^\ast= \frac{1}{\sqrt{1-(\vec{v}_k\cdot \hat{e}_{ab})^2}},
\vec{S}_k^\ast= \gamma^\ast_k \left(1+u_k+\frac{P_k}{n_k}\right) \vec{v}_k
\quad 
{\rm and} 
\quad
\vec{\epsilon}_k^\ast= \gamma^\ast_k \left(1+u_k+\frac{P_k}{n_k}\right) - \frac{P_k}{N_k},
\ee
where the asterisk denotes the projection to the line connecting two particles,
give good results, even in very challenging shock tests \citep{rosswog10b}.
A good choice for $v_\mathrm{sig}$ is \citep{rosswog10b} 
\be
v_\mathrm{sig,ab}= \max (\alpha_a,\alpha_b),\label{eq:vsig}
\ee
where
\be
\alpha_k^{\pm}= \max (0,\pm \lambda^\pm_k)
\ee
with $\lambda^\pm_k$ being the extreme local eigenvalues of the Euler equations, see e.g., \citet{marti03},
\be
\lambda^\pm_k= \frac{v_\parallel(1-c_\mathrm{s}^2) \pm c_\mathrm{s} \sqrt{(1-v^2)(1-v_\parallel^2 - 
v_\perp^2 c_\mathrm{s}^2)}}{1-v^2 c_\mathrm{s}^2}
\ee
and $c_{{\rm s},k}$ is the relativistic sound velocity of particle $k$,   
$c_{s,k}= \sqrt{\frac{(\Gamma-1) (\enth-1)}{\enth}}$. 
In 1 D, this simply reduces
to the usual velocity addition law, $\lambda^\pm_k= (v_k\pm c_{{\rm s},k})/(1\pm v_k c_{{\rm s},k}) $.
As in the non-relativistic case, the challenge lies in designing
triggers that switch on where needed, but not otherwise. The
strategies that can be applied here are straight forward translations
of those described in  Sect.~\ref{sec:AV}. We refer to
\citet{rosswog15b} for more details on this topic. A successful
special-relativistic Riemann solver approach has recently been developed,
implemented and tested in \citet{kitajima25}.

\subsection{General relativistic SPH in dynamical spacetimes}
\label{sec:fullGR_SPH}

The step from gas dynamics in a fixed metric to full numerical relativity comes with serious additional challenges.
Here the spacetime and the matter need to be evolved consistently together,
rather than fixing the metric to, say, the Kerr metric or assuming that
the spacetime is given by a sequence of waveless equilibrium slices,
like in the conformal flatness approach
\citep{isenberg08,wilson95a,wilson96,mathews97,mathews98}. Since now
the interesting quantities, such as the properties of a ``ringing''
spacetime, are no longer attached to the matter (like in standard
Newtonian fluid dynamics where gravity is treated as an elliptic
problem that is independent of the previous history), one needs, in
addition to the particles, a computational entity that evolves the
spacetime, even if it is devoid of matter. 
The most straight forward way to deal with this is to borrow the mesh-based techniques
that are used in conventional Eulerian numerical relativity \citep{alcubierre08,bona09,baumgarte10,rezzolla13a,shibata16}. 
In other words, one uses a mixed methodology with both particles (for matter) and meshes (for spacetime) with the 
technically most challenging part being the accurate information transfer back and forth between particles and mesh.

Given the maturity of relativistic Eulerian (magneto-)hydrodynamic methods, it is fair to ask: 
\enquote{Is it worth developing a new methodology with particles?} 
Particle methods have, for example, major advantages when it comes to dealing with vacuum.
Standard Eulerian methods need to model it as a low-density \enquote{atmosphere}
and need to pay attention that this background gas does not impact the results of the
\enquote{real fluid}, in numerical relativity usually neutron star material.
In SPH, in contrast, vacuum is just the absence of matter/SPH-particles, and no special 
treatment is needed. If the initial conditions are well-prepared, neutron star surfaces
remain (in contrast to Eulerian methods) very well-behaved,  see
Fig.~10 in \cite{rosswog21a} or Fig.~11 in \cite{diener22a} for illustrations.\\
Yet another advantage is exact advection, see Sec.~\ref{sec:advection_test},  and in particular the ease 
with which ejecta can be followed. In the
recently started era of multi-messenger gravitational wave astronomy, see Sec.~\ref{sec:GW_detections},
this is of particular importance: major breakthroughs have been achieved by observing
the first neutron star merger event in both gravitational {\em and} electromagnetic emission. 
While the gravitational wave emission is sourced by the (relatively easy to evolve) bulk 
matter motion, all of the electromagnetic emission is caused by a small fraction ($\sim 1$ \%) 
of the binary mass. Here SPH's ease in following the small ejecta mass 
without loss of information  is a major benefit of relativistic 
particle methods. This allows for a straight forward post-processing with nuclear networks
which, in turn, is the basis for predicting the electromagnetic signatures of neutron star mergers.\\
But these potential benefits do not come for free. Since the spacetime needs
to be evolved on (adaptive) meshes, the accurate mapping
between particles and spacetime mesh is both technically very complicated and computationally
rather expensive, since it involves a fair amount of matrix algebra, see 
Secs.~\ref{sec:LRE} and \ref{sec:matter_spacetime_coupling}. This computational burden comes
on top of the SPH- and the expensive spacetime evolution.\\
Another serious challenge are general relativistic initial conditions: they need to be an accurate representation
of the desired physical system,  a solution of the Einstein field equations and they
need to fulfill in particular the constraint equations (\ref{eq:ham_constr}) and (\ref{eq:mom_constr}),
see below. Several initial data library packages are available, for example 
 \lo~\citep{lor}, \fu~\citep{papenfort21}, \sgrid~\citep{tichy09}, \spells~\citep{pfeiffer03}, \Elliptica~\citep{rashti22}
or \nRPyElliptic~\citep{assumpcao22}, but once such consistent matter+spacetime solutions are found,
they still need to be accurately represented by the SPH particles, which can be done via GR
generalizations \citep{rosswog21a,diener22a,rosswog23a} of the APM method described in Sec.~\ref{sec:IC}. \\
In terms of which astrophysical systems can be tackled with full GR-SPH, one has to keep in mind
that in a dynamical spacetime evolution the "signal speed"
that sets the admissible numerical time step is the speed of light. If black holes are involved, one needs to resolve
the event horizon region to sufficient accuracy and this combination of small resolution scale and large signal speed
massively reduces the size of allowed time  steps. Thus, fully dynamical spacetime evolutions are typically restricted
to moderate multiples of dynamical time scales and 
before embarking on a challenging simulation project with spacetime evolution, say long-term
evolutions of black holes in gaseous disks, one should make sure that a) full GR is really needed to 
obtain a satisfactory answer to the question (and  approximate methods such as post-Newtonian 
expansions are not good enough; they would be computationally {\em much} faster) and b) that, given the 
serious time step restrictions, the relevant time scales can really be reached.

To date, there is to our knowledge only a single particle code that
solves the full set of Einstein equations and can handle the extreme
spacetime of neutron star mergers, the code \SpB
\citep{rosswog21a,diener22a,rosswog22b,rosswog23a,rosswog25c}, and we
follow below this implementation\footnote{There is a GR version of
  the PHANTOM code which has been applied in a (less extreme)
  cosmological context \citep{magnall23}.}.

\subsubsection{Spacetime evolution}

In general relativity\footnote{We follow here the text of
  \citet{rosswog25c} that was written by the author in close
  collaboration with Peter Diener.},  gravity manifests itself as curvature of spacetime and the
fundamental object that contains this geometric information is the spacetime metric,
$g_{\mu\nu}$, from which the spacetime interval between infinitesimally close points
can be calculated via $ds^2=g_{\mu\nu}dx^{\mu}dx^{\nu}$. From the metric one can calculate
the \emph{Christoffel symbols} $\Gamma^{\lambda}_{\mu\nu}$ from which one can calculate the
\emph{Riemann curvature tensor} ${R^{\lambda}}_{\sigma\mu\nu}$. 
Contracting the first and third index on the Riemann curvature tensor delivers the
\emph{Ricci tensor}, $R_{\mu\nu}={R^{\lambda}}_{\mu\lambda\nu}$ and by contracting the two
indices on the Ricci tensor one finds the \emph{Ricci scalar} $R=g^{\mu\nu}R_{\mu\nu}$.
These combine to define the \emph{Einstein tensor} that is related to the stress-energy
tensor in the full covariant 4-dimensional set of Einstein field equations
\be
G_{\mu\nu}=R_{\mu\nu}-\frac{1}{2}g_{\mu\nu}R=8\pi T_{\mu\nu},
\ee
which, in turn, is the starting point for any spacetime evolution code.

\begin{figure}
\centerline{
\includegraphics[width=8cm]{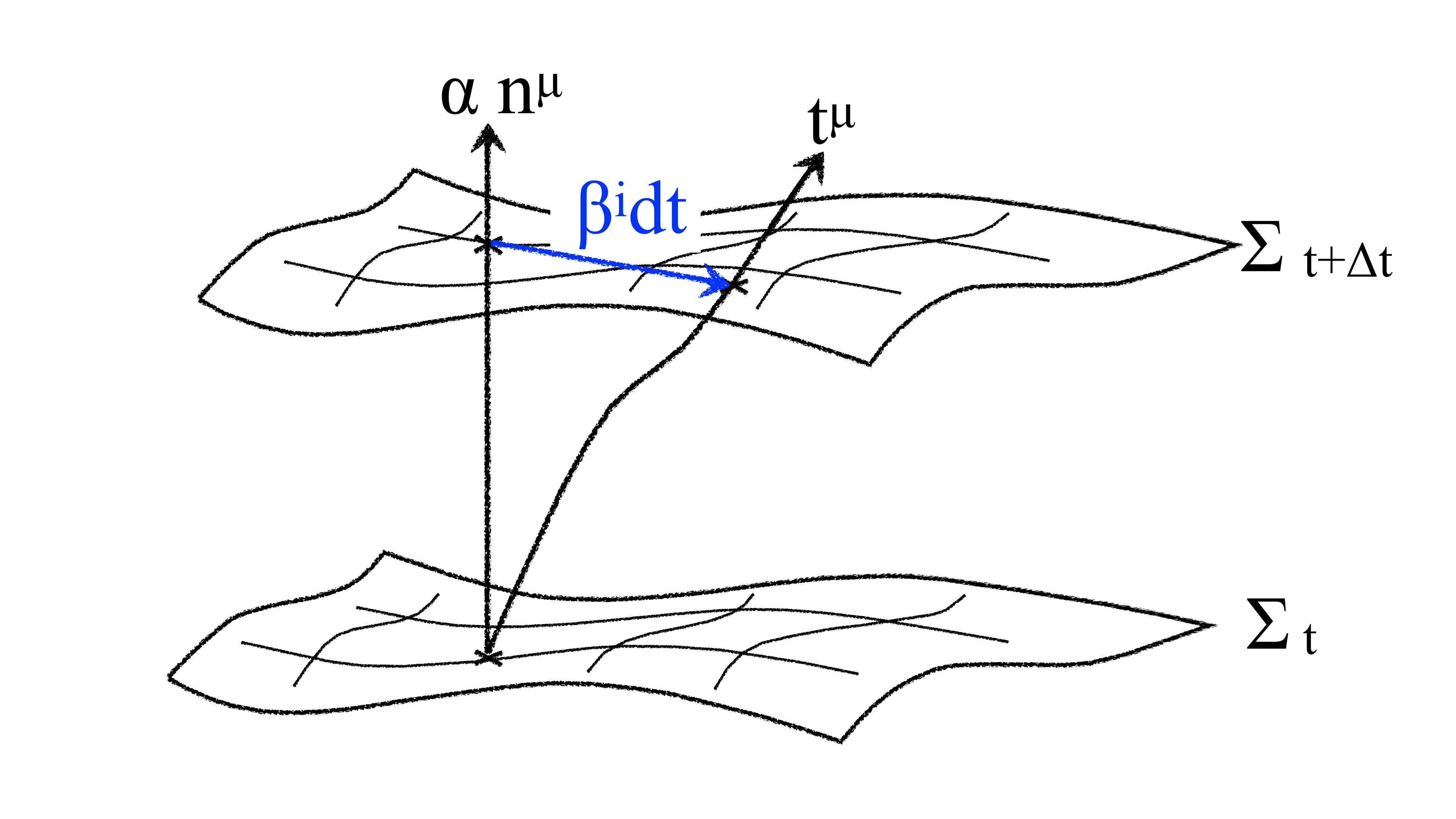}
}
\caption{3+1 foliation of the spacetime. The spacetime is sliced into spacelike hypersurfaces 
which are integrated forward in time. The lapse function $\alpha$ measures the proper time 
along the normal $n^{\mu}$, the shift vector $\beta^i$ measures the displacement, on consecutive 
hypersurfaces, between the observer time lines $t^{\mu}$ and the normal lines $n^{\mu}$.}\label{fig:3+1}
\end{figure}

These equations are 4-dimensional and to evolve 3-dimensional data
forward in time one usually splits spacetime into 3+1 dimensions. To this end,
one can write the line element as
\be
ds^2=g_{\mu\nu} dx^{\mu}dx^{\nu}= (-\alpha^2 +\beta_i\beta^i) dt^2+2\beta_i dt dx^i+
     \gamma_{ij}dx^i dx^j,
\ee
where $\alpha$ is the lapse function, $\beta^i$, the shift vector, and $\gamma_{ij}$ the
spatial 3-metric. This foliates spacetime into non-intersecting spacelike hypersurfaces
$\Sigma$ that are labeled by the coordinate time, $t$, as sketched in Fig.~\ref{fig:3+1}.
Each spatial slice has a future pointing normal vector, $n^{\mu}$, whose 
coordinates are $n^{\mu}=(1/\alpha, -\beta^i/\alpha)$. In this picture the lapse $\alpha$
measures the elapsed proper time between two nearby spatial slices (in the normal direction),
the shift vector, $\beta^i$, is the relative velocity of Eulerian observers and the lines of
constant spatial coordinates and the spatial metric, $\gamma_{ij}$, measure proper distance
within the slices.

The curvature that is intrinsic to the slices is given by the 3-dimensional Riemann
tensor defined in terms of the 3-metric $\gamma_{ij}$. However, there is also an
extrinsic curvature associated with how the 3-dimensional slices are embedded in the
overall 4-dimensional spacetime. Remarkably, this can be described in terms of another
3-dimensional tensor, $K_{ij}$, that can be defined as the projection of the gradient of
the normal vector onto the spatial slice
\be
K_{\mu\nu}:= -P^{\lambda}_{\mu}n_{\nu;\lambda}, \label{eq:Kij}
\ee
where $P^{\lambda}_{\mu}$ is the projection operator,
\be
P^{\lambda}_{\mu}:=\delta^{\lambda}_{\mu}+n^{\lambda}n_{\mu}.
\ee
Note that even though the
extrinsic curvature is defined as a 4-dimensional object in Eq.~(\ref{eq:Kij}), it is indeed
a 3-dimensional object with $K^{00}=K^{0i}=0$ and can be referred to as $K^{ij}$. 
$K_{00}$ and $K_{0i}$ on the other hand are not zero, but the components of $K_{ij}$ 
can still be obtained by using only the spatial metric, $\gamma_{ij}$, to lower the
indices of $K^{ij}$.

It is possible \citep{alcubierre08,bona09, baumgarte10,rezzolla13a,shibata16}
to split the 10 original second order Einstein equations into 12 first order
hyperbolic evolution equations
\begin{align}
\dt{\gamma_{ij}} & = -2\alpha K_{ij}+D_i\beta_j+D_j\beta_i, \label{eq:adm1}\\
\dt{K_{ij}} & = -D_i D_j\alpha + \alpha\left [ {}^{(3)}R_{ij}+K K_{ij}-2K_{ik}K^{k}_{j}\right ] \nonumber \\
            & + 4\pi\alpha[\gamma_{ij}(S-\rho)-2S_{ij}]+\beta^k\pdl{K}{ij}{k}+K_{ki}\pdu{\beta}{k}{j}
              + K_{kj}\pdu{\beta}{k}{i}, \label{eq:adm2}
\end{align}
where $D$ is the 3-dimensional covariant derivative related to $\gamma_{ij}$, $\rho=n^{\mu}n^{\nu}T_{\mu\nu}$, 
$S_{\mu\nu}=P^{\lambda}_{\mu}P^{\sigma}_{\nu}T_{\lambda\sigma}$ and
$S=S^{\mu}_{\mu}$. In addition, one has four elliptical constraint equations
\begin{align}
    {}^{(3)}R+K^2-K_{ij}K^{ij} & = 16\pi\rho \label{eq:ham_constr} \\
    D_j(K^{ij}-\gamma^{ij}K) & = 8\pi j^i, \label{eq:mom_constr}
\end{align}
where
$j^{\mu}=-P^{\mu\lambda}n^{\nu}T_{\lambda\nu}$. Equations~(\ref{eq:adm1})
and (\ref{eq:adm2}), as written here, are due to  York~\cite{york79}
and are known as the ADM equations after \citet{misner73}. Given initial data for $\gamma_{ij}$ and
$K_{ij}$ on a 3-dimensional spatial slice, the ADM equations can in
principle be used to evolve the spacetime forward in time. 
Eq.~(\ref{eq:ham_constr}) is the \emph{Hamiltonian or energy constraint} while 
Eq.~(\ref{eq:mom_constr}) is the so-called \emph{momentum constraint}. These do not involve any time
derivatives but rather provide a set of equations that any physical 
data has to satisfy at any point in time. This is similar, for example, to the case of 
magneto-hydrodynamics, where one has evolution equations
that have to preserve the $\nabla \cdot \vec{B}=0$-constraint.

The only problem with the ADM equations is that they do not actually work in practice. They can
be shown (see for example \citealt{alcubierre08}) to be only \emph{weakly 
hyperbolic} and hence they are not a well-posed\footnote{A well-posed
problem has the following properties: i) it has a solution, ii) the
solution is unique and iii) the solution changes continuously with the initial
conditions.} formulation of the Einstein equation.
An alternative formulation due to Baumgarte, Shapiro, Shibata and Nakamura, the 
so-called BSSN formulation \citep{baumgarte10,shibata16},
has proven to be a robust choice for many codes and has
therefore become rather popular. In the BSSN formulation 
a conformal rescaling is introduced
\be
  \tlg_{ij}=e^{-4\phi}\gamma_{ij},
\ee
where $\phi$ is defined in terms of the determinant of the physical three-metric,
$\gamma$, as $\phi = \frac{1}{12}\log\gamma$. With this choice the determinant of the
conformal metric becomes $\tlg=1$. In addition, the extrinsic curvature is separated
into its trace and its tracefree part and conformally rescaled 
\be
  \tlA_{ij}=e^{-4\phi}A_{ij}= e^{-4\phi}(K_{ij}- \frac{1}{3}\gamma_{ij}K).
\ee
The final addition is to also evolve the three quantities
\be
  \tlG^{i}=\tlg^{jk}\tlG^i_{jk}=-\partial_j\tlg^{ij},
\ee
where $\tlG^{i}_{jk}$ is the Christoffel symbols related to the conformal metric.
That is, the BSSN evolution variables are $\phi$, $\tlg$, $K$, $\tlA$ and $\Gamma^{i}$, and result
in the following set of evolution equations
\begin{align}
  \dt{\phi} & = -\frac{1}{6} \left ( \alpha K - \pdu{\beta}{i}{i} \right) + \upwindu{\phi}{}{i}, \label{eq:BSSN1}\\
 \dt{\tlg_{ij}} & = -2\alpha \tlA_{ij} + \tlg_{ik} \pdu{\beta}{k}{j}
                   + \tlg_{jk} \pdu{\beta}{k}{i}
                    -\frac{2}{3} \tlg_{ij} \pdu{\beta}{k}{k}
                         + \upwindl{\tlg}{ij}{k}, \\
  \dt{K} & = -\emfp \left ( \tlg^{ij} \left [ \pdpdu{\alpha}{}{i}{j}
               +2\pdu{\phi}{}{i}\pdu{\alpha}{}{j} \right ]
               - \tlGn^{i}\pdu{\alpha}{}{i} \right ) \nonumber \\
        & + \alpha \left ( \tlA^{i}_{j} \tlA^{j}_{i} +\frac{1}{3} K^2
               \right ) + \upwindu{K}{}{i} + 4 \pi \alpha ( \rho + s ), \label{eq:dtK} \\
  \dt{\tlA_{ij}} & = \emfp \left [ -\pdpdu{\alpha}{}{i}{j} + \tlG^{k}_{ij}
                       \pdu{\alpha}{}{k} + 2 \left ( \pdu{\alpha}{}{i}
                       \pdu{\phi}{}{j}+\pdu{\alpha}{}{j} \pdu{\phi}{}{i}\right ) +\alpha R_{ij} \right ]^{\mathrm{TF}} \nonumber \\
                &      +\alpha ( K \tlA_{ij}- 2 \tlA_{ik} \tlA^{k}_{j} ) + \tlA_{ik} \pdu{\beta}{k}{j}
                       + \tlA_{jk} \pdu{\beta}{k}{i}
                       - \frac{2}{3} \tlA_{ij} \pdu{\beta}{k}{k} \nonumber \\
                 & 
                 +\upwindl{\tlA}{ij}{k} - \emfp \alpha 8 \pi
                       \left (T_{ij}-\frac{1}{3} \gamma_{ij} s\right ), \label{eq:dtA} \\
  \dt{\tlG^{i}} & = -2 \tlA^{ij} \pdu{\alpha}{}{j} + 2 \alpha \left (
                    \tlG^{i}_{jk} \tlA^{jk} - \frac{2}{3} \tlg^{ij}
                    \pdu{K}{}{j} +
                  6 \tlA^{ij} \pdu{\phi}{}{j}\right )
                 \nonumber\\
                  &                     +\tlg^{jk} \pdpdu{\beta}{i}{j}{k} + \frac{1}{3}
                    \tlg^{ij} \pdpdu{\beta}{k}{j}{k} -\tlGn^{j}\pdu{\beta}{i}{j}
                    + \frac{2}{3} \tlGn^{i}\pdu{\beta}{j}{j} \nonumber \\
                    & 
                    + \upwindu{\tlG}{i}{j} -16 \pi \alpha \tlg^{ij} s_j, \label{eq:dtG}
\end{align}
where
\begin{align}
  \rho & =  \frac{1}{\alpha^2} ( T_{00} - 2 \beta^{i} T_{0i} +
             \beta^{i}\beta^{j} T_{ij} ),\label{eq:BSSN_rho} \\
  s & =  \gamma^{ij} T_{ij}, \\
  s_{i} & =  -\frac{1}{\alpha} ( T_{0i} - \beta^{j} T_{ij}),\label{eq:BSSN_Si}
\end{align}
and $\upwindu{}{}{i}$ denote partial derivatives that are ``upwinded'' based on the
shift vector. This means that the stencil used for
finite differencing is shifted by one in the direction of 
the shift. As an example of this, look at second order
finite differencing, where a derivative in the x-direction 
at grid point $x_i$ would normally be approximated by the 
centered finite difference
\be
\left.\frac{\partial f}{\partial x}\right|_{x=x_i}\approx \frac{-f_{i-1}+f_{i+1}}{2\Delta x} + O\left(\Delta x^2\right).
\ee
If the shift is positive we instead use the ``upwinded"
finite difference approximation
\be
\left.\frac{\bar{\partial} f}{\partial x}\right|_{x=x_i}\approx \frac{-3 f_{i}+4 f_{i+1}-f_{i+2}}{2\Delta x}+O\left(\Delta x^2\right), 
\ee
whereas if the shift is negative we use
\be
\left.\frac{\bar{\partial} f}{\partial x}\right|_{x=x_i}\approx \frac{f_{i-2}-4 f_{i-1}+3f_{i}}{2\Delta x}+O\left(\Delta x^2\right).
\ee
The superscript ``TF'' in the evolution equation of $\tlA_{ij}$ denotes 
the trace-free part of the bracketed term. Note that there is a
slight subtlety to the treatment of $\tlG^{i}$. We introduce
\be
\tlGn^i=\tlg^{jk}\tlG^i_{jk},
\ee
i.e.\ a numerical recalculation of
the contracted conformal Christoffel symbols from the current
conformal metric. This is used instead of $\tlG^{i}$ whenever
derivatives of $\tlG^{i}$ are not needed. In all other places,
i.e.\ when finite differences are needed, the evolved variables, $\tlG^{i}$, are used directly.
This helps with numerical stability and makes the constraint
$\tlG^{i}=-\partial_j\tlg^{ij}$ better behaved. 
Finally $R_{ij} = \tlR_{ij} + R^{\phi}_{ij}$, where
\begin{align}
  \tlG_{ijk} = & \frac{1}{2}\left ( \pdl{\tlg}{ij}{k} + \pdl{\tlg}{ik}{j}
               - \pdl{\tlg}{jk}{i} \right ), \\
  \tlGmixed{ij}{k} = & \tlg^{kl} \tlG_{ijl}, \\
  \tlG^{i}_{jk} = & \tlg^{il}\tlG_{ljk}, \\
  \tlGn^{i} = & \tlg^{jk} \tlG^{i}_{jk}, \\
  \tlR_{ij}  = &  -\frac{1}{2} \tlg^{kl} \pdpdl{\tlg}{ij}{k}{l}
                  +\tlg_{k(i} \pdu{\tlG}{k}{j)}
                  +\tlGn^{k} \tlG_{(ij)k} 
            +\tlG^{k}_{il} \tlGmixed{jk}{l} \nonumber \\
                &  +\tlG^{k}_{jl} \tlGmixed{ik}{l}
                  +\tlG^{k}_{il} \tlGmixed{kj}{l}, \\
  R^{\phi}_{ij} = &  -2\left (\pdpdu{\phi}{}{i}{j}
                 -\tlG^{k}_{ij}\pdu{\phi}{}{k}\right )
                 -2\tlg_{ij} \tlg^{kl} 
            \left ( \pdpdu{\phi}{}{k}{l}
                 -\tlG^{m}_{kl}\pdu{\phi}{}{m}\right ) \nonumber \\
            &  + 4\pdu{\phi}{}{i}\pdu{\phi}{}{j}
             - 4\tlg_{ij}\tlg^{kl}\pdu{\phi}{}{k}\pdu{\phi}{}{l},
\end{align}
where the standard notation for symmetrization, $A_{(ij)}=(A_{ij}+A_{ji})/2$, has been used.
The additional constraints associated with the conformal metric, $1-\tlg=0$, and
conformal traceless part of the extrinsic curvature, $\tlA=\tlg^{ij}\tlA_{ij}=0$,
are enforced at every sub-step of the time integrator.

The BSSN evolution equations above do not give a prescription for how to choose the
gauge variables, $\alpha$ and $\beta^{i}$, but a simple form are the so-called \emph{moving puncture gauges}, i.e.
\be
  \partial_t \alpha = -2 \alpha K
\ee
and
\be
  \partial_t \beta^i = \frac{3}{4}\left (\tlG^{i}-\eta \beta^i\right ).
\ee

\subsubsection{Coupling between matter and spacetime}
\label{sec:matter_spacetime_coupling}
Since in the \SpB methodology the hydrodynamic equations are solved with Lagrangian
particles, but the spacetime is solved on a mesh, the particles and the mesh
need to continuously and accurately exchange information. The spacetime
is evolved on a relatively simple adaptive mesh which is described in
detail in \citet{diener22a,rosswog23a,rosswog25c}. As one sees
from Eqs.~(\ref{eq:GR_momentum_evolution}) and
(\ref{eq:GR_energy_evolution}), the particle evolution equations need the metric and its derivatives,
which are known at grid points, while the spacetime/BSSN-evolution equations, see
Eqs~(\ref{eq:dtK}), (\ref{eq:dtA}) and (\ref{eq:dtG}), need the
matter's energy momentum tensor, Eq.~(\ref{eq:EM_tensor}),  which is
known at the particle positions. In  other words, the metric and its
derivatives need to be interpolated --at every Runge-Kutta sub-step--
to the particle positions ("mesh-to-particle"-, or M2P-step) and the
energy momentum tensor needs to be mapped to grid points based on the
values at the surrounding particles ("particle-to-mesh"-, or
P2M-step).

The simpler of the two is the M2P-step. In \SPHI it is performed via a
5$^{\rm th}$ order Hermite interpolation, that was developed in \citet{rosswog21a} 
and extends  the work of \citet{timmes00a}. Contrary to a standard
Lagrange polynomial interpolation, the Hermite  interpolation
guarantees that the metric remains twice differentiable as particles
pass from one grid cell to another and therefore this sophisticated
interpolation avoids the introduction of additional noise.  
The details of this approach are explained in Sect.~2.4 of 
\citet{rosswog21a} to which we refer the interested reader.

Since the particles are continuously moving and not arranged
in any particular lattice, the accurate interpolation of the energy-momentum tensor
$T_{\mu\nu}$, known at the particles, see Eq.~(\ref{eq:EM_tensor}), to
the mesh points is substantially more challenging than the
M2P-step. After many experiments \citep{rosswog21a,diener22a} involving moving
least square methods and  kernels that are used in the context of
vortex methods \citep{cottet00}, the \SpB team settled
\citep{rosswog23a} on the local regression estimate (LRE) approach
described in Sect.~\ref{sec:accurate_gradients} in combination with a
Multi-dimensional Optimal Order Detection (MOOD). The main idea is to
have a mesh point as ``point of interest'' and to optimize the
polynomial coefficients $\beta$, Eq.~(\ref{eq:coeff}), at this point
\emph{for a given polynomial order}. Typically one wants high
polynomial order in very well resolved regions such as the stellar
interior and lower order at, say, the stellar surface to avoid possible oscillations.  
To decide which polynomial order
is most accurate at a given mesh point, one calculates
the results for all polynomial orders up to a maximum (usually up to
cubic or quartic polynomials) and out of these possible solutions one
selects the one that minimizes an error criterion. For more information, 
the interested reader is referred to the detailed description in \citet{rosswog23a}.

\subsubsection{GR-SPH tests}
\begin{figure}
   \centering
   \includegraphics[width=1.\columnwidth]{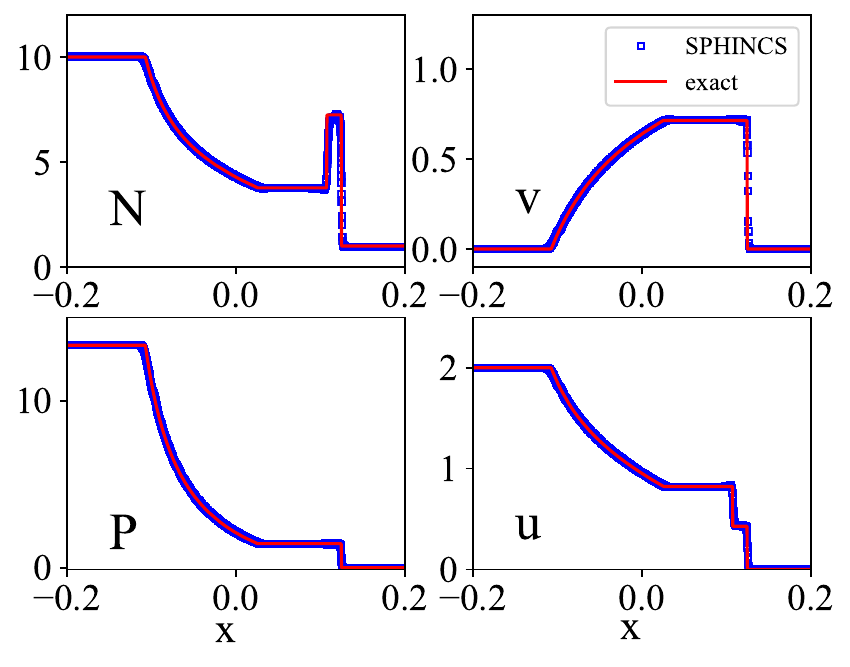} 
   \caption{Result of a 3D relativistic shock tube (initial particle spacing on the left $\Delta x_L=0.0005$) at t= 0.15, 
   numerical results are shown as blue squares, the exact solution is shown in red. From \citet{rosswog21a}.}
\label{fig:rel_Sod}
\end{figure}

\begin{figure}
\includegraphics[width=\textwidth]{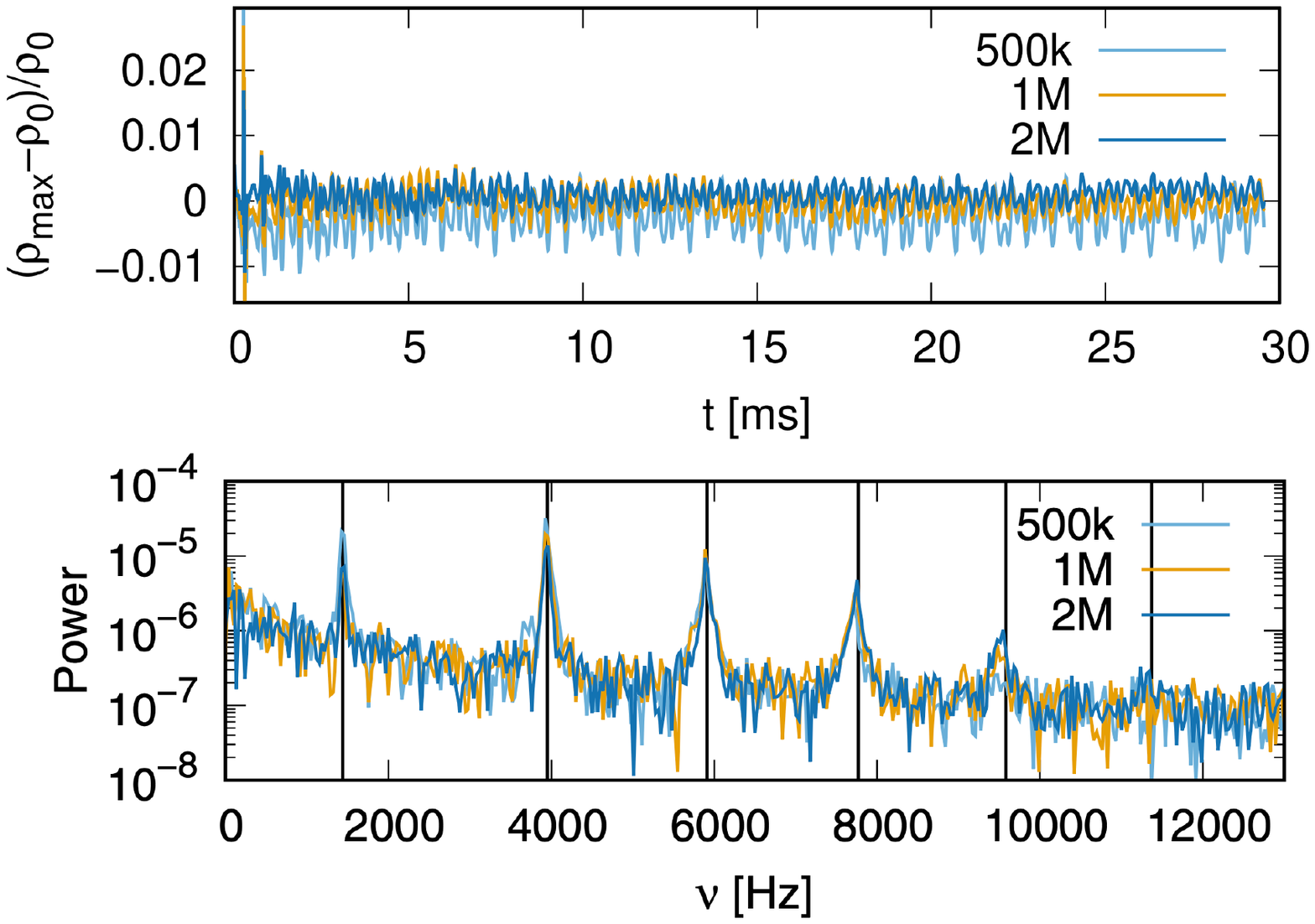}
\caption{Top plot: The relative difference between the maximal density 
$\rho_{\mathrm{max}}$ and the initial maximal density, $\rho_0$, for the 
neutron  star as function of time in units of ms. Bottom plot: The power in the 
Fourier spectrum as function of
frequency in units of Hz. In both plots the light blue line is for 500k, the
orange is for 1M and the dark blue is for 2M particles. In the bottom plot, 
the vertical lines indicate (from left to right) the position of the expected frequency for the fundamental mode at 1.442 kHz, the first harmonic at 3.952
kHz, the second harmonic at 5.913 kHz, the third harmonic at 7.771 kHz, the
fourth harmonic at 9.584 kHz and the fifth harmonic at 11.373 kHz. Even
though the fifth harmonic is barely visible, we can still extract the  frequency to better than 1\% accuracy (see Table~\ref{tab:tov}) from
the simulation data. Figure from \cite{rosswog25c}.}\label{fig:tov}
\end{figure}

Our  first numerical example scrutinizes the ability of the full-GR code \SpB to correctly
reproduce the special-relativistic hydrodynamics limit when a fixed Minkowski metric
is specified. 
The test is a relativistic version of ``Sod's shocktube'' \citep{sod78}, 
a widespread benchmark for relativistic hydrodynamics codes 
\citep{marti96,chow97,siegler00a,delzanna02,marti03}. It uses a polytropic 
exponent $\Gamma=5/3$ and starts from
\be
\left[ N, P \right]=   \left\{
    \begin{array}{ll}
         \left[10,\frac{40}{3}\right], & {\rm for \;}  x<0\\
   	 \left[1,10^{-6}\right]          & {\rm for \;} x \ge 0,
   \end{array}
  \right.
\ee
with zero velocities everywhere.
Particles with equal baryon numbers are initially placed on close-packed lattices as described in \cite{rosswog15b},
so that on the left side the particle spacing  is $\Delta x_L= 0.0005$ and there are 12 particles in both $y$- 
and $z$-direction. 
The numerical result \citep{rosswog21a} at $t= 0.15$ is shown in Fig.~\ref{fig:rel_Sod}  together with
 the exact solution \citep{marti03}. Overall there is very good 
agreement  with practically no spurious oscillations. \\
The second test is an oscillating star evolved in a fully general relativistic spacetime.
This test probes all ingredients, general relativistic hydrodynamics, 
the spacetime evolution and their mutual coupling via the mappings from the 
grid to the particles and from the particles to the grid. 
For this problem oscillation frequencies are known from independent, 
linear perturbation approaches, see e.g. \cite{font02}, which serve as a measure of the accuracy of the 
\SpB results. The initial neutron star is constructed for a polytropic EOS with $K=100$ and a
central density $\rho_c=1.28\times 10^{-3}$ ($=7.91\times 10^{14}$g/cm$^3$), which results
in a gravitational mass of 1.4~\Msun and a baryonic mass of 1.506~\msun.
Due to truncation error the star does
not remain in exact equilibrium and different eigenmodes  are
excited at eigenfrequencies that depend on the equation of state and the mass
of the star. 
Fig.~\ref{fig:tov}  shows 3 simulations (up to $\approx$ 30 ms) with varying resolution of both the number of
particles (low: 500k; medium: 1M; high: 2M particles) and the grid. In all cases 
the outer grid boundaries in all directions are at 160 
in code  units ($\approx$ 236 km),  4 refinement levels are used
with the star being completely contained within the finest grid. The grid resolutions are
0.4$\approx$ 590 m (low), 0.317$\approx$ 468 m (medium) and 0.25$\approx$ 369 m 
(high). The smallest SPH smoothing length in each case was about 310 m (low), 255 m 
(medium) and 208 m (high).
In the top plot we show the relative 
difference between the maximal density, $\rho_{\mathrm{max}}$, and the initial maximal 
density, $\rho_0$, as a function of time for the three runs. Here light blue is low, orange
is medium and dark blue is high resolution. As expected, the amplitudes of the oscillations 
decrease with higher resolution ($\approx 0.002$ for 2M particles) and  multiple modes at different 
frequencies are excited. Note also the absence of any systematic drift in the central density value.
The Fourier transform of
$(\rho_{\mathrm{max}}-\rho_0)/\rho_0$, displayed in the bottom plot of 
Fig.~\ref{fig:tov}, shows  one weak peak as well as five strong peaks, all
at frequencies that agree very well with the frequencies of the
fundamental mode and the five first harmonics \footnote{Kindly provided  by Kostas Kokkotas and Nick Stergioulas.}.
The reference values for the known frequencies are indicated
with black vertical lines in the plot, quantitative error measures
are provided in Table~\ref{tab:tov}. Even at
the lowest resolution, the frequencies agree to better than 1\% with the 
reference solution and they further improve with resolution.\\
\begin{table}
\caption{Oscillation frequencies: fundamental mode (F), first, second,
third, fourth and fifth harmonic (H1, H2, H3, H4 and H5) frequencies 
are given in Hz in the "Reference" column. For each
resolution the extracted frequency of all modes (number before 
$\pm$), as well as an estimate of the accuracy of its extraction (number
after $\pm$) are given. The relative error between the extracted frequency and
the reference frequency is given in the parenthesis. Even at the
lowest resolution the agreement with the reference solution 
is better than 1\%.}
\label{tab:tov}       
\begin{tabular}{p{1.1cm}p{2.0cm}p{2.5cm}p{2.5cm}p{2.5cm}}
\hline\noalign{\smallskip}
Mode & Reference (Hz) & 500k (Hz) & 1M (Hz) & 2M (Hz) \\
\hline\noalign{\smallskip}
F & 1441.9 & 1435.9 $\pm$ 2.5 (0.4) & 1441.5 $\pm$ 6.1 (0.03) &
1439.0 $\pm$ 4.0 (0.2) \\
H1 & 3952.4 & 3937.0 $\pm$ 3.3 (0.4) & 3942.1 $\pm$ 5.5 (0.3) &
3945.4 $\pm$ 4.2 (0.2) \\
H2 & 5912.5 & 5890.1 $\pm$ 2.7 (0.4) & 5895.7 $\pm$ 5.9 (0.3) &
5901.8 $\pm$ 4.4 (0.2) \\
H3 & 7770.6 & 7713.4 $\pm$ 5.0 (0.7) & 7733.7 $\pm$ 6.1 (0.5) &
7750.4 $\pm$ 4.0 (0.3) \\
H4 & 9583.9 & 9571.5 $\pm$ 32 (0.1) & 9532.4 $\pm$ 52 (0.5) &
9537.8 $\pm$ 21 (0.5) \\
H5 & 11373.1 & 11345.9 $\pm$ 19 (0.2) & 11357.5 $\pm$ 31 (0.1) &
11326.3 $\pm$ 36 (0.4) \\
\end{tabular}
\end{table}
Further successfully passed tests \citep{rosswog21a} include a) evolving an initial equilibrium neutron star
constructed in 1D with the full 3D code which shows that the equilibrium is nearly perfectly maintained,
see Fig.~10 ibid., b) oscillations in Cowling approximation which very accurately reproduce
the oscillation frequencies found with completely independent methods, see Fig.~7 ibid.,  and c)
the so-called "migration test" where a star is prepared on the unstable
branch of the mass radius relation.  The evolution of this star depends --according to Eulerian full 
GR simulations \citep{font02,bernuzzi10}-- very sensitively on the initial perturbation: if one just 
lets the star evolve (i.e. the initial perturbation comes just from truncation error) the star undergoes 
violent oscillations until it settles on the stable branch. If instead there is a tiny velocity perturbation 
inward (0.005c), the neutron star collapses into a black hole. This subtle response to different,
small disturbances could also be reproduced with \spB,  see Secs.~3.4 and 3.5 in \citet{rosswog21a}.

\newpage

\section{SPH simulations of compact object mergers}
\label{sec:astro_mergers}
The remaining part of this review is dedicated to  actual applications of SPH
to astrophysical studies of compact objects.  We will focus on three types of encounters:
\begin{itemize}
\item two white dwarfs (Sect.~\ref{sec:astro_WDWD}), 
\item two neutron stars (Sect.~\ref{sec:NSNS}) and 
\item a neutron star with a black hole  (Sect.~\ref{sec:NSBH}).
\end{itemize}
In each case the main focus will be on gravitational-wave-driven binary mergers, but also
dynamical collisions as they may occur in locations with large stellar number densities
may yield very interesting (though probably rarer) signatures and are therefore also 
briefly discussed. In each sub-field, important results have also been obtained by means 
of other methods. Naturally, since the scope of this review is on SPH-methods, we will focus
here on results that are at least partially based on SPH simulations.

\subsection{Encounters of two white dwarfs}
\label{sec:astro_WDWD}
\subsubsection{Relevance}
White Dwarfs (WDs) are the evolutionary end stages of most stars in the Universe,
for every solar mass of stars that forms $\sim 0.22$ WDs will be produced on average. As a result, the Milky Way 
contains $\sim 10^{10}$ WDs \citep{napiwotzki09} in total and  $\sim 10^8$ double WD systems
\citep{nelemans01a}. About half of these systems have separations that are small enough (orbital
periods $<10$ hrs) so that gravitational-wave emission will bring them into contact 
within a Hubble time, making them a major target for the LISA mission \citep{amaro23}. 
Once in contact, the binary system will merge in almost all cases. In the remaining
small fraction of cases mass transfer may stabilize the orbital decay and lead to long-lived 
interacting binaries such as AM CVn systems \citep{paczynski67,warner95,nelemans01b,nelemans05,solheim10}.

Those systems that merge may have a manifold of interesting possible outcomes. The  merger 
of two He WDs may produce a low-mass He star \citep{webbink84,iben86,saio00,han02}, He-CO 
mergers may form hydrogen-deficient giant or R CrB stars \citep{webbink84,iben96,clayton07} and 
if two CO WDs merge, the outcome may be a more massive, possibly hot and high B-field WD 
\citep{bergeron91,barstow95,segretain97}. A good fraction of the CO-CO merger remnants probably transforms into
ONeMg WDs which finally, due to electron capture on Ne and Mg, undergo an accretion-induced collapse (AIC)
to a neutron star \citep{saio85,nomoto91,saio98}.
Given that the nuclear binding energy that can still be released by burning to iron group elements 
(1.6 MeV from He, 1.1 MeV from C and 0.8 MeV from O) is large, it is not too surprising
that there are also various pathways to thermonuclear explosions. The ignition of helium
on the surface of a WD may lead to weak thermonuclear explosions \citep{bildsten07,foley09,perets10}, 
sometimes called ``.Ia'' supernovae. WDWD mergers might also trigger
type Ia supernovae (SN~Ia) \citep{webbink84,iben84} and, in some cases,
even particularly bright ``super-Chandrasekhar'' explosions, e.g. \citet{howell06,hicken07}. 

SN~Ia  are important as cosmological distance indicators, as factories for intermediate mass 
and iron-group nuclei, as cosmic ray accelerators, kinetic energy sources for galaxy evolution or 
simply in their own right as end points of  binary stellar evolution. After having been the second-best 
option behind the ``single degenerate'' model for decades, it now seems likely that double degenerate mergers 
are behind at least a sizeable fraction of SN~Ia.
See \citet{howel11}, \citet{maoz14}, and \citet{ruiter25} for excellent
reviews on this topic.

Below, we will briefly summarize the challenges in a numerical simulation of a WDWD merger 
(Sect.~\ref{sec:challenges_WDWD}) and then discuss recent results concerning mass transferring 
systems (Sect.~\ref{sec:WDWD_MT}) and, closely related, to the final merger of a WDWD binary 
and possibilities of triggering SN~Ia (Sect.~\ref{sec:WDWD_SNIa}). We will also briefly discuss dynamical
collisions of WDs (Sect.~\ref{sec:WDWD_collisions}). For SPH studies that explore the 
gravitational-wave signatures of WDWD mergers we refer to the literature 
\citep{loren05,dan11,vandenbroek12}.

Note that in this section we explicitly include the constants $G$ and $c$ in the equations to 
allow for a simple link to the astrophysical literature.

\subsubsection{Challenges}
\label{sec:challenges_WDWD}
WDWD merger simulations are challenging for a number of reasons not the least of which are
the onset of mass transfer and the self-consistent triggering of thermonuclear explosions.

While two white dwarfs revolve around their common centre of mass, gravitational-wave 
emission reduces the separation $a$ of a circular binary orbit at a rate of \citep{peters63,peters64}
\be
\dot{a}_\mathrm{GW}= -\frac{64 G^3}{5 c^5} \frac{m_1 m_2 M}{a^3},
\ee
where $m_1$ and $m_2$ are the component masses and $M$ is the total mass.
Although it is gravitational-wave emission that drives the binary towards mass transfer/merger
in the first place,  its dynamical impact at the merger stage is completely negligible since
\be
\frac{\tau_\mathrm{GW}}{P_\mathrm{orb}}= \frac{a/|\dot{a}_\mathrm{GW}|}{2\pi/\omega_\mathrm{K}}=
6.6 \cdot 10^8 \left( \frac{a}{2 \times 10^9 \, \mathrm{cm}}\right)^{5/2} \left(\frac{0.6 \; M_\odot}{m_1} \right) 
\left( \frac{0.6 \; M_\odot}{m_2}\right) \left( \frac{1.2 \;
    M_\odot}{M}\right)^{1/2}
\label{eq:tau_GW_tau_orb}
\ee
and therefore does not need to be
modelled explicitly in a WDWD merger. 
Mass transfer will set in once the size of the Roche lobe of one of the stars has become comparable to the 
enclosed star. Due to the inverted mass-radius relationship of WDs, it is always the less massive WD (``secondary'') that fills 
its Roche lobe first. From the Roche lobe size and Kepler's third law, the average density 
$\bar{\rho}$ of the donor star can be related to the orbital period \citep{paczynski71,frank02}
\be
\bar{\rho} \approx \frac{ 115 \, \mathrm{g\, cm}^{-3} } {P^2_\mathrm{hr}},
\ee
where $P_\mathrm{hr}$ is the orbital period measured in  hours. In other words: the shorter the
orbital period, the higher the density of the mass donating star. For periods below 1 hr the donor 
densities exceed those of  main sequence stars which shows that a compact star is involved. If
one uses the typical dynamical timescale of a WD, $\tau_\mathrm{dyn}\approx (G \bar{\rho})^{-1/2}$, one finds
\be
\frac{P_\mathrm{orb}}{\tau_\mathrm{dyn}}\approx 10,
\ee
so that a single orbit would already require $\sim 10000 (n_\mathrm{dyn}/1000)$ numerical time steps, if $n_\mathrm{dyn}$
denotes the number of numerical time steps per stellar dynamical time. This demonstrates that longlived mass transfer
over tens of orbital periods can become quite computationally expensive and may place limits on
the numerical resolution that can be afforded in such a simulation. On the other hand, when
numerically resolvable mass transfer sets in, it already has a rate of
\begin{equation}\label{eq:mmin}
\dot M_\mathrm{lim}\sim \frac{1\ \text{particle mass}}{\text{orbital period}}
\sim 2 \times 10^{-8}
\frac{{\rm M}_\odot}{\rm s}\left(\frac{10^6}{\mathrm{n_{part}}}\right)
\left(\frac{{\rm M}}{1\, {\rm M}_\odot}\right)^{3/2}
\left(\frac{2\cdot 10^9\, \mathrm{cm}}{a_0}\right)^{3/2},
\end{equation}
where $\mathrm{n_{part}}$ is the total number of SPH particles, ${\rm M}$ is the
total mass of the binary and $a_{0}$ is the separation between the stars
at the onset of mass transfer. This limit, which is due to finite
numerical resolution, is several
orders of magnitude above the Eddington limit of WDs. Therefore sub-Eddington 
accretion rates are hardly ever resolvable within a global 3D SPH simulation. The 
transferred matter comes initially
from the tenuous WD surface which, in SPH, is the poorest resolved region of
the star. Following this matter is also a challenge for Eulerian methods since it
needs to be disentangled from the ``vacuum'' background and, due to the 
resolution-dependent angular momentum conservation, it is difficult to obtain the 
correct feedback on the orbital evolution. In other words: {\em the consistent simulation
of mass transfer and its feedback on the binary dynamics is a serious challenge for 
every numerical method}. 

The onset of mass transfer represents a juncture in the life of WDWD binary, since now the stability
of mass transfer decides whether the binary can survive or will inevitably merge. 
The latter depends sensitively on the internal structure of the donor star, the binary mass 
ratio and the angular momentum transport mechanisms (e.g. \citealt{marsh04,gokhale07}).
Due to the inverse mass-radius relationship of WDs (less massive WDs are larger), the secondary will expand on mass loss and 
therefore tendentially speed up the mass loss further. On the other hand, since the mass is transferred 
to the higher mass object, momentum/centre of mass conservation will tend to 
widen the orbit and therefore tendentially reduce mass transfer. If the circularization radius 
of the transferred matter is smaller than the primary radius it will directly impact on the stellar 
surface and tend to spin up the accreting star. In this way, orbital angular momentum 
is lost to the spin of the primary which, in turn, decreases the orbital separation and accelerates 
the mass transfer. If, on the other hand, the circularization radius is larger than the primary radius 
and a disk can form, angular momentum can, via the large lever arm of the disk, be fed back into the orbital 
motion and stabilize the system \citep{iben98,piro11}. To make things even more complicated, if tidal interaction 
substantially heats up the mass donating star it may have an impact on
its internal structure and therefore change its  response to mass loss.

\begin{figure}[ht]
   \centering
   \includegraphics[width=\textwidth]{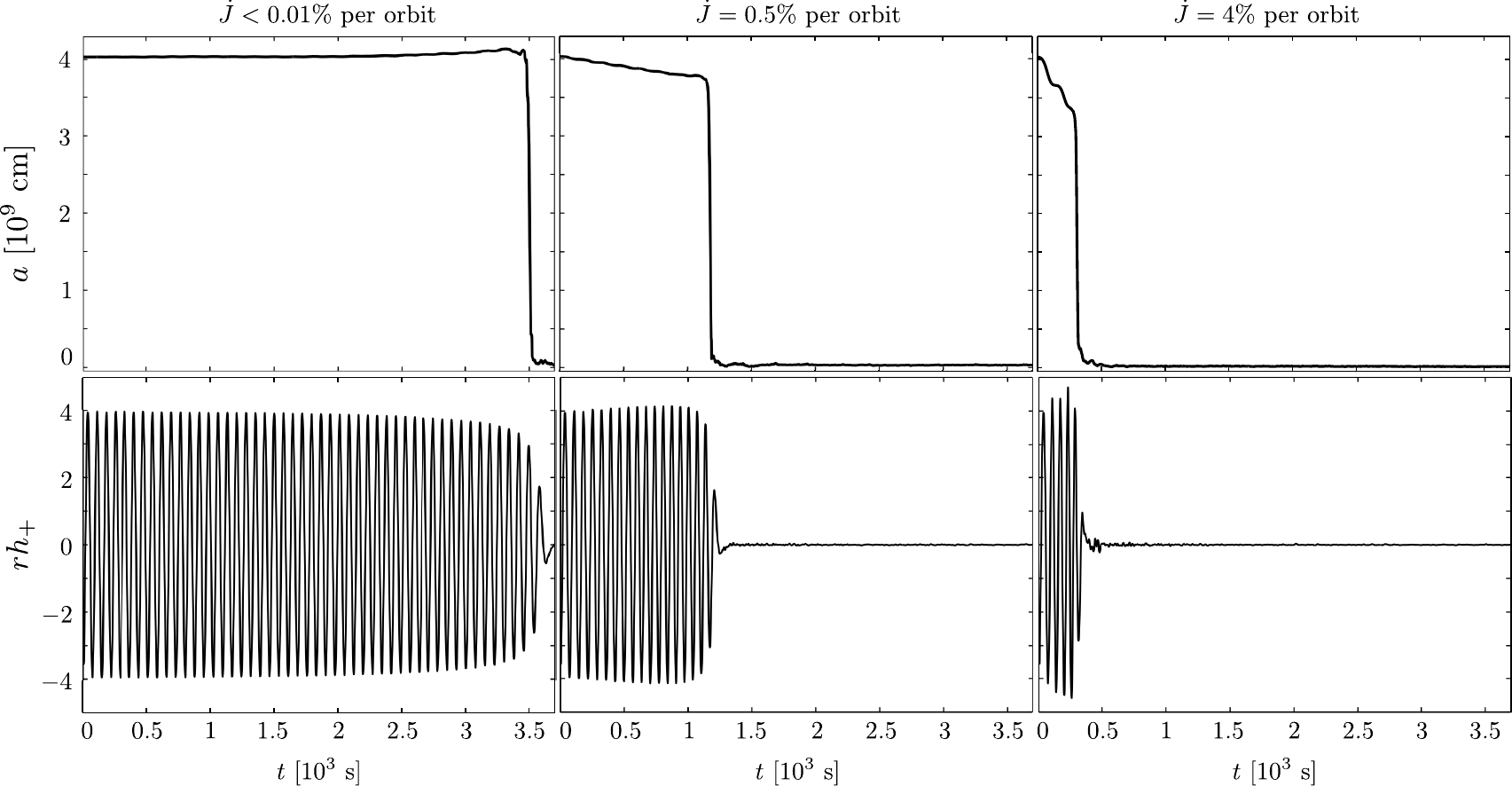}
 \caption{Numerical experiment to illustrate the sensitivity of a binary system to the non-conservation
               of angular momentum. A WDWD binary system (0.3 and 0.6 \msun) is adiabatically relaxed to the point where mass
               transfer is about to set in and then its orbital evolution is followed in a 3D hydrodynamic simulation. 
               To mimic numerical loss of angular momentum, an artificial force
               is applied that removes angular momentum at a rate of 0.5 \% (middle) and 4\% per orbit (right panel).
               The simulation shown in the leftmost panel  conserves angular momentum to better than 0.01\% per orbit. Shown
               is the binary separation $a$ (upper row) and the gravitational-wave amplitude $h_+$ times to
               distance to the source $r$. Image courtesy of Marius Dan.}
   \label{fig:conservation_GWs}
\end{figure}

To numerically reliably capture these complex angular momentum transfer mechanisms
requires a very accurate intrinsic angular momentum conservation of the used numerical method. We want to 
briefly illustrate this point with a small numerical experiment. A $0.3+0.6\,M_\odot$ WD binary system 
is prepared in a Keplerian orbit, so that mass transfer is about to set in. To mimic the effect of numerical 
angular momentum loss in a controllable way, we add small artificial forces similar to those emerging 
from gravitational-wave emission \citep{peters63,peters64,davies94} and adjust the overall value so 
that 4\% or 0.5\% of angular momentum per orbit are lost. These results are compared to a 
simulation without artificial loss terms where the angular momentum is conserved to better 
than 0.01\% per orbit, see Fig.~\ref{fig:conservation_GWs}. The effect on the mutual separation 
$a$ (in $10^9\,\mathrm{cm}$) is shown in the upper panels and the gravitational-wave amplitude, 
$h_+$ ($r$ is the distance to the observer) as calculated in the quadrupole formalism, are shown 
the lower panels. 

Even the moderate loss of 0.5\% angular momentum per orbit leads to a quick artificial merger and a mass transfer
duration that is reduced by more than a factor of three. These conservation requirements make SPH a natural 
choice for WDWD merger simulations and it has indeed been  the first  method used for these type of problems.

As outlined above, one of the most exciting possibilities is the triggering of thermonuclear 
explosions during the interaction of two WDs. Such explosions can  either be triggered by a shock wave
where the thermonuclear energy generation behind the shock wears down possible dissipative effects
or, spontaneously, if the local conditions for burning are favourable enough so that it occurs 
faster than the star can react by expanding. \citet{seitenzahl09} have studied detonation conditions
in detail via local simulations and found that critical detonation conditions can require that length 
scales down to centimetres are resolved which is, of course,  a serious challenge for global, 3D simulations of
objects with radii of $\sim 10^9$ cm.

There is also a huge disparity in terms of timescales. Whenever nuclear burning is important for
the dynamics of the gas flow, the nuclear timescales are many orders of magnitude shorter than
the admissible hydrodynamic time steps. Therefore, nuclear networks are usually implemented via
operator splitting methods, see e.g. \citet{benz89,rosswog09a, raskin10,garcia_senz13}. Because of 
the exact advection in SPH the post-processing of hydrodynamic trajectories with larger nuclear
networks to obtain detailed abundance patterns is straight forward.  For burning processes in tenuous
surface layers, however, SPH is seriously challenged since here the resolution is poorest. For
such problems hybrid approaches that combine SPH with, say, AMR methods \citep{guillochon10}
seem to be the best strategies.


\subsubsection{Dynamics and mass transfer in white dwarf binaries systems}
\label{sec:WDWD_MT}
Three-dimensional simulations of WDWD mergers were pioneered by 
\citet{benz90b}. Their major motivation was to understand the merger dynamics and the possible
role of double degenerate systems as SN~Ia progenitors. 
They used an SPH formulation as described in Sect.~\ref{sec:vanilla_ice} (``vanilla ice'') together with  7000 SPH 
particles, an equation of state for a non-degenerate ideal gas with a completely
degenerate, fully relativistic electron component and they restricted themselves to the study 
of a 0.9 - 1.2 \Msun system. No attempts were undertaken to include nuclear burning in this 
study (but see \citealt{benz89}). Each star was relaxed in isolation, see Sect.~\ref{sec:IC}, and subsequently placed in 
a circular Keplerian orbit so that the secondary was overfilling its critical lobe by $\sim8\%$.
Under these conditions the secondary star was disrupted within slightly more than two orbital periods, 
forming a three-component system of a rather unperturbed primary, a hot pressure 
supported spherical envelope and a rotationally supported outer disk. About 0.6 \% of a  
solar mass were able to escape, the remaining $\sim 1.7$ \msun, supported mainly by 
pressure gradients, showed no sign of collapse.

\citet{rasio95} were more interested in the equilibrium and the (secular,
dynamical and mass transfer) stability properties of close binary systems. They studied
systems both with stiff ($\Gamma> 5/3$) as models for neutron stars and soft ($\Gamma= 5/3$)
polytropic equations of state, as approximations for the EOS of  (not too massive) WDs 
and low-mass main sequence star binaries.
They put particular emphasis on constructing accurate, \emph{synchronized} initial conditions 
\citep{rasio94}. These were obtained by relaxing the binary system in a corotating frame where,
in equilibrium, all velocities should vanish. The resulting configurations satisfied the virial 
theorem to an accuracy of about one part in $10^3$. With these initial 
conditions they found a more gradual increase in the mass transfer rate in comparison to 
\citet{benz90b}, but nevertheless the binary was disrupted after only a few  orbital periods.

\citet{segretain97} focussed on the question whether 
particularly massive and hot WDs could be the result of binary mergers \citep{bergeron91}. They applied
a simulation technology similar to \citet{benz90b} and concentrated on a binary system
with non-spinning WDs of 0.6 and 0.9 \msun.  They showed, for example, that such a merger remnant would need
to loose about 90\% of its angular momentum in order to reproduce  properties of the observed 
candidate WDs.

\begin{figure}[ht]
   \centering
   \includegraphics[width=10cm]{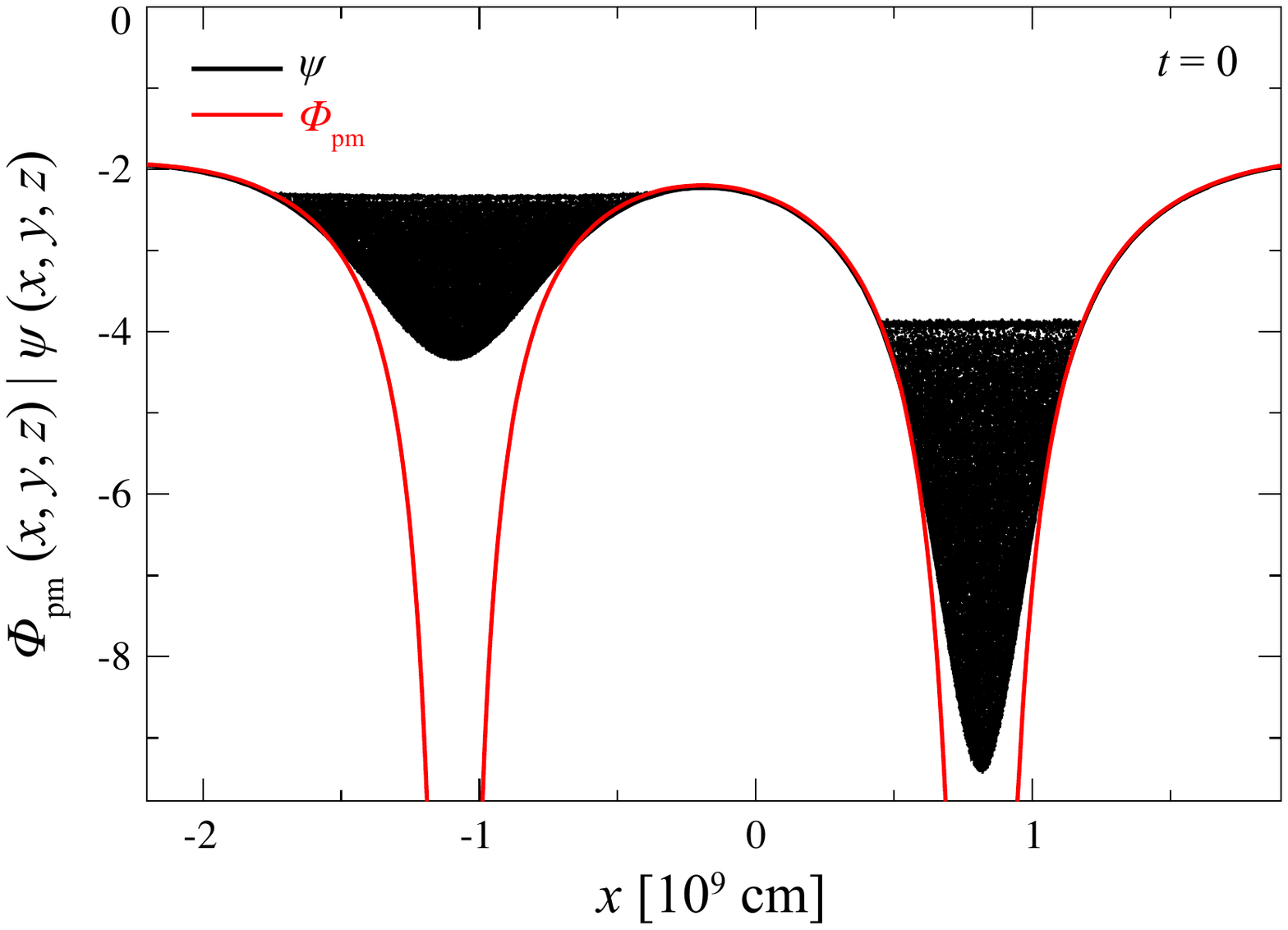}
 \caption{Initial conditions for a synchronized WDWD binary system at the onset
               of resolvable mass transfer. The  red line is the point mass Roche potential, 
               the SPH particle values are shown as filled black circles. Once the first SPH
               particle has crossed the $L_1$ point, the system is transformed from the corotating
               binary frame to a space-fixed frame where it is hydrodynamically evolved. 
                Image reproduced with permission from \citet{dan11}, copyright by AAS.}
   \label{fig:effect_potential}
\end{figure}

Although \citet{rasio95} had already  explored the construction of accurate initial
conditions, essentially all subsequent simulations 
\citep{guerrero04,loren05,yoon07a,loren09,pakmor10,pakmor11,zhu13a} 
were carried out with rather approximate initial conditions consisting of spherical stars placed
in orbits where, according to simple Roche-lobe geometry estimates, mass transfer should
set in. \citet{marsh04} had identified, in  a detailed orbital stability analysis,
definitely stable regions (roughly for primary masses substantially larger than the companion mass), 
definitely unstable (mass ratios between 2/3 and 1) and an intermediate region where the 
stability of mass transfer is less clear. 
Motivated by large discrepancies in the mass transfer 
duration that had been observed between careful grid-based \citep{motl02,dsouza06,motl07} 
and earlier SPH simulations \citep{benz90b,rasio95,segretain97,guerrero04,yoon07a,pakmor10}
\citet{dan09,dan11} focussed on the mass transfer in this unclear 
regime.  They  very carefully relaxed the binary system in a corotating frame 
and thereby adiabatically reduced the mutual separation until the first particle climbed up 
to saddle point $L_1$ in the effective potential, see Fig.~\ref{fig:effect_potential}. 

In their study such carefully constructed initial conditions were compared to the previously 
commonly used approximate initial conditions.  Apart from the inaccuracies inherent to the 
analytical Roche lobe estimates, approximate initial conditions also
neglect the tidal deformations of the stars and therefore seriously underestimate 
the initial separation at the onset of mass transfer. Therefore, such initial conditions have  
up to 15\% too little angular momentum and, as a result, merge more violently on a 
much too short timescale. As a result, temperatures 
and densities in the final remnant are overestimated and the size of tidal tails are underestimated.
The carefully prepared binary systems of \citet{dan11} all showed dozens of orbits of numerically resolvable 
mass transfer. Given that, due to the finite resolution, the mass transfer is already highly 
super-Eddington when it starts being resolvable, all the results on mass transfer duration 
have to be considered as strict lower limits. A 0.2 \Msun He-WD and a 0.8 CO-WD, merged  within
two orbital periods (comparable to earlier SPH results) when
approximate initial conditions were used, but the same system took as
long as 84 orbital periods when the initial conditions were prepared carefully. This particular
example also illustrated the suitability of SPH for such investigations: during the orbital
evolution, which corresponds to $\approx 17 \; 000$ dynamical timescales, energy and angular
momentum were conserved to better than 1\%! All of the investigated (according to the Marsh
et al. analysis) unstable binary systems merged in the end although only after several dozens 
of orbital periods. Some systems showed a systematic widening of the orbits after the onset
of mass transfer. Although they were still disrupted in the end, this indicated that
systems in the parameter space vicinity of this 0.2--0.8 \Msun system may evolve into short-period
AM CVn systems.

\citet{dan12,dan14a} systematically explored the  parameter space 
by simulating 225 different binary systems with masses ranging from 0.2 to 1.2 \msun. All of
the initial conditions were prepared as carefully as in \citet{dan11}. Despite the only moderate resolution 
(40 K particles) that could be afforded in such a broad study, they found 
excellent agreement with the orbital evolution predicted by mass transfer stability analysis 
\citep{marsh04,gokhale07}.

\subsubsection{Double white dwarf mergers and pathways to thermonuclear supernovae}
\label{sec:WDWD_SNIa}

  \begin{figure}[ht]
    \centerline{\includegraphics[width=0.8\textwidth]{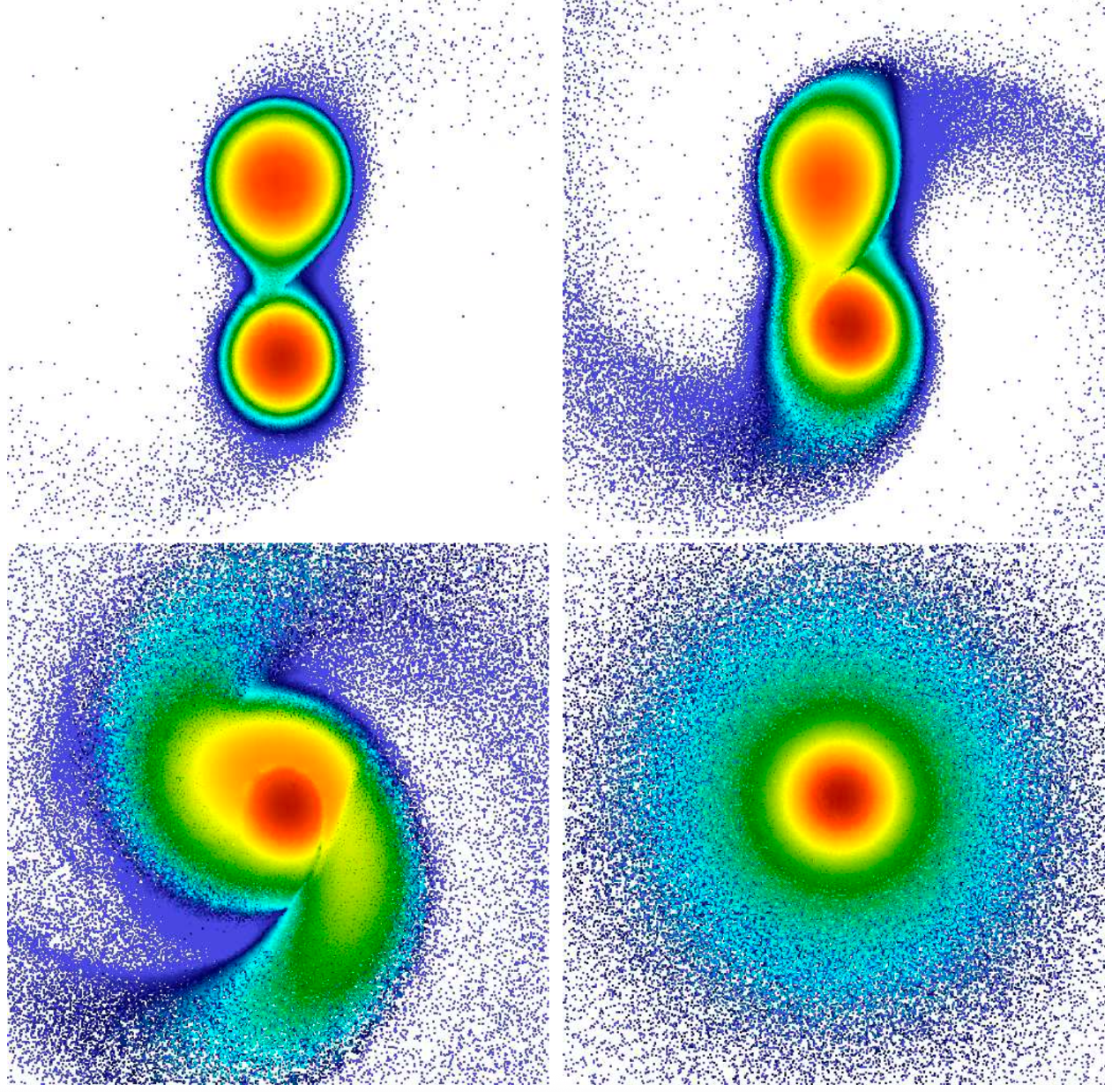}}
    \caption{Illustration of the morphology of a WDWD merger (mass ratio  $q= 0.78$). 
          Image reproduced with permission from \citet{diehl08}, copyright by the author(s).}
    \label{fig:WDWD_merger}
\end{figure}

The merger of two white dwarfs, the so-called ``double degenerate scenario'', had already 
been suggested relatively early on \citep{webbink84,iben84} as a promising type Ia progenitor channel.
It was initially modelled as CO-rich matter being accreted from a thick disk onto a central, 
cold WD \citep{nomoto85,saio85,saio98,piersanti03a,piersanti03b,saio04}.
Since accretion rates close to the Eddington limit ($\dot{M} \sim 10^{-5}$ \Msun yr$^{-1}$) 
are expected for such thick disks, most studies concluded that carbon
ignition would start in the envelope of the central WD and, as the
burning flame propagates inwards, it would
transform the WD  from CO into ONeMg within $\sim$ 5000 years \citep{saio85,saio98}. When
approaching the Chandrasekhar mass, Ne and Mg  would undergo electron
capture and the final result would be an accretion-induced collapse
to a  neutron star rather than a SN~Ia. Partially based on these studies, the double degenerate
model was long regarded as only the second best model, that had some good motivation (consistent
rates, lack of hydrogen in SN~Ia spectra, Chandrasekhar mass as motivation for the uniformity of 
type Ia properties), but lacked a convincing pathway to an explosion.

\citet{guerrero04} were the first to explore the effect
of nuclear burning  during a WD merger event. They implemented the
reduced 14-isotope $\alpha$-network of \citet{benz89} with updated
reaction rates into a ``vanilla-ice'' SPH code with artificial
viscosity enhanced by the Balsara factor, see Sect.~\ref{sec:AV}. They
typically used 40 K particles, approximate initial conditions 
as described above and explored six different combinations of masses/chemical compositions.
They found an orbital dynamics similar to \citet{benz89,segretain97}
and, although, temperatures around $10^9$ K were encountered in the outer
partially degenerate layers of the central core,
no dynamically important nuclear burning was observed. Whenever it set in, the remnant had time 
to quench it by expansion, both for the He and CO accreting systems. Therefore they concluded 
that direct SN~Ia explosions were unlikely, but some remnants could evolve into subdwarf B objects 
as suggested in \citet{iben90}.

\citet{yoon07a} challenged the ``classical picture'' of the cold WD accreting from a 
thick disk as an oversimplification. Instead, they suggested that the subsequent 
secular evolution of the remnant would be better studied by treating the central 
object as a differentially rotating CO star with a central, slowly rotating,
cold core engulfed by a rapidly rotating hot envelope, which, in turn, is embedded 
and fed by a centrifugally supported Keplerian accretion disk. The further evolution 
of such a system is then governed by the thermal cooling of the hot envelope and 
the redistribution of angular momentum inside of the central remnant and the 
accretion of the matter from the disk into the envelope.
They based their study of the secular remnant evolution on a dynamical merger 
calculation of two CO WDs with 0.6 and 0.9 \msun. To this end they used an SPH code originally
developed for neutron star merger calculations \citep{rosswog99,rosswog00,rosswog02a,rosswog03a,rosswog08b} 
extended by the Helmholtz EOS \citep{timmes99} and a quasi-equilibrium reduced $\alpha$-network
\citep{hix98}. Particular care was taken to avoid artefacts from the artificial viscosity treatment
and time-dependent viscosity parameters \citep{morris97} and a Balsara-switch \citep{balsara95}, see
Sect.~\ref{sec:AV}, were used in the simulation. As suggested by the work
of \citet{segretain97}, they assumed non-synchronized stars and started the simulations from
approximate initial conditions, see Sect.~\ref{sec:WDWD_MT}. Once a stationary remnant had 
been formed, the results were mapped into a 1D hydrodynamic stellar evolution code \citep{yoon04a} and its 
secular evolution was followed including the effects of rotation and angular momentum transport. 
They found that the growth of the stellar core is controlled by the neutrino 
cooling at the interface between the core and the envelope and that carbon ignition could be avoided 
provided that a) once the merger reaches a quasi-static equilibrium temperatures are below the 
carbon ignition threshold, b) the angular momentum loss occurs on a timescale longer than the neutrino 
cooling timescale and c) the mass accretion from the centrifugally supported disk is low enough
($\dot{M} \le 5 \times 10^{-6} - 10^{-5}$ \Msun yr$^{-1}$). From such remnants an explosion may be
triggered $\sim 10^5$ years after the merger.

\citep{shen12a} started from two remnants of  CO WD mergers from \citet{dan11} and followed 
their viscous longterm evolution. Their conclusion, however, was more in line with earlier studies: they
expected that the long-term result would be an ONe WD or an accretion-induced collapse to a neutron star rather
than a SN~Ia.

In recent years, WD mergers have been extensively explored as possible pathways to SN~Ia. Not too
surprisingly, a number of pathways have been discovered that likely lead to a thermonuclear
explosion directly prior to or during the merger. Whether these explosions are responsible for 
(some fraction of) normal SNe~Ia or for peculiar subtypes needs to be further explored 
in future work. Many of the recent studies used different numerical tools to 
explore WDWD mergers, but a general review of such studies is beyond
the focus of this article. We therefore focus here on those studies where SPH simulations 
were involved.

\paragraph{Explosions prior to merger}

\citet{dan11} had carefully studied the impact of mass transfer on the orbital dynamics.
In these SPH simulations the feedback on the orbit is accurately modelled, but due to SPH's automatic 
``refinement  on density'' the properties of the transferred matter are not well resolved.
Therefore, they \citep{guillochon10,dan11} decided
for a ``best-of-both-worlds'' approach where they simulated the impact of the mass transfer on 
orbital dynamics with SPH, but recorded the orbital evolution and the mass transfer rate for
a second set of simulations that were performed with the FLASH code \citep{fryxell00}. This second study 
focussed on the detailed hydrodynamic interaction of the transferred mass with the accretor star. For 
He-CO binary systems where He directly impacts on a primary of a mass $>$ 0.9 \msun,
they found that helium surface explosions can be self-consistently triggered via Kelvin--Helmholtz (KH)
instabilities. These instabilities occur at the interface between the incoming helium stream 
and an already formed helium torus around the primary. ``Knots'' produced by the KH instabilities can
lead to local ignition points once the triple-alpha timescale becomes shorter than the 
local dynamical timescale. The resulting detonations travel around the primary surface 
to collide on the side opposite to the ignition point. Such helium surface detonations
may resemble weak type Ia SNe \citep{bildsten07,foley09,perets10} and they may
drive shock waves into the CO core which concentrate in one or more focal 
points, similar to what was found in the 2D study of \citet{fink07}. This 
could possibly lead to an explosion via a ``double-detonation'' mechanism.
In a subsequent large-scale parameter study \citep{dan12,dan14a}, Dan et al. found,
based on a comparison between nuclear burning and hydrodynamical timescales,
that a large fraction of the He-accreting systems do produce explosions early on: 
all dynamically unstable systems with primary masses $< 1.1$ \Msun together with 
secondary masses $>0.4$ \Msun triggered He-detonations at surface contact. A 
good fraction of these systems could also, in addition, produce KH-instability-induced 
detonations as described in detail in \citet{guillochon10}. 
There was no definitive evidence for explosions prior to  contact for any of the 
studied CO-transferring systems.

The scenario where accretion of helium from the secondary WD triggers
a helium explosion on the surface of the primary WD has been further
studied by \citet{pakmor13} with the moving mesh code
AREPO \citep{springel10b}. They found that this detonation propagates
around the primary CO WD and subsequently sends a converging shock
wave into its core, which is known to robustly trigger the detonation
of the CO core (\enquote{double detonation}). \citet{tanikawa18,tanikawa19} also systematically studied the 
double detonation mechanism for the double degenerate model with
high-resolution SPH simulations. They used the Helmholtz EOS 
\citep{timmes00a} together with a 13 isotope nuclear reaction network 
\citep{timmes00b}  and modelled the white dwarfs with up to 80 million
SPH particles. They explored under which conditions the
so-called ``Dynamically-Driven Double-Degenerate Double-Detonation
(D$^6$)'' scenario \citep{shen18} occurs, in which helium is transferred
to a sub-Chandrasekhar WD and --via double-detonation-- explodes the
CO primary. With the more massive WD being abruptly removed through
the explosion, the secondary survives as a ``hypervelocity WD". The
authors further explored the ``triple-detonation'' scenario
\citep{papish15} where the helium accretion triggers the explosion of
the CO primary in a double detonation and this explosion, impacting on
the mass donating helium WD, can trigger its explosion for small
enough binary separations. They also explored yet another variation of
the theme, the so-called ``quadruple-detonation'' scenario, where again
accreted helium triggers a detonation which explodes the CO core of
the primary (first double-detonation) and the explosion interacts with
the secondary, a CO WD surrounded by a helium shell, which then
ignites the secondaries helium which subsequently triggers the CO
detonation of the secondaries core (second double-detonation).

\paragraph{Explosions during merger}
\citet{pakmor10} studied double degenerate mergers, but --contrary to earlier
studies-- they focussed on very massive WDs with masses close to 0.9 \msun.
They used the GADGET code \citep{springel05a}  with some modifications \citep{pakmor12a}, for example, the 
Timmes EOS \citep{timmes99} and a 13 isotope network were implemented for their study.  
In order to facilitate the network implementation, the energy equation (instead of,
as usually in GADGET, the entropy equation) was evolved. No efforts were undertaken
to reduce the constant, untriggered artificial viscosity. They placed
the stars on an orbit with the
approximate initial conditions described above and found the secondary to be disrupted 
within two orbital periods. In a second step, several hot ($>2.9 \times 10^9$K) particles
were identified and the remnant was artificially ignited in these hot spots. The explosion
was followed with a grid-based hydrodynamics code \citep{fink07,roepke07} that had
been used in earlier SN~Ia studies. In a third step, the nucleosynthesis was post-processed
and synthetic light curves were calculated \citep{kromer09}. The explosion resulted in 
a moderate amount of \Nifs (0.1 \msun), large amounts (1.1 \msun) of intermediate 
mass elements and oxygen (0.5 \msun) and less than 0.1 \Msun of unburnt
carbon. The kinetic energy of the explosion ($1.3 \times 10^{51}$ erg) was typical for 
a SN~Ia, but the resulting velocities were relatively small, so that the explosion
resembled a sub-luminous 1991bg-like supernova. An important condition for reaching ignition 
is  a mass ratio close to unity. Some variation in total mass is expected, but cases with less 
than 0.9 \Msun of a primary mass would struggle to reach the ignition temperatures and --if successful-- 
the lower densities would lead to even lower resulting \Nifs   masses and therefore lower luminosities. 
For substantially higher masses, in contrast, the burning would proceed at larger densities
and therefore result in  much larger amounts of \nifs. Based on population synthesis
models \citep{ruiter09} they estimated that mergers of this type of system could account for
2--11 \% of the observed SN~Ia rate.
In a follow up study \citep{pakmor11} the sensitivity of the proposed model to the mass ratio
was studied. The authors concluded that binaries with a primary mass near 0.9 \Msun ignite
a detonation immediately at contact, provided that the mass ratio $q$ exceeds 0.8. Both the 
abundance tomography  and the lower-than-standard velocities provided support for
the idea of this type of merger producing sub-luminous, 1991bg-type supernovae.

As a variation of the theme \citep{pakmor12b} they also explored the merger of a higher 
mass system with 0.9 and 1.1 \msun. Using the
same assumption about the triggering of detonations as before, they found a substantially
larger mass of \Nifs (0.6 \msun), 0.5 \Msun of intermediate mass elements, 0.5 \Msun of Oxygen 
and about 0.15 \Msun of unburnt carbon. Due to its higher density, only the primary is able
to burn \Nifs and therefore the brightness of the SN~Ia would be closely related to the primary
mass. The secondary is only incompletely burnt and thus provides the bulk of the intermediate mass 
elements. Overall, the authors concluded that such a merger reproduces the observational 
properties of normal SN~Ia reasonably well.
In \citet{kromer13}  the results of a 0.9 and 0.76 \Msun CO-CO merger were analyzed
and unburned oxygen was found close to the centre of the ejecta, which produces narrow emission
lines of [O1] in the late-time spectrum, similar to what is observed in the sub-luminous 
SN 2010lp \citep{leibundgut93}.

\citet{fryer10} applied a sequence of computational tools to study the spectra that 
can be expected from a supernova triggered by a double-degenerate merger. Motivated by population synthesis
calculations, they simulated a CO-CO binary of 0.9 and 1.2 \Msun  with the SNSPH
code \citep{fryer06}. They assumed that the remnant would explode at the Chandrasekhar
mass limit, into a gas cloud consisting of the remaining merger debris. Since they found a density
profile with $\rho \propto r^{-4}$, they  inserted an explosion
\citep{meakin09} into such a matter distribution and calculated signatures with the radiation-hydrodynamics 
code RAGE \citep{fryer07,fryer09}.  They found that in such ``enshrouded''  SNe~Ia, the debris extends 
and delays the X-ray flux from a shock breakout and produces a signal that is closer to a SN Ib/c. 
Also the  V-band peak was extended and much broader than in a normal SN~Ia  with
the early spectra being dominated by CO lines only.  They concluded that, within their model, 
a CO-CO merger with a total mass $>1.5$ \Msun would not produce spectra and light curves 
that resemble normal type Ia supernovae.

One of the insights gained from the study of \citet{pakmor10} was that (massive) mergers 
with similar masses are more likely progenitors of SNe~Ia than the mergers with larger mass differences that
were studied earlier. This motivated \citet{zhu13a} to perform a large parameter study where they
systematically scanned the parameter space from 0.4-0.4 \Msun up to 1.0-1.0 \msun. In their study,
they used the GASOLINE code \citep{wadsley04} together with the Helmholtz EOS \citep{timmes99,timmes00a}, 
no nuclear reactions were included. All their simulations were performed with non-spinning stars
and approximate initial conditions as outlined above. Mergers with ``similar'' masses produced a 
well-mixed, hot central core while ``dissimilar'' masses produced a rather unaffected cold core 
surrounded by a hot envelope and disk, consistent with earlier studies.  They found that the central density 
ratio of the accreting and donating star of $\rho_a/\rho_d > 0.6$ is a good criterion for those 
systems that produce hot cores (i.e. to define ``similar'' masses).

\citet{dan14a}  also performed a very broad scan of the parameter space. They studied 
the temperature distribution inside the remnant for different stellar spins: tidally locked initial conditions 
produce hot spots (which are the most likely locations for detonations
to be initiated) in the outer layers of the core, while irrotational
systems produce them deep inside of the core,  consistent with the
results of \citet{zhu13a}.  Thus, the spin state of the WDs may possibly 
be decisive for the question where the ignition is triggered. This, in
turn, may impact the properties of the resulting supernova. Dan et
al. found essentially no chemical mixing between the stars for mass
ratios below $q \approx 0.45$, but maximum mixing for a mass ratio of
unity. Contrary to \citet{zhu13a}, complete mixing was no found in any
case, but this difference can be convincingly attributed to 
the different stellar spins that were investigated (tidal locking in \citealt{dan14a}, no spins in \citealt{zhu13a}).
In addition to the He-accreting systems that likely undergo a detonation for a total mass beyond 
1.1 \Msun they also found CO binaries with total masses beyond 2.1 \Msun to
be prone to a CO explosion. Such systems may be candidates for the so-called super-Chandrasekhar
SN~Ia explosions.  They also  discussed the possibility of ``hybrid supernovae''
where a ONeMg core with a significant He layer  collapses and forms a neutron star. While
technically being a (probably weak) core-collapse supernova 
\citep{podsiadlowski04,kitaura06}, most of the explosion energy may come from helium
burning. Such hybrid supernovae may be candidates for the class of ``Ca-rich'' SNe Ib
\citep{perets10} as the burning conditions seem to favour the production of intermediate mass
elements.

\citet{raskin12a} explored 10 merging WDWD binary systems with total masses between 
1.28 and 2.12 \Msun with  the SNSPH code  \citep{fryer06} coupled to the Helmholtz EOS and a 13-isotope 
nuclear network. They used a heuristic procedure to construct tidally deformed, synchronized binaries
by letting the stars fall towards each other in free fall, and repeatedly set the fluid velocities to zero.
They had coated their CO WDs with atmospheres of helium (smaller than 2.5 \% of the total mass) and found 
that, in all cases where the primary had a mass of 1.06 \msun, the helium detonated.  No Carbon detonation 
was encountered, though.

Given the difficulty in identifying the central engine of a SNe~Ia on purely theoretical grounds, \citet{raskin13a} 
explored possible observational signatures stemming from the tidal tails of a WD 
merger. They followed the ejecta from a SNSPH simulation \citep{fryer06} with n-body methods and --
assuming spherical symmetry-- they explored how a supernova would
interact with such a medium by means of a 1D Lagrangian code. Provided
the time lag between merger and supernova is short enough ($<100$ s),  
detectable shock emission at radio, optical, and/or X-ray wavelengths is expected. For delay times 
between $10^8$ s and 100 years one expects broad NaID absorption features, and, since this has not been 
observed to date, they concluded that if (some) type Ia supernovae are indeed caused by WDWD mergers, the delay 
times need to be either short ($<100$ s) or rather long ($>100$ years). If the tails can expand and cool for 
$\sim 10^4$ years, they produce the observable narrow NaID and Ca II K\& K lines which are seen in some 
fraction of type Ia supernovae.

An interesting study from a Santa Cruz--Berkeley collaboration
\citep{moll14a,raskin14a}, again,  combined
the strengths of different numerical methods. The merger process was calculated with the SNSPH code
\citep{fryer06}, the subsequent explosion with the grid-based code 
CASTRO \citep{almgren10,zhang11} and synthetic light curves and spectra were calculated with SEDONA  \citep{kasen06}.
For some cases without immediate explosion the viscous remnant evolution was followed further with ZEUS-MP2 \citep{hayes06}. 
The first part of the study by \citet{moll14a} focussed on prompt detonations, while the second part explored 
the properties of detonations that emerge in later phases, after the secondary has been completely disrupted. 
For the first part, three mergers (1.20 - 1.06, 1.06 - 1.06 and 0.96 - 0.81 \msun) were simulated with
 the SPH code and subsequently mapped into CASTRO, where soon after simulation start ($<0.1$ s)
detonations emerged. They found the best agreement with common SNe~Ia  (0.58 \Msun of \nifs) for the
binary with 0.96 - 0.81 \msun. More massive systems  lead to 
more \Nifs and therefore unusually bright SNe~Ia. The remnant asymmetry at the moment of detonation leads
to large asymmetries in the elemental distributions and therefore to strong viewing angle effects for the resulting supernova. 
Depending on viewing angle, the peak bolometric luminosity varied by a factor of two and the flux in the ultraviolet 
even varied by an order of magnitude. All of the three models approximately fulfilled the width-luminosity 
relation (``broader means brighter''; \citealt{phillips99}).

The companion study \citep{raskin14a} explored cases where the secondary has become completely disrupted before a detonation
sets in, so that the primary explodes into a disk-like CO structure. To initiate detonations, the SPH simulations were
stopped once a stationary structure had formed, all the material with $\rho>10^6$ \Gcc was burnt in a separate simulation 
and, subsequently, the generated energy was deposited back as thermal energy into the merger remnant and the SPH simulation 
was resumed. As a double-check of the robustness of this approach, two simulations were also mapped into CASTRO and detonated
there as well. Overall there was good agreement both in morphology and the nucleosynthetic yields. The explosion 
inside the disk produced an hourglass-shaped  remnant geometry with strong viewing angle effects. The 
disk scale height, initially set by the mass ratio and the resulting different merger dynamics and burning processes,
turned out to be an important factor for the viewing angle dependence of the later supernova. The other 
crucial factor was the primary mass that  determines the resulting amount of
\nifs. Interestingly, the location of the detonation spot (whether at the surface or in the core of the primary) was found
to have a relatively small effect compared to the presence of an accretion disk. While qualitatively in agreement
with the width-luminosity relation, the lightcurves lasted much longer than 
standard SNe~Ia. The surrounding CO disk from the secondary remained
essentially unburnt, but it lead to relatively small intermediate mass
element absorption velocities by impeding the expansion. The large
asymmetries in the abundance distribution 
can lead to a large overestimate of the involved \Nifs masses if spherical symmetry is assumed in interpreting observations.
The lightcurves and spectra were peculiar with weak features 
from intermediate mass elements but relatively strong carbon absorption. The study also explored how 
longer-term viscous evolution before a detonation sets in would affect the supernova. Longer delay times 
were found to likely produce larger \Nifs masses and more symmetrical remnants.  Such systems might be 
candidates for super-Chandra SNe~Ia.

\citet{sato15} systematically explored mergers of CO WDs in the mass range between 0.5 and 1.1 \msun. 
They used a ``vanilla ice'' SPH formulation with time-dependent artificial viscosity parameters \`a la \citet{morris97}
together with the Balsara limiter \citep{balsara95} and a cubic spline kernel with a modified kernel derivative that is 
peaked at the centre\footnote{As shown in \citet{rosswog15b} peaked kernels perform rather poorly in estimating densities.} 
and on average 75 neighbour particles. They use the Helmholtz EOS \citep{timmes00a}, but do not include nuclear burning
and instead use a  comparison of the dynamic timescale with an analytically given burning timescale as a condition for dynamical carbon burning.
Based on this parameter study, they suggest that binaries where both CO WDs are in the mass range between 0.9 \Msun and 1.1 \Msun should result in a SNIA during the merger
phase. If instead the more massive WD is between 0.7 and 0.9 \Msun and the total mass exceeds 1.38 \Msun an explosion can still occur during the stationary post-merger phase. Overall, they estimate that CO WD mergers contribute less than 9\% to the
total galactic SNIa rate.

In a follow-up study \citep{sato16}, they used essentially the same methodology, but
included a simple energy release prescription to mimic the effect of carbon burning
and found a moderate enhancement of the peak temperatures during the
merger process. They further determined the critical mass ratio
that results in a violent merger as function of the primary mass. 

\subsubsection{Simulations of white dwarf--white dwarf collisions}
\label{sec:WDWD_collisions}
\begin{figure}[ht]
   \centering
      \includegraphics[angle=90,width=\textwidth]{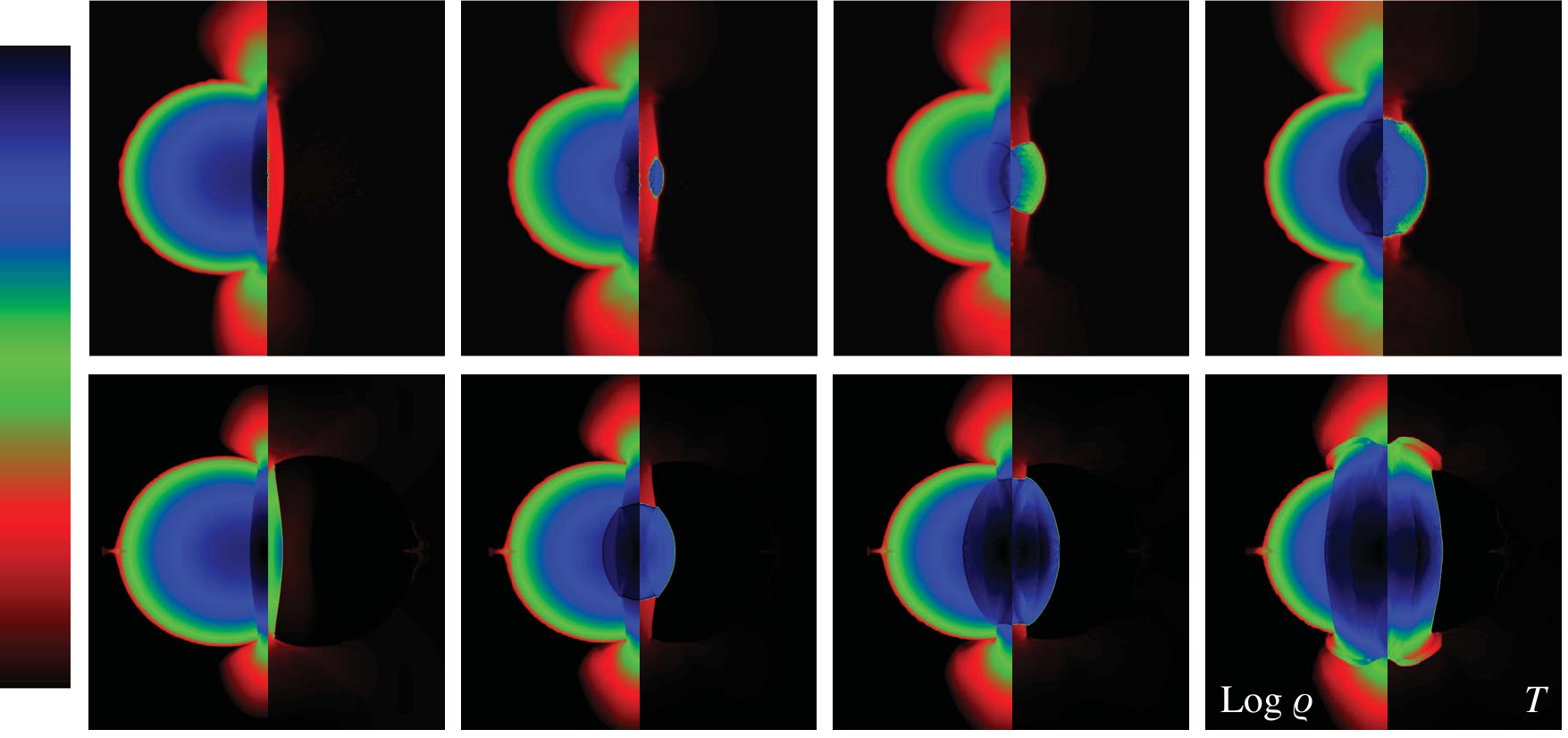}
 \caption{Comparison of the head-on collision of two WDs once calculated with an 
               SPH code \citep{rosswog08b} (upper row) and once with the Adaptive Mesh Refinement code FLASH
               \citep{fryxell00} (lower row). Each of the panels is split into density (left) and 
               temperature (right). Given the enormous temperature sensitivity of explosive nuclear
               burning the agreement is remarkably good. The produced energy from the SPH calculation
               is $\log(E_\mathrm{erg})= 51.21$ and  $\log(E_\mathrm{erg})= 51.11$ for the FLASH simulation. 
              Image reproduced with permission from \citet{rosswog09c}, copyright by AAS.}
   \label{fig:WDWD_collision}
\end{figure}

A physically interesting alternative to gravitational-wave-driven mergers are dynamical 
collisions between two WDs, as they are expected in globular clusters and galactic cores.
They have first been explored by \citet{benz89}, probably
in one of the first hydrodynamic simulations of WDs that included a nuclear reaction 
network. At that time the authors found that nuclear burning could, in central collisions,
help to unbind a substantial amount of matter, but this amount was found to be
irrelevant for the chemical evolution of galaxies. 
The topic has been taken up 
again by \citet{rosswog09c} and, independently, by \citet{raskin09}. Both of these studies employed SPH simulations ($2 \times 10^6$
SPH particles in \citealt{rosswog09c}, $8 \times 10^5$ particles in \citealt{raskin09})
coupled to small nuclear reactions networks. Both groups concluded that an amount of radioactive
\Nifs could be produced that is comparable to what is observed in normal type Ia supernovae
($\sim 0.5$ \msun). In \citet{rosswog09c} one of the simulations was repeated 
with  the Flash code \citep{fryxell00} to judge the robustness of the result, and
given the enormous sensitivity of the nuclear reactions to temperature,
good agreement was found overall, see Fig.~\ref{fig:WDWD_collision}. 
Rosswog et al. also post-processed the nucleosynthesis with larger nuclear networks
and calculated synthetic light curves and spectra  with the Sedona code \citep{kasen06}.
Interestingly, the resulting light-curves and spectra \citep{rosswog09c}
look like normal type Ia supernova, even the width-luminosity relationship \citep{phillips99}
is fulfilled well. These results have been confirmed 
in further studies
\citep{raskin10,hawley12,kushnir13,garcia_senz13}.

This result is interesting for a number of reasons. First, the detonation mechanism
is parameter-free and extremely robust: the free-fall velocity between WDs naturally
produces relative velocities in excess of the WD sound speeds. Therefore strong shocks
are inevitable. Moreover, the most likely involved WD masses are near the peak of the mass
distribution, $\sim 0.6$ \msun, or, due to mass segregation effects in
globular clusters, possibly slightly larger, but well
below the Chandrasekhar mass. This has the benefit that the nuclear burning occurs
at moderate densities ($\rho \sim 10^7$ \gcc) and thus naturally produces the observed
mix of $\sim 0.5$ \Msun  \Nifs and intermediate-mass elements, without any fine-tuning
such as the deflagration-detonation transition that is required in the single-degenerate
scenario \citep{hillebrandt00}. However, based on simple order of magnitude estimates,
the original studies \citep{raskin09,rosswog09c} concluded that, while being very interesting
explosions, the rates would likely be too low to make a substantial contribution to the observed 
supernova sample. There have, however, been claims \citep{thompson11,kushnir13}
that the Kozai--Lidov mechanism in triple stellar systems may substantially boost the rates of 
WDWD collisions so that they could constitute a sizeable fraction of the SN~Ia rate. Contrary
to these claims, a study by \citet{hamers13} finds  that the 
contribution from the triple-induced channels to SN~Ia is small. A
dedicated study by \citet{toonen18} concludes that
the contribution of dynamical collisions to the SNIa rate is of order
0.1\%, \citet{rajamuthukumar23} estimated that it could range
between 0.4 and 4 \%. In summary, dynamical collisions could potentially
provide a (small) contribution to type Ia supernovae.

\subsubsection{Summary double white dwarf encounters}
SPH has very often been used to model mergers of, and later on also collisions between, two WDs.
This is mainly since SPH is not restricted by any predefined geometry and has excellent conservation
properties. As illustrated in the numerical experiment shown in Fig.~\ref{fig:conservation_GWs},
even small (artificial) losses of angular momentum can lead to very large errors in the prediction
of the mass transfer duration and the gravitational-wave signal. SPH's tendency to ``follow the density'' 
makes it ideal to predict, for example, the gravitational-wave signatures of WD mergers. But it is 
exactly this tendency which makes it very difficult for SPH to study
thermonuclear ignition processes in low-density regions that are very important, for example, for the 
double-detonation scenario.  This suggests that in such cases, a
combination of numerical tools should be applied: 
SPH for bulk motion and orbital dynamics and Eulerian (Adaptive Mesh) hydrodynamics for 
low-density regions that need high resolution. As 
outlined above, there have recently been a number of successful studies that have followed such strategies.
SPH simulations have played a pivotal role in ``re-discovering'' the importance of white dwarf mergers
(and possibly, to a smaller extent, collisions) as progenitor systems of SNe~Ia. In the last few years, a number
of new possible pathways to thermonuclear explosions prior to or during a WD merger have been discovered.

\subsection{Encounters between two neutron stars and neutron stars and black holes}
\label{sec:astro_NSNS_NSBH}

\subsubsection{Relevance}
\label{sec:relevance_CompactBinaries}
The relevance of compact binary inspirals and mergers is hard to
overrate: they are related to many grand challenges of contemporary
physics and astrophysics.

\paragraph{Dark matter}

Thinking about dark matter has a history going back for centuries
\citep{bertone18}. The nature of dark matter, however,  is still unknown 
and the proposed candidates span 90 orders of magnitude in mass, from
ultra-light bosons to massive black holes
\citep{fereira21,green21,cirelli24}. Compact binary mergers may serve
as extreme environment probes to potentially detect the effects of dark
matter.  Since the dynamics of binary black holes in vacuum 
can be calculated to very high precision, see
e.g. \citet{blanchet14,kaelin20,bluemlein20}, they are ``clean
systems''  where deviations from the expected behaviour could be
tell-tales of beyond-standard-model physics \citep{barack19,cardoso20}. Dark
matter may also impact the structure of neutron stars with potentially
interesting consequences for their observational signatures during a
merger, see e.g. \citet{das21,fortin21,emma22,baryakhtar22}.

\paragraph{Strong gravity}

General Relativity cannot be the ultimate theory of gravity since it
does not meld well with quantum theory. Also, since dark matter and
dark energy are not well-understood concepts, one has to remain
open-minded about modifications of general relativity and
test gravity wherever possible. Einstein's theory of
gravity has been substantially constrained in the quasi-linear,
quasi-stationary regime, where velocities are low and gravitational
fields are weak, see e.g. \citet{stairs03,will06,psaltis08}. The late
inspiral, merger and ring-down of compact binary mergers, however,
offer the possibility to probe the extreme gravity regime
with very strong, dynamically changing fields and substantial velocities
($\sim0.5c$). As of today,  no significant disagreement with
Einstein's theory has been detected
\citep{abbott16i,yunes16,berti18a,berti18b,abbott19b,abbott19c,abbott21a,abbott21b,abac25b,abac25c}.
Still, black hole spectroscopy in the aftermath of compact mergers \citep{dreyer04,brito18,berti25}
seems to be our best chance to detect potential cracks in Einstein's theory.
 
\paragraph{Dense matter physics}

The density inside a neutron star reaches several times
nuclear matter density, $\rho_\mathrm{nuc}\approx 2.7 \times 10^{14}$
\gcc, during a neutron star merger it can be increased further by tens
of percent. At densities of $\sim 3 \rho_\mathrm{nuc}$ the distances
between nucleons become comparable to their size and phase transitions
to new states of matter that contain deconfined quarks may appear
\citep{baym18}.  Such high-density matter properties at essentially
zero temperature, however, are close to impossible to probe in
terrestrial experiments. In the last inspiral stages, however, at  GW
frequencies of several hundred Hz and separations a few neutron star
radii, the ``cold'' nuclear matter properties determine how neutron
stars deform due to tidal accelerations \citep{hinderer08,flanagan08}. The tidal
interaction acts like an attractive potential between the neutron
stars and speeds up the inspiral compared to a point-mass orbit, see e.g. \citet{bernuzzi20a}.
Therefore, the last inspiral stages carry the imprint of cold nuclear
matter.  After the merger, when temperatures have reached $>10^{11}$
K, the matter properties of ``hot'' nuclear matter determine the
post-merger GW spectrum that is a main target of the next
generation  GW detectors \citep{abac25d}. 

\paragraph{Cosmic nucleosynthesis}

The basics of the  rapid neutron capture process or ``r-process''
\citep{cowan21} were understood already in the 1950s
\citep{cameron57,burbidge57}, but the cosmic origin of r-process
elements was a matter of intense debates for decades, and identified as
one the ``11 science questions for the millennium''
\citep{11questions}. Mergers of neutron star black hole and double
neutron star binaries had been discussed as potentially promising sites
\citep{lattimer74,lattimer76,symbalisty82,eichler89}, and the first
neutron star merger plus nucleosynthesis calculations
\citep{rosswog99,freiburghaus99b} showed clearly that the ejecta consist of
r-process elements and contains enough material to, at least,
make neutron star mergers a
major cosmic source. More evidence for neutron star mergers
subsequently piled up from the theoretical side, see the excellent
review on the r-process of \citet{cowan21} and references
therein, but it took until the first neutron star merger 
detection in 2017 to observationally confirm neutron star mergers as
an r-process site
\citep{kasen17,drout17,evans17,kasliwal17,tanvir17,rosswog18a}. Although
the production of r-process in neutron star mergers is, now, beyond
reasonable  doubt, detailed information about which elements
are actually formed in  neutron star mergers 
has remained circumstantial. While neutron star mergers are the only
confirmed production site to date, additional r-process sources are potentially 
needed \citep{cote18,siegel19,siegel22,farouqi22,arcones23}.

\paragraph{Gamma-ray bursts}

That neutron star mergers could power gamma-ray bursts was suggested
already more than three decades ago \citep{eichler89}. While the
detection of the first afterglows of short GRBs
\citep{gehrels05,fox05,berger05} supported this idea, there was only indirect evidence for
a merger origin, see \citet{nakar07} and references 
therein, until GW170817 and its related sGRB
\citep{goldstein17,savchenko17}. For more than a decade there was a
consensus that neutron star mergers should only be responsible for 
\emph{short} GRBs, but there is now growing evidence that they may also
power some \emph{long} bursts
\citep{gehrels06,zhang07,troja22,rastinejad22,yang24,levan24}, but how
a system with such intrinsically short time scales (milli-seconds)
should produce such long-lasting bursts is currently not well
understood. 

\paragraph{Cosmic expansion}

To date there is a tension between the values of the Hubble
parameter $H_0$ determined via observations of the cosmic microwave
background and those from using a cosmic distance latter
\citep{freedman21,kamionkowski23}. This tension could be due to not
well enough understood systematic errors, or, maybe, be the imprint of
new physics. DNS and NSBH mergers are ``standard sirens'' and offer the
possibility to independently determine  $H_o$. From the
gravitational-wave amplitudes alone, one can measure the luminosity
distance (modulo inclination angle) and if a redshift can be
measured electromagnetically, it becomes possible to determine $H_0$
\citep{schutz86,holz05,nissanke10}. For an error in $H_0$ of $\sim 2\%$
one expects to need about 50 GW-EM events \citep{feeney19}.

\subsubsection{Gravitational-wave detections: the first decade}
\label{sec:GW_detections}
In 1916, only one year after the final version of his General Theory
of Relativity (GR), Einstein realized \citep{einstein16} that his
theory admits wave-like solutions that propagate at the speed of
light. But it took until the  discovery of the Hulse--Taylor pulsar in
1974 \citep{hulse75} to have, at least, an \emph{indirect} proof for the
existence of GWs. This provided a strong impetus to drive forward the
detector development aiming at  a  \emph{direct} detection of GWs. In
2015,  GWs were detected for the first time \citep{abbott16a}: both
LIGO detectors recorded the last 0.2 s of the inspiral and the final
merger of a $\sim$29 and a $\sim$36 \Msun black hole (GW150914),
leaving behind a $\sim$62 \Msun black hole. In other words, the
equivalent of $\sim$3 \msun, or $\sim 5 \times 10^{54}$ erg, was
radiated away as GWs.

The detection of the first observed binary neutron star merger followed soon:
in August 2017 a $\sim$ minute-long GW-signal from a
merging neutron star binary (GW170817) was detected \citep{abbott17b}. Most excitingly, this
event was also observed in electromagnetic (EM) waves, first as a short gamma-ray burst
(sGRB) \citep{abbott17d,savchenko17,goldstein17} that followed the peak of the GW emission with a delay of 1.7 s, days later
as a radioactively powered thermal transient peaking at optical/near-infrared wave-
lengths frequently called a “kilonova” \citep{abbott17c,arcavi17,coulter17,pian17,smartt17,tanvir17,evans17,cowperthwaite17},
and much later ($\sim$months) also in radio wave-lengths.
The event also displayed a rising X-ray flux starting nine days after merger
\citep{troja17}, peaking after 160 days, that then started to decline steeply
afterwards \citep{hallinan17,margutti17,lyman18,troja18}. 
This was interpreted as the imprints of a structured jet
observed at an angle of $\sim 25^\circ$ from the jet core \citep{lamb17,fong17,lamb18,hotokezaka18a}. Several years
later, broad-band synchrotron afterglow was detected \citep{hajela22,troja22,hajela22}
that was interpreted as a ``kilonova afterglow'' due to a mildly relativistic ejection
component interacting with the interstellar medium \citep{nakar11a,hotokezaka15a,hotokezaka18a}. For an excellent review
summarizing the observation related to GW170817 see \citet{margutti21}.

This ``golden event'' allowed us to make major leaps forward for several long-standing questions.  For example, the
electromagnetic follow-up observations of GW170817 revealed the host galaxy and thus connected the merger event to potential stellar evolution paths 
\citep{levan17}. Since the distance to the host galaxy is known, it was possible to use the 1.7~s delay between GWs and the sGRB to
show that GWs propagate at the speed of light to a relative accuracy
of $10^{-15}$ \citep{abbott17d}, placing strong constraints on
alternative theories of gravity. The event also demonstrated the
exciting possibility to constrain the properties of cold nuclear matter via the effects of
tidal deformability and thereby ruled out some very stiff equations of state \citep{abbott17c}.
The kilonova that was observed in the aftermath of the event demonstrated that neutron star
mergers are indeed major cosmic sources of rapid neutron capture elements \citep{cowan21}
as expected on theoretical grounds \citep{lattimer77,symbalisty82,eichler89,rosswog99,freiburghaus99b}.
The decay of the bolometric light curve agreed well  \citep{kasen17,rosswog18a,metzger20} with the predicted characteristic
power-law decay of the r-process heating rate \citep{metzger10b}, therefore the r-process nature of the ejected matter was established beyond reasonable doubt.
There is, however, still a fair amount of uncertainty in terms of which range of
elements was actually  produced. Interestingly, there is evidence for strontium \citep{watson19}, a lighter r-process element (also co-produced
by the s-process) that can only be formed for moderately large electron fractions $Y_e$.
There is hardly any such matter in cold neutron stars, which demonstrates that weak interactions have significantly changed the neutron-to-proton ratio during
during the merger process. There is also evidence \citep{wu19,kasliwal22}
for heavier r-process elements, consistent with second- and third-peak r-process material. Moreover, the detection of both gravitational and electromagnetic waves allowed
for an independent estimate of the Hubble parameter, yielding a value between
the Planck and the SHoeES collaboration results \citep{abbott17a}.

The fourth gravitational-wave transient catalog GWTC-4.0
\citep{abac25a} contains more than 200 binary black hole mergers, two neutron star mergers
(but only one with an EM counterpart) and two neutron star black hole mergers have been firmly detected.
Based on the events detected so far, the merger rate is estimated as 7.6 - 250 Gpc$^{-3}$ yr$^{-1}$ for binary
neutron stars and 9.1 - 84 Gpc$^{-3}$ yr$^{-1}$ for neutron star-black hole binaries.
As the gravitational-wave  amplitude decreases with the inverse of the distance to the
source, an increase in detector sensitivity by a factor of $k$ will increase the accessible volume by a factor of $k^3$.
Therefore, with an expected  range increase of LIGO from currently $\sim$ 150 Mpc to $\sim$ 300 Mpc in O5 (expected start 2028, \citealt{LVK_observing_plans}),
one expects an order of magnitude increase in the detection rate with excellent prospects for exciting new discoveries.

\subsubsection{The modeling challenge}
\label{sec:modelling_challenge}
\begin{figure}[ht]
   \centering
   \includegraphics[width=\textwidth]{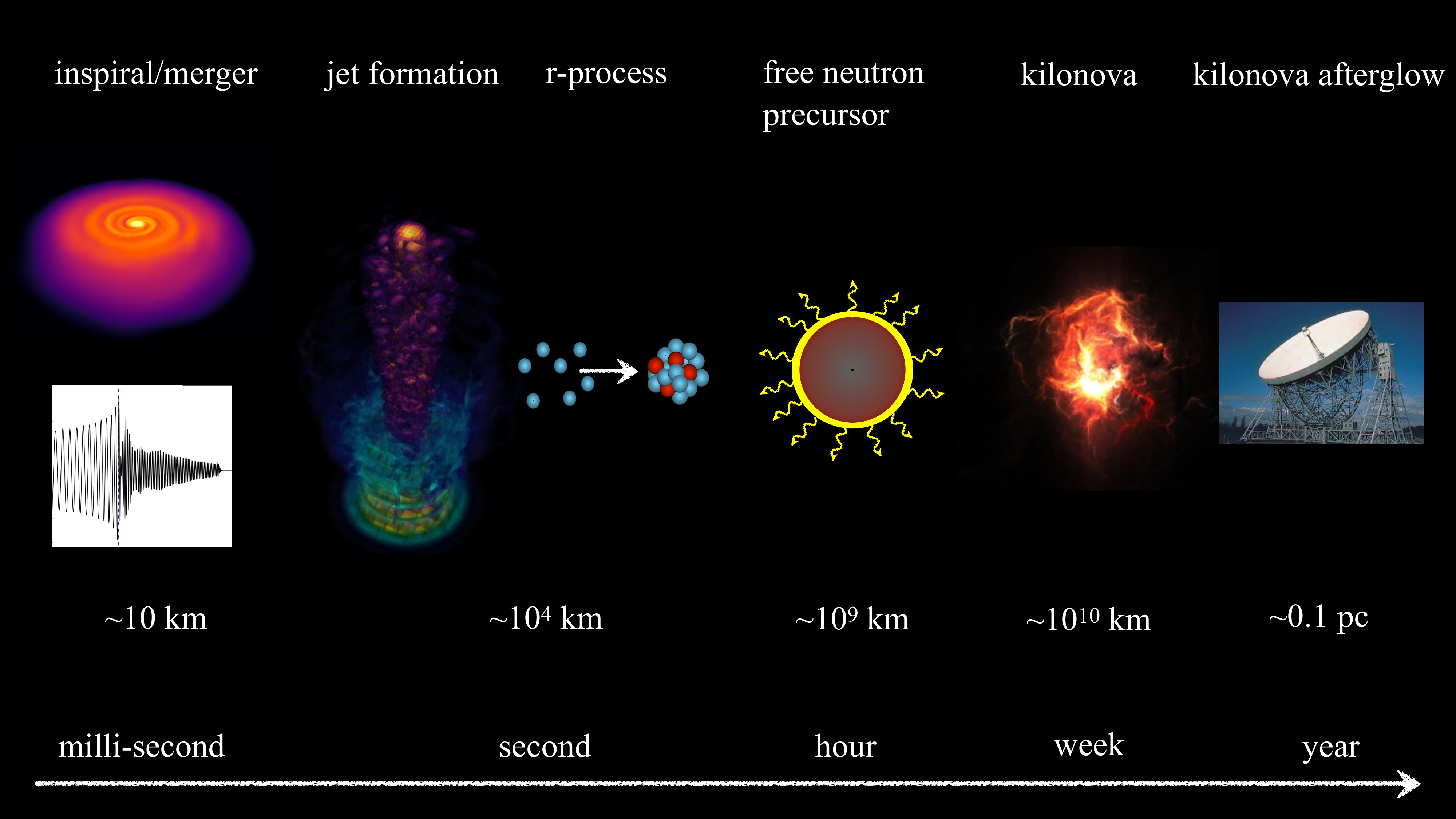} 
   \caption{The questions related to mergers of neutron stars and
   black holes involve many orders of magnitude in length and time scales.}
   \label{fig:NSNS_challenges}
 \end{figure}

The modelling of the merger of two compact objects is a serious
computational physics challenge, see Fig.~\ref{fig:NSNS_challenges}, both in terms of the involved
physical processes and in terms of the length and time scales that one
ideally should cover.

\paragraph{Physics at the extremes}

Mergers between neutron stars and black holes require a treatment
with full 3+1 \emph{numerical relativity} of Einstein's (or some alternative)
theory. This is crucial for predicting e.g. the collapse to black holes,
but also to obtain (within the framework of a given equation of state)
reliable temperatures which, in turn, determine the neutrino emission
and weak interaction rates and therefore the nucleosynthesis and the
resulting electromagnetic signal.

Since the high-density \emph{nuclear matter} equation of state is
currently not known \citep{baym18}, one needs to 
explore the potentially observable signatures of physically plausible
nuclear matter models in order to prepare for future
multi-messenger observations. 
In numerical relativity simulations, one often uses piecewise
polytropic approximations to cold nuclear 
matter equations of state \citep{read09} together with thermal
contributions, e.g. either via a simple ideal-gas type contribution with a
fixed thermal polytropic exponent $\Gamma_\mathrm{th}$ \citep{janka93},
or, more physics-motivated thermal contributions \citep{raithel19}. Many
microscopic-theory-based equations of state are available in a tabular
form \citep{compose}. While replacing one table with another in a
working code is trivial, the evolution code needs to be able to
robustly handle the conservative-to-primitive transformation with a
tabulated EOS in the
first place. This requires usually rather sophisticated root-finding algorithms
\citep{cerda_duran08,galeazzi13,siegel18a,kastaun21,shankar26a} that 
deep in their core perform interpolations inside the potentially non-smooth
table. This may not be bullet-proof and root-finding may fail in some
cases. Moreover, using tables is guaranteed to be more expensive than
dealing with analytical expressions. Another drawback of a tabulated EOSs is
that they are usually restricted to densities $\gtrsim 10^3$ \gcc. This
may be sufficient  for core-collapse supernova simulations, but this value is
way too large to reliably follow the ejecta in a merger. For
comparison, the long-term simulation of \citet{rosswog14a} followed the
ejecta up to 100 years post-merger, during this time the density in
the simulation dropped below the value of the interstellar medium
($\sim 10^{-24}$ \gcc).

With densities of
several times $10^{14}$ \Gcc and temperatures exceeding $10^{11}$ K,
\emph{neutrino reactions} can become very fast, $Y_e/\dot{Y}_e\sim 1$
ms. The $\sim 10$ MeV neutrinos in a neutron star merger are
efficiently trapped  in the central  remnant,  while the outer disk
and ejecta regions are essentially 
neutrino-transparent. Much of the physics that determines the
observable electromagnetic signal, however, occurs  in the
difficult-to-handle semi-transparent regime. Ideally, one should solve
the GR Boltzmann equation \citep{lindquist66,mezzacappa20}, but this seven-dimensional problem (six
phase-space dimensions plus time) currently cannot be solved ``on the
fly'', and one has to retreat to various transport approximations such as
leakage schemes, truncated moment expansions or Monte Carlo
approaches, see \citet{foucart23} for a recent review on neutrino
transport in a neutron star merger context. Another important unsolved, but
relevant problem, is neutrino oscillations \citep{tamborra21,li21,grohs23,volpe24}
which also cannot be dealt with during dynamical merger simulations. 

Neutron stars are naturally endowed with strong \emph{magnetic fields} and
the dynamics of a neutron star merger offers a plethora of
possibilities for amplifying the magnetic field to gigantic field strengths exceeding
$10^{17}$ G
\citep{price06,anderson08b,rezzolla11,zrake13,kiuchi15,palenzuela22,aguilera22,aguilera24,kiuchi24,aguilera25,neuweiler26,cook26,gutierrez26}. 
Challenges related to magnetic field evolution include fulfilling the
$\nabla\cdot\vec{B}=0$ constraint, the stable evolution of high
magnetic fields in low density regions and, crucially, to resolve the
length scales that are relevant for the magnetic field
amplification. This is challenging since the field is often amplified
via instabilities where the shortest wavelengths grow fastest and by
small scale turbulence.

\paragraph{Length and time scales}

Also the length and time scales related to compact binary mergers span
huge ranges. The merger involves time scales of milliseconds and
a system size of $\sim 10$ km. The relevant scales for magnetic
field amplification are much smaller  and can usually only
be reliably treated via sub-grid models that account for the effects
of unresolved scales \citep{radice24,schmidt25}.
While the dynamic mass ejection occurs on ms-scales,
 neutrino-/MHD winds occur on $\sim 100$ ms, and torus unbinding 
takes $\sim 1$ s.  During the r-process in the ejecta, the initial neutrons
are captured within $\sim 1 $s. If there are fast ejecta they may
produce a ``free neutron'' precursor to the kilonova \citep{metzger15a}
that peaks $\sim 1$ hour post-merger. 
The kilonova emission (depending on the nuclear composition) peaks at time scales
of $\sim 1$ week, when the ejecta have expanded to $\sim10^{10}$~km,
and the densities have dropped to $\sim 10^{-14}$ \Gcc
(see, e.g. Fig.~1 in \citealt{rosswog14a}). The physics
that determines this stage is the heating by radioactive decays,
the thermalization of the decay products, atomic line opacities and,
at late stages, NLTE effects in the rapidly expanding gas cloud.
So between the observed signatures,
gravitational-waves and kilonova, there is a huge gap in scales,
physics and observations and also in our understanding
of what is actually happening.
This gap currently needs to be bridged by making 
simplifying assumptions such as homology and/or spherical
symmetry, but reality is likely substantially more
complicated, for example geometric effects, nuclear heating
etc. can still impact the evolution significantly
\citep{rosswog14a,korobkin21,wu22}.

\subsubsection{Double neutron star mergers}
\label{sec:NSNS}

\paragraph{First Newtonian calculations with polytropic equation of state}

The earliest NSNS merger calculations \citep{rasio92,davies94,zhuge94,rasio94,rasio95,zhuge96} were performed with Newtonian 
gravity and a polytropic equation of state, sometimes a simple gravitational-wave backreaction force was added. While
these pioneering studies were, of course, rather simple, they set the stage for future efforts and began to explore
the qualitative merger dynamics. They already established
the emergence of a Kelvin--Helmholtz unstable vortex sheet at the interface between the two stars, which, due to the larger shear, is 
more pronounced for initially non-rotating stars. This vortex sheet has subsequently been understood to be crucial for the
early magnetic field amplification in a neutron stars merger \citep{price06,anderson08b,rezzolla11,zrake13,kiuchi15,palenzuela22,aguilera22,aguilera24,kiuchi24,aguilera25}.
These early simulations also established the basic morphological differences between tidally locked and irrotational binaries and
between binaries of different  mass ratios. In addition, these studies also drove some technical developments that became very useful in later studies
such as the relaxation techniques \citep{rasio94} to construct synchronized binary systems or the procedures to 
analyze the GW energy spectrum in the frequency band \citep{zhuge94,zhuge96}. See also \citet{rasio99} for a excellent
review of earlier work.

\paragraph{Early studies with focus on microphysics}

Newtonian studies with a focus on the involved microphysics were pioneered by \citet{ruffert96} 
who implemented a nuclear equation of state and a neutrino-leakage scheme into their Eulerian (PPM) hydrodynamics code.
In SPH, the effects of a nuclear equation of state (EOS)  were first explored in \citet{rosswog99}. The authors implemented
the Lattimer--Swesty EOS \citep{lattimer91} and neutrino cooling in the simple free-streaming limit in order to bracket the
effects that neutrino emission may possibly have.  To avoid artefacts from excessive artificial dissipation, they also included 
the time-dependent viscosity parameters \citep{morris97}, see Sect.~\ref{sec:AV}. 
They found torus masses between 0.1 and 0.3 \msun, and, maybe most importantly, that between $4 \times 10^{-3}$ 
and $4 \times 10^{-2}$ \Msun of neutron-rich matter becomes ejected \citep{rosswog98b,rosswog99}. This was realized
to be enough for neutron star mergers to be the major cosmic r-process source, \emph{provided} that the ejecta are
predominantly r-process. In a companion paper \citep{freiburghaus99b} the authors performed nuclear network calculations
on the  hydrodynamic trajectories found in the simulations and showed that the ejected matter indeed consists of r-process elements and shows an abundance 
pattern close to the one observed in the solar system, provided that the initial electron fraction is $Y_e\approx 0.1$.
These studies showed in particular, that the neutron-rich tidal ejecta effortlessly produce the third r-process peak (``platinum peak'')
around $A=195$, that supernova simulations had struggled to produce for decades.\footnote{They found essentially no light r-process,
  which, according to modern simulations with better neutrino transport, is also produced in a neutron star merger, see e.g. \citet{wanajo14}.}
In a follow-up study \citep{rosswog00} the effects of different initial stellar spins and mass ratios $q\neq 1$
on the ejecta masses  were explored.

The simulation ingredients were further refined in \citet{rosswog02a,rosswog03a,rosswog03c}. Here the Shen et al. 
\citep{shen98a,shen98b} EOS, extended down to very low densities, was used and a detailed multi-flavour neutrino leakage scheme was
developed \citep{rosswog03a} that integrated over the spectral energy distribution of the neutrinos. These studies were also performed
at a substantially higher resolution (up to $10^6$ particles) than previous SPH studies of the topic. The typical neutrino luminosities
turned out to be $\sim 2 \times 10^{53}$ erg/s  with typical average neutrino energies of 8/15/20 MeV for $\nu_e, \bar{\nu}_e$ and
the heavy lepton neutrinos. This leakage scheme, together with the early work of \citet{ruffert96,ruffert97a}, served as blueprint
for later, general relativistic neutrino leakage schemes \citep{galeazzi13,radice16a}.
A particular focus of \citet{rosswog03b} was the launch of GRBs, therefore, the authors calculated the annihilation of
neutrino-anti-neutrino pairs in a post-processing step and, barring possible complications from
baryonic pollution, it was concluded that $\nu \bar{\nu}$ annihilation should lead to 
relativistic outflows and could produce moderately energetic sGRBs. Simple estimates indicated, however, that strong 
neutrino-driven winds likely occur which could threaten the emergence of the ultra-relativistic outflow that is needed for a sGRB.
Later dedicated studies, see e.g. \citet{dessart09,perego14b,radice18a,fujibayashi20c,bernuzzi25} have indeed
confirmed that strong baryonic winds are blown out in polar directions.

In a set of SPH studies the neutron star mass parameter space was scanned by  systematically exploring  mass
combinations from 1.0 to 2.0 \Msun \citep{korobkin12a,rosswog13a,rosswog13b}. The main focus here was the dynamically 
ejected mass and its possible observational signatures. One interesting result \citep{korobkin12a} was that the nucleosynthetic
abundance pattern for the tidal dynamical ejecta is essentially identical for all mass combinations and
even NSBH systems yield practically an identical pattern. While extremely robust to  variations of the astrophysical parameters,
the pattern showed some sensitivity to the involved nuclear physics, for example to a change of the mass formula or the
distribution of the fission fragments. The authors concluded that the dynamic ejecta of neutron star mergers are excellent
candidates for the source of the heavy, so-called ``strong r-process'' that is observed in a variety of metal-poor stars and, each time,
shows  the same relative abundance pattern for nuclei beyond barium \citep{sneden08}.

Several studies on the electromagnetic transients powered by the radioactive decay of freshly
synthesized r-process matter were based on SPH simulations \citep{piran13a,rosswog13a}, mostly
because they allow to track ejecta properties in a straight-forward way, and to perform nuclear
network calculations based on such trajectories. In two back-to-back studies \citep{rosswog14a,grossman14a}
the authors explored the ejecta evolution, including the energy release due to radioactive  decay,
up 100 years after the merger event.  This study substantially profited from SPH's geometric
flexibility and its treatment of vacuum as just empty (i.e. SPH particle-free) space. The ejecta expansion was followed for as many
as 40 orders of magnitude in density, from nuclear matter down to the densities of interstellar matter. Since from the 
latter calculations the 3D remnant structure was known, also viewing angle effects for kilonovae could be explored 
\citep{grossman14a}. In the ramp-up to GW170817 and its aftermath, studies related to the kilonova emission
from neutron star mergers have proliferated and it is far beyond the scope of this study to summarize
this very broad and rapidly growing field. For recent reviews on the topic we refer to
\citet{nakar20,metzger20,margutti21,rosswog24a}.

\paragraph{Studies with approximate GR gravity}

A natural next step beyond Newtonian gravity is the application of  Post-Newtonian expansions.
\citet{blanchet90} developed an approximate formalism for moderately relativistic, self-gravitating 
fluids.  This formalism allows writing all the equations in a quasi-Newtonian form and casts all relativistic non-localities
in terms of Poisson equations with compactly supported sources. The 1PN equations require the solution of eight Poisson
equations and to account for the
lowest order radiation reaction terms requires the solution of yet another Poisson equation. While --with nine Poisson equations--
computationally already quite heavy, the efforts to implement the scheme into SPH by two groups \citep{faber00,ayal01,faber01,faber02b}
turned out not to be particularly useful, mainly since the corrective
1PN terms are comparable to the Newtonian ones for realistic neutron stars with compactness $\mathcal{C}\approx 0.17$, which can lead to instabilities. As a result, one of the groups \citep{ayal01}
decided to study ``neutron stars'' of small compactness ($M < 1$ \msun, $R\approx 30$ km), while the other \citep{faber00,faber01,faber02b}
artificially downsized the 1PN effects by choosing a different speed of light for the corresponding terms. While both approaches
represented admissible first steps, the corresponding results are astrophysically difficult to interpret.

A second, more successful approach, was to use the so-called conformal flatness approximation (CFA)
\citep{isenberg08,wilson95a,wilson96,mathews97,mathews98}. Here the basic assumption is that the spatial part of the metric
is conformally flat, i.e.  it can be written as a multiple (the prefactor depends on space and time and absorbs the overall 
scale of the metric) of the Kronecker symbol $\gamma_{ij}= \Psi^4 \delta_{ij}$, and that
it remains so during the subsequent evolution. The latter, however, is an assumption and by no means guaranteed. 
Physically this corresponds to a gravitational-wave-free spacetime. Consequently, the inspiral of a binary system has 
to be achieved by adding an ad hoc radiation reaction force. The CFA also cannot handle frame dragging effects.
On the positive side, for spherically symmetric spacetimes the CFA coincides exactly with GR and for small deviations
from spherical symmetry, say for rapidly rotating neutron stars, it has been shown \citep{cook96} to deliver very accurate results.
For more general cases such as a binary merger, the accuracy is difficult to assess. Nevertheless, the CFA
has turned out to be a useful approximation, although new results should be carefully scrutinized against
full GR simulations.

The CFA was implemented into SPH by \citet{oechslin02} and slightly later by \citet{faber04}. The 
major difference between the two approaches was that Oechslin et al. solved the set of six coupled, non-linear elliptic
field equations by means of a multi-grid solver \citep{press92}, while Faber et al. used spectral methods from the LORENE
library \citep{lor} on two spherically symmetric grids around the stars. Both studies used polytropic equations of state (\citealt{oechslin02} used
$\Gamma= 2.0, 2.6$ and 3.0, while \citet{faber04} used $\Gamma= 2.0$) and approximate radiation reaction  terms
based on the Burke--Thorne potential \citep{burke71,thorne69b}. Oechslin et al. used a combination of a bulk and a 
von-Neumann--Richtmyer artificial viscosity steered similarly as in the Morris and Monaghan approach \citep{morris97},
see Sect.~\ref{sec:AV}, while Faber et al. argued that shocks would not be important and artificial dissipation could be ignored.\footnote{This is not correct, shocks play an important role for ejecting matter in the GR case.}

In a subsequent study, \citet{oechslin04} explored how the presence of quark matter in neutron stars would 
impact a NSNS merger and its gravitational-wave signal. They combined a relativistic mean field model (above
$\rho = 2 \times 10^{14}$ \gcc) with a stiff polytrope as a model for the hadronic EOS and added in an MIT bag model
so that quark matter would appear at $5 \times 10^{14}$ and would completely dominate the EOS
for high densities ($> 10^{15}$ \gcc). While the impact on the GW frequencies at the ISCO remained moderate ($<10\%$), 
the post-merger GW signal was substantially influenced in those cases where the central object did not collapse
into a bh immediately. In a subsequent study
\citep{oechslin06} they implemented the Shen et al. EOS \citep{shen98a,shen98b} and adopted a range of NS mass ratios to explore
how large post-merger accretion tori would be. Based on the values found they concluded that double neutron star mergers
could plausibly be the central engines of short GRBs.

In \citet{oechslin07a} the same group also explored the Lattimer--Swesty EOS \citep{lattimer91}, the cold
EOS of \citet{akmal98} and the ideal gas equations of state with parameters fitted to nuclear EOSs. The merger outcome
was  rather sensitive to the nuclear matter EOS: the remnant collapsed either immediately or very soon after merger 
for the soft Lattimer--Swesty EOS and for all other cases it did not show signs of collapse for tens of dynamical time scales.
Both ejecta and disk masses were found to increase with an increasing deviation of the mass ratio from unity. The ejecta
masses were in a range between $10^{-3}$ and $10^{-2}$ \msun, comparable, but slightly lower than the earlier, Newtonian
estimates \citep{rosswog99}. In terms of  their GW signature, it 
turned out, that the peak in the GW wave energy spectrum that is related to the formation of the hypermassive merger remnant,
has a frequency that is sensitive to the nuclear EOS \citep{oechslin07b}. In comparison, the mass ratio and neutron star spin 
only had a weak impact. 

\begin{figure}[htp]
    \centerline{
    \includegraphics[width=0.49\textwidth]{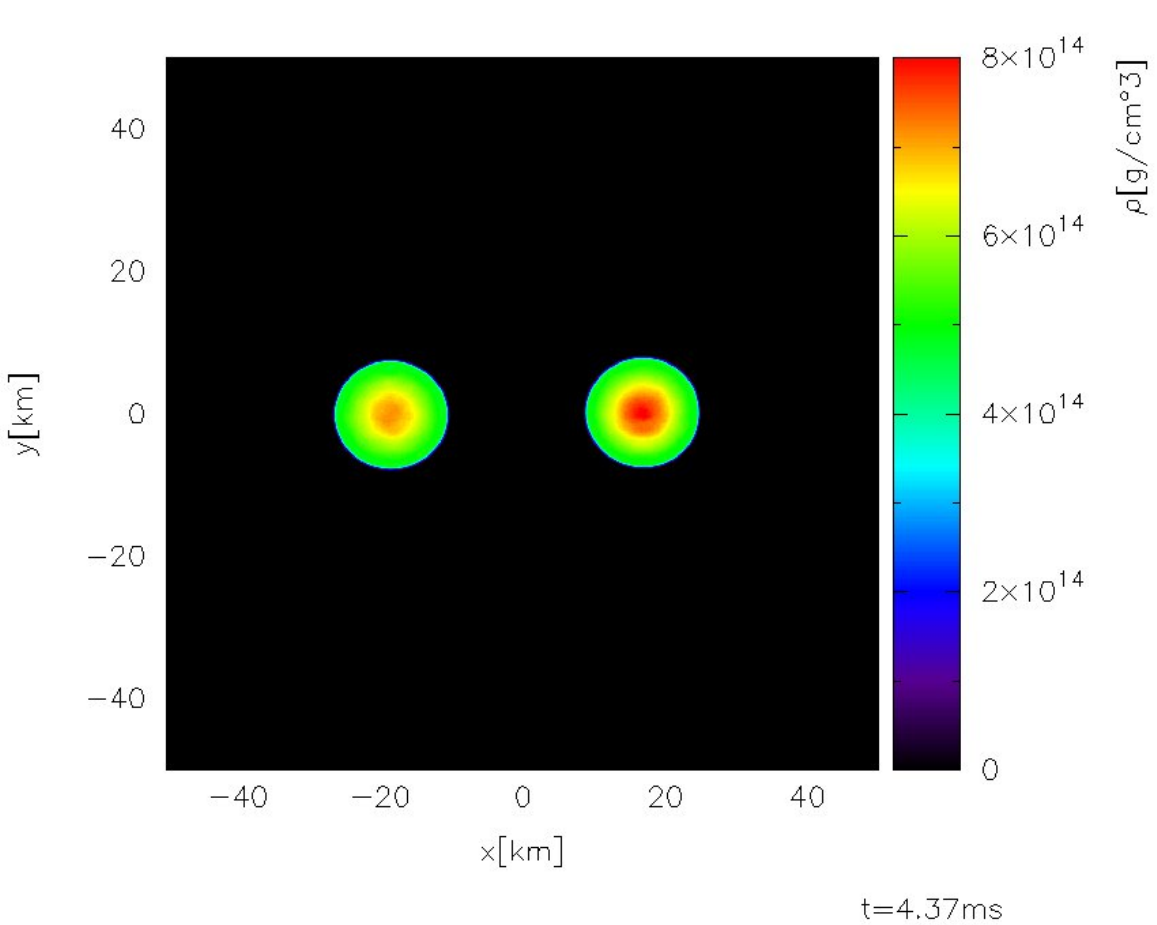} 
    \includegraphics[width=0.49\textwidth]{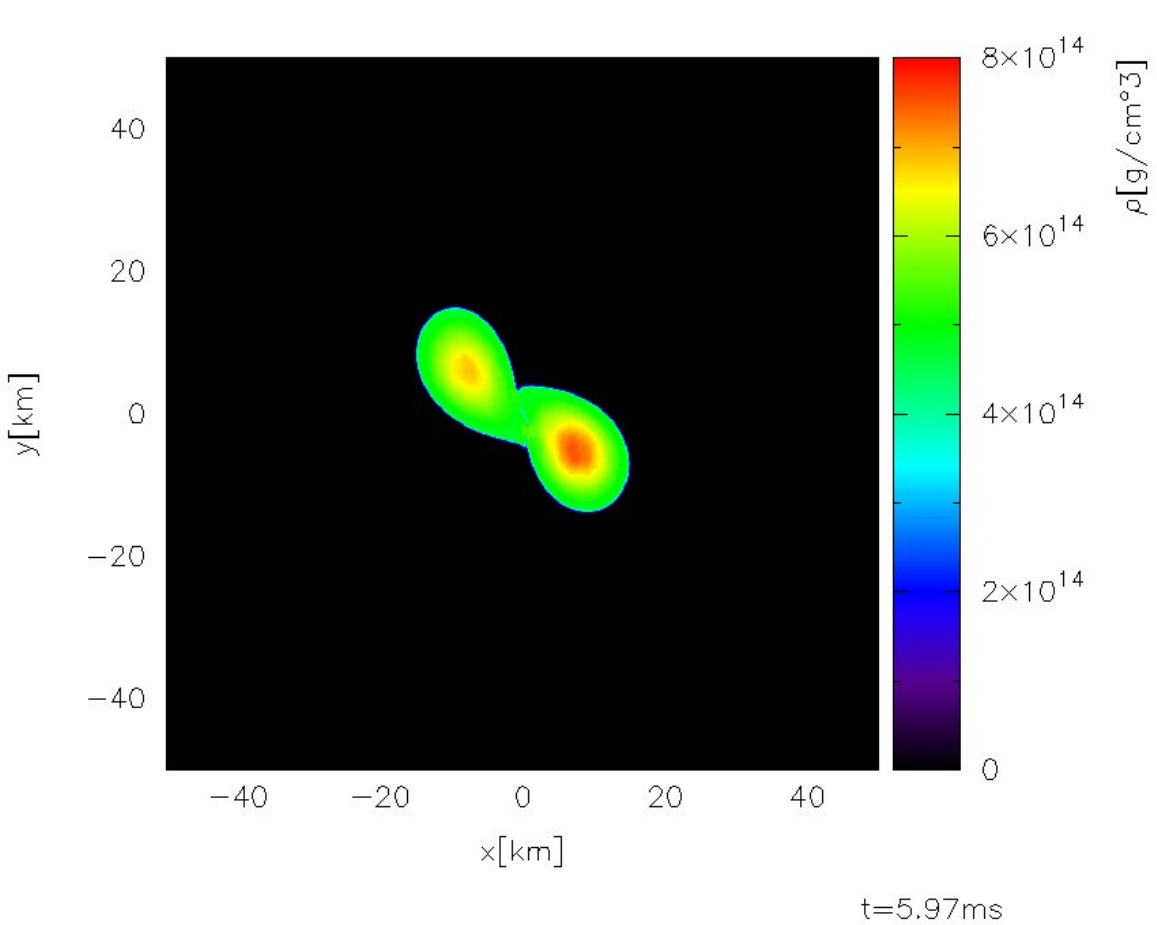} 
    }
    \centerline{
    \includegraphics[width=0.49\textwidth]{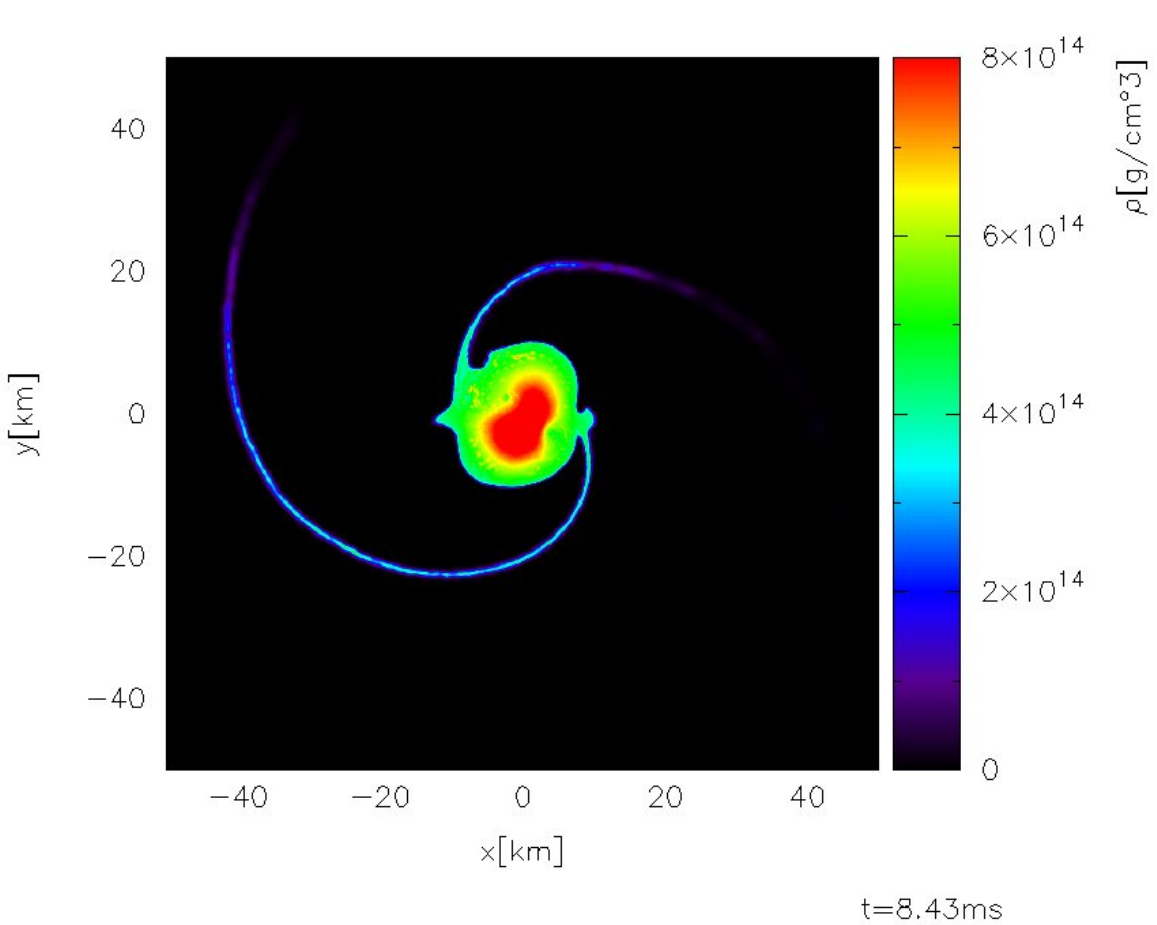}
    \includegraphics[width=0.49\textwidth]{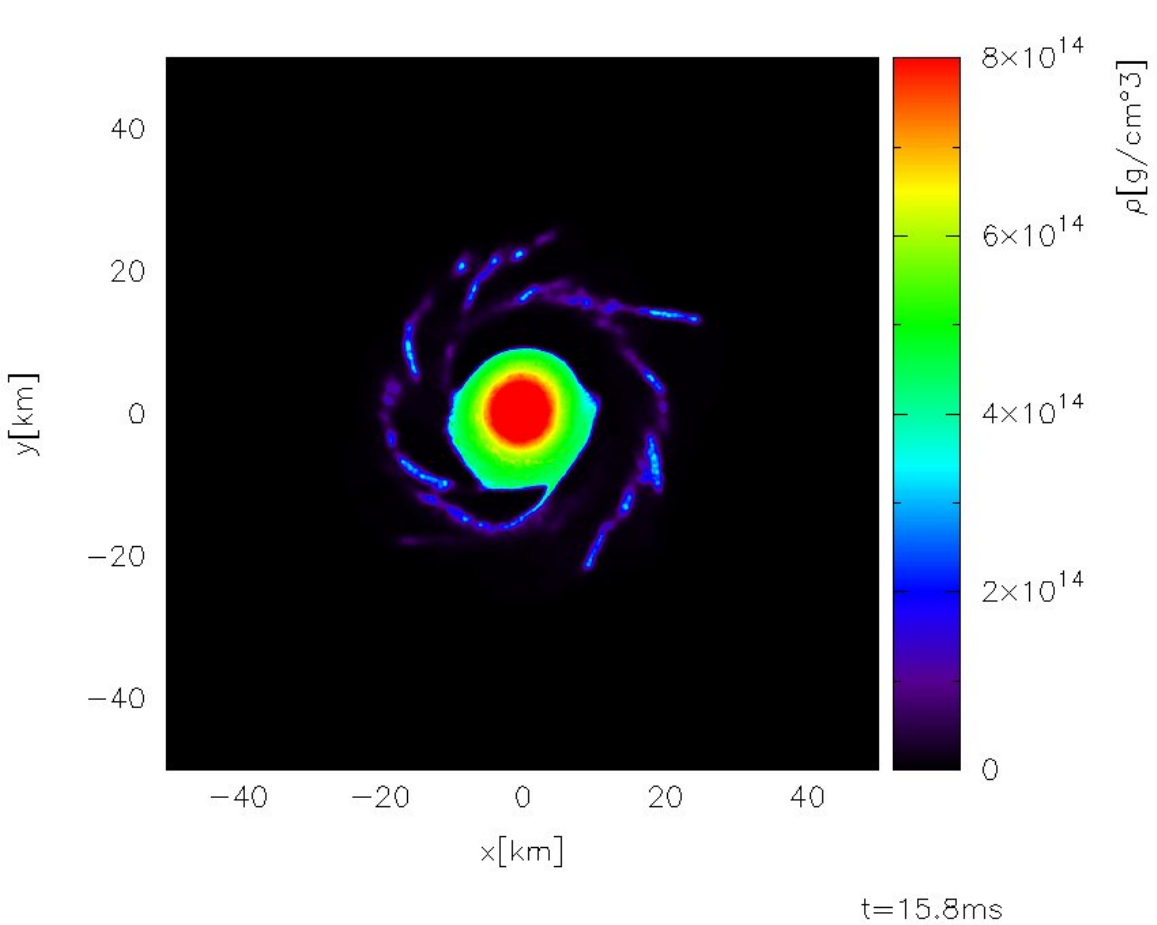}
    }
 \caption{Merger of two strange stars with masses of 1.2 and 1.35 \msun. Shown is the density in the orbital plane.
  Image reproduced with permission from \citet{bauswein10a}, copyright by APS.}
   \label{fig:nsns_strange}
\end{figure}

This line of work was subqsequently continued by Bauswein et al. in a series of papers \citep{bauswein09,bauswein10a,bauswein10b,
bauswein12a,bauswein12b,bauswein13a,bauswein13b}. In their first study, they explored the merger of two strange stars 
\citep{bauswein09,bauswein10a}. If strange quark matter is really the ground state of matter as hypothesized in 
\citet{bodmer71} and \citet{witten84}, compact stars made of strange quark matter might exist. Such stars would differ from neutron 
stars in the sense that they are self-bound, they do not have an overall inverse mass-radius relationship and they can be more
compact. Therefore the gravitational torques close to merger are different and it is more difficult to tidally
tear apart a strange star. 
In their study \citep{bauswein10a} the quark matter EOS was modelled via the MIT bag model \citep{chodos74,farhi84}
for two different bag constants (60 and 80 MeV/fm$^3$) and a large number of binary systems (each time with 130 000 SPH particles) 
was explored.  The coalescence of two strange stars is indeed morphologically different from a neutron star merger:
the result is a differentially rotating hypermassive object with sharp boundary 
layers surrounded by a thin and clumpy strange matter disk, see  Fig.~\ref{fig:nsns_strange} for an example of a strange 
star merger with 1.2 and 1.35 \Msun \citep{bauswein10a}. Moreover, due to the greater compactness, the peak GW 
frequencies were larger during both inspiral and the subsequent ringdown phase. 
If the merger of two strange stars would eject matter in the form of ``strangelets'' these should contribute to the cosmic
ray flux. The ejected mass and therefore the contribution to the cosmic ray flux strongly depends on the chosen bag 
constant and for large values no mass loss could be resolved. For such values neutron stars and 
strange stars could coexist, since neutron stars would not be converted into strange stars by capturing  
cosmic strangelets.
In this and another study \citep{bauswein10b} thermal effects (and their consistent treatment) were shown
to have a substantial impact on the remnant structure.

In subsequent work a  large number of microphysical EOSs was explored 
\citep{bauswein12a,bauswein12b,bauswein13a,bauswein13b} to study the effect on the GW signal.They found that the peak
frequency of the post-merger signal correlates well
with the radii of the non-rotating neutron stars \citep{bauswein12a,bauswein12b} and they concluded that a GW
detection would allow to constrain the NS radius within a few hundred meters.

In a separate study \citep{bauswein13a} they used their large range of equations of state and several mass ratios to systematically explore 
dynamic mass ejection. According to their study, softer equations of state  with correspondingly smaller radii eject
a larger amount of mass. In the explored cases they found the mass of
the dynamical ejecta to be in the  range from $10^{-3}$ to $2 \times
10^{-2}$ \msun. For the arguably most likely case with two 1.35 \Msun
stars they found, that the ejecta masses varied by  about one order of
magnitude depending on the equation of state. Moreover,  consistent
with other  studies, they found  a robust r-process that produces a
close-to-solar abundance pattern beyond  nucleon number of $A= 130$
and they discussed the implications for kilonovae and possibly emerging radio remnants due to the ejecta.

Bauswein et al. have also explored, with their approximate-GR SPH approach, various empirical relations
related to the prompt collapse of a neutron star binary, where a black hole forms immediately at contact
without the central remnant ``bouncing back'' \citep{bauswein13b,bauswein17,bauswein20,bauswein21}.
Overall, their empirical relations agree rather well with full-GR simulations on the same topic
\citep{shibata05b,hotokezaka11,koeppel19,tootle21,koelsch22,kashyap22}.
In \citet{bauswein17} two such empirical relations were used to constrain NS mass and radius. The authors
interpreted the large ejecta mass of GW170817 as an argument that no prompt collapse had occured which
constrained $M_\mathrm{thr} > M_\mathrm{GW170817} \approx 2.74$ \msun, and, via the empirical relations, they constrained
a 1.6 \Msun neutron star to have radius $R_{1.6} > 10.68$ km and the radius of the maximum mass configuration to
be $R_{\max} > 9.60$ km. In later work, the authors also explored the effects of mass ratios deviating from unity
and of potential first-order phase transitions on empirical relations, for which they provide various fits
\citep{bauswein20,bauswein21}.

\paragraph{The first SPH simulations in full General Relativity}

The first fully general relativistic SPH code, \SpB \citep{rosswog21a,diener22a,rosswog22b,rosswog23a,rosswog25c},
integrates, similar to many Eulerian numerical relativity codes, the spacetime evolution equation set of
Baumgarte-Shapiro-Shibata-Nakamura-Oohara (BSSN-OK) which are described in detail in several excellent textbooks
\citep{alcubierre08,bona09,baumgarte10,rezzolla13a,shibata16,baumgarte21}, see also Sect.~\ref{sec:fullGR_SPH}.
The spacetime is evolved numerically on an
adaptive mesh, while, contrary to conventional numerical relativity codes, matter is evolved by means of freely moving particles
according to a modern version of SPH. The corresponding evolution equations are described in detail in Sect.~\ref{sec:fullGR_SPH}.
The hydrodynamic equations are enhanced by artificial dissipation terms, that make use of both slope-limited reconstruction and
time-dependent artificial dissipation parameters, see Sect.~\ref{sec:AV}. For the full details of the equations and their implemenation
we refer the interested reader to \citet{rosswog21a,rosswog22b}.

The construction of initial conditions for \SpB is a two-stage process. In the first step constraint-satisfying initial conditions need to
be found, that are an accurate representation of the desired physical initial conditions. To this end standard
initial data solvers such as \lo \citep{lor}, \fu \citep{papenfort21}, \sgrid \citep{tichy09}, \spells \citep{pfeiffer03}, \Elliptica \citep{rashti22}
or \nRPyElliptic \citep{assumpcao22} can be used. For a particle hydrodynamics code such as \SpB one is, however,
not yet done after this step: one still has to
place the SPH particles so that they represent as accurately as possible the solution found by the initial data solver. This is done
for the \SpB simulations via the so-called ``Artificial Pressure Method'' (APM), see \citet{rosswog20a} for the original method and
\citet{rosswog21a,diener22a,rosswog22b} for applications of the APM to relativistic neutron stars.
As outlined earlier, see Sect.~\ref{sec:IC}, the main idea of the method is to start from a guessed particle distribution, measure
how well the current density agrees with the desired density and calculate some ``artificial pressure'' based on this density
error, which is then used to  move the particles, in an iterative process, to positions where they minimize the error.

%
%
Studying observational data of pulsars observed in globular clusters, the authors of \citet{rosswog24b} realized
that $\sim$ 5 \% of the merging neutron star binaries should contain one rapidly spinning neutron star together
with a slowly/not spinning neutron star companion. Such systems can form naturally in the high density environments of
globular cluster cores. Starting from a binary system containing a neutron star and a non-degenerate companion (which
may have formed either via isolated binary evolution or dynamically), the non-degenerate companion will at some
point fill its Roche-lobe and transfer mass to the neutron star.  In this phase the binary is observable as a low-mass
X-ray binary. The end result of this phase is a spun-up milli-second pulsar with a WD companion. Close encounters occur frequently in dense cluster
cores and if another neutron star flies by closely enough, the likely
outcome is an exchange of the lighter WD for a heavier neutron star, which is typically  slowly  rotating.
As for other neutron star binaries, gravitational-wave emission will then eventually lead to a merger of such single-spin binaries.

\begin{figure}[ht]
   \centering
   \includegraphics[width=0.49\textwidth]{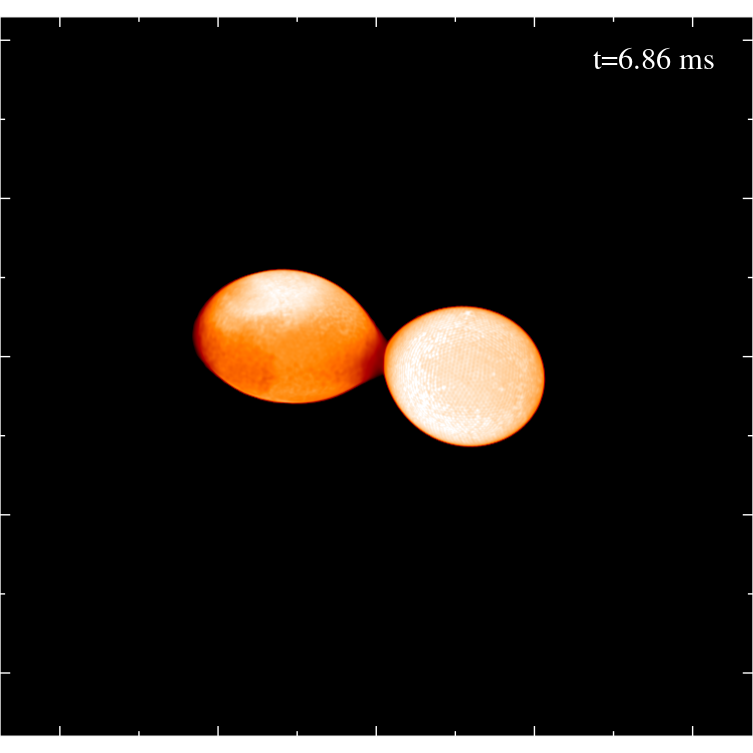}
   \includegraphics[width=0.49\textwidth]{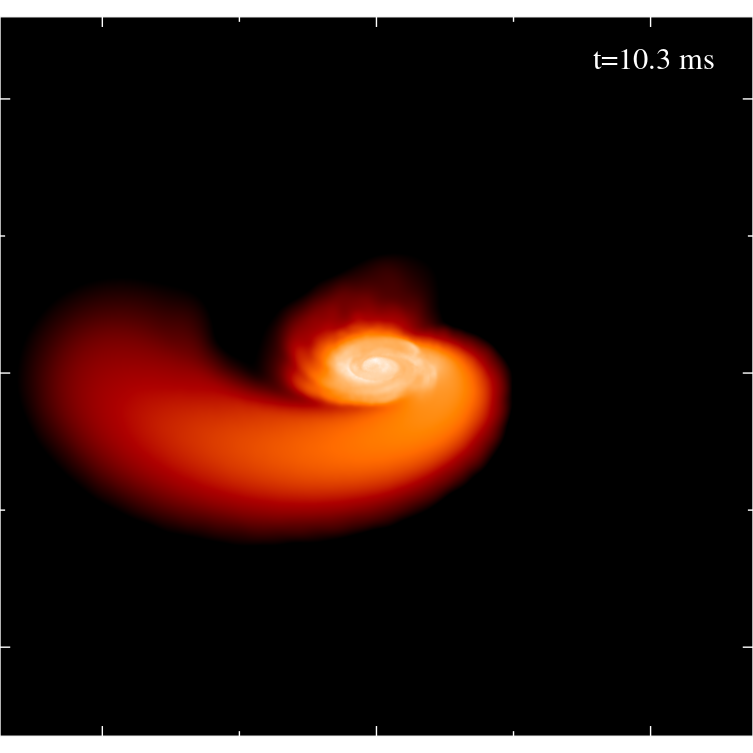}  
   \caption{Volume rendering of two merging neutron stars with 1.3 \msun. One of the stars is spun-up by accretion
   (spin parameter $\chi=0.5$; left star in the left panel) while the other star is non-spinning. Overall the morphology resembles
   a non-spinning binary with a mass ratio substantially deviating from unity. Image reproduced with permission from \citet{rosswog25c}, copyright by the author(s).}
   \label{fig:volren_single_spin}
\end{figure}

To explore how different the merger of such a binary is compared to the ``standard case'' of non-spinning neutron stars,
\citet{rosswog24b} used the \Fu library to construct initial data for one neutron star with a spin parameter
of $\chi=0.5$ and the other having $\chi=0$. They used 1.3 \Msun for each of the stars and explored three piecewise polytropic equations of state:
SLy \citep{SLY_eos} as an example of a soft EOS, MS1b \citep{mueller96} as an example of a (likely too) stiff EOS and APR3 \citep{akmal98}
as an intermediate case. Such mergers are significantly different from mergers with $\chi_1= \chi_2=0$. First, the spinning
equal mass cases resemble mergers with a mass ratio that is significantly different from unity, see Fig.~\ref{fig:volren_single_spin}.
This is because the rapidly spinning stars are substantially more extended due to the rotational flattening, 
more vulnerable to tidal forces and therefore produce a single massive tidal tail. This enhances the dynamical mass ejection
by an order of magnitude compared to the non-spinning case and also produces more massive accretion disks.
Since ejecta are launched tidally, they escape before weak interactions can change their electron fraction $Y_e$ substantially
and due to their very low $Y_e$ they result in lanthanide-/actinide-rich heavy r-process. The resulting kilonovae are
substantially brighter and peak later than for the non-spinning cases.
Since they are rotationally flatted, the tides act earlier on the spinning stars and lead to an overall less violent
collision. This results in smaller amounts of fast ejecta, see below, but also
in less efficient GW emission. Also, since more matter is launched into highly eccentric,  but bound orbits,
single-spin mergers also result in longer fallback activity. In addition, the less efficient emission of angular momentum via
gravitational waves will prolong the lifetime of the central remnant before it collapses to a black hole. Both of these effects
therefore lead to ``prolonged central engine activity''. This is an interesting property in the light of recent GRB observations.
After decades with a clear distribution of roles, ``compact binary mergers produce short GRBs, collapsar events
produce long GRBs'', the recent past has seen detections of kilonovae from \emph{long} GRBs \citep{rastinejad22,levan24}.
Such bursts are uncomfortably long compared to the expected time scales in a neutron star merger, therefore mergers
that produce longer-lived central engines may be an interesting way forward.
Whether single-spin mergers can explain such long GRBs or not will need to
be further studied in the future.

%
%
\begin{figure}[ht]
   \centering
   \includegraphics[width=\textwidth]{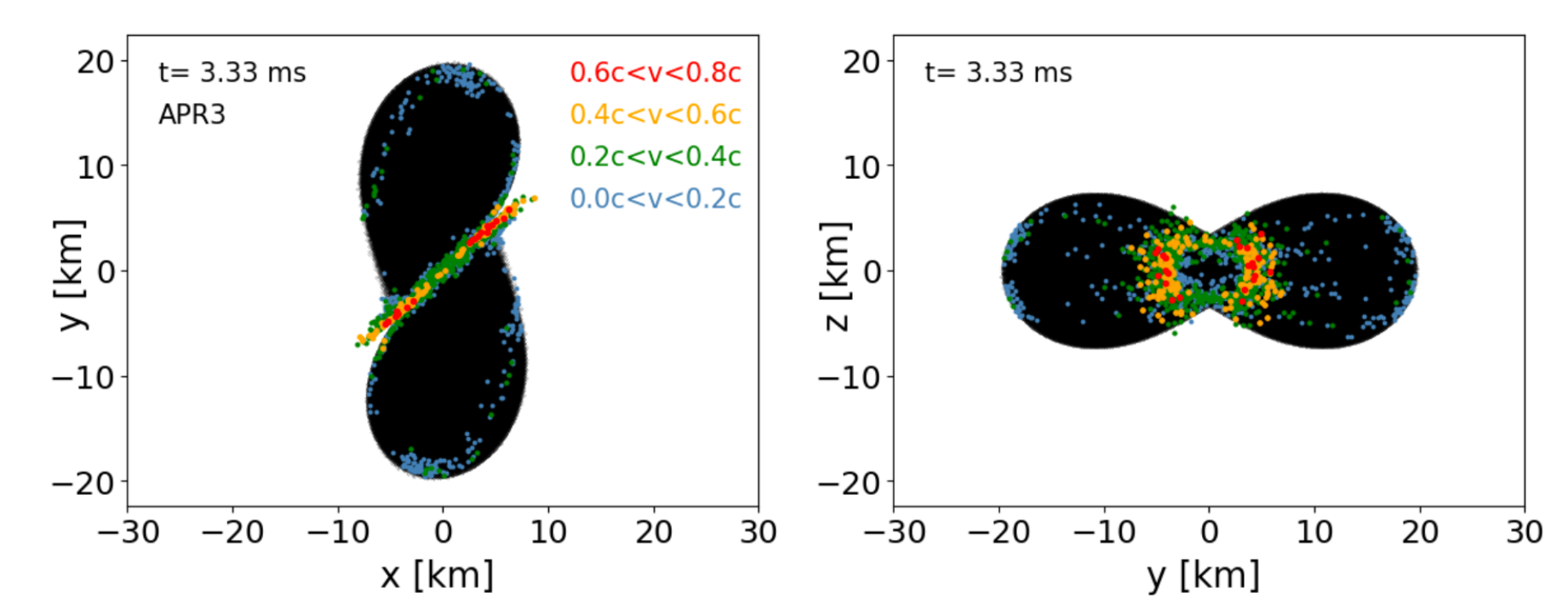} 
   \caption{Particle positions (black) during the time of merger
  (2$\times$ 1.3 \msun, APR3 equation of state). Overlaid are
  the positions of particles that are finally ejected with asymptotic
  velocities $v_\infty$ between 0 and 0.2$c$ (blue), between 0.2 and
  0.4 $c$ (green), between 0.4 and 0.6$c$ (orange) and above 0.6$c$ (red).
  The simulation was performed with the Lagrangian Numerical
  Relativity code \spB. Image reproduced with permission from \citet{rosswog24a}, copyright by the author(s).}
   \label{fig:fast_ejecta_positions}
 \end{figure}
 
Small amounts of ejecta extending  to large velocities ($>0.5 c$) 
have been seen in Newtonian \citep{dean21}, approximate GR \citep{bauswein13a} and
full GR simulations \citep{hotokezaka13,kyutoku14,kiuchi17,radice18a,hotokezaka18a,dean21,nedora21,rosswog22b,combi23,bernuzzi25}.
\citet{metzger15a} had discussed that, if such fast ejecta are real, they may have interesting consequences. The
leading edge of the homologously expanding ejecta may be so fast that many free neutrons can escape without being captured
by nuclei. The subsequent decay of this leading edge of free neutrons would then results in a blue/ultraviolett signal $\sim 1$ hour 
after the merger event. A recent detailed study of the nucleosynthesis in fast ejecta \citep{schnabel26a} found indeed
that free neutrons survive for a range of conditions.
Apart from the blue/UV signal from the free neutron decay such fast ejecta may also be responsible for
producing a flare of gamma-rays. Once a jet is launched following a neutron star merger, it will inflate a cocoon in the ejecta.
This cocoon drives a shock that breaks out of the fast ejecta, thereby generating a flare of gamma rays, which may be detected
over a substantially larger angle than the jet-opening angle \citep{kasliwal17,gottlieb18}.

Due to the interesting observational consequences, even small amounts of such fast ejecta may be impactful.
While such tiny amounts of fast ejecta are very difficult to resolve, the large number of simulations with different numerical
methods that have seen them gives us some confidence that these fast ejecta are not just a mere numerical artifact. Still, the
ejection mechanism was not entirely clear, mostly because the majority of the simulations was performed in a Eulerian framework,
where it is non-trivial to follow the ejecta history. In a recent set of \SpB simulations \citep{rosswog25b} the ejecta history was
tracked to understand \emph{how} these fast ejecta are launched. As suggested in earlier work, these fast ejecta come indeed
mostly from the interface between the two merging neutron stars, see also Fig.~\ref{fig:fast_ejecta_positions}. They are ejected
by two mechanisms: a) there is a ``spray component'' that is sheared out from the interface and immediately escapes, predominantly
along the orbital plane and b) a ``bounce component'', which is also sheared out from the interface, but remains initially close
to the central remnant. The central remnant subsequently becomes deeply compressed and on bouncing back it ejects several,
close-to-spherical pulses of fast ejecta that partially catch up with the spray component.

\subsubsection{Neutron star--black hole mergers}
\label{sec:NSBH}
A number of issues that have complicated the merger dynamics in the WDWD case, such as stability of mass transfer
or the formation of a disk, see Sect.~\ref{sec:WDWD_MT}, are also very important for  NSBH mergers.\footnote{If the neutron star
mass distribution allows for small enough mass ratios, this applies as well to nsns binaries.} Here, however,
they are further complicated by i) a incompletely known high-density equation of state which determines the mass-radius relationship
and therefore the reaction of the neutron star on mass loss, ii) general-relativistic effects such as the appearance of an
innermost stable circular orbit or effects from the BH spin and iii) the fact that now the GW radiation-reaction time 
scale can now become comparable to the dynamical time scale, see Eq.~(\ref{eq:tau_GW_tau_orb}).

\begin{figure}[ht]
   \centering
   \includegraphics[width=11cm,angle=0]{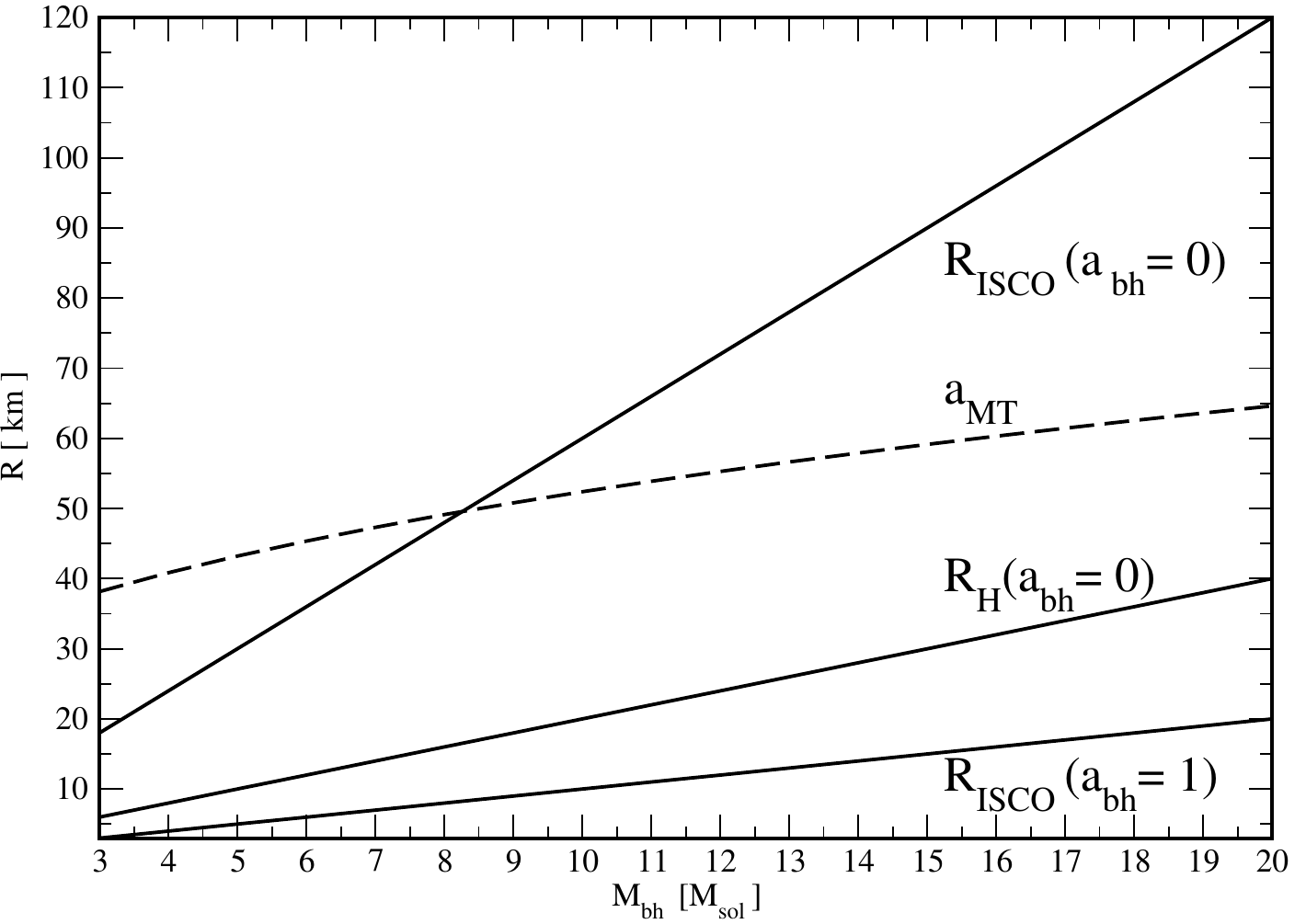}
 \caption{Illustration of the mass transfer separation $a_\mathrm{MT}$ with respect to horizon and innermost stable
               circular orbits based on Newtonian estimates (for fiducial neutron star properties, $M_\mathrm{ns}= 1.4$ \Msun and $R_\mathrm{ns}= 12$ km).}
   \label{fig:BH_radii}
 \end{figure}

Of particular relevance is the question for which binary systems sizable accretion 
tori form since they are thought to be the crucial ``transformation engines'' that channel available energy into (relativistic)
outflow. The final answer to this question requires 3D numerical simulations with the relevant physics, but a qualitative
idea can be gained already from simple estimates (but, based on the experience from WDWD binaries, keep in mind that even plausible 
approximations can yield rather poor results). Mass transfer is expected to set in when the Roche volume becomes 
comparable to the volume of the neutron star. By applying Paczynski's estimate for the Roche lobe radius \citep{paczynski71} 
and equating it with the NS radius, one finds that the onset of mass transfer (which we use here as a 
proxy for the tidal disruption radius) can be expected near a separation of
\be
a_\mathrm{MT}= 2.17 R_\mathrm{ns} \left( \frac{1 + q}{q} \right)^{1/3}. 
\ee
Since in the limit of $q \ll 1$ the mass transfer separation $a_\mathrm{MT}$ grows only proportional to $M_\mathrm{BH}^{1/3}$, but the ISCO and the event horizon
grow $\propto M_\mathrm{BH}$, the onset of mass transfer/disruption can  take place inside the ISCO 
for large BH masses. At the high end of BH masses, the neutron star is swallowed as whole 
without being disrupted at all. A qualitative illustration (for fiducial neutron star properties, $M_\mathrm{ns}= 1.4$ 
\Msun and $R_\mathrm{ns}= 12$ km) is shown in Fig.~\ref{fig:BH_radii}.
Roughly, already for black holes near $M_\mathrm{BH}\approx 8$ \Msun the mass transfer/disruption occurs near the ISCO 
which makes it potentially difficult to form a massive torus from ns
debris. So, low-mass black holes are clearly preferred as GRB
engines. The 4th Gravitational Wave Transient Catalogue (GWTC-4.0)
\citep{abac25e} shows a clear peak around 10 \Msun in the primary black hole
mass spectrum, but also contains a NSBH candidate (GW230529), where the black
hole mass is between 2.5 and 4.5 \Msun while the neutron star
candidate is between 1.2 and 2.0 \msun. 
Numerical simulations \citep{faber06b} have shown,
however,  that even if the disruption occurs deep inside the ISCO this does not necessarily mean that all the matter is 
doomed to fall straight into the hole and  a torus can still possibly
form.

When discussing disk formation in a GRB context, it is worth keeping in mind that even seemingly small disk masses
allow, at least in principle, for the extraction of energies,
\be
E_\mathrm{extr} \sim 1.8\times 10^{51} \; \mathrm{erg} \; \left( \frac{\epsilon}{0.1}\right) \left( \frac{M_\mathrm{disk}}{0.01 M_\odot}\right),
\ee
that are large enough to accommodate the isotropic gamma-ray energies, $E_{\gamma, \mathrm{iso}} \sim 10^{50}$ erg, that have been inferred
for short bursts \citep{berger14a}. If short bursts are collimated into a half-opening angle $\theta$, their 
true energies are substantially lower than this number, $E_{\gamma, \mathrm{true}} = \left( E_{\gamma, \rm iso}/65 \right) \left( 
\theta/10^\circ\right)^2$.

\paragraph{Early Newtonian calculations with polytropic equations of state}

The first NSBH simulations were performed by Lee and collaborators \citep{lee99a,lee99b,lee00,lee01a}
using Newtonian physics and polytropic equations of state. Although simple and missing 
some qualitative features of black holes (such as a last stable orbit), these simulations
provided insight and a qualitative understanding of the system dynamics, the impact
of neutron star spin, the mass ratio and the equation of state. The first set of simulations \citep{lee99a}
were carried out with an SPH formulation similar to the one of \citet{hernquist89}, but 
an alternative kernel gradient symmetrization, artificial viscosity 
was implemented with fixed dissipation parameters $\alpha$ and $\beta$ and $\sim 10^4$ SPH particles were used. 
The black hole was modelled as a Newtonian point mass with an absorbing boundary 
condition at the Schwarzschild radius $R_\mathrm{S}$, no backreaction from gravitational-wave emission was accounted for.
In the simulations of, initially tidally locked, binaries with a stiff EOS ($\Gamma=3$; \citealt{lee99a}), 
they found the neutron star to survive the onset of mass transfer and to keep orbiting, at a reduced 
mass, for several orbital periods. A similar behaviour had been realized in subsequent work with a 
stiff nuclear equation of state \citep{rosswog04b,rosswog07b}. In a follow-up paper \citep{lee99b},
a simple point-mass backreaction force was applied, and, in one case, the 
\citet{paczynski80} potential was used (now with absorbing
boundary at $1.5 R_\mathrm{S}$). But the main focus of this study was to explore the effect of a softer  equation of state 
($\Gamma=5/3$). In all explored cases the system dynamics was very different from the previous 
study, the neutron star was completely disrupted and formed a massive disk of $\sim 0.2 $ \msun,
with $\sim 0.01$ \Msun being dynamically ejected. The sensitivity to the EOS stiffness is not entirely surprising,
since the solutions to the Lane--Emden equations give the mass-radius relationship 
\be
\frac{d\log R}{d \log M}= \frac{\Gamma - 2}{3\Gamma - 4}
\label{eq:LE_MR}
\ee 
for a polytropic star, so that the neutron star reacts differently on mass loss: it shrinks for 
$\Gamma > 2$, so mass loss is quenched, and expands for $\Gamma < 2$ and therefore further enhances
mass transfer.

In a second set of calculations ($\approx 80$ K SPH particles), they explored non-rotating neutron 
stars that were modelled as compressible triaxial ellipsoids according to the semi-analytic work of 
\citet*{lai93a,lai93b,lai94b}, both with stiff ($\Gamma=2.5$ and 3) \citep{lee00} 
and soft ($\Gamma=5/3$ and 2) \citep{lee01a} polytropic equations of state. They used the same simulation
technology, but also applied a Balsara-limiter, see Eq.~(\ref{eq:Balsara}), in their artificial viscosity treatment
and only a purely Newtonian interaction between NS and BH was considered.  For the $\Gamma=3$ case,
the neutron star survived again until the end of the simulation, with $\Gamma=2.5$ it survived
the first mass transfer episode, but was subsequently completely disrupted and formed a disk of
nearly 0.2 \msun, about 0.03 \Msun were dynamically ejected.

\citet{lee01b} also simulated mergers between a black hole and a strange star which was modelled
with a simple quark-matter EOS. The dynamical evolution for these systems was quite different from
the  polytropic case: the strange star was stretched into a thin
matter stream that wound around the  
black hole and was finally swallowed. Although ``starlets'' of $
\approx 0.03$ \Msun formed during the disruption process, all of them
were in the end swallowed by the hole within milli-seconds, no mass loss
could be resolved.

\paragraph{Studies with focus on microphysics}

The first NSBH studies based on Newtonian gravity, but including detailed microphysics were performed 
by \citet{ruffert99,janka99} using a Eulerian PPM code on a Cartesian 
grid.\footnote{See Sect.~\ref{sec:challenges_WDWD} for a discussion of the complications of such an approach.} 
The first Newtonian-gravity-plus-microphysics SPH simulations of NSBH mergers were discussed in
\citet{rosswog04b,rosswog05a}.
Here the black hole was simulated as a Newtonian point mass with an absorbing boundary and a simple GW backreaction
force was applied. For the neutron star the \citet{shen98a,shen98b} temperature-dependent nuclear EOS was 
used and the star was modelled with $3\times 10^5  - 10^6$ SPH particles. In addition, neutrino cooling and electron/positron 
captures were followed with a detailed multi-flavour leakage scheme \citep{rosswog03a}. The initial study 
focussed on systems with low mass black holes ($q=0.5 - 0.1$) since this way there are greater chances to 
disrupt the neutron star outside of the ISCO, see the discussion above. Moreover, both (carefully constructed) corotating 
and irrotational neutron stars were studied. In all cases the core of the neutron star ($0.15 - 0.68$ \msun) survived the initial 
mass transfer episodes until the end of the simulations (22 - 64 ms). If disks formed at all during the simulated time, 
they had only moderate masses ($\sim0.005$ \msun). One of the NSBH binary 
($M_\mathrm{ns}=1.4$ \msun, $M_\mathrm{bh}=3$ \msun) systems was followed through the whole mass transfer episode \citep{rosswog07b}
which lasted for 220 ms or 47 orbital revolutions and only ended when
the neutron star finally became disrupted and
formed a disk of 0.05 \msun. A set of test calculations with a stiff ($\Gamma=3$) and a softer polytropic EOS ($\Gamma=2$) 
indicated that such episodic mass transfer is related to the stiffness of the NS EOS and only occurs for 
 stiff cases, consistent with the results of  \citet{lee00}. Subsequent studies with better approximations to 
relativistic gravity, e.g. \citet{faber06a}, have seen qualitatively 
similar effects for stiffer EOSs, but after a few orbital period the neutron was always disrupted. Shibata discussed such episodic,
long-lived mass transfer in a GR context \citep{shibata11} and concluded that, while possible, it has so far never been
seen in fully relativistic studies.

A follow-up study \citep{rosswog05a} with similar simulation tools focussed on higher mass, non-spinning black holes 
(M$_\mathrm{BH}= 14 ... 20$ \msun) that were approximated by  pseudo-relativistic potentials \citep{paczynski80}. While being
very simple to implement, this approach mimics some GR effects reasonably well and in particular it has an innermost stable
circular particle orbit at the correct location ($6 GM_\mathrm{BH}/c^2$), see \citet{tejeda13a} for quantitative assessment of
various properties. In none of these high black hole mass cases was episodic mass transfer observed, the neutron star was 
always completely disrupted shortly after the onset of mass transfer. However, although disks  formed for
systems below 18 \msun, a large part of them was inside the ISCO and  falling practically radially into the hole on a dynamical
time scale. As a result, they were thin and cold and not considered promising GRB engines. It was suggested, however, that 
even black holes at the high end of the mass distribution could possibly be GRB engines, provided they spin rapidly 
enough, since then both ISCO and  horizon move closer to the BH. The investigated systems ejected 
between 0.01 and 0.2 \Msun at large velocities ($\sim 0.5$ c) and analytical estimates suggested that such systems
should produce  bright optical/near-infrared transients (``macronovae'') powered by the radioactive decay of the freshly 
produced r-process elements within the ejecta, as originally suggested \citep{li98}.

\paragraph{Studies with approximate GR gravity around a non-spinning black hole}

\begin{figure}[ht]
 \centerline{\includegraphics[width=\textwidth]{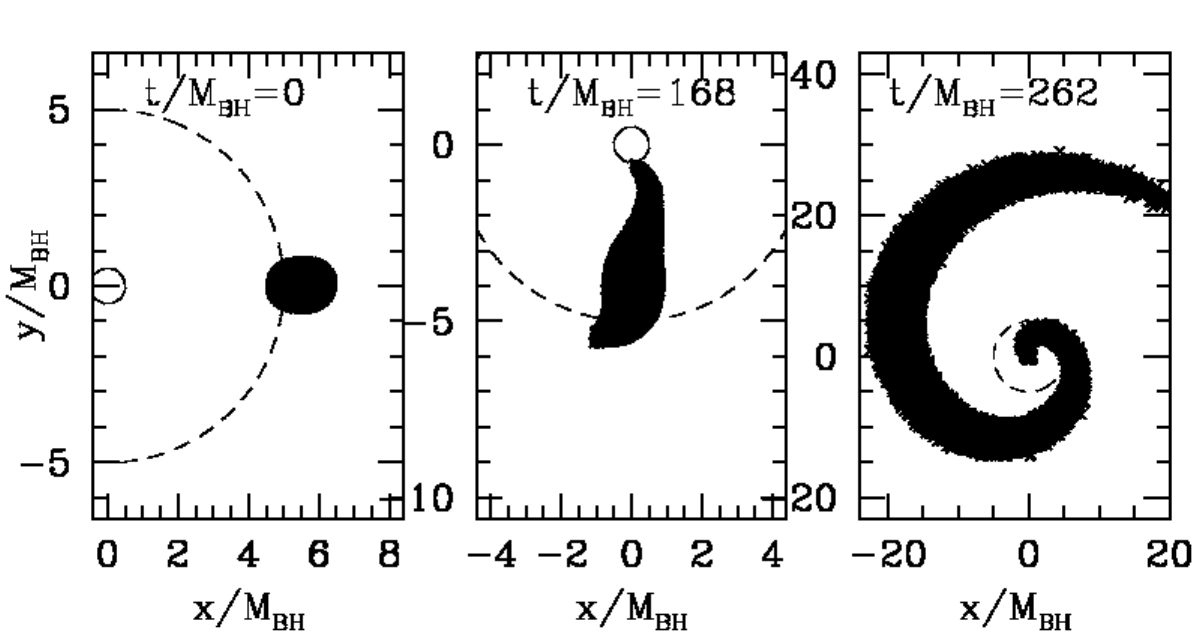}}
 \caption{Disruption of an ``undercompact'' neutron star (compactness =0.09),
               simulated within the conformal flatness approximation 
               by a non-rotating black hole ($q=0.1$). Although at some stage 
               98\% of the neutron star are within the innermost stable circular orbit
               (dashed circle),  rapid redistribution of angular momentum
               leads to the ejection of a one-armed spiral. Finally a disk made of 
               0.12 M$_\mathrm{ns}$ forms and 0.13 M$_\mathrm{ns}$ are dynamically ejected.
                Image reproduced with permission from \citet{faber06b}, copyright by AAS.}
   \label{fig:NSBH_faber}
\end{figure}

\citet{faber06a,faber06b} studied the merger of a non-rotating black hole with a polytropic neutron star
in approximate GR gravity. While the hydrodynamics code was fully relativistic, the self-gravity of the neutron star was
treated within the conformal flatness approximation. Since the black hole was kept at a fixed position its
mass needed to be substantially larger than the one of the neutron star, therefore a mass ratio of $q=0.1$ was chosen.
The neutron star matter was modelled with two (by nuclear matter standards) relatively soft polytropes ($\Gamma= 1.5$ 
and 2). In the first study \citep{faber06a} they focussed on tidally locked neutron stars and solved the five linked non-linear
field equations of the CF approach by means of the LORENE libraries. The second study \citep{faber06b} used
irrotational neutron stars and solved the CF equations by means of a Fast Fourier transform solver. In a first case
they considered a neutron star of compactness $\mathcal{C}= 0.15$, and, to simulate a case
where the disruption of the neutron star occurs near the ISCO, they also considered a second case where
$\mathcal{C}$ was only 0.09. The first case turned out to be astrophysically unspectacular: the entire
neutron star was swallowed as a whole without leaving matter behind. The second, ``undercompact'' case, however, see 
Fig.~\ref{fig:NSBH_faber}, showed some very interesting behaviour: the neutron star spiralled towards the black hole,
became tidally stretched and although at some point 98\% of the NS
mass were inside of the ISCO, see panel two, a substantial fraction of
the matter was ejected as a one-sided spiral via
a rapid redistribution of angular momentum. Approximately 12\% of the initial neutron star formed a disk and an additional 13\% of the initial neutron star 
were tidally ejected into unbound orbits. Such systems, they concluded, would be interesting sGRB engines.

\paragraph{Studies in the fixed metric of a spinning black hole}

\citet{rantsiou08} explored how the outcome of a neutron star black hole merger depends on
the spin of the black hole and on the inclination angle of the binary orbit with respect to the equatorial plane
of the black hole. They used the relativistic SPH code originally developed by 
\citet{laguna93a,laguna93b} to study the tidal disruption of a main sequence star by a massive black hole.
The new code version employed Kerr--Schild coordinates to avoid
coordinate singularities at the horizon that appear in the frequently
used Boyer-Lindquist coordinates. Since the spacetime was kept fixed,
they focused on a small mass ratio $q= 0.1$, where the impact of the neutron star on the spacetime is sub-dominant.
Both the black hole mass and spin were frozen at their initial values during the simulation and the GW backreaction
was implemented via the quadrupole approximation in the point mass limit, similar to the one used by \citet{lee99b}.
The neutron star itself was modelled as a $\Gamma=2.0$ polytrope with Newtonian self-gravity, the artificial
dissipation parameters were fixed to 0.2 (instead of values near unity which are needed to properly deal with 
shocks). Note also, that $\Gamma=2$ is a special choice, since a Newtonian star does not change its 
radius if mass is added or lost, see Eq.~(\ref{eq:LE_MR}). The bulk of the simulations was calculated
with 10$^4$ SPH particles, in one case 10$^5$ particles were used to confirm the robustness of the
results.

\begin{figure}[ht]
        \centerline{
        \includegraphics[angle=-90,width=0.49\textwidth]{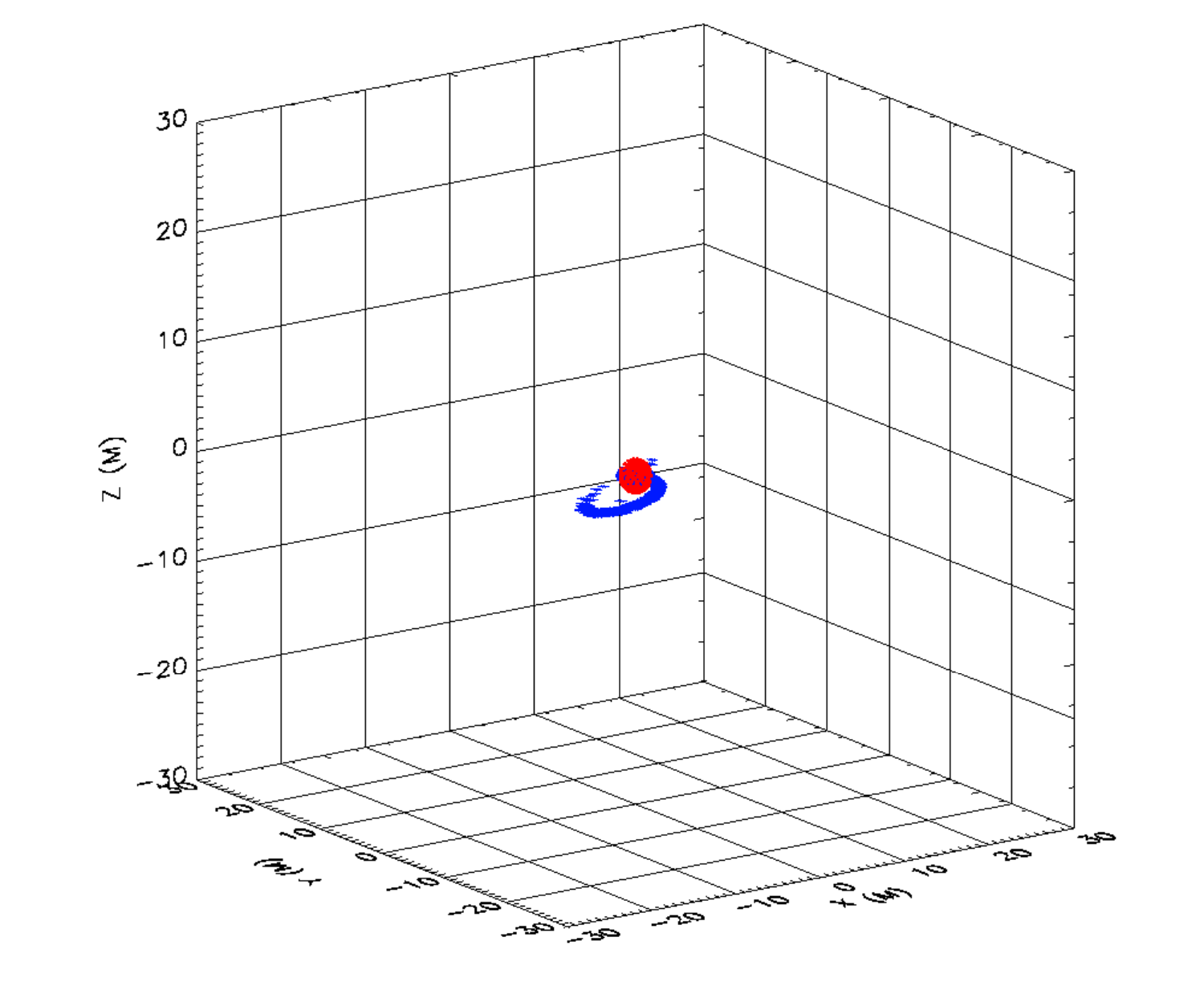}
        \includegraphics[angle=-90,width=0.49\textwidth]{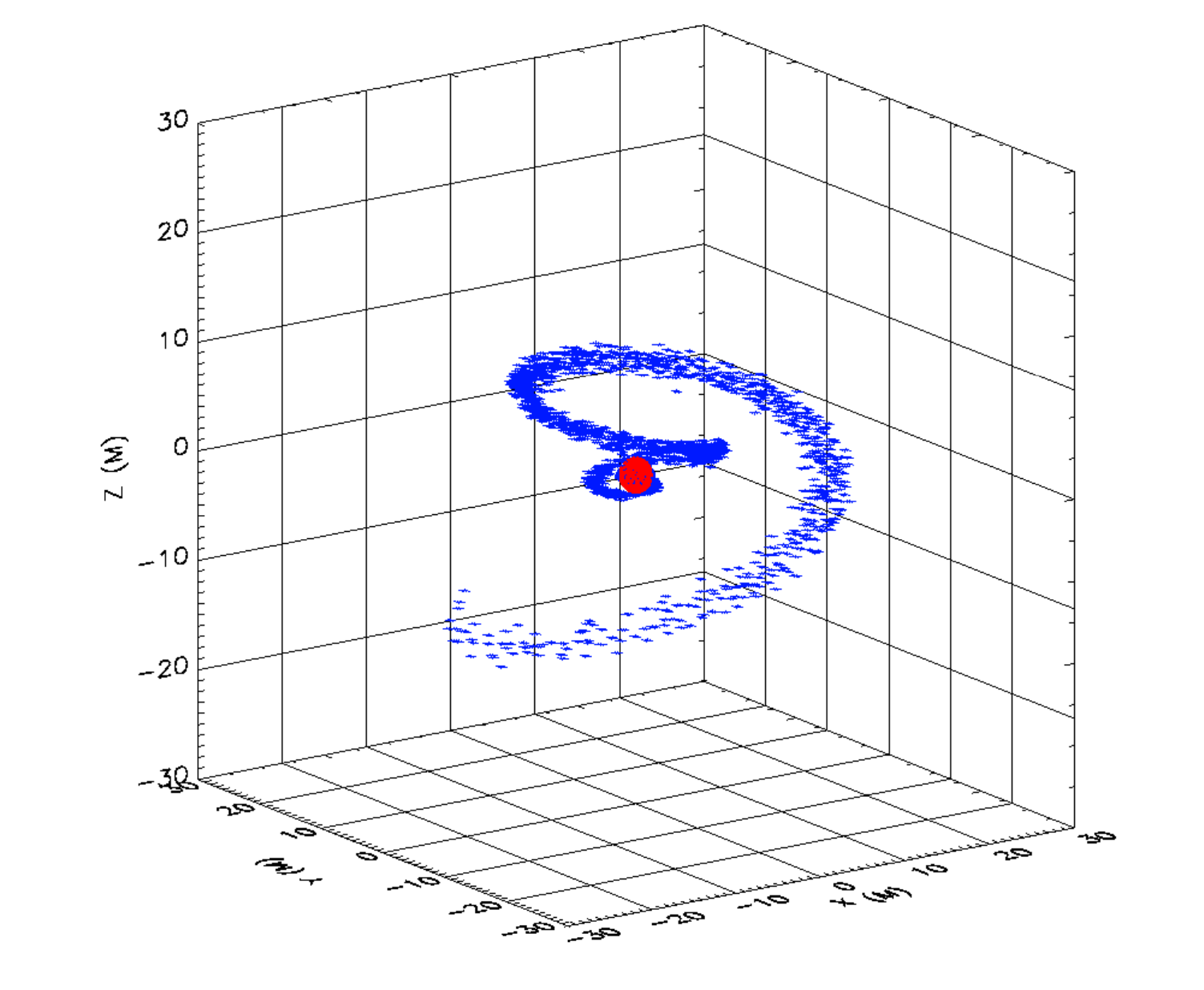}
    }
 \caption{Disruption of a polytropic star ($\Gamma=2.0$) by a close-to-maximally 
               spinning black hole ($a_\mathrm{BH}=0.99$) where the binary orbit has an initial inclination
               of 45$^\circ$ with respect to the black hole spin. The ``helix'' of unbound material contains 25\% of 
               the original neutron star mass. Image reproduced with permission from \citet{rantsiou08}, copyright by AAS.}
   \label{fig:NSBH_rantsiou}
\end{figure}

For the case of a Schwarzschild black hole ($a_\mathrm{BH}=0$) they found that neither a disk 
formed nor any material was ejected.  For equatorial mergers with spinning black holes, it 
required a spin parameter of $a_\mathrm{BH} > 0.7$ for any mass to form a disk or to become ejected. 
For a rapidly spinning BH  ($a_\mathrm{BH}=0.75$) an amount of matter of order
0.01 \Msun became unbound, for a close-to-maximally spinning BH ($a_\mathrm{BH}=0.99$) a huge amount of matter 
($> 0.4$ \msun) was ejected. Mergers with inclination angles $> 60^\circ$ lead to the complete swallowing of the neutron
star. An example of close-to-maximally spinning black hole ($a_\mathrm{BH}=0.99$) and a neutron star whose orbital plane is
inclined by 45$^\circ$ with respect to the black hole spin is shown in Fig.~\ref{fig:NSBH_rantsiou}. Here, as much as
25\% of the neutron star, in the shape of a helix, become unbound.

%

At the time of writing this review (spring 2026) we are not aware of
any fully relativistic simulation of neutron star black hole mergers.

\subsubsection{Dynamical collisions between neutron stars and black holes}
\label{sec:compact_collisions}
Traditionally, the focus of compact object encounters has been on GW-driven binary systems such as the Hulse--Taylor pulsar
\citep{taylor89,weisberg10}. 
However, also dynamical collisions/high-eccentricity encounters between two compact objects should occur in nature
\citep{kocsis06a,oleary09,lee10a,east12a,kocsis12} although their rates are difficult to estimate reliably.

Collisions differ from gravitational-wave-driven mergers in a number of ways. Since gravitational-wave emission
of eccentric binaries  removes angular momentum more efficiently than energy, primordial binaries will have radiated away
their eccentricity and will finally merge from a nearly circular orbit.  Binaries that have formed dynamically, in contrast, 
say in a globular cluster, start from a small orbital separation, but with large eccentricities and may not have enough time 
to circularize before merger. This leads to pronouncedly different gravitational-wave signatures, ``chirping'' signals of 
increasing frequency and amplitude for mergers and initially well-separated, repeated GW bursts that continue from 
minutes to days in the case of collisions. Moreover, 
compact binaries are strongly gravitationally bound at the onset of the dynamical merger phase while collisions, in contrast, 
have total orbital energies close to zero and need to get rid of energy and angular momentum via GW emission 
and  mass shedding episodes before they can form a single remnant. Due to the strong dependence on the
impact parameter and the lack of strong constraints on it, one expects an even  larger variety of dynamical
behaviour for collisions than for mergers, and some will be substantially more violent.

\citet{lee10a}  provided detailed rate estimates of compact object collisions and concluded that such encounters could 
possibly produce an interesting contribution to the observed GRB rate. They also performed the first SPH simulations 
of such encounters.   They used the SPH code from their earlier studies \citep{lee99a,lee99b,lee00,lee01a} and
explored the dynamics and remnant structure of encounters with different strengths between all types of compact stellar objects
(WD/NS/BH; typically with 100k SPH particles) using polytropic equations of state and Newtonian point masses with absorbing
boundaries at the Schwarzschild radii as models for black holes. Their calculations indicated in particular that
such encounters would produce interesting GRB engines with massive disks and additional  external
reservoirs (tidal tails, one for each close encounter) where large amounts of matter ($>0.1$ \msun) could be stored to 
possibly prolong the central engine activity, as observed in some bursts. In addition, a substantial amount of mass 
was dynamically ejected (0.03 \Msun for NSNS and up to 0.2 \Msun for NSBH systems).

\begin{figure}[htp]
   \centerline{
    \includegraphics[width=0.5\textwidth]{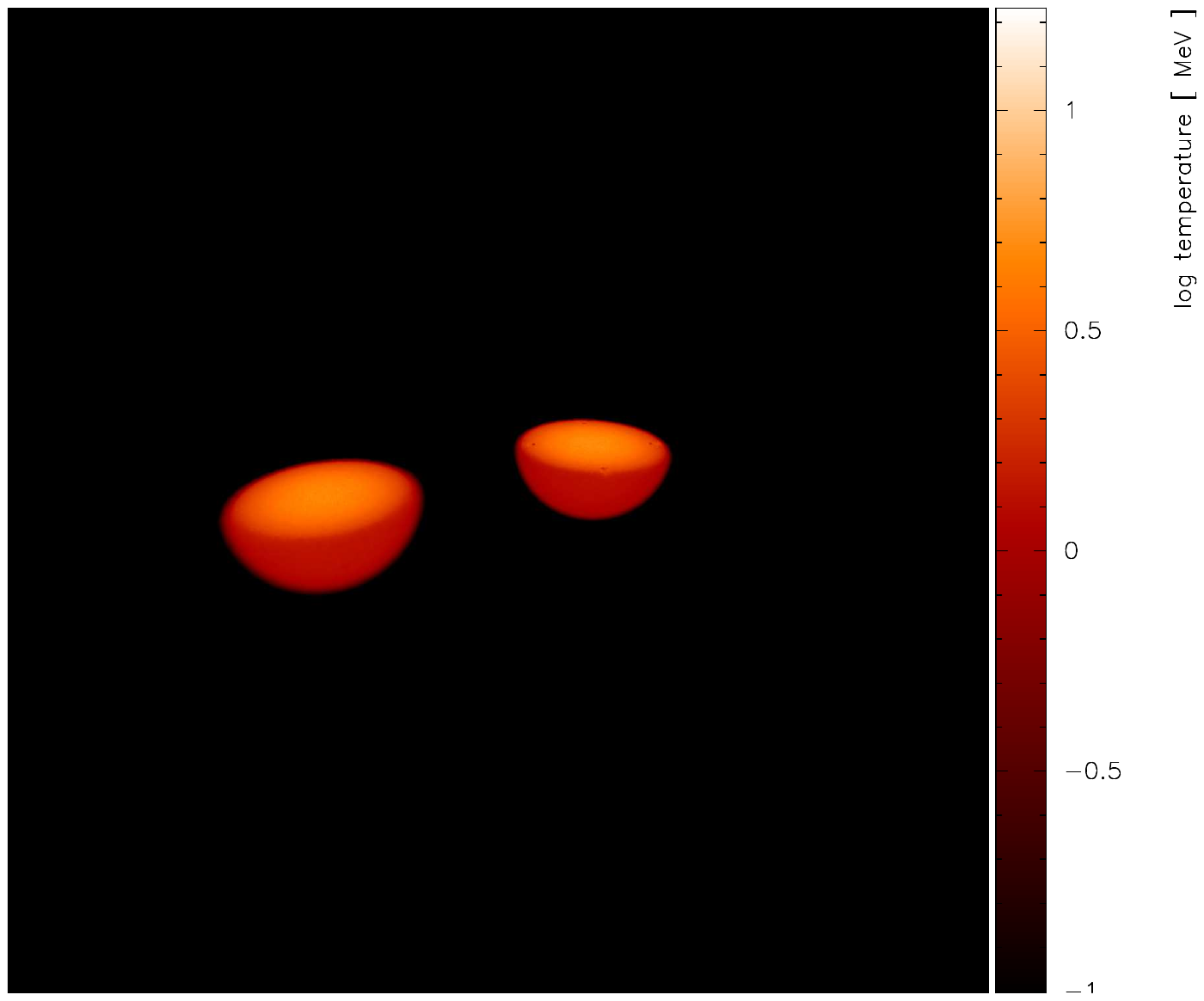}
    }
    \centerline{
    \includegraphics[width=0.5\textwidth]{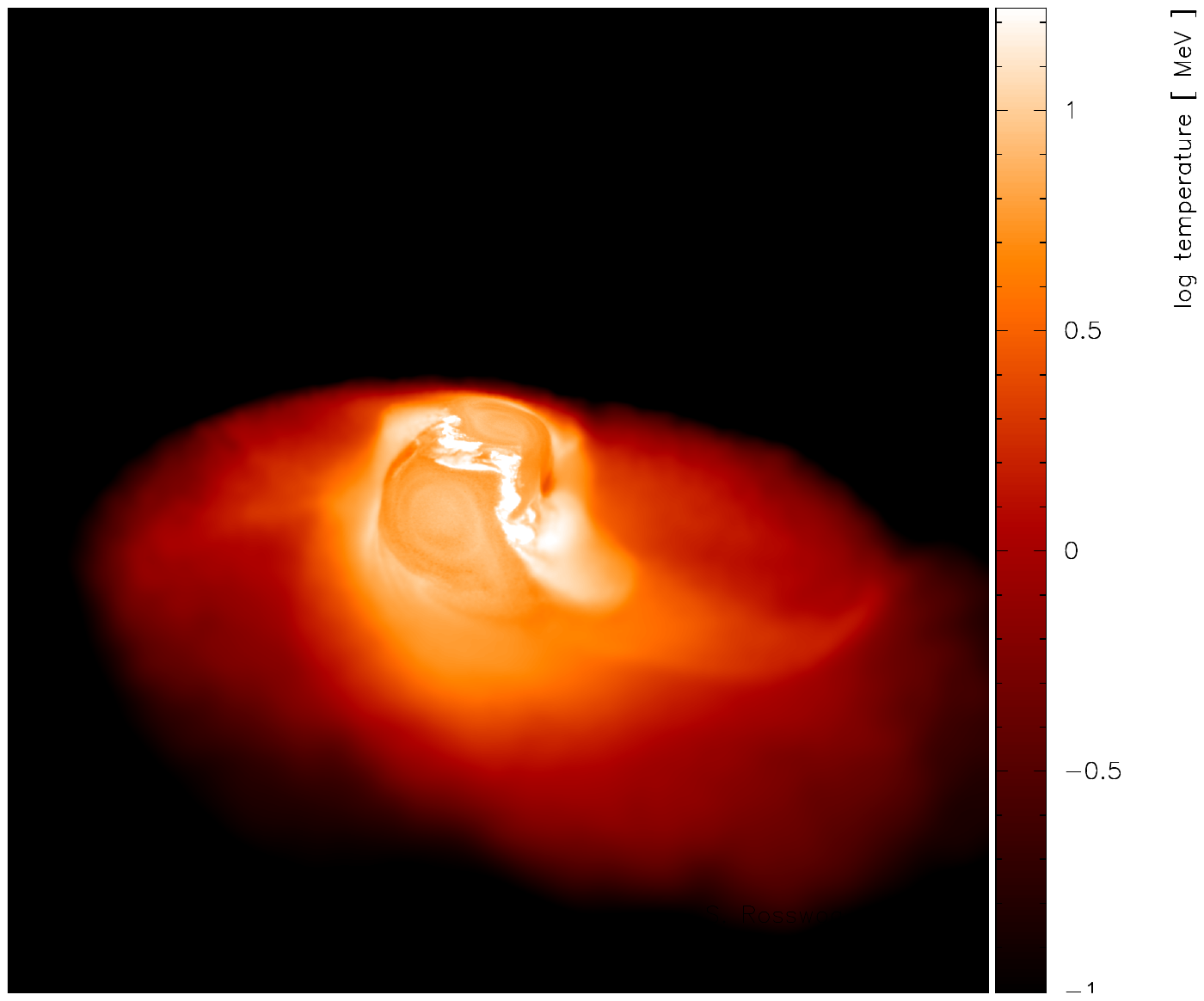}
    \includegraphics[width=0.5\textwidth]{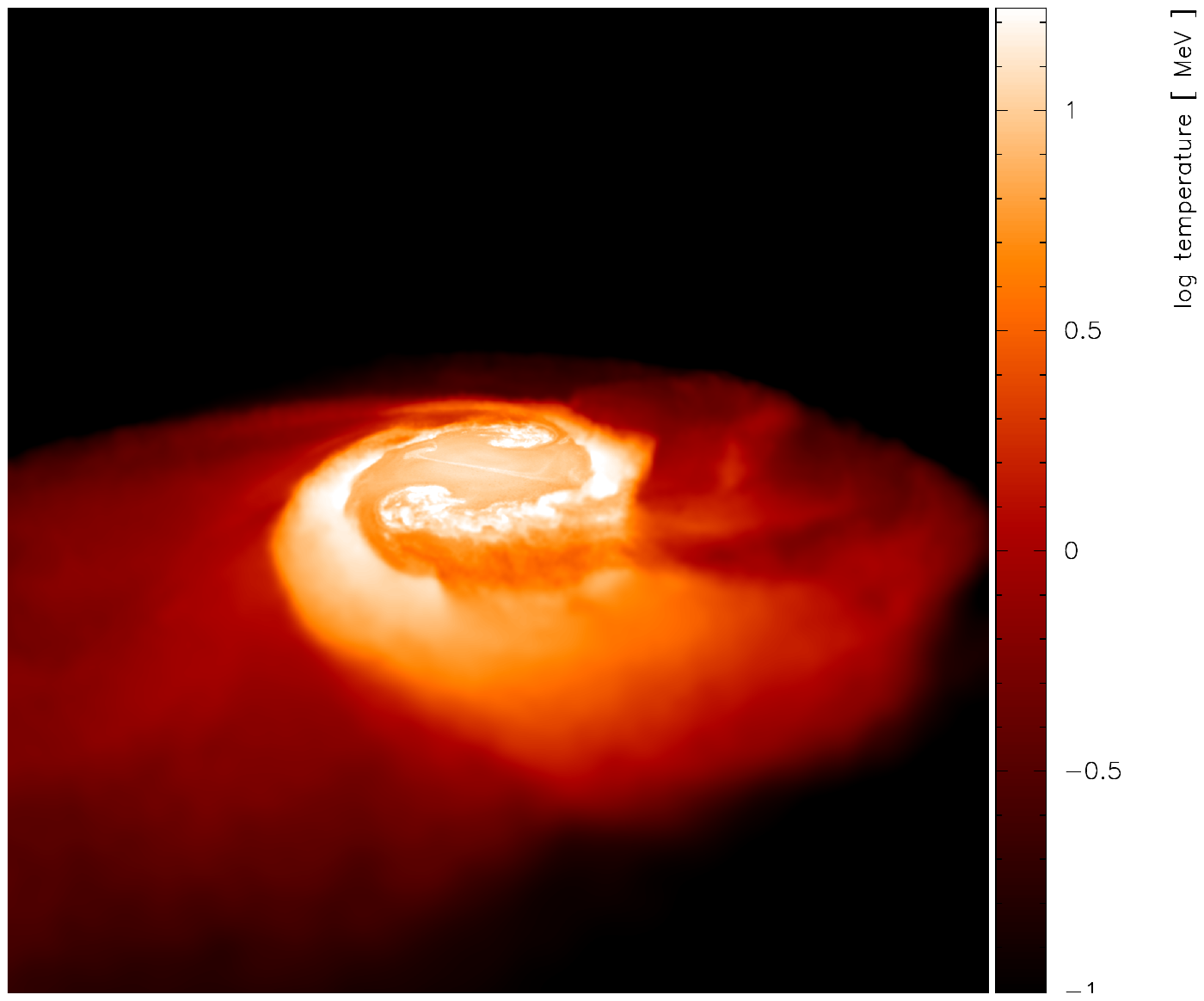}
    }
 \caption{Collision between a 1.3 \Msun and a 1.4 \Msun neutron stars
               (modelled by $8 \times 10^6$ SPH particles; pericentre distance equal to 
                the average of the two neutron star radii). 
               Only matter below the orbital plane is shown, color-coded is temperature
               in MeV. The corresponding simulations are discussed in detail in \citet{rosswog13a}. 
               (For a movie, please see the Supplementary Material of this article.)}
   \label{fig:nsns_collision}
\end{figure}

\begin{figure}[ht]
   \centerline{
   \includegraphics[width=0.5\textwidth]{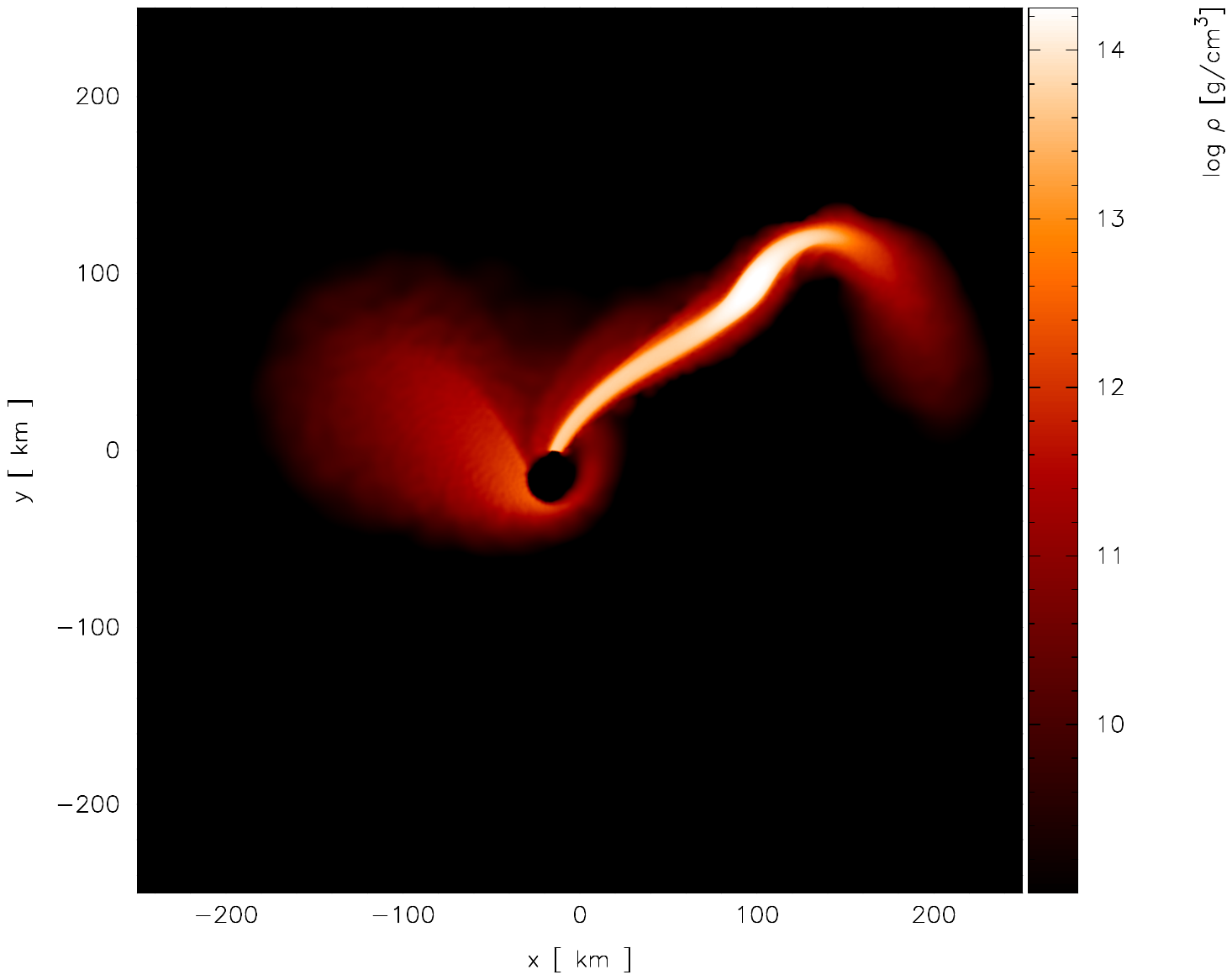}
   }   
   \centerline{
    \includegraphics[width=0.5\textwidth]{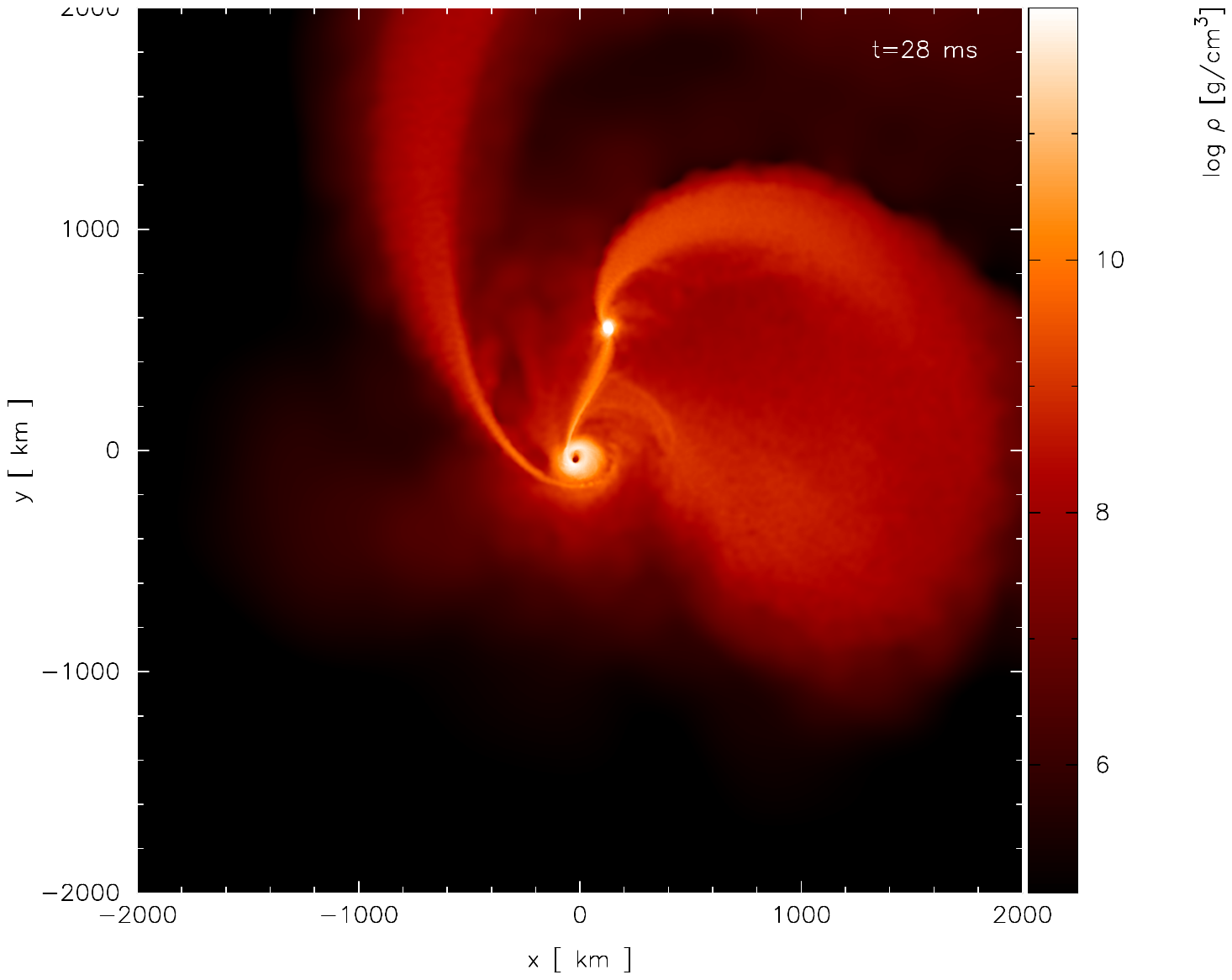}
    \includegraphics[width=0.5\textwidth]{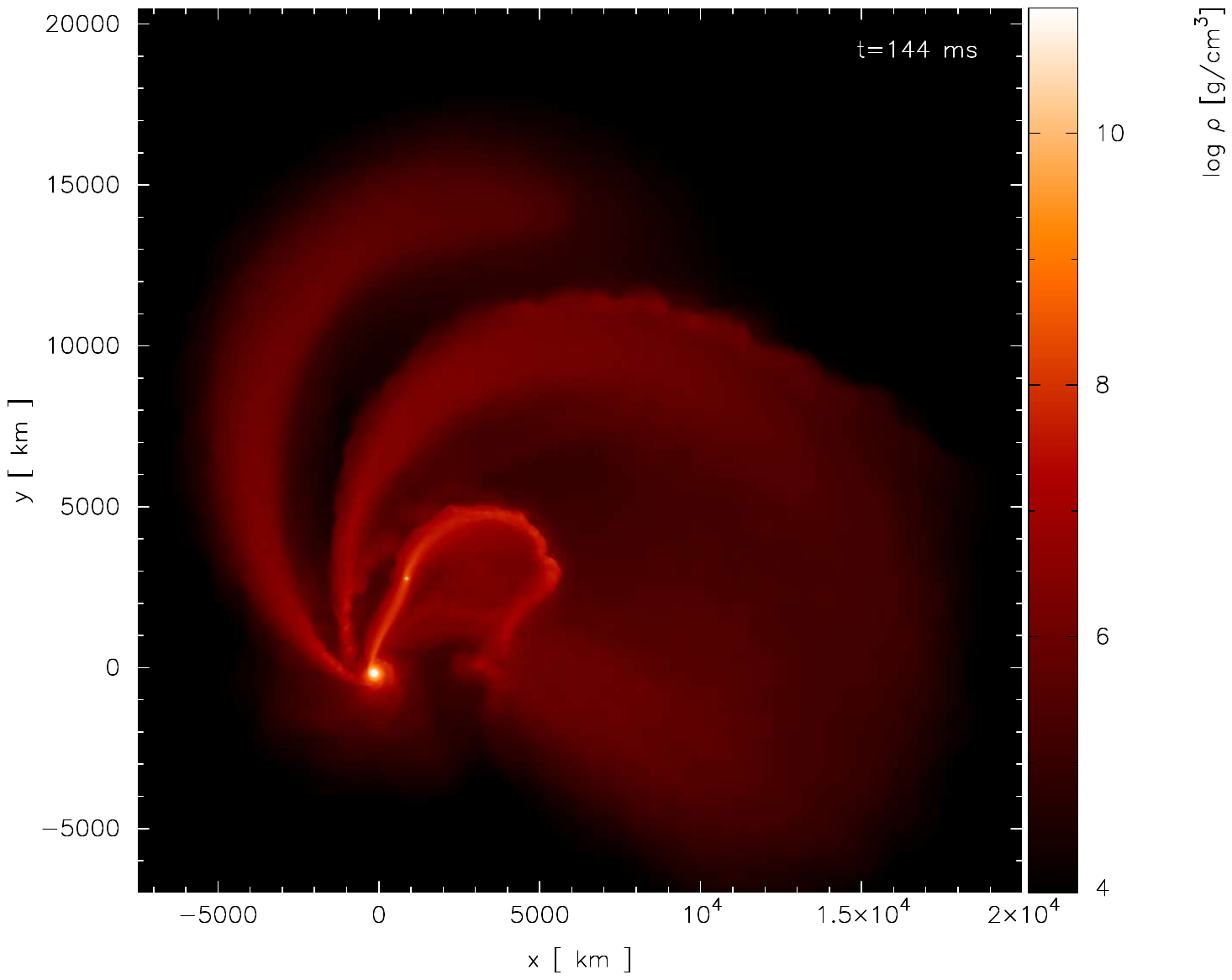}
    }
 \caption{Grazing collision between a 5 \Msun black hole and a 1.3 \Msun neutron star. 
               Shown are the densities in the orbital plane after the first, second and third close encounter.
               In this particular case the neutron star's core survives all three encounters,
               each time producing a tidal tail, and is finally ejected at $\sim 0.1$ \msun.
               The corresponding simulations are discussed in detail in \citet{rosswog13a}.
               (For a movie, please see the Supplementary Material of this article.)}
   \label{fig:nsbh_collision}
\end{figure}

In \citet{rosswog13a} various signatures of gravitational-wave-driven mergers and dynamical collisions were compared,
both for NSNS and NSBH encounters. The study applied Newtonian SPH (with up to $8 \times 10^6$ particles) together with a nuclear equation
of state \citep{shen98a,shen98b} and a multi-flavour neutrino leakage scheme \citep{rosswog03a}. As above, black holes were modelled as Newtonian
point masses with absorbing boundaries at the $R_S$. A simulation result of a strong encounter between a 1.3 and a 1.4 \Msun neutron
star (pericentre distance equal to the average of the two neutron star radii) is shown in Fig.~\ref{fig:nsns_collision}. Due to the strong
shear at their interface, a string of Kelvin--Helmholtz vortices forms in each of the  close encounters before a final central 
remnant is formed. Such conditions offer plenty of opportunity for magnetic field amplification 
\citep{price06,anderson08b,obergaulinger10,rezzolla11,zrake13,kiuchi15,palenzuela22,aguilera22,aguilera24,kiuchi24,aguilera25,neuweiler26,cook26,gutierrez26}.
In all explored cases the neutrino luminosity was at least comparable to
the merger case, $L_\nu \approx 10^{53}$ erg/s, but for the more extreme cases exceeded this value by an order of magnitude. Thus, if
neutrino annihilation should be the main agent driving a GRB outflow, chances for collisions should be at least as good as in the
merger case. But both scenarios also share the same caveat: neutrino-driven, baryonic pollution could prevent, in at least a fraction of cases,
the emergence of relativistic outflows. In the explored NSBH collisions, see for example Fig.~\ref{fig:nsbh_collision}, the neutron star usually suffered several encounters before being completely
disrupted. In some cases its core survived several encounters and was finally ejected  with a mass of $\sim 0.1$ \msun. Of course,
this offers a number of interesting possibilities (production of low-mass neutron stars, explosion at the minimal mass \citep{colpi89} etc.). But first
of all, such events may be very rare and it needs to be seen  whether such behaviour can occur in the general-relativistic case.

Generally, both  NSNS and NSBH collisions ejected  large quantities of unbound debris. Collisions between neutron stars ejected a few percent
(dependent on the impact strength) of a solar mass, while all investigated NSBH collisions ejected $\sim 0.15$ \msun, consistent with the
findings of \citet{lee10a}. Since NSBH encounters should dominate the rates \citep{lee10a}, it was concluded in \citet{rosswog13a} 
that collisions must be (possibly much) less frequent than 10\% of the NSNS merger rate to avoid  a conflict with constraints from the
chemical evolution of galaxies.

Since the ejecta masses are substantially larger ($m_\mathrm{ej}\sim 0.1$ \msun) than in the neutron  
star merger case ($m_\mathrm{ej}\sim 0.03$ \msun) simple scaling relations \citep{grossman14a}
suggest that a resulting radioactively powered kilonova should roughly peak after 
\be
t_\mathrm{P}= 11 {\rm d} \left( \frac{\kappa}{10 \; \mathrm{cm^2/g}}  \;  \frac{m_\mathrm{ej}}{0.1 \; \mathrm{M_\odot}} \; \frac{0.2 {\rm c}}{v_\mathrm{ej}}\right)^{1/2}
\ee
with a luminosity of
\be
 L_\mathrm{P}= 8.8 \times 10^{40} \; \mathrm{erg/s} \left( \frac{v_\mathrm{ej}}{0.2{\rm c}}  \;  \frac{10 \; \mathrm{cm^2/g}}{\kappa} \right)^{0.65}
 \left( \frac{m_\mathrm{ej}}{0.1 \; \mathrm{M_\odot}} \right)^{0.35}.
\ee
Here $\kappa$ is the r-process material opacity \citep{kasen13a} and a radioactive heating rate 
$\dot{\epsilon} \propto t^{-1.3}$ \citep{metzger10b,korobkin12a,lippuner15} has been assumed. 

At the moment of writing this review (spring 2026) there are no full GR simulations
of such collisions with SPH. For full GR simulations with Eulerian hydrodynamics see, 
for example, \citet{radice16a,fontbute25a}.

\subsection{Outlook on future challenges}
On the methodological side, both for Newtonian and relativistic codes, many of the recent improvements such as the enhanced gradient 
accuracy, see Sec.~\ref{sec:accurate_gradients} or more sophisticated dissipation approaches, see Sec.~\ref{sec:shocks}, have been explored on 
the level of benchmark tests, but they have not yet found their ways into the  codes that are routinely used to produce the astrophysical 
results.  Moreover, one of the near-future challenges is  to port existing production codes to up-to-date multi-GPU 
computing infrastructures. \\
In terms of including new/better physics ingredients for white dwarf simulations, the coupling to larger nuclear reaction networks would be desirable for an 
accurate prediction of nuclear abundances and the implementation of sophisticated particle-based radiative transfer schemes
would facilitate the self-consistent modeling of e.g. type Ia supernova progenitors.\\
On the relativistic SPH side, there is a wealth of important challenges that need to be addressed.
%
%
For example, although it is now possible to evolve the particle-based fluid in a self-consistent, dynamical general relativistic space-time,
see Sec.~\ref{sec:GR_SPH}, this may still not be viable approach for some problems. If black holes are to be evolved in a fully dynamical 
space-time, the event horizon needs to be resolved with a sufficient number of grid points, otherwise very serious unphysical artifacts may 
result. This translates into very serious constraints on the grid size,  the integration time step and therefore on the overall 
possible length of a simulation. For cases where the interesting time scales are much longer than the dynamical time near the black hole horizon,
a full space-time evolution approach will become unfeasible. 
For systems such as relativistic binaries embedded in a gaseous environment, say a black hole merger in
an AGN disk \citep{grobner20}, hybrid schemes may be highly advantageous. For example, if the black hole dynamics is known, e.g. from a
separate numerical relativity simulation or from a Post-Newtonian approach, one could just evolve relativistic gas dynamics in a dynamical, but a-priori
known space-time.
For other systems in addition to a "nearly known space-time" additional perturbations due to self-gravity may be important. Such approaches
are used, for example, in tidal disruption studies \citep{liptai19b,calderon26,mahapatra26,banerjee26}, but they might be generalized and refined
further. Alternative theories of gravity in the strong-field regime within an SPH approach are to our knowledge so far un-explored territory.\\
%
%
There also remains  much to do in terms of neutrino transport approximations in relativistic SPH.
Neutrinos play a very crucial role for the fate of  compact mergers and their observational signatures. Neutrino cooling as a sink of thermal energy
may determine the exact point when a merger remnant  collapses to a BH, neutrino absorption may drive additional matter outflows, it changes 
the electron fraction of ejecta, thus determines the nucleosynthetic yields and the resulting electromagnetic emission \citep{rosswog24a}. So far, there are hardly 
any neutrino transport approximations available in GR SPH, with the only exception being  ILEAS \citep{ardevol19}, which is an advanced 
leakage scheme for conformal flatness approximation SPH that accounts also for neutrino absorption/heating effects. There are plenty of 
possibilities for future neutrino transport approximations in relativistic SPH, many exist already in a Eulerian context, but would need to be translated 
for particle approaches. For example, one could develop GR M0/Flux-limited diffusion schemes, see e.g.
\cite{wilson07},  gray/spectral M1 moment schemes, see e.g. \cite{oconnor15}, \cite{radice22} and \cite{cheong23}, Monte Carlo transport,
see \cite{foucart23} for a recent comprehensive review, or hybrid schemes such as moment schemes that calculate closure relations 
by means of Monte Carlo transport \citep{foucart18,izquierdo24}. More such hybrid schemes that combine the best parts of different transport 
approximations  are certainly possible.
All of the above mentioned neutrino transport approximations refer to the case of mass-less neutrinos, but including neutrino oscillations 
would be yet another, though very challenging, step forward.\\
%
%
Another  ingredient of paramount importance are magnetic fields. Initial neutron star seed magnetic fields become massively amplified during the
merger process, see e.g. \cite{price06}, \cite{kiuchi15}, \cite{aguilera25}, \cite{neuweiler26} or \cite{gutierrez26}, and they are understood to play a major role in 
launching the relativistic jets that are needed for gamma-ray bursts. There is to our knowledge no GR-MHD implementation in SPH yet,
but it would for sure be both interesting and challenging. One of the challenges comes from a major advantage of SPH: the
particles follow the density. This means that the region where jets are likely formed, along the binary rotation axis where the density is particularly low, 
is among the worst resolved regions in a compact binary merger simulation. Another challenge, similar to the Newtonian case, would be fulfilling the 
no-monopole constraint. Here,  one could think along two lines of development, moving either towards a purely particle-based scheme
along the ideas outlined in the introduction, or, alternatively, one could borrow established grid-based methods for a particle-mesh approach,
since the fully dynamical spacetime evolution needs a mesh anyway.\\
All these challenges will for sure be addressed in the not too far future.

\backmatter

\bmhead{Acknowledgements}
\label{sec:acknowledgements}
We thank the anonymous referee for her/his careful reading and for constructive comments on  
how to improve the paper.\\
This work is based on a long term-effort and it has enormously profited
from many teachers, colleagues and friends and I want to collectively thank 
all of them. Particular thanks goes to Willy Benz, Marius Dan, Peter Diener, Melvyn Davies, 
Oleg Korobkin,  Joe J. Monaghan, Daniel J. Price, Enrico Ramirez-Ruiz,  Emilio Tejeda 
and Friedel Thielemann. Especially the development of a fully general relativistic SPH
method has enormously profited from my close collaboration with Peter Diener
and from the dedicated funding from the Swedish Research Council (VR) that gave
us the time and travel possibilities to spend useful amounts of time at each others
institutions. I also want to thank Bhaskar Biswas, Jan-Erik Christian, Peter Diener,  Lukas Schnabel 
and Swapnil Shankar for their comments on an earlier version of this review.

SR has been supported by the Swedish Research Council (VR) under 
grant numbers 2012-4870, 2016-03657 and 2020-05044. This work has further 
profited from  the Research Environment Grant
``Gravitational Radiation and Electromagnetic Astrophysical
Transients'' (GREAT) funded by VR  under Dnr 2016-06012, by the Knut and Alice Wallenberg Foundation
under grant Dnr. KAW 2019.0112, by Deutsche Forschungsgemeinschaft 
(DFG, German Research Foundation) under Germany's Excellence Strategy 
- EXC 2121 \enquote{Quantum Universe} - 390833306 and by the European Research 
Council (ERC) Advanced Grant INSPIRATION under the European Union's 
Horizon 2020 research and innovation programme (Grant agreement No. 
101053985).

\bmhead{Competing Interests} The authors declare no competing interests.

\begin{appendices}

\section{Common SPH Kernels}
\label{sec:kernels}

We write normalized SPH kernels in the form\footnote{We use the convention that $W$ 
refers to the full normalized kernel while $w$ is the un-normalized shape of the kernel.}
\be
W(|\vec{r}-\vec{r}'|,h)= \frac{\sigma}{h^D} w(q), \label{eq:normalized_SPH_kernel}
\ee
where $h$ is the smoothing length that determines the support of $W$, $q= |\vec{r}-\vec{r}'|/h$ 
and $D$ is the number of spatial dimensions. The normalizations are obtained from
\be
\sigma^{-1}= \left\{\begin{array}{ll}  
                       2 \int_0^{Q} w(q) dq  & \text{in 1D}\\\\
                       2 \pi \; \int_0^{Q} w(q)  \; q\; dq & \text{in 2D}\\\\
                       4 \pi \; \int_0^{Q} w(q)  \; q^2 \; dq & \text{in 3D},
                        \end{array}\right.
\label{eq:normalization}
\ee
where $Q$ is the kernel support for which, for historical reasons, often $Q=2$ is used.
In the following we will give the kernels in the form that is usually used in the literature.
This also means that the kernels usually have different support sizes e.g. some vanish outside
1 smoothing length, or outside of several smoothing lengths, see below.

For a fair comparison in terms of computational effort/neighbour numbers, the kernels  can be stretched to
different support sizes, but then, of course, the normalization constants $\sigma$ need to be adapted.
If a kernel has a normalization $\sigma_{lh}$ for a support size of $l h$,  its  normalization changes to
\be
\sigma_{kh}= \left(\frac{l}{k}\right)^D \sigma_{lh} \label{eq:normalization_change}
\ee
if it is stretched to a support of $k h$, where $D$ is the number of spatial dimensions.

In the following box, we collect some simple relations that are very useful when deriving SPH equations
and that are referred to frequently throughout this review.

\medskip
\fbox{\parbox{0.9\textwidth}{
\vspace*{0.5cm}
\centerline{\bf Derivatives of radial kernels}
\vspace*{0.5cm}

We use the notation $\vec{r}_{bk}= \vec{r}_{b} - \vec{r}_{k}$, $r_{bk}=
|\vec{r}_{bk}|$ and $\vec{v}_{bk}= \vec{v}_{b} - \vec{v}_{k}$. 
Derivatives resulting from the smoothing lengths are ignored for the 
moment, but they can be easily calculated if needed. 
By straight-forward component-wise differentiation one finds
\begin{align}
\frac{\partial}{\partial \vec{r}_a} |\vec{r}_b-\vec{r}_k|
= \frac{(\vec{r}_b-\vec{r}_k)
  (\delta_{ba}-\delta_{ka})}{|\vec{r}_b-\vec{r}_k|}
= \hat{e}_{bk} (\delta_{ba}-\delta_{ka}),\label{eq:k1}
\end{align}
where $\hat{e}_{bk}$ is the unit vector from particle $k$ to particle $b$,
\begin{align}
\frac{\partial}{\partial \vec{r}_a} \frac{1}{|\vec{r}_b-\vec{r}_k|}=
- \frac{\hat{e}_{bk} (\delta_{ba}-\delta_{ka})}{|\vec{r}_b-\vec{r}_k|^2}
\label{eq:k2}
\end{align}
and $\delta_{ij}$ is the usual Kronecker symbol.
Another frequently needed expression is
\begin{align}
\frac{d r_{ab}}{dt}=& \frac{\partial r_{ab}}{\partial x_a} \frac{d x_a}{dt} 
                     + \frac{\partial r_{ab}}{\partial y_a} \frac{d y_a}{dt} 
                     + \frac{\partial r_{ab}}{\partial z_a} \frac{d z_a}{dt} 
                     +  \frac{\partial r_{ab}}{\partial x_b} \frac{d x_b}{dt} 
                     + \frac{\partial r_{ab}}{\partial y_b} \frac{d y_b}{dt} 
                     + \frac{\partial r_{ab}}{\partial z_b} \frac{d z_b}{dt} 
                     \nonumber \\
                   =& \nabla_a r_{ab} \cdot \vec{v}_a + \nabla_b r_{ab}
                   \cdot \vec{v}_b  = \nabla_a r_{ab} \cdot \vec{v}_a - 
                   \nabla_a r_{ab}
                   \cdot \vec{v}_b
                   = \nabla_a r_{ab} \cdot \vec{v}_{ab} 
                   = \hat{e}_{ab} \cdot \vec{v}_{ab},
\label{eq:basic:drab_dt} 
\end{align}
where 
$\partial r_{ab}/\partial x_b= - \partial r_{ab}/\partial x_a$ etc. was used.
For kernels that only depend on the magnitude of the separation,
$W(\vec{r}_b-\vec{r}_k)= W(|\vec{r}_b-\vec{r}_k|)\equiv W_{bk}$, the derivative with respect
to the coordinate of an arbitrary particle $a$  is
\begin{align}
\nabla_a W_{bk}=\frac{\partial}{\partial \vec{r}_a} W_{bk}= \frac{\partial  W_{bk}}{\partial
  r_{bk}}  \frac{\partial r_{bk}} {\partial \vec{r}_a}= 
\frac{\partial  W_{bk}}{\partial r_{bk}} \hat{e}_{bk}
(\delta_{ba}-\delta_{ka})= \nabla_b W_{kb} (\delta_{ba}-\delta_{ka}) \label{eq:k3},
\end{align}
where Eq.~(\ref{eq:k1}) was used. This yields in particular the important property
\begin{align}
\nabla_a W_{ab}=\frac{\partial}{\partial \vec{r}_a}W_{ab}
=& \frac{\partial W_{ab}}{\partial r_{ab}} \frac{\partial r_{ab}} {\partial
  \vec{r}_a} 
 = \frac{\partial W_{ab}}{\partial r_{ab}} \hat{e}_{ab}
= - \frac{\partial W_{ab}}{\partial r_{ab}} \frac{\partial r_{ab}} {\partial
  \vec{r}_b} \nonumber\\
=& - \frac{\partial}{\partial \vec{r}_b}W_{ab}= - \nabla_b W_{ab}\label{eq:k4}
\end{align}
and also shows that the gradient points in the direction of the line connecting
two particles. This is important for ensuring exact angular momentum conservation.
The time derivative of the kernel is given by
\begin{align}
\frac{d W_{ab}}{dt}= \frac{\partial W_{ab}}{\partial r_{ab}} \frac{d r_{ab}}
{dt} 
= \frac{\partial W_{ab}}{\partial r_{ab}}
\frac{(\vec{r}_a-\vec{r}_b)\cdot (\vec{v}_a-\vec{v}_b)}{r_{ab}} 
= \frac{\partial W_{ab}}{\partial r_{ab}} \hat{e}_{ab} \cdot \vec{v}_{ab}
= \vec{v}_{ab} \cdot  \nabla_a W_{ab}. \label{eq:k5}
\end{align}
}}

\medskip

In the following subsections we summarize frequently used SPH kernels. It is
worth noting, however, that kernels can actually also be combined, say via linear combinations.
For example, \citet{wissing26a} fine-tune such linear combinations to the specific set of
evolution equations and they find improved convergence without  any extra 
computational costs.

\subsection{B-Spline kernels}
\label{sec:spline_kernel}
So-called B-spline  functions \citep{schoenberg46}, $M_n$, which are generated as Fourier transforms,
\be
M_n(x,h)= \frac{1}{2 \pi} \int_{-\infty}^\infty \left[ \frac{\sin(kh/2)}{kh/2}\right]^n \cos(kx) dk,
\ee
have long been the most common choice for SPH kernels.
The functions $M_n$ are approximations to Gaussians and they consist
of $\lceil n/2 \rceil$ piece-wise polynomials of degree $n-1$ and they
are $n-2$ times continuously differentiable.

Since SPH requires at the very least the continuity in the 
first and second derivative, the cubic spline kernel M$_4$ is the
lowest-order SPH kernel that is a possible choice and it has, for
a long time, been the default in SPH. In Table~\ref{tab:spline_kernels} we
give the innermost pieces from the cubic up to the nonic spline
kernels.
From these innermost pieces, the full kernels can be
constructed by omitting the last summand from the previous piece
when going one piece further out.  As an illustration, we give the
septic kernel which reads
\be
M_8=  \frac{6}{8! \pi h^3} \left\{\begin{array}{ll}  
         (4 - q)^7 - 8 (3 - q)^7 + 28 (2 - q)^7 - 56 (1 - q)^7 & 0
         \leq q  < 1\\
         (4 - q)^7 - 8 (3 - q)^7 + 28 (2 - q)^7  & 1
         \leq q  < 2\\
         (4 - q)^7 - 8 (3 - q)^7  & 2
         \leq q  < 3\\
         (4 - q)^7  & 3
         \leq q  < 4\\
          0      & {\rm elsewhere}.
         \end{array}\right.
\ee

\begin{landscape}
    
\begin{table}[htp]
\caption{B-spline kernels with the innermost piece, the normalisation factor
$\sigma$, and radial extent $Q$. From the innermost piece the full piecewise function can
be constructed.}

\small

\label{tab:spline_kernels}
\begin{tabular}{llllll}
\toprule
Name & innermost piece & $\sigma$ (1D) & $\sigma$ (2D) & $\sigma$ (3D) & $Q$ \\
\midrule
Cubic M$_4$ 	& $(2 - q)^3 - 4 (1 - q)^3$ & $3!$ & $5/(14\pi)$ & $6/(4!\pi)$ & 2 \\[2pt]
Quartic M$_5$ 	& $(\tfrac{5}{2} - q)^4 - 5 (\tfrac{3}{2} - q)^4 + 10 (\tfrac{1}{2} - q)^4$ & $4!$ & $96/(1~199\pi)$ & $6/(5!\pi)$ & 2.5 \\[2pt]
Quintic M$_6$ 	& $(3 - q)^5 - 6 (2 - q)^5 + 15 (1 - q)^5$ & $5!$ & $7/(478\pi)$ & $6/(6!\pi)$ & 3 \\[2pt]
Sextic M$_7$ 	& $(\tfrac{7}{2} - q)^6 - 7 (\tfrac{5}{2} - q)^6 + 21 (\tfrac{3}{2} - q)^6 - 35 (\tfrac{1}{2} - q)^6$ & $6!$ & $256/(113~149\pi)$& $6/(7!\pi)$  & 3.5 \\[2pt]
Septic  M$_8$ 	& $(4 - q)^7 - 8 (3 - q)^7 + 28 (2 - q)^7 - 56 (1 - q)^7$ & $7!$ & $9/(29~749\pi)$ & $6/(8!\pi)$ & 4 \\[2pt]
Octic  M$_9$ 	& $(\tfrac{9}{2} - q)^8 - 9 (\tfrac{7}{2} - q)^8 + 36 (\tfrac{5}{2} - q)^8 - 84 (\tfrac{3}{2} - q)^8 + 126 (\tfrac{1}{2} - q)^8$ & $8!$ & $512/(14~345~663\pi)$ & $6/(9!\pi)$ & 4.5 \\[2pt]
Nonic  M$_{10}$ 	& $(5 - q)^9 - 10 (4 - q)^9 + 45 (3 - q)^9 - 120 (2 - q)^9 + 210 (1- q)^9$ & $9!$ & $11/(2~922~230\pi)$ & $6/(10!\pi)$  & 5 \\[2pt]
\bottomrule 
\end{tabular}
\end{table}

\end{landscape}

\begin{figure}[ht]
   \centering
   \includegraphics[width=10cm]{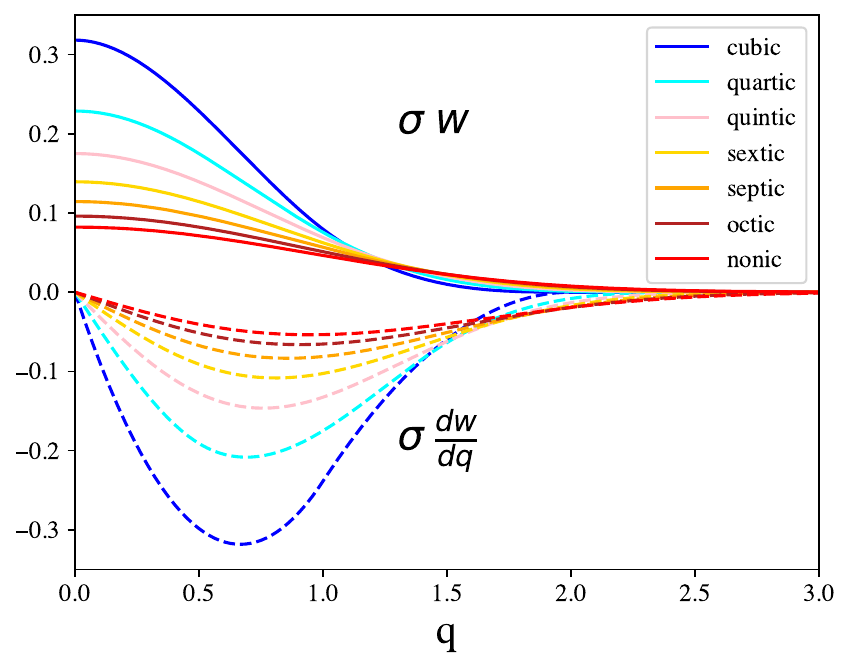} 
   \caption{Shown are the kernels  (solid lines) and their derivatives (dashed lines) for the cubic to nonic B-spline kernels with their support sizes $Q$ as indicated in Table~\ref{tab:spline_kernels}, the $q$-axis has been truncated at $q=3$ for visibility reasons.}
   \label{fig:Bsplines}
\end{figure}

\subsection{Harmonic-like kernels}
\label{sec:WHn_kernel}
A one parameter family of harmonic-like kernels has been suggested in \citet{cabezon08}  
\be
W_{{\rm H},n}= \frac{\sigma_{{{\rm H},n}}}{h^D} \left\{\begin{array}{ll}  
         1 & q = 0\\
         \left( \frac{\sin [\frac{\pi}{2} q]}{\frac{\pi}{2} q}\right)^n    & 0 < q \le 2\\
         0      & \text{else,}
         \end{array}\right. 
\ee
where the kernels become smoother and more centrally peaked for larger
values of $n$.
The normalization of this kernel family, $\sigma_{{\rm H},n}$,  for integer $n$ from 3 to 9 is given in 
Table~\ref{tab:kernel_params}. In \citet{cabezon08}, their Table~2, a fifth order polynomial is 
given that provides the normalization for continuous $n$ between 3 and 7.
The $W_{{\rm H},3}$ kernel is very similar to the  M$_4$,
while $W_{{\rm H},5}$ is a close approximation of M$_6$,
 provided they have the same support. 

\begin{table}[ht]
\caption{Normalization $\sigma_{{\rm H},n}$ for the harmonic-like
  kernels, all vanish at $Q=2$.}
\label{tab:kernel_params}

\begin{tabular}{ l | l l l l l l l}
               & n= 3    & n= 4             & n= 5            & n= 6            & n= 7             & n= 8             & n= 9 \\
\hline \\
1D  & 0.66020338 & 0.75221501 & 0.83435371 & 0.90920480 & 0.97840221 & 1.04305235  & 1.10394401\\
2D  & 0.45073324 & 0.58031218 & 0.71037946 & 0.84070999 & 0.97119717 & 1.10178466  & 1.23244006\\
3D  & 0.31787809 & 0.45891752 & 0.61701265 & 0.79044959 & 0.97794935 & 1.17851074  & 1.39132215\\
\end{tabular}
\end{table}

\begin{figure}[ht]
   \centering
   \includegraphics[width=10cm]{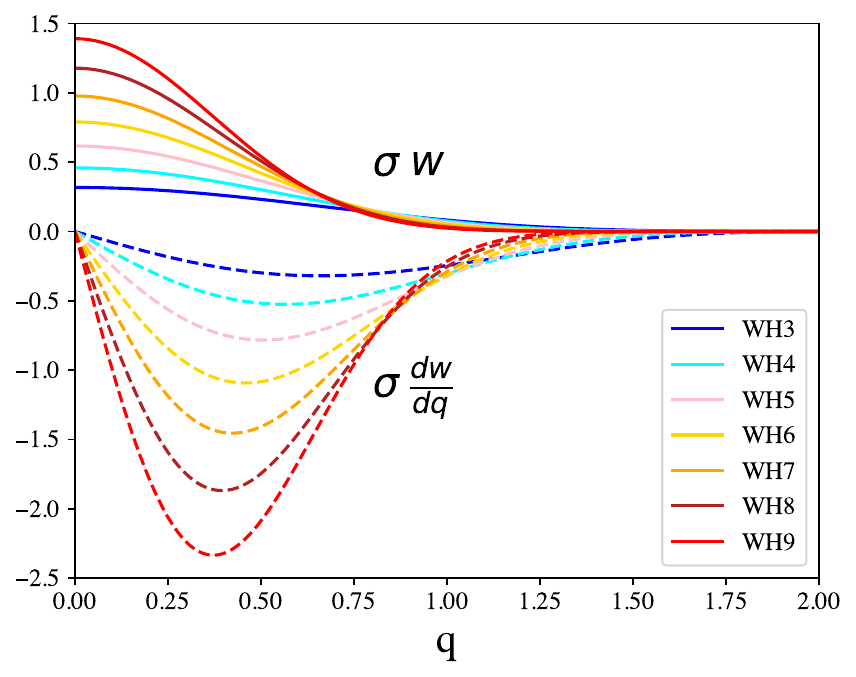} 
   \caption{Shown are the kernels (solid lines) and their
     derivatives  (dashed lines) for the sinc-kernel family for
     $n=3$ to $n=9$.}
   \label{fig:WH_kernels}
\end{figure}

\subsection{Wendland kernels}
\label{sec:Wendland_kernel}
The Wendland kernels \citep{wendland95,wendland05}  are radial basis
functions that are constructed to have compact support and
non-negative Fourier transforms. At first sight, their shape is
similar to other bell-shaped kernels with vanishing derivatives at the
origin. Most other such kernels, however, become ``pairing-unstable'' for
many contributing neighbour particles. Practically, that means that
particles form pairs rather than being spread quasi-uniformly. This
instability is rather benign in the sense that nothing really bad
happens, but since one particle is essentially replaced by two,
there is an effective loss of resolution and the non-perfect
distribution of particles may lead to a compromised gradient accuracy.
Contrary, to what was believed for a long time, however, this
instability is \emph{not} related to the kernel gradient vanishing at
the centre, but instead a linear instability analysis shows
\citep{dehnen12} that a non-negative Fourier transform is needed to
prevent the pairing instability. Keep in mind that kernels with a
non-vanishing central derivative avoid the pairing instability, but
they are not twice continuously differentiable at the centre and they
are very bad density estimators, see, for example, Fig.~1 in
\citet{dehnen12} or Fig.~4 in \citet{rosswog15b}.
Scrutinizing Wendland kernels in a number of benchmark tests \citep{rosswog15b} has shown that these
kernels maintain a highly ordered particle distribution during the
dynamical evolution, see e.g. Fig.~\ref{fig:particle_dist},  and they therefore yield very good interpolation
properties.\footnote{Despite the fact that for static particle distributions on a
grid other kernels may deliver more accurate density estimates.}

We list in Table~\ref{tab:Wendland_kernels} a number of  Wendland 3D kernels that
are suitable for SPH simulations. We write them as they are often
found in the mathematical literature with a support of $Q=1$, but they
can straight forwardly be stretched to different supports, if desired,
by means of Eq.~(\ref{eq:normalization_change}). Note, however, that the
strongly centrally peaked high-order kernels need more neighbours in
their support for a good density estimation, see Fig.~\ref{fig:density_error}.

\begin{table}[ht]
\caption{Some useful 3D Wendland kernel functions, in a form where $Q=1$.}
\small
\label{tab:Wendland_kernels}
\begin{tabular}{llcccc}
\toprule
Name & $w(q)$ & $\sigma$ (3D)  \\
\midrule
$W_{C^2}$ 	& $(1-q)^4_{+} \; (1 + 4 q)$ & $\frac{21}{2 \; \pi}$ \\[4pt]
$W_{C^4}$ 	& $(1-q)^6_{+} \; (1 + 6 q + \frac{35}{3}q^2)$ &
$\frac{495}{32 \; \pi}$ \\[4pt]
$W_{C^6}$ 	&$(1 - q)^8_{+} \; (1 + 8q + 25 q^2 + 32 q^3$) &
$\frac{1365}{64 \; \pi}$ \\[4pt]
$W_{C^8}$ 	& $(1-q)^{10}_{+} \; (1 + 10 q + 42 q^2 + 90 q^3 +
             \frac{429}{5} q^4)$ &  $\frac{1785}{64 \; \pi}$ \\[4pt]
$W_{C^{10}}$ 	& $(1-q)^{12}_{+} \; (9+ 108 q + 566 q^2 + 1644 q^3 +
             2697 q^4 + 2048 q^5)$ &  $\frac{1955}{512 \; \pi}$ \\
\bottomrule
\end{tabular}
\end{table}

\begin{figure}[ht]
   \centering
   \includegraphics[width=10cm]{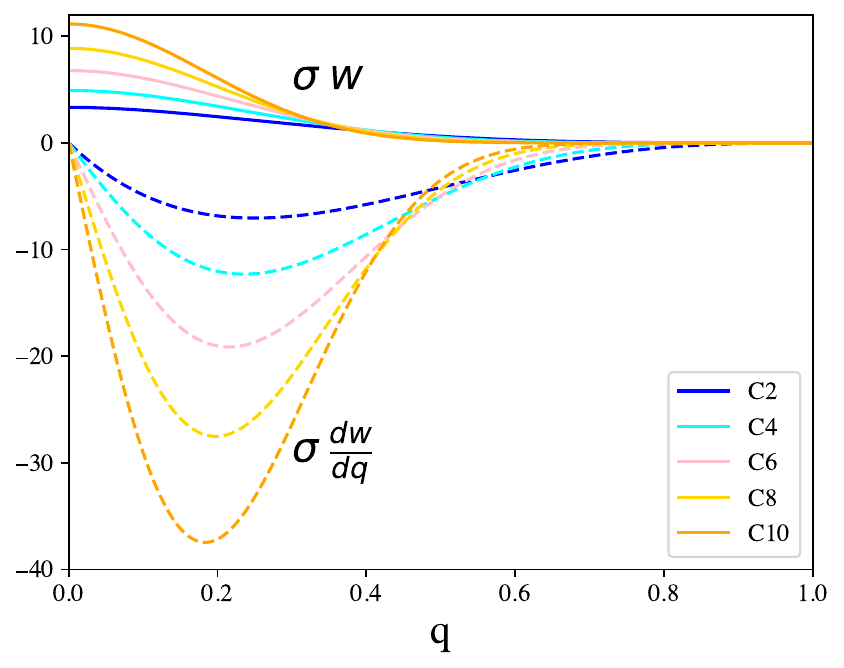} 
   \caption{Shown are the Wendland kernels (solid lines) of
     Table~\ref{tab:Wendland_kernels} and their derivatives
      (dashed lines).}
   \label{fig:Wendland_kernels}
\end{figure}

\begin{figure}[ht]
   \centerline{\includegraphics[width=0.63\textwidth]{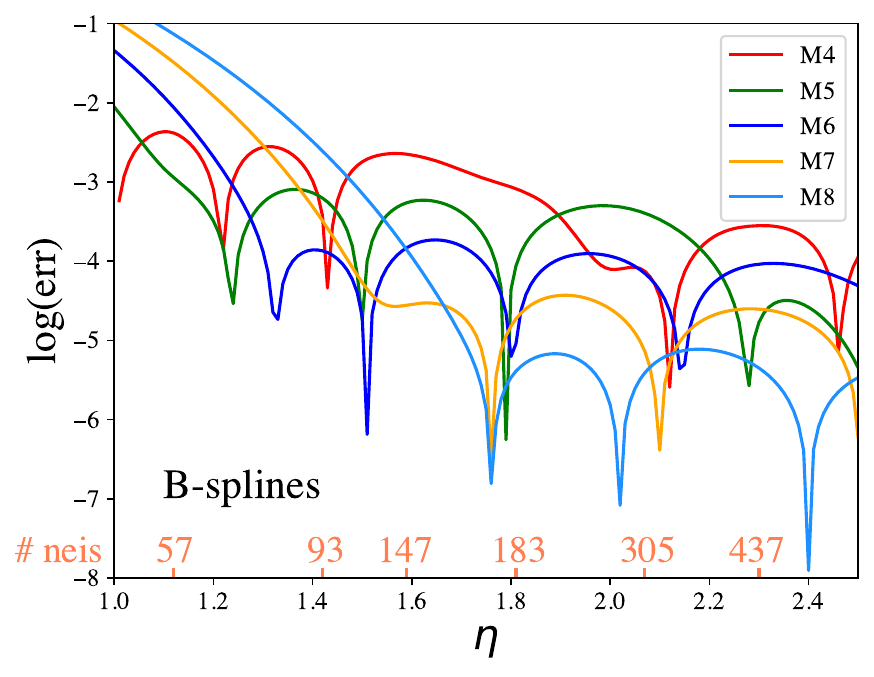}} \vspace*{-0.2cm}
   \centerline{\includegraphics[width=0.63\textwidth]{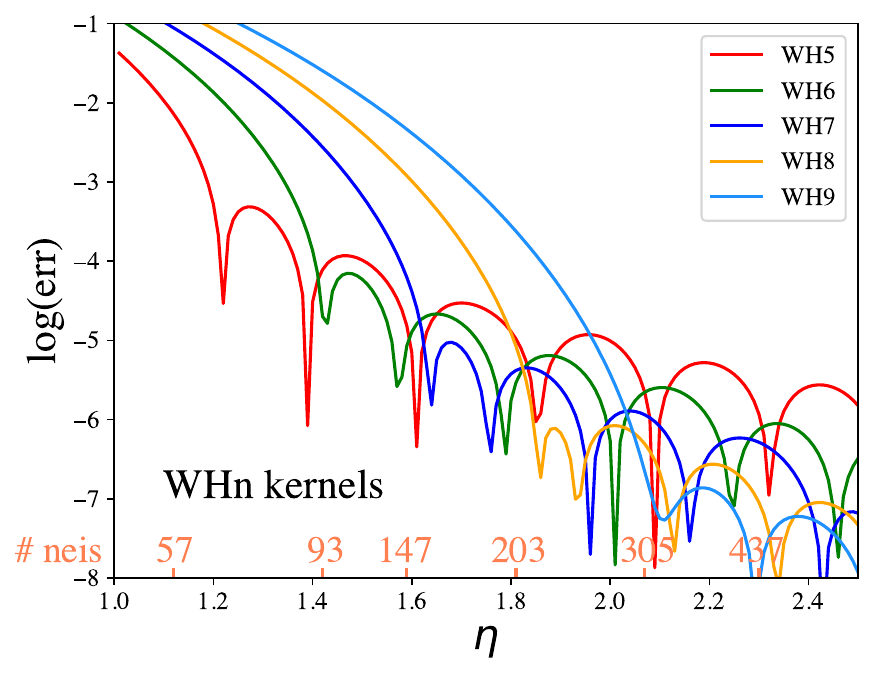}}  \vspace*{-0.2cm}
    \centerline{\includegraphics[width=0.63\textwidth]{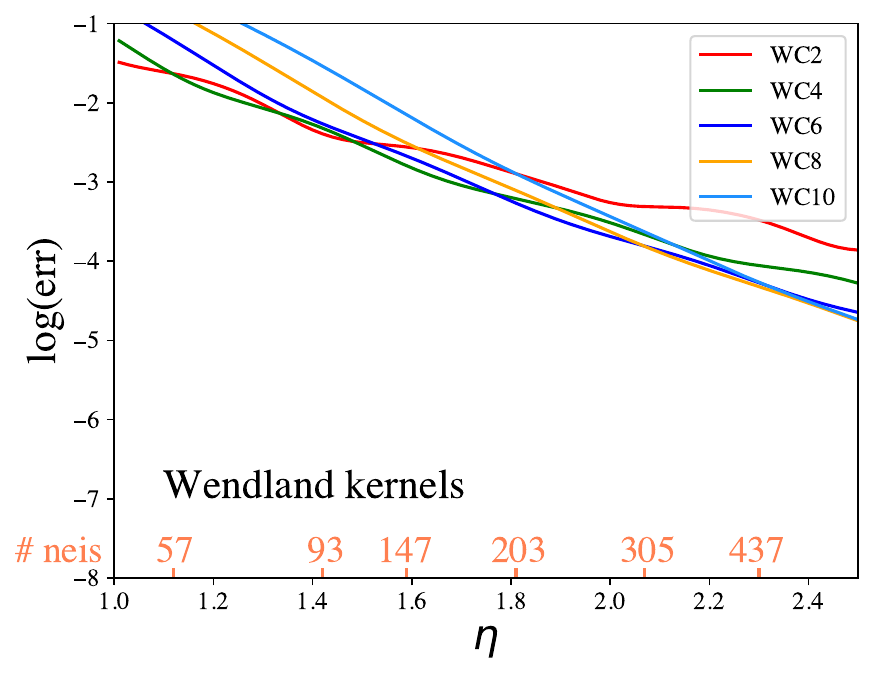}} 
    \caption{Density accuracy measurements for three kernel families. Shown is the average
    density error as a function of the quantity $\eta$ in Eq.(\ref{eq:h_update2}). The orange numbers
    above the abscissa indicate the numbers of neighbours.}
   \label{fig:density_error}
\end{figure}
\subsection{Practical density measurement}
We set up a particle distribution on a cubic lattice so that each particle
has theoretical density $\rho_{\rm theo}=1$ and we measure what 
different kernels find, as a function of  support size, 
for the density using Eq.(\ref{eq:rho_sum}). We show 
in Fig.~\ref{fig:density_error} the average relative density error
\be
{\rm err}= \frac{1}{N \rho_{\rm theo}} \sum_{a=1}^N |\rho_a - \rho_{\rm theo} |,
\ee
where $\eta$ is the prefactor from Eq.(\ref{eq:h_update2}) which determines
smoothing length and neighbour number. Note that, for a fair comparison of the neighbour numbers/the computational effort,
we scale all the kernels via Eq.(\ref{eq:normalization_change}) so that they have a support size of $2 h$. The neighbour numbers
for the chosen cubic lattice are indicated just above the abscissa.\\
Note that the Wendland kernels do not show "dips", which correspond to a change of sign 
of the density error: they always {\em overestimate} the density, but continuously improve
with increasing neighbour number. The B-spline and harmonic-like kernels
deliver in this idealized test a better density estimate, in particular for low neighbour numbers.
Another noteworthy feature is that the high-order, sharply peaked kernels, see Figs.~\ref{fig:Bsplines}, \ref{fig:WH_kernels} and \ref{fig:Wendland_kernels},
are poor density estimators at low neighbour numbers, but strongly improve with more neighbours.\\
The caveat, however, is that this density estimate  has been performed under idealized 
conditions, with particles being distributed in a perfectly regular way. In a practical simulation,
however, one of the major questions is whether particles {\em remain} in a regular configuration
during the dynamical evolution. The members of the Wendland kernel family have shown in a large number
of tests very good results in this respect \citep{rosswog15b}, therefore only  very little noise/
local velocity fluctuations occur.

\subsection{The impact of kernel and neighbour number choice on Kelvin-Helmholtz instabilities}
\begin{figure}[t]
\centerline{
\includegraphics[width=1.1\textwidth]{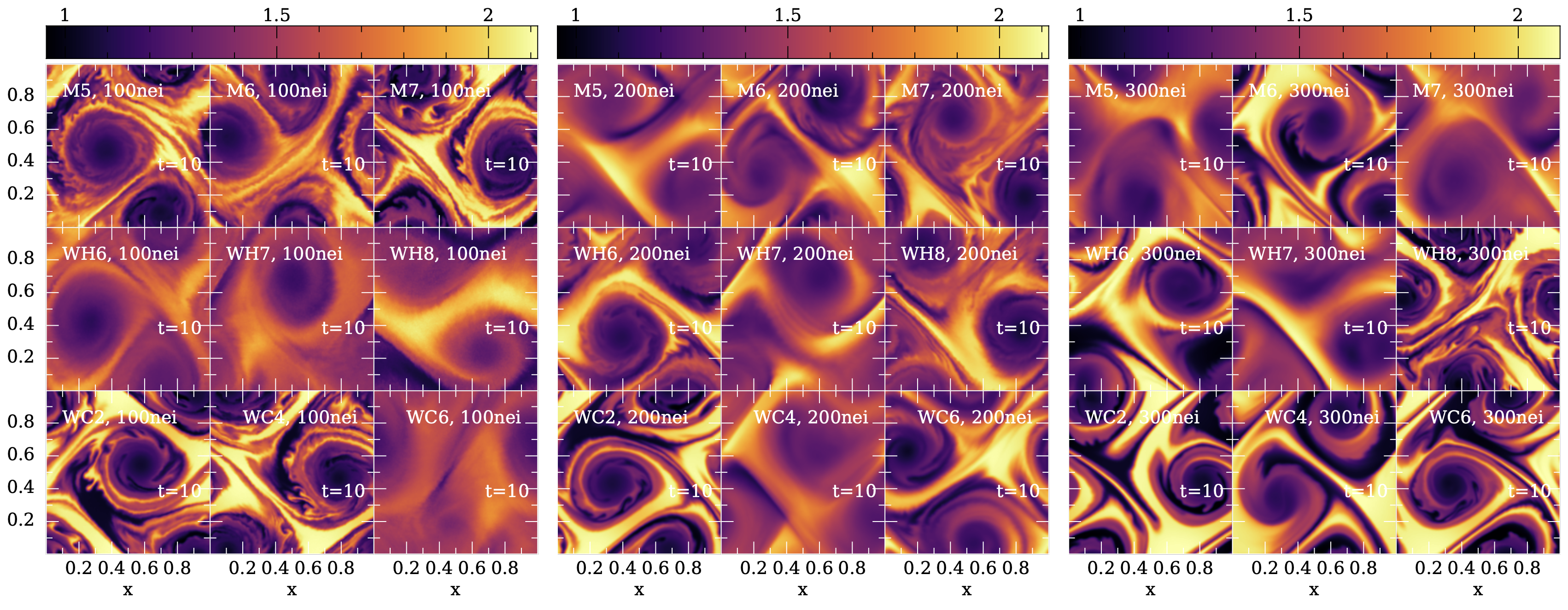}
}
\caption{Long-term Kelvin-Helmholtz evolution ($256^2$ particles with fixed dissipation parameter and linear reconstruction)
for different kernels (M$_5$ ... M$_7$, W$_{H,6}$ ... W$_{H,8}$ and W$_{C^2}$ ... W$_{C^6}$) at $t=10$. The left $3\times3$-panel
shows the case for exactly 100 particles ("neighbours") within the reach of each particle, the middle panel shows the case for exactly 200 neighbours
and the right panel shows the 300 neighbour case.}
\label{fig:KeHe_kernel_neis}   
\end{figure}

In this experiment we explore the impact of the kernel choice/neighbour number on the
long-term evolution of a Kelvin-Helmholtz instability. More specifically, we ask: 
\enquote{How dissipative are different kernels for a given number of neighbour particles?}
The answer depends in a non-trivial way on the combination of kernel
and neighbour number.\\
We focus entirely on the kernel impact and set up a Kelvin-Helmholtz 
test ($256^2$ particles) exactly as described in Sec.~\ref{sec:KeHe_test}. We use the
SPH version $\mathcal{V}_4$, see Sec.~\ref{sec:numerical_examples}, we use only the
quadratic von-Neumann--Richtmyer artificial pressure term in the form
\be
Q_a= 10 \rho_a \mu_a^2, \label{eq:Qvis_quad}
\ee
where $\mu_a$ is defined in Eq.~(\ref{eq:mu_vis}). Moreover, we keep the prefactor in $Q_a$ constant 
and perform a linear reconstruction,  in which we use the reproducing kernels described in 
Sec.~\ref{sec:RPK} and a small amount of conductivity as in Sec.~\ref{sec:numerical_examples}.
We perform experiments for the kernels M$_5$, M$_6$ and M$_7$ from the B-spline family, for
W$_{H,6}$, W$_{H,7}$ and W$_{H,8}$ from the harmonic-like family and for W$_{C^2}$, W$_{C^4}$
and W$_{C^6}$ from the Wendland kernel family, each time for exactly 100, 200 and 300 neighbour
particles within the support of each particle. We note that the B-spline and W$_{H,n}$ kernels become
pairing unstable in the 300 neighbour case, but this has no dramatic effect here. \\
Ideally, one should find the sweet spot
between low smoothing length (to have formally low discretization error) and large smoothing  length to have low numerical noise. Also keep in mind, 
that the above viscous pressure  (ignoring the divergence-avoiding $\epsilon$-term in $\mu$)
is quadratic in the smoothing length, so with everything else being the same, a larger smooting length
should lead to more dissipation. In practice, however, too few neighbours can lead
to noise, which in turn increases the dissipation as well. So it is a non-trivial problem
to figure out ideal (or at least very good) combinations of kernel function and neighbour numbers 
and the outcome is expected to differ from the idealized experiment of the previous section.
The results are shown in Fig.~\ref{fig:KeHe_kernel_neis}.  \\
These plots illustrate the non-trivial dependence on the neigbour number. While the differences in the initial growth rates 
(like Fig.~\ref{fig:KH_growth_rate}) are very small, 
in such long-term evolutions of an instability, tiny changes in the dissipation make a substantial difference in the appearance at late times. For example, the W$_{C^4}$
kernel produces somewhat noisy, but otherwise crisp results for 100 neighbours,
appears rather diffusive for 200 neighbours, and sharp again
for 300 neighbours.

\section{Coefficients of the linearly reproducing kernel method}
\label{sec:RPK_expressions}
The linearly reproducing kernel method, see Sect.~\ref{sec:RPK}, is designed to ensure the 
partition of unity and the partition of nullity at each particle position.
As a first step one needs the first discrete moments (at position $a$)
\bea
(M_0)_a &\equiv& \sum_b V_b \; \bar{W}_{ab} \\
(M_1^i)_a &\equiv& \sum_b (\vec{r}_{ab})^i \; V_b \;  \bar{W}_{ab} \\
(M_2^{ij})_a &\equiv& \sum_b (\vec{r}_{ab})^i  \; (\vec{r}_{ab})^j  \; V_b \; \bar{W}_{ab}
\eea
and their derivatives
\bea
(\p_k M_0)_a &=& \sum_b V_b \; \nabla_a^k  \bar{W}_{ab} \\
(\p_k M_1^i)_a &=& \sum_b V_b \left[ (\vec{r}_{ab})^i (\nabla_a^k \bar{W}_{ab}) + \delta^{ki} \bar{W}_{ab} \right]\\
(\p_k M_2^{ij})_a &=& \sum_b V_b \; \left[ (\vec{r}_{ab})^i  \; (\vec{r}_{ab})^j  \; (\nabla_a^k \bar{W}_{ab}) + 
(\vec{r}_{ab})^i \; \delta^{jk} \; \bar{W}_{ab} +  (\vec{r}_{ab})^j \; \delta^{ik} \; \bar{W}_{ab} \right].
\eea
With these expressions at hand,  one can straight forwardly calculate the kernel parameters
\bea
A_a &=& \frac{1}{(M_0)_a - (M_2^{ij})_a^{-1} \; (M_1^i)_a \; (M_1^j)_a}\\
B_a^i&=& - (M_2^{ij})_a^{-1} \; (M_1^j)_a
\eea
and their somewhat lengthy, but otherwise straight forwardly calculable, derivatives (needed for $\p_k \mathcal{W}_{ab}$, see Eq.~(\ref{eq:RPK_kernel_gradient}))
\begin{align}
\p_k A_a =& -A^2_a \left[ (\p_k M_0)_a - 2 (M_2^{ij})_a^{-1}   (M_1^j)_a  \left( \p_k M_1^i \right)_a   \right. \nonumber\\
 & \left. + (M_2^{il})_a^{-1}   (\p_k M_2^{lm})_a (M_2^{mj})_a^{-1} (M_1^j)_a (M_1)_a^i \right]
\end{align}
and
\be
\p_k B_a^i= -(M_2^{ij})_a^{-1}  \; (\p_k M_1^j)_a + (M_2^{il})^{-1} \; (\p_k M_2^{lm})_a  (M_2^{mj})_a^{-1} (M_1^j)_a. 
\ee

\clearpage

\end{appendices}

\newpage

\phantomsection
\addcontentsline{toc}{section}{References}
\bibliography{refs}

\end{document}

%% file: own_latex_commands_i.tex
\newcommand{\nifs}{\ensuremath{^{56}\mathrm{Ni}}}
\newcommand{\Nifs}{\ensuremath{^{56}\mathrm{Ni}} $\;$}

\def\p{\partial}
\def\msun{M$_{\odot}$}
\def\Msun{M$_{\odot}$ }
\def\be{\begin{equation}}
\def\ee{\end{equation}}
\def\bea{\begin{eqnarray}}
\def\eea{\end{eqnarray}}
\def\bt{\begin{tabbing}}
\def\et{\end{tabbing}}
\def\gcc{g cm$^{-3}$}
\def\Gcc{g cm$^{-3} \;$}

\def\edo{\end{document}}

\newcommand{\enth}{\mathcal{E}}

\newcommand{\tlg}{\tilde{\gamma}}
\newcommand{\tlG}{\tilde{\Gamma}}
\newcommand{\emfp}{e^{-4\phi}}
\newcommand{\tlA}{\tilde{A}}
\newcommand{\tlR}{\tilde{R}}

\newcommand{\dt}[1]{\partial_t #1}
\newcommand{\pdu}[3]{\partial_{#3} #1^{#2}}
\newcommand{\pdl}[3]{\partial_{#3} #1_{#2}}
\newcommand{\pdpdu}[4]{\partial_{#3}\partial_{#4} #1^{#2}}
\newcommand{\pdpdl}[4]{\partial_{#3}\partial_{#4} #1_{#2}}
\newcommand{\upwindu}[3]{\beta^{#3}\bar{\partial}_{#3} #1^{#2}}
\newcommand{\upwindl}[3]{\beta^{#3}\bar{\partial}_{#3} #1_{#2}}
\newcommand{\tlGn}{\tilde{\Gamma}_{\mathrm{(n)}}}
\newcommand{\tlGmixed}[2]{\tlG_{#1}^{\;\;\;#2}}

\def\vr{\vec{r}}
\def\ra{\vec{r}_a}
\def\rb{\vec{r}_b}
\def\rab{\vec{r}_{ab}}
\def\rba{(\vec{r}_{ba})}

\def\f{{\bf f}}

\def\M4{M$_4$}
\def\M6{M$_6$}
\def\lcm6{LCM$_6$}
\def\qcm6{QCM$_6$}

\def\wh3{W$_{\rm H,3}$}
\def\wh4{W$_{\rm H,4}$}
\def\wh5{W$_{\rm H,5}$}
\def\wh6{W$_{\rm H,6}$}
\def\wh7{W$_{\rm H,7}$}

\def\divv{\nabla \cdot \vec{v}}

\newcommand{\SpB}{\texttt{SPHINCS\_BSSN}\; }
\newcommand{\spB}{\texttt{SPHINCS\_BSSN}}

\newcommand{\Ma}{\texttt{MAGMA2}\;}

\newcommand{\lo}{\texttt{LORENE}}
\newcommand{\Fu}{\texttt{FUKA}\;}
\newcommand{\fu}{\texttt{FUKA}}

\newcommand{\sgrid}{\texttt{SGRID}}

\newcommand{\spells}{\texttt{Spells}}
\newcommand{\Elliptica}{\texttt{Elliptica \;}}

\newcommand{\nRPyElliptic}{\texttt{NRPyElliptic}}